Applying a Label Propagation Algorithm to
Detect Communities in Graph Databases

# CLUSTERING GRAPHS

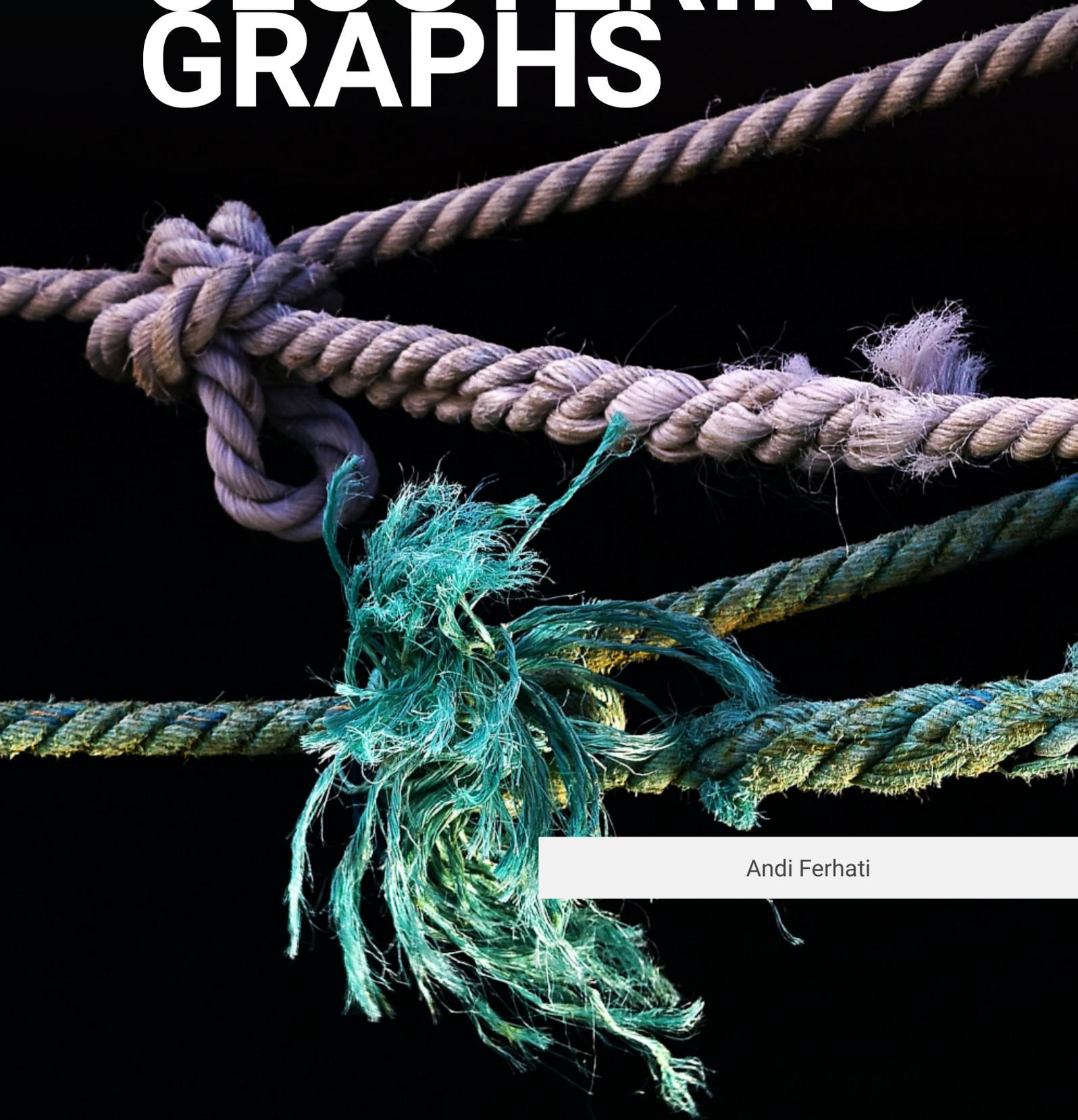

Andi Ferhati

# Clustering Graphs

Applying a Label Propagation Algorithm to Detect Communities in Graph Databases

**Andi Ferhati**

Bergamo, Italy - September 2021

**Author:**  Andi Ferhati
**Reg. Number:**  1012310

**Supervisor:**  Prof. Stefano Paraboschi

Subject: Master's Thesis
Course: Master's Degree in Computer Science & Engineering
Institution: University of Bergamo

# Index

## Index

# Dedication

*To my family.*

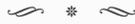

# Abstract

In the last few decades, Database Management Systems (DBMSs) became powerful tools for storing large amount of data and executing complex queries over them. In the recent years, the growing amount of unstructured or semi-structured data has seen a shift from representing data in the relational model towards alternative data models. Graph Databases and Graph Database Management Systems (GDBMSs) have seen an increase in use due to their ability to manage highly-interconnected, continuously evolving data.

In a typical graph data model, a node represents an entity and an edge represents a connection between two nodes, describing the relationship between them. In the application domains where relationships have relevant importance, Graph Database Management Systems (GDBMSs) are having great popularity since the relationships can be explicitly modeled and easily visualized in a graph data model. This kind of model is suitable for storing data without rigid schema for use cases like network processing or data integration. In addition to the storage flexibility, graph databases provide new querying possibilities in the form of Path Queries, detection of Connected Components, Pattern Matching, etc.

This thesis is a documentation of the work done in implementing a system to identify clusters in graph modeled data using a Label Propagation Community Detection Algorithm. The graph was built using datasets of academic publications in the field of Computer Science obtained from dblp.org . The system developed is a FullStack WebApp consisting of a web-based user interface, an API and the data (nodes, edges, graph) stored in a Graph Database Management System (GDBMS).

Described in this document are:

- the process of manipulation pre-import and import of the data in a Graph Database Management System (GDBMS) such as ArangoDB, creation of nodes, relations (edges) between the nodes and a graph composed of these nodes and edges;
- the GraphQL API implemented in NodeJS to request data from the Graph Database Management System (GDBMS);
- the frontend interface made with TypeScript and React consisting of the search functionalities and ability to visualize results in Cytoscape Network Graphs;
- the Label Propagation Community Detection Algorithm execution on the graph, the found clusters which are stored and visualized to the user whenever requested.

This thesis hopes to contribute with a practical hands-on approach on the graph representation, integration and analysis of interconnected data.

# Acknowledgments

I would like to thank my teacher and supervisor Prof. Stefano Paraboschi for his guidance during my studies and their finalization with this thesis.

I would also like to thank:

- Mbarsila Kadiu @Comm-it Software (commitsoftware.it) for helping me with the implementation of the Frontend;
- Simone Massimo Zucchi @SORINT.lab (sorint.it) for his help with the DevOps of the project.

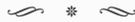

# List of figures

# List of tables

# List of code listings

# List of definitions, theorems, proofs and remarks

# Acronyms

All acronyms and their pages of appearance in the thesis.

**Symbols**

**2i** Secondary Indexing

**A**

**ACID** Atomicity, Consistency, Isolation, Durability
**ANNs** Artificial Neural Networks
**API** Application Program Interface
**AQL** ArangoDB Query Language
**ASCII** American Standard Code for Information Interchange

**C**

**C++** C++ Programming language, a superset of the C language
**CNN** Convolutional Neural Network
**CPU** Central Processing Unit
**CRUD** Create Retrieve Update Delete

**D**

**DB** DataBase
**DBMS** DataBase Management System
**DCL** Data Control Language
**DDL** Data Definition Language
**DML** Data Manipulation Language

**E**

**EBA** Edge Betweenness Algorithm
**ER Diagram** Entity Relationship Diagram

**G**

**GA** Genetic Algorithm
**GDBMS** Graph DataBase Management System
**GIMP** GNU Image Manipulation Program
**GNA** Girvan-Newman Algorithm
**GNU** GNU's Not Unix
**GOMA** Greedy Optimization Of Modularity Algorithm
**GPU** Graphical Processing Unit
**GUI** Graphical User Interface

**H**

**HTML** HyperText Markup Language

**HTTP** Hypertext Transfer Protocol
**HTTPS** HTTP over SSL

**I**

**I/O** Input/Output
**IaaS** Infrastructure as a Service
**ID** Identifier
**IDE** Integrated Development Environment
**IMA** InfoMap Algorithm
**IP** Internet Protocol
**IPv6** IP version 6

**J**

**JSON** JavaScript Object Notation

**K**

**kNN** k Nearest Neighbors

**L**

**LEA** Leading Eigenvector Algorithm
**LMA** Louvain Modularity Algorithm
**LPA** Label Propagation Algorithm
**LPCDA** Label Propagation Community Detection Algorithm. See LPA

**M**

**MVC** Model View Controller

**N**

**NAT** Network Address Translation
**NoSQL** Not only SQL
**NPM** Node Packet Manager

**O**

**OLAP** Online Analytical Processing
**OLTP** Online Transactional Processing
**OODBMS** Object Oriented DataBase Management System
**OOP** Object Oriented Programming
**ORDBMS** Object-Relational DataBase Management System
**OS** Operating System

**P**

**PaaS** Platform as a Service
**PC** Personal Computer
**PDF** Portable Document Format

**R**

**RAM** Random Access Memory
**RDBMS** Relational DataBase Management System
**RDF** Resource Description Framework
**REST** REpresentational State Transfer
**RGB** Red+Green+Blue color model
**RNN** Recursive Neural Network
**ROM** Read Only Memory
**RPC** Remote Procedure Call

**S**

**SaaS** Software as a Service
**SCC** Strongly Connected Components
**SCP** Secure CoPy
**SFTP** Secure File Transfer Protocol
**SPARQL** SPARQL Protocol And RDF Query Language
**SQL** Structured Query Language

**SSH** Secure Shell
**SVM** Support Vector Machine

**T**

**TCP** Transmission Control Protocol

**U**

**UDP** User Datagram Protocol
**UI** User Interface
**UML** Unified Modeling Language
**URI** Uniform Resource Identifier
**URL** Uniform Resource Locator

**W**

**W3C** World Wide Web Consortium
**WCC** Weakly Connected Components
**WTA** WalkTrap Algorithm

**X**

**XML** eXtensible Markup Language

# Nomenclature

**Bachmann-Landau or asymptotic notations**

| | |
|---|---|
| $O$ | Big O notation |
| $\Theta$ | Big Theta notation |

**Constants**

| | |
|---|---|
| $\pi$ | Pi |

**Graph notation**

| | |
|---|---|
| $G$ | Graph |
| $v$ | Vertex |
| $e$ | Edge |
| $V$ | Set of vertices |
| $E$ | Set of edges |
| $P$ | Path |
| $k$ | Length of a path |
| $A$ | Adjacency matrix |

**Logic notation**

| | |
|---|---|
| $\exists$ | Exists |
| $\nexists$ | Does not exist |
| $\forall$ | For all, for each |
| $\vee$ | Logical disjunction, or |
| $\wedge$ | Logical conjunction, and |
| $\neg$ | Logical negation, not |
| $\rightarrow$ | Logical consequence |
| $\leftarrow$ | Converse implication |
| $\leftrightarrow$ | Logical equivalence |

**Set notation**

| | |
|---|---|
| $\in$ | Member of, belongs to |
| $\notin$ | Not member of, does not belong to |
| $\emptyset$ | Empty, empty set |
| $\subseteq$ | Subset of |
| $\subset$ | Proper subset of |
| $\not\subset$ | Not proper subset of |
| $\nsubseteq$ | Not subset of |
| $\cup$ | Union |
| $\cap$ | Intersection |
| $\setminus$ | Difference |

**Other symbols**

| | |
|---|---|
| $\infty$ | Infinity |
| $\vert\vert$ | Cardinality |
| $\ldots$ | Horizontal ellipsis |
| $\vdots$ | Vertical ellipsis |

# 0. Preface

This thesis is the final chapter of my Computer Science and Engineering Master's Degree. Deciding the thesis subject was not an easy choice. I was really into the development of some software in distributed clouds related topics. There were a few different options proposed to me, worth noting: one was akin to compilers; another one was related to systems security with SELinux.

In order to bring together the interest for distributed systems and the options on the table, I went for an exploratory work in Container Isolation with SELinux in Kubernetes, OpenShift Clusters on the supervision of Prof. Stefano Paraboschi. While the topic was engaging, dense of new knowledge and intensively under continuous development, these properties, at least for me, made its learning curve pretty steep. I had to pivot.

The new chosen topic was on the usage of relatively new database systems to solve a real life problem. The problem on which to work was the individuation of the collaboration communities in academia. Using graph databases and a community detection algorithm made perfect sense - this way, this thesis was framed in its present shape.

## 0.1. About this thesis

This document describes the work on the thesis done during the ending of the Master's Degree. The work started before mid 2021, was initially exploratory and was carried out under the supervision of Prof. Stefano Paraboschi.

### 0.1.1. What this thesis covers

The topic this thesis treats is the application of a Label Propagation Community Detection algorithm to detect the collaboration communities in the academia, between researchers, their publications and so on...

The data on academic publications was obtained from a complete dataset of *The dblp computer science bibliography* of Schloss Dagstuhl - Leibniz Center for Informatics[1]. The version of the dataset is of the $1^{st}$ of July 2021.

In order to host the data and do the necessary on graph calculations, a Graph Database Management System was used, specifically ArangoDB.

To query the data and visualize the results, an API and a frontend interface were built.

For more details, head to § 1 - Introduction.

### 0.1.2. Get and run the code

**Pull the source code**

The instructions to get the source code used for the data manipulation, the importing in the database, the further collection updates, the GraphQL API and the frontend - are in appendix § A - Source Code, specifically § A.1 - Project repositories on page 117.

**Run, build and deploy**

The instructions on how to get everything running are in § A.2 - Instructions on how to run, build and deploy on page 117.

---

[1] Schloss Dagstuhl - Leibniz Center for Informatics (2021)

Schloss Dagstuhl - Leibniz Center for Informatics. *The dblp computer science bibliography*. Online. dblp.org. July 2021. URL: `http://dblp.uni-trier.de/xml/`.

### 0.1.3. Notation conventions used

*Italic*

Indicates new terms or terms that are non part of standard vocabulary.

`Constant width, monospace` (in black color)

Used for code listings, as well as within paragraphs to refer to program elements such as variable or function names, databases, data types and instructions.

`Constant width, monospace` or normal fontstyle (in blue color)

Used for links, urls, websites

[2]

Used for citations. Clicking on the number sends to Bibliography on page 133. In Bibliography are also indicated all the pages where each reference has been cited. The bibliographic information of a reference is shown in footnotes of a page (see the bottom of the page) only the first time it is cited. Numbering and sorting of references in the bibliography is in ascending order of their citation.

This is[a]

Used for simple footnotes[b]. Differently from bibliographic citations, footnotes use superscripted smallcase letters[c] in green color inside round brackets.

§ 2.1

Silcrow (section sign), used to refer to a section or subsection of the document.

≔ Figure 1.2: A random title

FontAwesome list icon, used to show a figure (or table, code listing, definition, theorem, proof or remark) is included in the list or figures (or list of tables, of code listings and so on ...). It is hyperlinked, clicking on it sends to the list of that element type.

Used for breadcrumbs in the header of every page except for the frontmatter and chapters, appendices and bibliography's first page. Other then general orientational information on the current topic and position in the chapter, breadcrumbs offer also interactive browsing of the whole document.

Clicking on Index sends to the general contents index.

Clicking on 0 Preface sends to the current chapters' first page, where its subcontents can be consulted.

Clicking on 0.1 About this thesis sends to the current sections' beginning.

Clicking on 0.1.3 Notation conventions used sends to the current subsections' beginning.

### 0.1.4. About the cover illustration

On the cover and the backcover of the thesis is a picture from David Clode taken in 2018 with caption "*A couple of interesting frayed ropes, subtly lit in the shadow between two fishing boats at the Cairns prawn trawler base*".[3]

Symbolically knots can be intended as nodes of a graph, ropes as the edges and the different coloring of these intended as the different communities of that graph.

---

[a] this is a footnote
[b] not a bibliographic citation
[c] uses letters, not numbers

[2] Università degli Studi di Bergamo (2021)
Università degli Studi di Bergamo. *Logo UniBG*. Online. July 2021. URL: https://unibg.it/.
[3] Clode (2018)
David Clode. *A couple of interesting frayed ropes, subtly lit in the shadow between two fishing boats at the Cairns prawn trawler base*. Online. Photographer's profile: https://unsplash.com/@davidclode. Jan. 2018. URL: https://unsplash.com/photos/hpQAUR9jkaM.

### 0.1.5. About illustrations on the first page of each chapter or appendix

Icons placed on the first page of each chapter or appendix were obtained from `https://uxwing.com/`.[4]

### 0.1.6. Graph of order of reading the chapters

Let Chapter X be the begin and Chapter Y the end node of an edge displayed below.

If the edge is a *continuous arrow*, in order to read (and well understand) Chapter Y, is *required* to have read Chapter X beforehand.

If the edge is a *dashed arrow*, in order to read (and well understand) Chapter Y, is *recommended* to have read Chapter X beforehand.

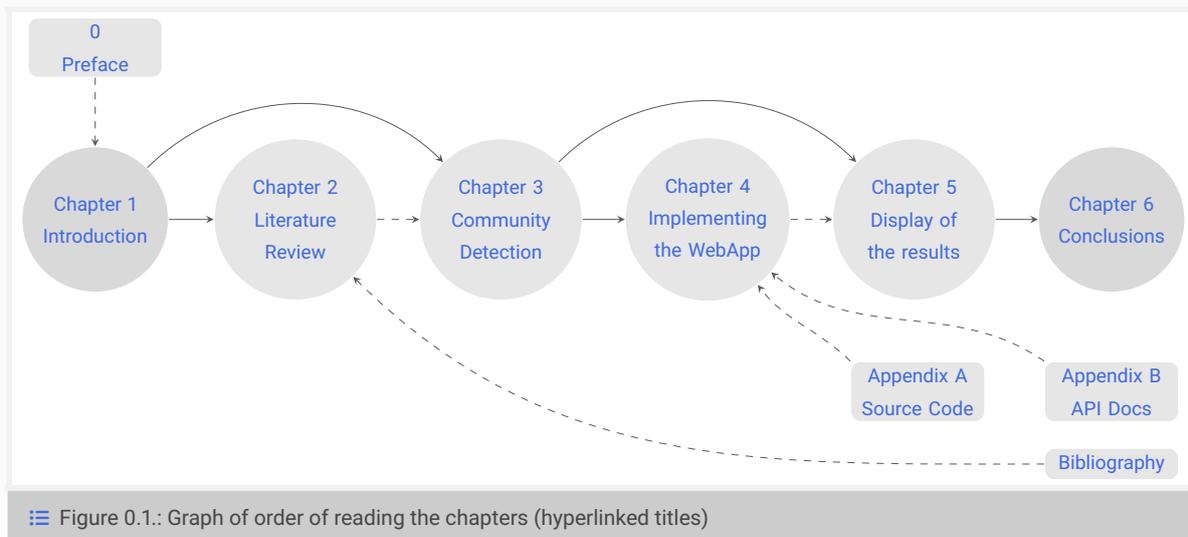

≔ Figure 0.1.: Graph of order of reading the chapters (hyperlinked titles)

### 0.1.7. Technologies and tools used

During the work for this thesis, the following technologies and tools were used:

- CentOS and Ubuntu;
- ArangoDB and temporarily Neo4j;
- JavaScript, Python, AQL, NodeJS, React, improperly TypeScript, Bootstrap, Express, GraphQL, Apollo, Cytoscape.JS and many more;
- Visual Studio Code, WebStorm, PyCharm, Postman, `vim` and `gedit`;
- `git`, `konsole` and GitHub;
- Evince Document Viewer and Calibre;
- Firefox and Chromium browsers;

During the writing of this document (and the presentation afterwards, the following technologies and tools were used:

- Ubuntu;
- LATEX, Overleaf and Latex Beamer;
- `vim` and `gedit`;
- `git`, `konsole` and GitHub;
- Inkscape, Gimp and Spectacle;
- Evince Document Viewer and Calibre;
- Firefox and Chromium browsers;

[4] UXWing (2020)

UXWing. *Icons*. Online. May 2020. URL: `https://uxwing.com/`.

### 0.1.8.  Some statistics on this document

This thesis is 156 pages long. In absolute terms the document is 182 pages long, including frontmatter's roman numbered pages.

This document contains:

- 77 figures;
- 15 tables;
- 14 code listings;
- 24 definitions, 16 remarks, 2 theorems, 2 proofs, 1 corollary, 3 formulas and 6 mathematical expressions; 0 todo notes.
- 78 acronyms, 34 nomenclature entries and 58 glossary terms;
- 2 parts, 6 chapters, 31 sections, 71 subsections and 40 subsubsections;
- 249 bibliography references. Of these:
  - 116 are books, in conference proceedings, scientific articles, tech reports, thesis documents and similar. Total sum of the number of pages of these references is 9396 pages.
  - 133 are online documentation webpages, blog posts, wiki pages and similar. The total sum of pages varies according to the webpage-to-PDF exportation options.

## 0.2.  About the author

### 0.2.1.  `whoami`

Andi Ferhati, during the work for this thesis (2021) is (was) a graduating student at University of Bergamo in a Master's Degree in Computer Science & Engineering.

The supervisor professor during the work process is (was) Prof. Stefano Paraboschi.

### 0.2.2.  Contacts

To get in touch with the author, write to `a.ferhati@engineer.com` or connect on LinkedIn at `https://www.linkedin.com/in/a-fe/`.

**Part I.**

# Graph Theory, Graph Databases and Graph Algorithms

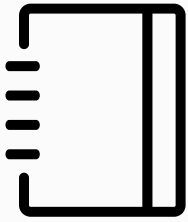

# 1. Introduction

**This chapter's contents:**

## 1.1. Overview

The end of the last century saw the birth of the web. It was one of most disruptive technological events (since maybe the invention of the transistor or the TCP/IP protocol) that radically changed the way people communicate, express and relate to each other.

> "Although Nelson was polite, charming, and smooth, I was too slow for his fast talk. But I got an aha! from his marvelous notion of hypertext. He was certain that every document in the world should be a footnote to some other document, and computers could make the links between them visible and permanent. But that was just the beginning! Scribbling on index cards, he sketched out complicated notions of transferring authorship back to creators and tracking payments as readers hopped along networks of documents, what he called the docuverse. He spoke of "transclusion" and "intertwingularity" as he described the grand utopian benefits of his embedded structure. It was going to save the world from stupidity."[5]

Apart from that last line, today, the web (and the internet in general) has made it possible to have highly interconnected communication, transactions and data transfer in general. Everyday online is produced and exchanged a gigantic amount of information, data.[6] The need to store, manage, transfer and efficiently analyze this information is more present than ever today.

---

[5] KELLY (2005)
Kevin Kelly. *We Are the Web*. Wired article. Jan. 2005. URL: https://www.wired.com/2005/08/tech/.
[6] MARX (2013)
Vivien Marx. *The big challenges of big data*. In: Nature 498 (7453 June 2013), pages 255–260. DOI: 10.1038/498255a. URL: https://www.nature.com/articles/498255a.

In addition, the study of networks has brought significant advances to the understanding of complex systems. Graphs, extremely useful in representing a wide variety of networks, and graph analysis have become crucial to understand the features of these systems.

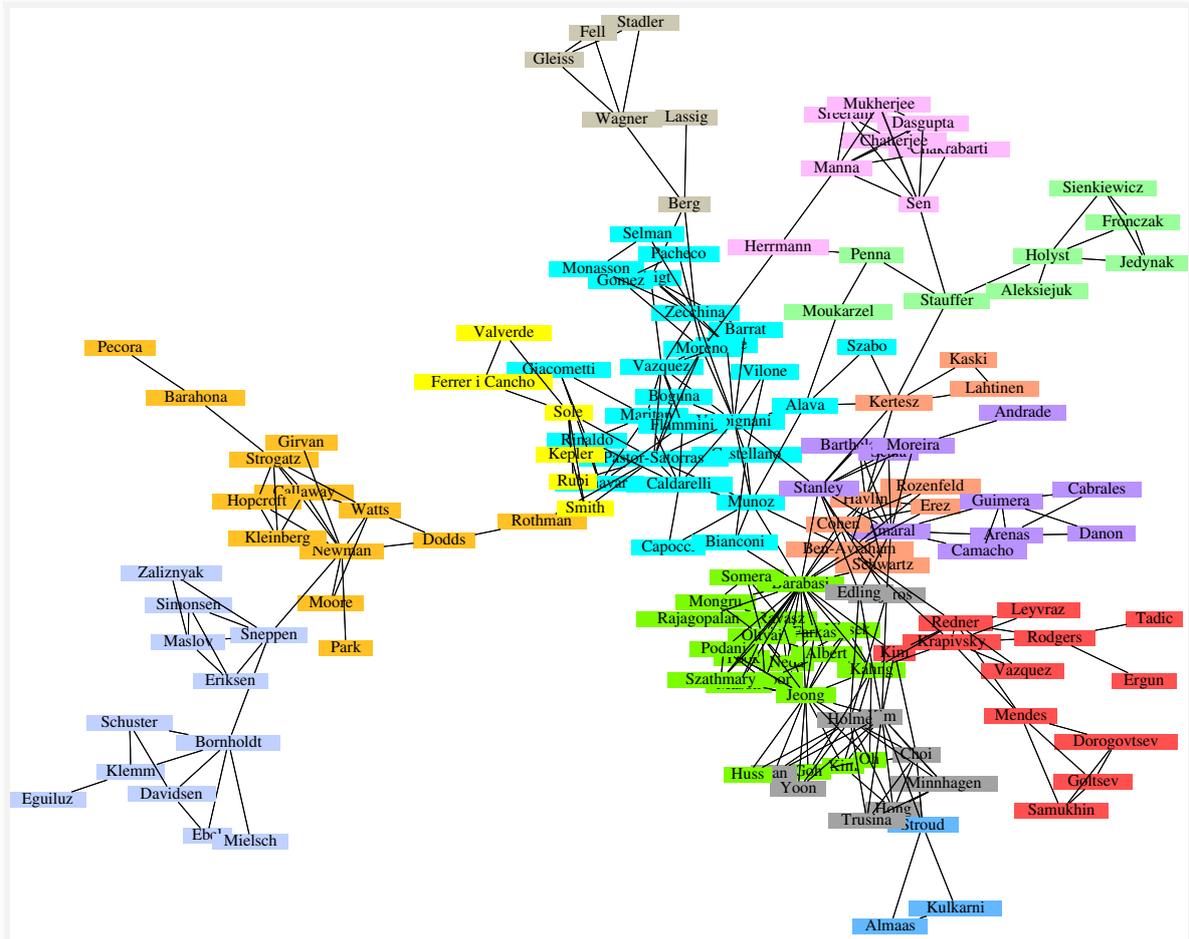

▤ Figure 1.2.: Illustration of the community structure of a network of coauthorships between physicists, scientists, taken from "Finding and evaluating community structure in networks" - Newman and Girvan (2004)[7]

Community structure[7], or clustering is one of the most relevant features of graphs representing real systems. Individuating such organization structure, detecting communities - is of great importance in various disciplines like:

- biology[8],
- sociology[9],
- computer science

[7] Newman and Girvan (2004)

M. E. J. Newman and M. Girvan. *Finding and evaluating community structure in networks*. In: Physical Review E 69.2 (Feb. 2004). ISSN: 1550-2376. DOI: 10.1103/physreve.69.026113. URL: https://arxiv.org/abs/cond-mat/0308217.

[8] Girvan and Newman (2002)

M. Girvan and M. E. J. Newman. *Community structure in social and biological networks*. In: Proceedings of the National Academy of Sciences 99.12 (June 2002). https://arxiv.org/abs/cond-mat/0112110, pages 7821–7826. ISSN: 1091-6490. DOI: 10.1073/pnas.122653799. URL: https://www.pnas.org/content/99/12/7821.

[9] Parthasarathy, Shah and Zaman (2019)

Dhruv Parthasarathy, Devavrat Shah and Tauhid Zaman. *Leaders, Followers, and Community Detection*. 2019. eprint: 1011.0774. URL: https://arxiv.org/abs/1011.0774.

• and many more.[10]

The nature of the problem makes it hard to solve. Despite the huge academic effort, it is not yet satisfactorily solved.[11]

## 1.2. Focus of the thesis

In this thesis shall be performed a practical analysis of large interconnected data representable by graphs (in the mathematical sense of vertices linked by edges) to the end of detecting communities, clusters of more densely connected nodes than others. For the storage and management of the data Graph Databases are used. By making use of Graph Databases and the advantages they offer in performing complex computations on graphs, with a hands-on approach - clusters (collaboration communities) shall be detected in a graph built on data from a dataset of academic researcher publications. The dataset used is obtained from dblp.org[1].

Using a graph database management system, a Label Propagation Community Detection algorithm shall be executed on the graph, thus detecting its clusters and by implementing a Web Application, the results of the detection shall be shown to the user whenever queried.

## 1.3. Context

The work for this thesis comes, on one hand, in a moment when NoSQL databases have peaked their hype cycle for some time now. It has been more than a decade now the time graph database management systems started getting developed and used. These softwares have now been tested in real environments with production data and processes. They may thus be defined mature enough.

Moreover, the cost for developers, researchers, students to explore this new world and make use of it for their needs - is almost zero. Community and Educational Editions of the softwares of graph DBMSs are in most cases free and/or open source, with almost all features and license permissions to use in complex projects.

On the other hand, the nowdays higher interconnectivity between people - be it because of faster transportation means, networking infrastructure, online social networks, instant messaging or just easier access and distribution of information - has made it possible to develop software in pairs remotely, write collaboratively, meet (in the sense of work meetings) online or at the nearby coffeeshop with fiber speed internet connection.

[10] PALLA, DERÉNYI, FARKAS and VICSEK (2005)

Gergely Palla, Imre Derényi, Illés Farkas and Tamás Vicsek. *Uncovering the overlapping community structure of complex networks in nature and society*. In: Nature 435.7043 (June 2005), pages 814–818. ISSN: 1476-4687. DOI: 10.1038/nature03607. URL: https://www.nature.com/articles/nature03607.

[11] FORTUNATO (2010)

Santo Fortunato. *Community detection in graphs*. In: Physics Reports 486.3-5 (Feb. 2010), pages 75–174. ISSN: 0370-1573. DOI: 10.1016/j.physrep.2009.11.002. URL: https://arxiv.org/abs/0906.0612.

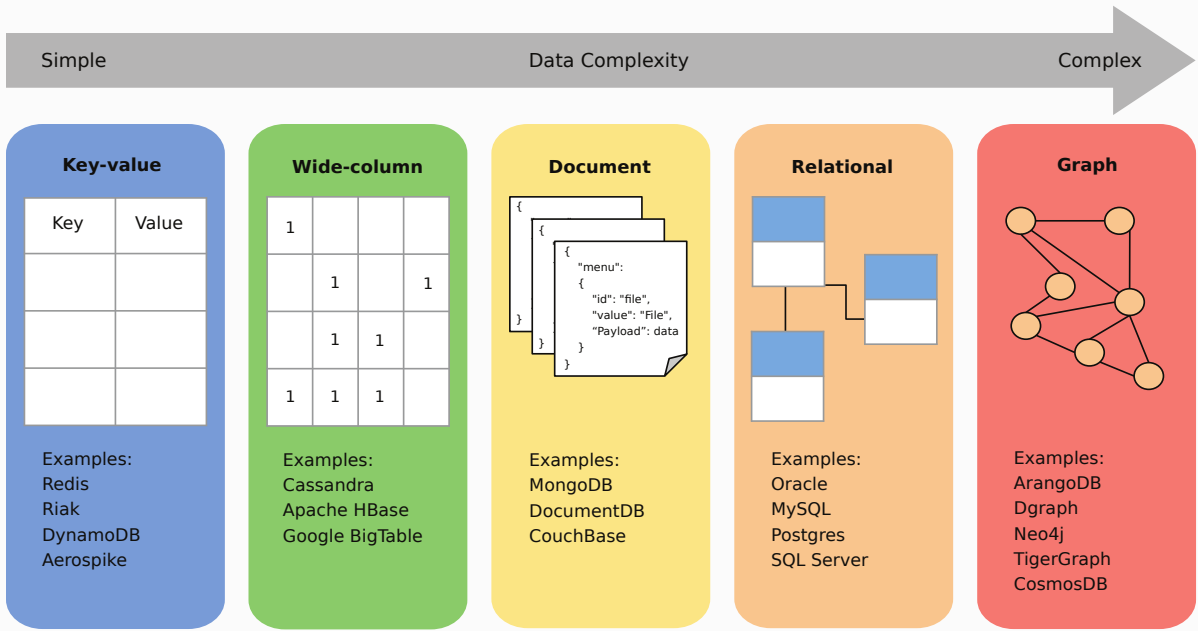

Figure 1.3.: Database engine types ordered by data complexity, taken from *Graph Databases in Action* - Bechberger and Perryman (2020)[12]

It probably is more easy today (in the general sense of these years, non-pandemic years) to take a plane, book a hotel, participate in a conference on the other side of the continent and come back with tons of new ideas, new people met and new relations instaurated. This ease of collaboration makes it possible for newer, larger and geographically more dispersed scientific collaboration communities to flourish.

In this context, capturing a snapshot of who collaborates with whom, which researchers work in smaller groups and which in wider collaboration communities - might be an intriguing exercise to do. Furthermore, as one of the most elegant features of graphs is the possibility to express them graphically in very fancy and expressive diagrams - it would be a huge plus if it was possible to implement an application that displayed the results of the detected clusters. So, besides detecting the communities, a WebApp shall be developed to display the results of the communities in response to user queries.

### 1.3.1. What is known

This study is based on two relatively heterogeneous macrotopics, the graph theory, graph databases and the community detection algorithm on one hand - and the development of a FullStack Web Application involving an API and a client interface where graphs are rendered on the other hand.

Specifically, it is based on:

- graphs
    - graph theory literature[13]

---

[13] Bollobás (1998)
Béla Bollobás. *Modern Graph Theory*. 1st edition. Graduate Texts in Mathematics 184. Springer-Verlag New York, 1998. ISBN: 978-0-387-98488-9. URL: https://www.springer.com/gp/book/9780387984889.

- – graph databases literature[12,14]
- – community detection literature from a theoretical point of view[15–22] and in practical case studies[23,24]
- – Google's Pregel algorithm[25] and ArangoDB's custom Pregel implementation[26]
- · and general web development
  - – JavaScript books and documentation
  - – NodeJS books and documentation
  - – ArangoDB documentation
  - – React books and documentation
  - – TypeScript books and documentation

[12] Bechberger and Perryman (2020)

Dave Bechberger and Josh Perryman. *Graph Databases in Action*. In Action. Manning Publications, 2020. 338 pages. ISBN: 9781617296376. URL: https://books.google.it/books?id=kWIFEAAAQBAJ.

[14] Besta, Peter, Gerstenberger, Fischer, Podstawski, Barthels, Alonso and Hoefler (2019)

Maciej Besta, Emanuel Peter, Robert Gerstenberger, Marc Fischer, Michał Podstawski, Claude Barthels, Gustavo Alonso and Torsten Hoefler. *Demystifying Graph Databases: Analysis and Taxonomy of Data Organization, System Designs, and Graph Queries - Towards Understanding Modern Graph Processing, Storage, and Analytics*. In: (Oct. 2019). URL: https://arxiv.org/abs/1910.09017.

[15] Brandes, Gaertler and Wagner (2003)

Ulrik Brandes, Marco Gaertler and Dorothea Wagner. *Experiments on Graph Clustering Algorithms*. In: volume 2832. Nov. 2003. ISBN: 978-3-540-20064-2. DOI: 10.1007/978-3-540-39658-1_52. URL: https://www.researchgate.net/publication/2939172.

[16] Dao, Bothorel and Lenca (2020)

Vinh Loc Dao, Cécile Bothorel and Philippe Lenca. *Community structure: A comparative evaluation of community detection methods*. In: Network Science 8.1 (Jan. 2020), pages 1–41. ISSN: 2050-1250. DOI: 10.1017/nws.2019.59. URL: https://arxiv.org/abs/1812.06598.

[17] Fortunato and Castellano (2007)

Santo Fortunato and Claudio Castellano. *Community Structure in Graphs*. 2007. eprint: 0712.2716. URL: https://arxiv.org/abs/0712.2716.

[18] Lancichinetti and Fortunato (2009)

Andrea Lancichinetti and Santo Fortunato. *Community detection algorithms: A comparative analysis*. In: Physical Review E 80.5 (Nov. 2009). ISSN: 1550-2376. DOI: 10.1103/physreve.80.056117. URL: https://arxiv.org/abs/0908.1062.

[19] Liu, Cheng and Zhang (2019)

Xin Liu, Hui-Min Cheng and Zhong-Yuan Zhang. *Evaluation of Community Detection Methods*. 2019. eprint: 1807.01130. URL: https://arxiv.org/abs/1807.01130.

[20] Newman (2004)

M. E. J. Newman. *Detecting community structure in networks*. In: The European Physical Journal B 38.2 (Mar. 2004), pages 321–330. ISSN: 1434-6036. DOI: 10.1140/epjb/e2004-00124-y. URL: http://www-personal.umich.edu/~mejn/papers/epjb.pdf.

[21] Orman and Labatut (2009)

Günce Orman and Vincent Labatut. *A Comparison of Community Detection Algorithms on Artificial Networks*. In: volume 5808. Nov. 2009, pages 242–256. DOI: 10.1007/978-3-642-04747-3_20. URL: https://www.researchgate.net/publication/224921426.

[22] Rosvall, Delvenne, Schaub and Lambiotte (2019)

Martin Rosvall, Jean-Charles Delvenne, Michael T. Schaub and Renaud Lambiotte. *Different Approaches to Community Detection*. In: Advances in Network Clustering and Blockmodeling (Nov. 2019), pages 105–119. DOI: 10.1002/9781119483298.ch4. URL: https://arxiv.org/abs/1712.06468.

[23] Shai, Stanley, Granell, Taylor and Mucha (2017)

Saray Shai, Natalie Stanley, Clara Granell, Dane Taylor and Peter J. Mucha. *Case studies in network community detection*. 2017. eprint: 1705.02305. URL: https://arxiv.org/abs/1705.02305.

[24] Wu, Wu, Chen, Li and Zhang (2021)

Sissi Xiaoxiao Wu, Zixian Wu, Shihui Chen, Gangqiang Li and Shengli Zhang. *Community Detection in Blockchain Social Networks*. 2021. eprint: 2101.06406. URL: https://arxiv.org/abs/2101.06406.

[25] Yan, Cheng, Xing, Lu, Ng and Bu (2014)

Da Yan, James Cheng, Kai Xing, Yi Lu, Wee Keong Ng and Yingyi Bu. *Pregel Algorithms for Graph Connectivity Problems with Performance Guarantees*. In: Proceedings of the VLDB Endowment 7(14). Volume 7. Oct. 2014, pages 1821–1832. DOI: 10.14778/2733085.2733089. URL: https://www.researchgate.net/publication/271020290_Pregel_Algorithms_for_Graph_Connectivity_Problems_with_Performance_Guarantees.

[26] ArangoDB (2021)

ArangoDB. *ArangoDB - Distributed Iterative Graph Processing (Pregel), Community Detection*. Online. Documentation. Aug. 2021. URL: https://www.arangodb.com/docs/devel/graphs-pregel.html#community-detection.

– Cytoscape.JS[27] documentation

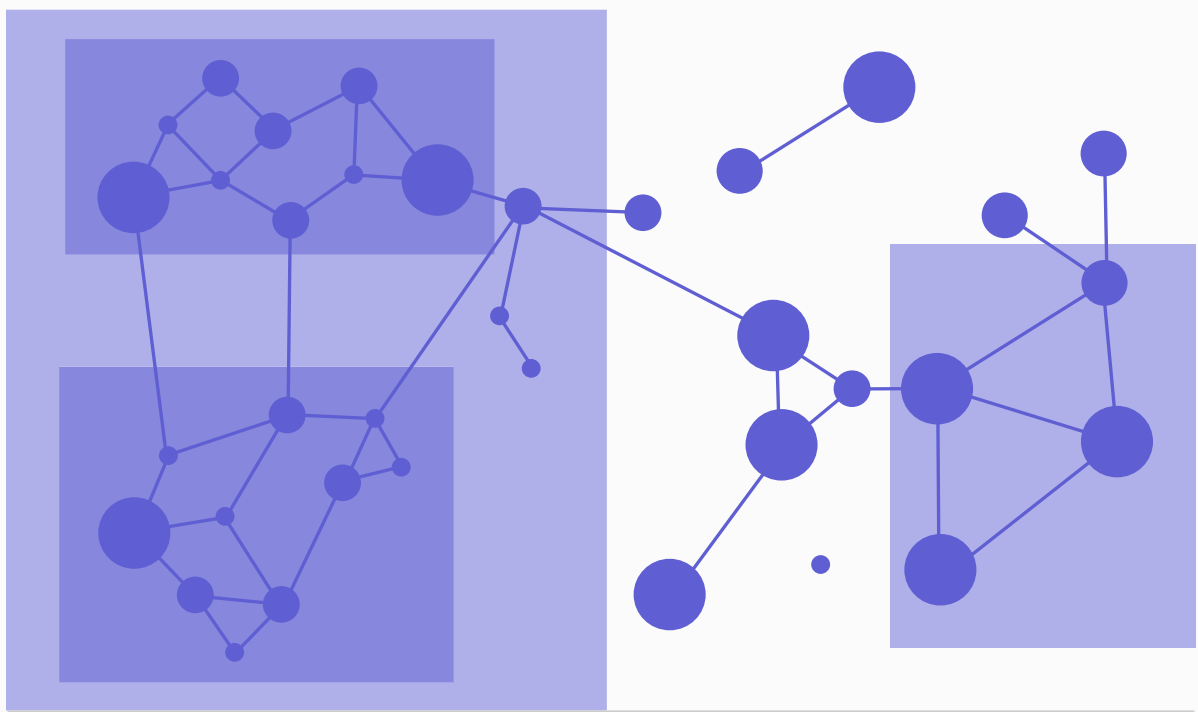

≔ Figure 1.4.: Example of a graph containing compound nodes rendered with Cytoscape.JS[27] using Cose Bilkent Layout

## 1.3.2. Boundaries of the study

This study is limited to the contours of a practical label propagation community detection algorithm execution on the dblp.org[1] dataset of computer science researcher's publications and on the display of the clustering results in a Web Application. No empirical analysis or comparison is made on the efficiency of the algorithm. No security enforcement mechanisms are implemented in the Web Application. The whole work is kind of a proof-of-concept of what can be done having access to a dataset of interconnected data, a good graph database management system with different graph algorithms offered, libraries to beautifully present the results and lots of imagination.

## 1.3.3. Past similar studies

The problem of detecting communities in scientific collaboration networks has been studied thoroughly for many decades now. Initially patterns of bibliographic reference relations, like citations, were studied.[28] In recent years, because of advances in information technology and the massive digitalization of documents, data collection and mining capabilities allow for system-level analysis of huge bibliometric datasets that are regularly collected in digital

[27] DONNELLY CENTRE - UNIVERSITY OF TORONTO (2021)

Donnelly Centre - University of Toronto. *Cytoscape cytoscape*. Online. Documentation. Aug. 2021. URL: https://js.cytoscape.org/.

[28] PRICE (1965)

Derek J. de Solla Price. *Networks of scientific papers*. In: Science (New York, N.Y.) 149.3683 (July 1965), pages 510–515. ISSN: 0036-8075. DOI: 10.1126/science.149.3683.510. URL: https://doi.org/10.1126/science.149.3683.510.

format. For the first time the structure of scientific collaboration networks was researched.[7,29–33]

These studies along with some other specifically related to graph databases and community detection[34] are the starting point for the work on this thesis.

## 1.4. Relevance

Bibliographic databases constitute the most important point for any empirical analysis of the dynamics involved in scientific activities, citation patterns and in defining the importance of specific contributions, journals, researchers.

There are a variety of compelling reasons for studying the community structure in complex networks. The partitioning of acquaintances into clusters in social networks represents valuable information of human interactions. Community structure in metabolic networks helps detect reaction modules. Communities in the web reveal connections to and from web pages, on related topics. The underlying scientific community structure in citation networks helps understand both obvious and more subtle relations between groups, subfields, publication entities as well as the birth or death of subfields or communities.[29]

**Benefits of this study**

By conducting this work, a step-by-step A to Z description of the of all the implementation process is given. As a result of this study, it is possible to easily get insights on the collaboration communities in the CS academia, their members and the interactions between them.

## 1.5. Aim and objectives

The goals of the work on this thesis are:

- usage of a graph database system with a large dataset of real data, specifically computer science academic publications from dblp.org[1];
- application of a complex graph algorithm such as the Pregel Label Propagation Community Detection algorithm

[29] CHEN and REDNER (2010)

P. Chen and S. Redner. *Community structure of the physical review citation network*. In: Journal of Informetrics 4.3 (July 2010), pages 278–290. ISSN: 1751-1577. DOI: 10.1016/j.joi.2010.01.001. URL: https://arxiv.org/abs/0911.0694.

[30] NEWMAN (2001)

M. E. J. Newman. *The structure of scientific collaboration networks*. In: Proceedings of the National Academy of Sciences 98.2 (Jan. 2001). https://arxiv.org/abs/cond-mat/0007214, pages 404–409. ISSN: 1091-6490. DOI: 10.1073/pnas.98.2.404. URL: https://www.pnas.org/content/98/2/404.

[31] NEWMAN (2001)

M. E. J. Newman. *Scientific collaboration networks. I. Network construction and fundamental results*. In: Physical review. E, Statistical, nonlinear, and soft matter physics 64 (Aug. 2001). DOI: 10.1103/PhysRevE.64.016131.

[32] NEWMAN (2001)

M. E. J. Newman. *Scientific collaboration networks. II. Shortest paths, weighted networks, and centrality*. In: Phys. Rev. E 64 (1 June 2001). DOI: 10.1103/PhysRevE.64.016132. URL: https://link.aps.org/doi/10.1103/PhysRevE.64.016132.

[33] RADICCHI, FORTUNATO and VESPIGNANI (2012)

Filippo Radicchi, Santo Fortunato and Alessandro Vespignani. *Citation Networks*. In: Models of Science Dynamics: Encounters Between Complexity Theory and Information Sciences. Edited by Andrea Scharnhorst, Katy Börner and Peter van den Besselaar. Berlin, Heidelberg - Germany: Springer Berlin Heidelberg, 2012, pages 233–257. ISBN: 978-3-642-23068-4. DOI: 10.1007/978-3-642-23068-4_7. URL: https://link.springer.com/chapter/10.1007/978-3-642-23068-4_7.

[34] BEIS, PAPADOPOULOS and KOMPATSIARIS (2015)

Sotirios Beis, Symeon Papadopoulos and Yiannis Kompatsiaris. *Benchmarking Graph Databases on the Problem of Community Detection*. In: New Trends in Database and Information Systems II. edited by Nick Bassiliades, Mirjana Ivanovic, Margita Kon-Popovska, Yannis Manolopoulos, Themis Palpanas, Goce Trajcevski and Athena Vakali. https://github.com/socialsensor/graphdb-benchmarks. Springer International Publishing, 2015, pages 3–14. ISBN: 978-3-319-10518-5. DOI: 10.1007/978-3-319-10518-5_1. URL: https://link.springer.com/chapter/10.1007/978-3-319-10518-5_1.

- and therefore detection of collaboration clusters of the researchers on that dataset;
- implementation of a fullstack Web Application to display the results of the detected communities.

### 1.5.1. What is developed and why

Different pieces of code are written during the work on this thesis. Development in the sense of programming, with reasons why, was done on:

- Python scripts to split and convert XML dataset file to ArangoDB importable JSON files;
- Bash scripts to import the data into ArangoDB graph database;
- AQL queries to manipulate the data and transform it to be used as vertices and edges in a graph;
- Fullstack self-hosted Web Application development made of:
    - NodeJS, Express and Apollo Server backend to serve a GraphQL API;
    - React, Typescript and Apollo Client frontend interface to display Cytoscape graphs of the detected communities.
- LaTeX and Beamer functions used in writing this thesis and the presentation.

### 1.5.2. What is determined

As stated in § 1.5 - Aim and objectives, one of the objectives of this work is to determine the collaboration communities between academic researchers in the field of computer science. For each of the entities involved, like authors, publications, editors, publishers, a community of which they are part of - is determined. The same is done with affiliation institutions, schools, journal and series.

### 1.5.3. The process

The work for this thesis is divided in three distinct macroparts, with many iterations between them (not to be confused with the parts into which this document is organized):

1. The first part of the work is on getting familiarized with the different topics involved. So, a profound literature review is conducted.
2. The second part is about getting the hands dirty with graph databases, using them with the chosen dataset and applying the selected cluster detection algorithm to find the communities.
3. The third and last part involves the development of the Web Application to display the results of the second part with fancy graphs in a user friendly interface.

## 1.6. Contributions in this thesis

While the work on this thesis is built on the math concepts of graph theory (commonly known as a difficult subject for students) and on relatively new technologies in computer science such as graph databases and GraphQL, React (for web app development) - its advancements of the scientific knowledge are humble. It might be considered to be a new footprint, a new trail, open for more in-depth future works.

It is shown how new information was produced from the application of a community detection algorithm on data of CS academic researchers. Furthermore it is shown how it is possible to visually express the clustering results using beautiful graphs embedded in the implemented WebApp.

This whole project is a nice practical exercise and might also be a good starting point for other future studies.

## 1.7.  Document layout

The document is divided into two parts. The first part is about a literature review of graph theory, graph databases and algorithms, including the process of detecting the communities. The second part of the thesis is about the implementation of a Web Application with graph databases, GraphQL API and graph rendering libraries in order to display the results of the detected communities.

In § 0 - Preface is given some general information on the thesis, what it covers, indications on how to get and run the code, description of the notations used, a word on cover and chapter illustrations, a graph of how is best or possible to read the chapters of the document and some basic statistics on the numbers of the document.

Follows some brief information on the author.

**Part I** is composed of the following chapters:

In § 1 - Introduction (this chapter) a general introduction to the subject is presented. Firstly the focus of the thesis is stated, with the relative context and relevance. Then aim and objectives of the thesis are described, followed by the contributions of the work and finally this general layout of the document.

In § 2 - Literature Review an overview or the current state of knowledge is presented. Firstly a brief review of Graph Theory concepts and algorithms are reported. Afterwards, Graph Database Systems and Querying Languages are considered, closing the chapter with a comparison between them.

In § 3 - Community Detection an in depth review of the community detection methodologies is presented. The chosen algorithm is described, its execution is brought and some preliminary results are shown.

**Part II**'s layout is the following:

In § 4 - Implementing the WebApp initially the hardware and OS used is described. Then the data preparation, importing and manipulation is presented. Follows the GraphQL API implementation and the frontend UI interface afterwards. Closing the chapter are the updates made after the clusters were detected.

In § 5 - Display of the results are brought a few snapshots from the results of all the project. Different kind of detected communities are shown and a brief description in given on each of them.

After that, conclusions are drawn:

In § 6 - Conclusions a summary of what was achieved is synthesized, possible future improvements are presented. The chapter ends with some further avenues of exploration.

In the end are attached two appendices, § A - Source Code and § B - API Docs followed by the Glossary and the Bibliography.

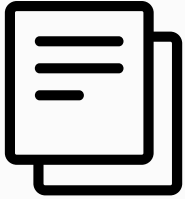

# 2. Literature Review

"To effectively leverage the social graph, every company needs to understand that they need to make their information easily transferable."

*— Erik Qualman*

**This chapter's contents:**

## 2.1.  Short review of Graph Theory concepts

To get a better understanding of Graph Databases and algorithms runnable on them, a revision of general graph theory concepts might be recommended. In the following subsections these concepts will be presented.

### 2.1.1.  Graphs

#### 2.1.1.1.  What is a graph

▤ **Definition 2.1 (of graphs):**  *A graph is a pair $G$ of disjoint sets $(V, E)$ satisfying $E \subseteq [V]^2$. The elements of $E$ are 2-element subsets of $V$. The elements of $V$ are the vertices (or nodes, or points) of the graph $G$, the elements of $E$ are its edges (or arcs, or links, lines).*[13]

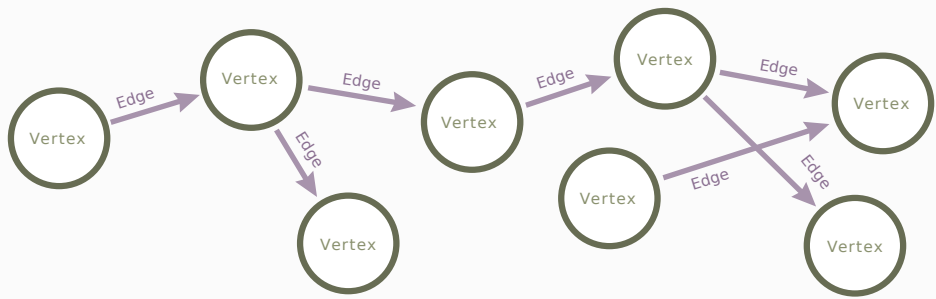

▤ Figure 2.2.: Example of a graph

The usual way to picture a graph is by drawing a dot for each vertex and joining two of these dots by a line if the corresponding two vertices form an edge. It is considered to be irrelevant how the dots and lines are drawn; Which pairs of vertices form an edge and which do not is all that matters.[35]

▤ **Definition 2.2 (of adjacent, neighbor vertices):**  *A graph is a pair $G = (V, E)$ of sets satisfying $E \subseteq [V]^2$. Two vertices $x, y$ of $G$ are adjacent, or neighbors, if $xy$ is an edge of $G$. Two edges $e \neq f$ are adjacent if they have an end in common. If all the vertices of $G$ are pairwise adjacent, then $G$ is complete.*[35]

▤ **Definition 2.3 (of the order of a graph):**  *The number of vertices of a graph $G$ is its order, written as $|G|$; its number of edges is denoted by $\|G\|$. Graphs are finite or infinite according to their order.*[35]

The graphs considered in this thesis are all finite.

[35] D<span>IESTEL</span> (2000)

Reinhard Diestel. *Graph Theory*. Volume 173. Graduate Texts in Mathematics. Heidelberg, Germany: Springer-Verlag, 2000. 447 pages. ISBN: 0-387-98976-5. URL: https://diestel-graph-theory.com/index.html.

## 2.1.1.2. Directed graphs (Digraphs) and Undirected graphs

☰ **Definition 2.4 (of directed edges):** *A directed edge is an edge $e$ one of whose endpoints is designated as the tail, and whose other endpoint is designated as the head. They are denoted $head(e)$ and $tail(e)$, respectively.*

*A directed edge is said to be directed from its tail and directed to its head. In a line drawing, the arrow points toward the head.*[36]

☰ **Definition 2.5 (of digraphs or directed graphs):** *A digraph (or directed graph) $G = (V, E)$ is a graph each one of its edges is directed. It consists of a finite, nonempty set of vertices $V$ and a set of edges $E$, each edge is an ordered pair $(v, w)$ of vertices.*[36]

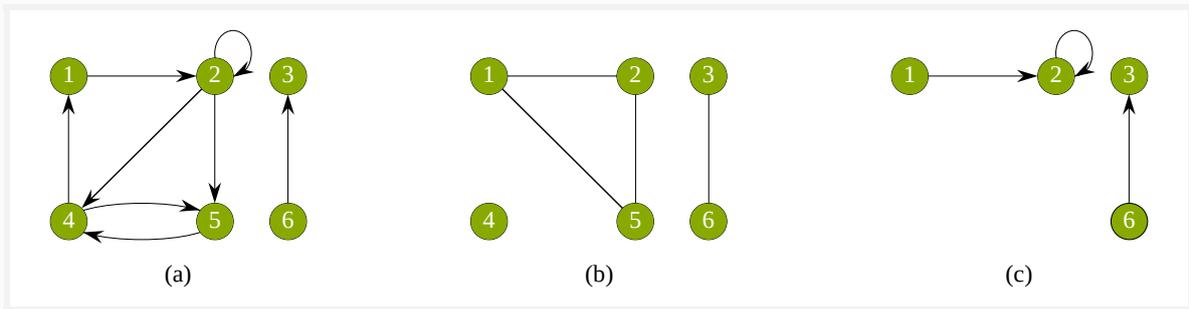

☰ Figure 2.3.: Directed and undirected graphs. (a) A digraph. (b) An undirected graph. (c) The subgraph of the graph in (a) induced by the set $\{1, 2, 3, 6\}$[37]

☰ **Definition 2.6 (of undirected graphs):** *An undirected graph $G = (V, E)$ consists of a finite, nonempty set of vertices $V$ and a set of edges $E$, each edge is a set $\{v, w\}$ of vertices.*[36]

## 2.1.1.3. Walks, paths, trails, cycles and connected graphs

☰ **Definition 2.7 (of walks):** *A walk of length $k$ in a graph is a succession of $k$ edges of the form*

$$x_0 x_1, \ x_1 x_3, \ x_3 x_7, \ \ldots, x_{k-1} x_k.$$ 

Expression 2.1

*and is referred to as a walk between $x_0$ and $x_k$.*[38]

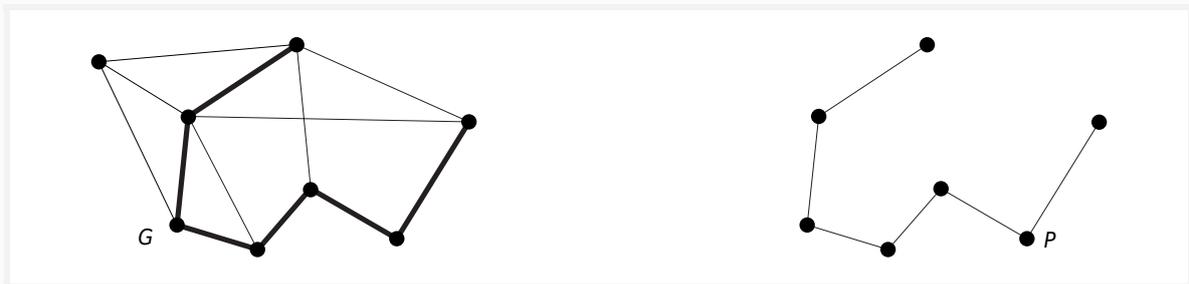

☰ Figure 2.4.: A path $P = P^6$ in $G$[35]

[36] Gross, Yellen and Zhang (2014)

Jonathan L. Gross, Jay Yellen and Ping Zhang, editors. *Handbook of Graph Theory*. Discrete Mathematics and it's Applications. 6000 Broken Sound Parkway NW, Suite 300: CRC Press, 2014. 1633 pages. ISBN: 9781439880180. DOI: https://doi.org/10.1201/b16132. URL: https://www.taylorfrancis.com/books/mono/10.1201/b16132/handbook-graph-theory-jonathan-gross-jay-yellen-ping-zhang.

[38] Aldous and Wilson J. (2004)

Joan M. Aldous and Robin Wilson J. *Graph and Applications*. Walton Hall, Milton Keynes MK7 6AA, UK: Springer-Verlag, 2004. 458 pages. ISBN: 1-85233-259-X. URL: https://www.springer.com/gp/book/9781852332594.

**Definition 2.8 (of walks,** another one)**:** *Another definition is in the form of a non-empty alternating sequence*

$$x_0 e_0 x_1 e_1 \dots e_{k-1} x_k$$

Expression 2.2

*of vertices and edges in $G$ such that $e_i = \{v_i,\ v_{i+1}\}$ for all $i < k$. If $v_0 = v_k$ the walk is closed.*[35]

**Definition 2.9 (of trails)**: *A trail is a walk in which all the edges, but not necessarily all the vertices, are different.*[38]

**Definition 2.10 (of paths)**: *A path is a non-empty graph $P = (V, E)$ of the form*

$$V = \{x_0,\ x_1,\ \dots,\ x_k\}, \quad E = \{x_0 x_1,\ x_1 x_2,\ \dots,\ x_{k-1} x_k\}$$

Expression 2.3

*where the vertices $x_i$ and the edges $x_i x_{i+1}$ are all distinct. The vertices $x_0$ and $x_k$ are linked by $P$ and are called its ends; the vertices $x_1,\ \dots,\ x_{k-1}$ are the inner vertices of $P$. The number of edges of a path is its length and the path of length $k$ is denoted by $P^k$. As $k$ is allowed to be zero, $P^0 = K^1$.*[35]

**Definition 2.11 (of cycles)**: *As with paths, a cycle is often denoted by its (cyclic) sequence of vertices; A cycle $C$ might be written as $x_0 x_1 \dots x_{k-1} x_0$. The length of a cycle is its number of edges (or vertices); A cycle of length $k$ is called a k-cycle and denoted by $C^k$.*

*The minimum length of a cycle in a graph $G$ is the girth $g(G)$ of $G$; the maximum length of a cycle in $G$ is its circumference. If $G$ does not contain a cycle, the former is set to $\infty$, the latter to zero. An edge that joins two vertices of a cycle but is not itself an edge of the cycle is a chord of that cycle.*[35]

**Theorem 2.1 (on edge set partitioning into cycles)**: *The edge set of a graph can be partitioned into cycles if, and only if, every vertex has even degree.*[13]

**Proof 2.1 (of Theorem 2.1)**: *See* Modern Graph Theory - *Bollobás (1998), page 5.*

**Definition 2.12 (of connected graphs)**: *A graph is connected if there is a path between each pair of vertices, and is disconnected otherwise.*

*An edge in a connected graph is a bridge if its removal leaves a disconnected graph. Every disconnected graph can be split up into a number of connected subgraphs, called components.*[38]

**Theorem 2.2 (on the presence of triangles in a graph)**: *Every graph of order $n$ and size greater than $\lfloor frac{n^2}{4} \rfloor$ contains a triangle.*[13]

**Proof 2.2 (of Theorem 2.2)**: *See* Modern Graph Theory - *Bollobás (1998), page 6.*

**Corollary 2.1 (on spanning trees of connected graphs)**: *Every connected graph contains a spanning tree.*[13]

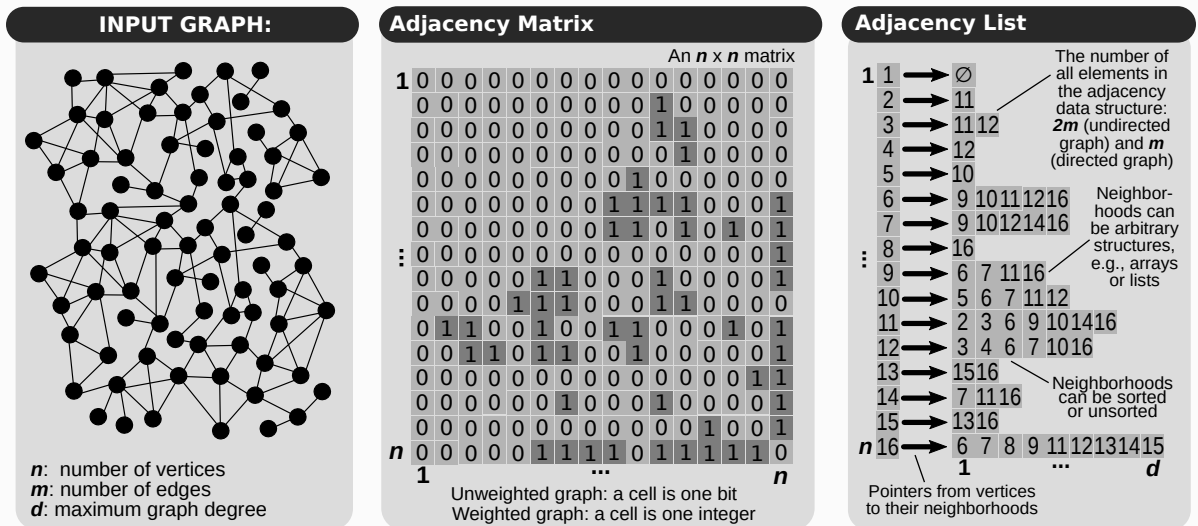

≡ Figure 2.5.: Illustration of Adjacency Matrix and Adjacency List[14]

**Adjacency lists and adjacency matrix**

There exist two standard ways to represent a graph: as a collection of *adjacency lists* or as an *adjacency matrix*; both valid for either directed and undirected graphs. The adjacency-list representation provides a compact way to represent *sparse* graphs, that is those for which $|E|$ is much less than $|V|^2$; when instead a graph is *dense* (i.e. $|E|$ is close to $|V|^2$), the adjacency matrix could be used for optimization purposes. The adjacency-matrix representation of a graph $G = (V, E)$ assumes that vertices are numbered $1, 2, \ldots, |V|$ in some arbitrary manner - consists of a $|V| \times |V|$ matrix $A = (a_{ij})$ such that $a_{ij} = 1$ if $(i, j) \in E$; 0 otherwise.[37]

An example is shown in Figure 2.5.

## 2.1.1.4. Notes on terminology used

≡ **Remark 2.1 (on nodes):** *Nodes and vertices are used interchangeably and represent the constituting discrete elements of a graph.*

≡ **Remark 2.2 (on edges):** *Edges, hops, links and arcs are used interchangeably and represent relationships and connections among vertices.*

≡ **Remark 2.3 (on depth and number of hops):** *Depth, level and distance are used interchangeably and represent the number of hops from a vertex to another given reference vertex.*

≡ **Remark 2.4 (on connected nodes):** *If a single edge connects two nodes, they are considered to be directly connected. If the pathways connecting two nodes have more than one edge, they are indirectly connected; that is, there is at least another vertex on the road between them.*

≡ **Remark 2.5 (on paths):** *A route is a path that begins at one source node and leads to a destination node by traversing consecutive edges (either in the same direction or not).*

≡ **Remark 2.6 (on outbound and inbound directions):** *If an edge's direction points away from the node being analyzed, it is said to have outbound or outgoing direction. If an edge's direction points exactly to the reference node, it has inbound or incoming direction.*

[37] Cormen, Leiserson, Rivest and Stein (2009)
Thomas H. Cormen, Charles E. Leiserson, Ronald L. Rivest and Clifford Stein. *Introduction to Algorithms*. 3rd Edition. MIT Press, 2009. 1313 pages. ISBN: 978-0-262-03384-8. URL: http://mitpress.mit.edu/books/introduction-algorithms.

≣ **Remark 2.7 (on directed paths):** *A path made up of just edges with the same concordant direction will be called a directed path; an undirected path, on the other hand, is a path made up of edges with no constraints on their relative directions. An outgoing path, for example, begins at the beginning node and consists solely of outgoing edges with respect to the path's vertices; an ingoing path, on the other hand, consists solely of ingoing edges.*

≣ **Remark 2.8 (on descendants):** *The nodes that can be reached by the original node via outgoing paths will be referred to as descendants. This set includes an initial node's children, as well as their children, and so on.*

≣ **Remark 2.9 (on neighbors):** *Those nodes that are directly related to the reference node, regardless of the direction of the edges involved, will be named neighbors.*

≣ **Remark 2.10 (on cycles):** *A directed path that starts and finishes with the same node is called a cycle. Therefore, if a path has the same vertex more than once it is said to contain a cycle. It is worth noting that the definition is based on avoiding multiple visits of the same vertex rather than multiple walks on the same arc. Furthermore, an undirected path that passes across the same node more than once does not mean it contains a cycle, because it is only an aggregation of different directed paths (though one of them could be a cycle; in such case, the path effectively contains a cycle).*

≣ **Remark 2.11 (on fathers and children):** *Will be called children the nodes that are directly connected to the given node and that can be reached by means of an outbound edge. In a dual way, will be called fathers the nodes that are directly connected to the given node and that can be reached by means of an inbound edge.*

≣ **Remark 2.12 (on ancestors and descendants):** *The nodes that are reachable from the beginning node via ingoing paths will be referred to as ancestors. The set of vertices that can be reached by the given node via outgoing paths can also be defined as ancestors. The same is true for descendants.*

≣ **Remark 2.13 (on exploration and traversals):** *Graph exploration and traversal are used interchangeably.*

In the next subsection are presented a few examples of usage of graphs to represent real-life interconnected entities.

## 2.1.2. Graph examples

The graph examples taken into consideration in this subsection are:

- road networks,
- computer networks,
- and social networks.

### 2.1.2.1. Road Networks

A great example of a graph is a road network as shown in Figure 2.6.[39].

[39] WIKIMEDIA FOUNDATION, INC. (2021)

Wikimedia Foundation, Inc. *Wikipedia - List of Milan Metro stations.* Online. Sept. 2021. URL: https://en.wikipedia.org/wiki/List_of_Milan_Metro_stations.

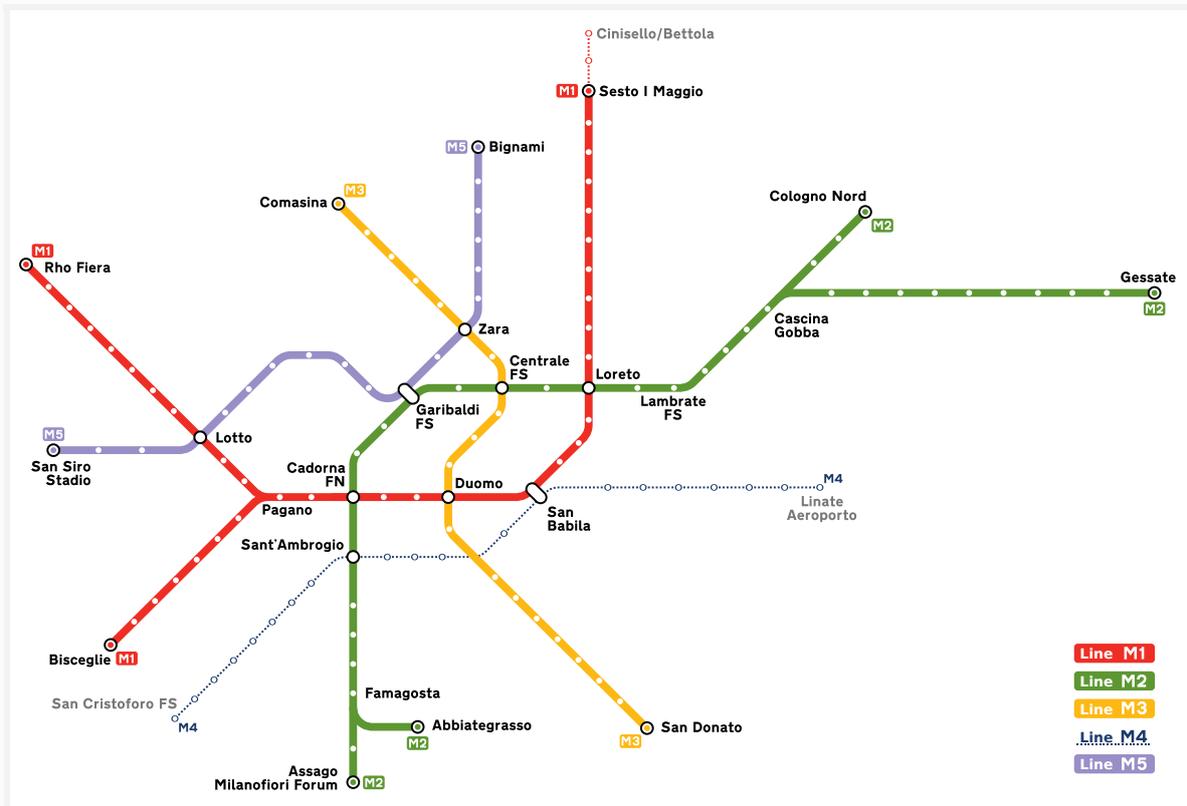

≣ Figure 2.6.: Network map of Milan Metro stations and lines[39]

   The nodes in such networks are road intersections or stations, stops, while the edges are the actual roadways, railways, airways or seaways. Graph analysis can address a variety of problems about road networks, including:

1. finding various alternative routes?
2. the shortest path's length;
3. a way to visit all the vertices in a list in the shortest time possible;
4. the shortest path from one location to another.

The third topic is particularly significant for postal delivery, as it helps reduce the distance traveled while increasing the number of shipments delivered.[40]

### 2.1.2.2. Computer networks

   In computer networks, computers and routers can be considered as vertices and cables, routing lines connecting them are indeed the links (of a graph). The following are some of the common questions that emerge with this type of network structure:

- A shortest path problem: how quickly can data be transferred from point A to point B?
- Node centrality algorithms: which of the vertices is the most important or critical one?

For more details on the algorithms, see § 2.1.3 - Graph analysis and algorithms on page 26.[40]

---

[40] Scifo (2020)

Estelle Scifo. *Hands-On Graph Analytics with Neo4j - Perform graph processing and visualization techniques using connected data across you entreprise*. Livery Place, 35 Livery Street, Birmingham, B3 2PB, UK: Packt, Aug. 2020. 510 pages. ISBN: 978-1-83921-261-1. URL: https://www.packtpub.com/product/hands-on-analytics-with-neo4j/9781839212611.

#### 2.1.2.3. Social networks

Graphs are used by Facebook, LinkedIn, and many other social media platforms to model their users and interactions. Nodes represent persons and edges show their *friendship* or *professional* relationship in the most basic example of a social graph, as shown in figure Figure 2.7.

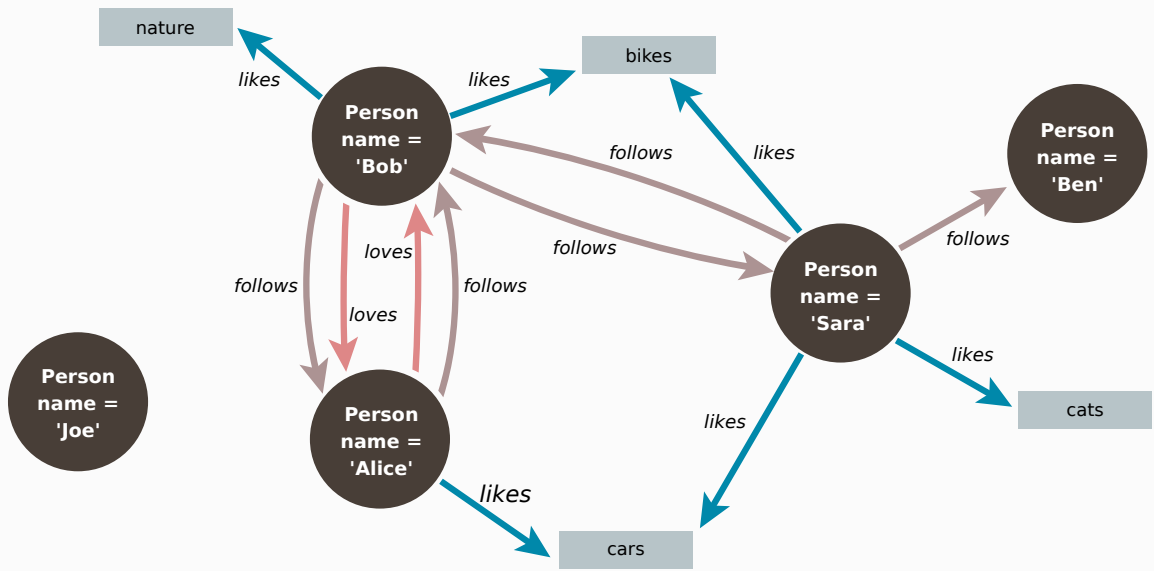

Figure 2.7.: Example of a social network graph

For instance, when looking at someone's profile on LinkedIn, it is possible to see the degree of connection of that profile from the logged-in user. If *3rd* is shown on that profile, it means that the logged-in user is just two connections away from the viewed profile. In other words, one person in the logged-in user's network is already connected to a person who is connected to the viewed profile.[40]

### 2.1.3. Graph analysis and algorithms

In this section are presented some methods of analysis that are at the very core of graph algorithms.

#### 2.1.3.1. Centrality (Node importance)

Centrality focuses on understanding which nodes in a network are more important than others. What exactly does 'importance' mean? Different forms of centrality algorithms have been developed to evaluate diverse things,[41] such as the ability to quickly spread information versus the ability to distinguish critical nodes in computer networks. By critical, it is meant that if a specific node is not working for some reason, the whole network will be impacted. Not every node affects the network in the same way. It comes naturally that backbone nodes are crucially important.

[41] Needham and Hodler (2021)

Mark Needham and Amy E. Hodler. *Graph Algorithms - Practical Examples in Apache Spark & Neo4j*. 1005 Gravenstein Highway North, Sebastopol, CA 95472, USA: O'Reilly, 2021. 300 pages. ISBN: 978-1-492-05781-9. URL: https://neo4j.com/graph-algorithms-book/.

| Algorithm type | What it does | Usecases |
|---|---|---|
| **Degree Centrality** | Calculates how many edges a vertex is connected to | Quantification of a person's popularity |
| **Closeness Centrality, Wasserman and Faust, Harmonic Centrality** | Determines which vertices have the shortest paths to all other vertices | Choosing optimal location of new public facilities for maximum accessibility |
| **Betweenness Centrality, Randomized-Approximate Brandes** | Calculates how many shortest paths pass through a vertex | Finding the control genes for specific diseases for better drug targeting |
| **All Pairs Shortest Path** | Computes the shortest path between all pairs of nodes in the graph | Finding alternate routes around traffic jams |
| **PageRank** | Quantifies a vertex's importance from its connections with the neighbors and their neighbors (popularized by Google) | Rating text for entity relevance in natural language processing and identifying the most relevant features for extraction in machine learning. |

Table 2.1.: Overview of centrality algorithms[41]

In the case of social media networks, while it may be possible a cascade reaction, generally it is very unlikely that a single person's retirement from social media makes that whole social community retire.[40]

#### 2.1.3.2. Pathfinding

For graph analytics and algorithms, paths are fundamental. Finding the shortest paths is perhaps the most common task performed with graph algorithms and it is a prerequisite for a variety of analyses. The shortest path is an exploration route with the least hops or lowest weight. If the graph is directed, the shortest path between two nodes is determined by the direction of the relationships.[41]

An example is shown in Figure 2.8.[42]

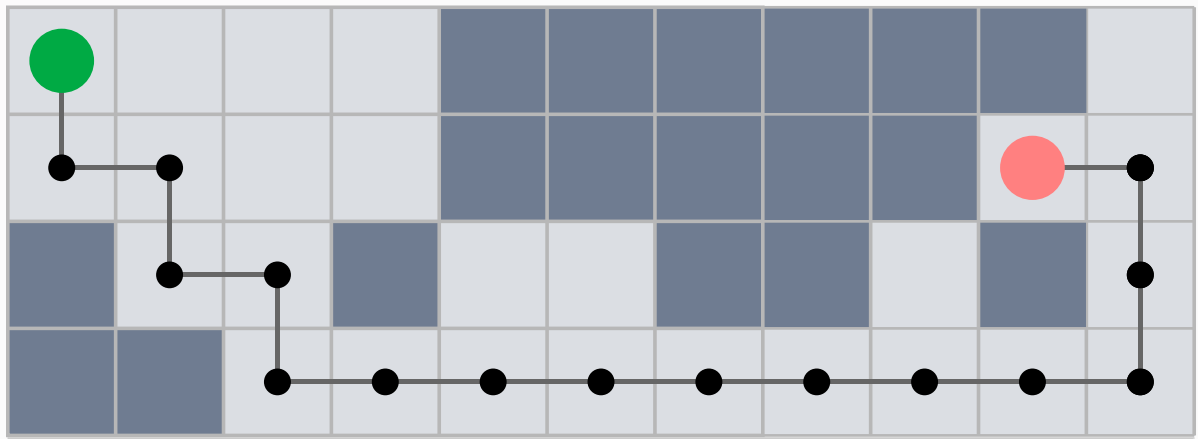

Figure 2.8.: An illustration of the directional movements during the execution of a shortest path algorithm, in a maze[42]

#### Path types

The *average shortest path* is used to evaluate a network's overall efficiency and resiliency, for example in determining the average distance between subway stations. Sometimes it may be necessary to get the longest optimized route for cases where is requested to know which subway stations are farthest apart or have the most number of stops between them, even when the best route is chosen. The *diameter* would be used in this case to discover the longest shortest path between all node pairs.[41]

[42] Growing with the Web and Imms (2021)
Growing with the Web and Daniel Imms. *growingwiththeweb.com website*. Online. Sept. 2021. URL: https://www.growingwiththeweb.com/.

| Algorithm type | What it does | Usecases |
|---|---|---|
| **Breadth First Search**[43] | Tree traversal by exploring right the nearest neighbors first and then the sublevel neighbors | GPS systems locating neighbor vertices to identify nearby places of interest |
| **Depth First Search**[44] | Tree traversal by exploring down the sublevel neighbors first before backtracking | Optimal solution path discovery with hierarchical choices |
| **Shortest Path** | Computes the shortest path between two vertices | Driving route discovery between two locations |
| **All Pairs Shortest Path** | Computes the shortest path between every two vertices of the graph | Finding alternate routes around a traffic jam |
| **Single Source Shortest Path** | Computes the shortest path between a root vertex and all other vertices | Least cost routing of phone calls |
| **Minimum Spanning Tree** | Computes the path in a connected tree structure that has the smallest cost for visiting all vertices | Routing optimization or garbage collection |
| **Random Walk** | Returns a list of vertices along a path of specified size by randomly choosing relationships to traverse. | Augmenting training for machine learning. |

Table 2.2.: Overview of pathfinding and graph search algorithms[41]

### 2.1.3.3. Community detection

Community detection, often referred to as clustering, is a method of finding a group of nodes sharing certain characteristics.[40]

Usually, real-world networks present substructures (often quasi-fractal) of almost independent subgraphs. Connectivity is utilized to detect communities and quantify the quality of clustering groupings. Analyzing different types of communities within a graph can uncover structures, like hubs, hierarchies and tendencies of groups to attract or repel others. These techniques are also used to study emergent phenomena such as echo chambers and filter bubble effects.[41]

This topic is thoroughly presented in § 3 - Community Detection on page 65 where also the execution of the Label Propagation Community Detection algorithm is step-by-step explained.

### 2.1.3.4. Link prediction

Using intelligent models, it is possible to predict whether two vertices of a graph shall be connected in the future or not.

Once again, perfect examples of applications of link prediction are recommendation engines.[40]

### 2.1.3.5. Other algorithms and problems in Graph Theory

While the problems and algorithms treated above are just a taste of only the most commonly known and often considered problems in graph theory literature, there is however a vast majority of other problems that are not presented here. The focus of this thesis is limited to a precise problem. Thus, for more on these topics, refer to graph theory books and scientific literature.

## 2.2. Review of Graph Database Systems

Graph databases address one of today's most important macroeconomic trends: generating insight and competitive advantage by harnessing complex and dynamic relationships in highly connected data. Whether it is understanding network elements, customer relationships, genes and proteins, or entertainment producers and consumers, the ability to understand and analyze vast graphs of highly connected data will be critical in determining which companies outperform their competitors over the next decade.

Graph databases are the most effective way to represent and query connected data. In order to interpret and understand connected data's value, first is needed to understand how its constituent elements are interconnected. Most of the time, naming and qualifying the relationships between things is needed to produce this understanding.

Despite the fact that huge corporations got this a long time ago and began developing their own proprietary graph processing technologies, today, technology has rapidly become democratized. General-purpose graph databases are now available, allowing ordinary people to profit from connected data without having to invest in their own graph infrastructure.[45]

Graphs have long been recognized as a natural manner of representing information and knowledge and the concept of a "graph database" has been around since at least the 1980s. However, because of the technological advancements, it is only lately possible to take full advantage of this abstract concept.[46]

Graph databases have since been used to tackle major challenges in social networking, recommendation systems, geospatial/GIS, master data management and many other domains. The massive commercial success of companies like Facebook, Google, and Twitter, all of which have built their business models around their own proprietary graph technologies, and the introduction of general-purpose graph databases into the technology landscape, are driving this increased focus on graph databases. This is clearly visible in the popularity change chart of different database systems categories shown in Figure 2.9.[47]

[45] Robinson, Webber and Eifrem (2015)

Ian Robinson, Jim Webber and Emil Eifrem. *Graph Databases*. 2nd Edition. O'Reilly Media, Inc., June 2015. 224 pages. ISBN: 9781491930892. URL: https://www.oreilly.com/library/view/graph-databases-2nd/9781491930885/.

[46] Angles and Gutierrez (2018)

Renzo Angles and Claudio Gutierrez. *An Introduction to Graph Data Management*. In: Graph Data Management (2018), pages 1–32. ISSN: 2197-974X. DOI: 10.1007/978-3-319-96193-4_1. URL: https://arxiv.org/abs/1801.00036.

[47] solid IT GmbH (2021)

solid IT GmbH. *db-engines.com website*. Online. Created and maintained by solid IT GmbH, an Austrian IT consulting company with a special focus on software development, consulting and training for database-centric applications. Aug. 2021. URL: https://db-engines.com.

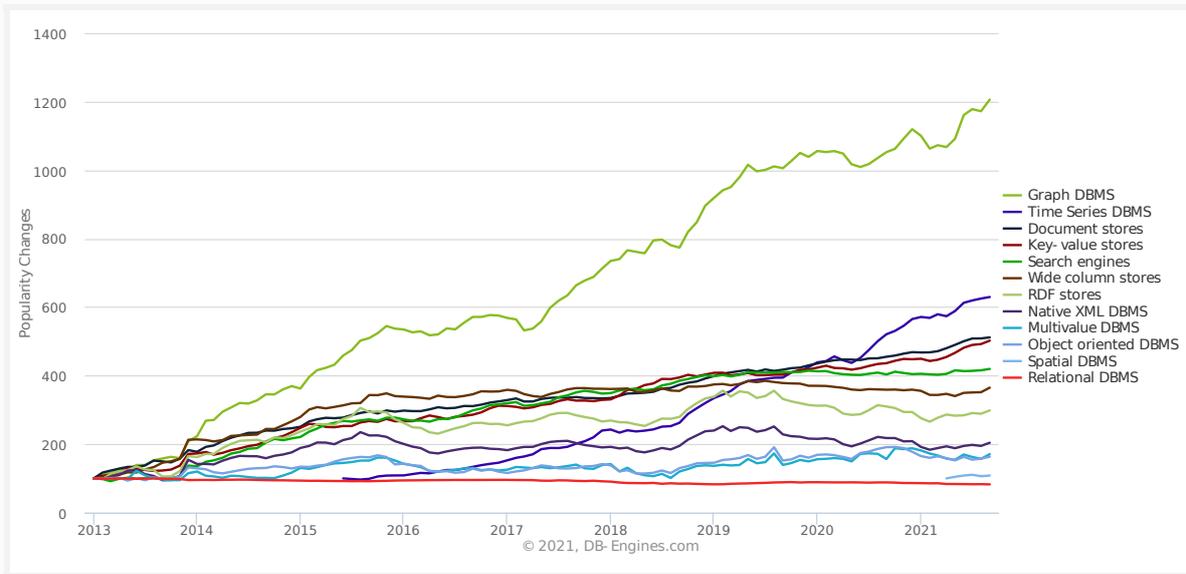

Figure 2.9.: Trend of DBMS popularity changes per category[47]

It is interesting to see what makes graph databases so compelling. To that end in the following subsection are presented some general notions, definitions of graph databases. Follows in § 2.2.2 - Graph data models a panoramic description of graph data models. After that in § 2.2.3 - Graph database characteristics the main characteristics of graph databases are investigated and in § 2.2.4 - Graph Query Languages an overview of the various graph query languages is presented. During the work of this thesis isn't done any benchmarking activity *per sé*. But to better understand the differences between graph databases, some comparisons, benchmarks are be presented between GDBMSs in § 2.2.5 - GDBMSs comparison. That shall be the closing of this chapter.

## 2.2.1. Different definitions of graph databases

There does not seem to be a universally accepted definition of what constitutes a graph database.

A common definition is the following:

**Definition 2.13 (of graph databases,** common**):** *A graph database is a database that organizes data by modeling it by means of the concepts of vertices and edges, where vertices represent the entities to be stored and edges represent the relationships that exist among nodes.*[48]

A more specific definition is:

**Definition 2.14 (of graph databases,** specific**):** *A graph database is any storage system that has the following properties:*

- *is designed to store graph-like data;*
- *organizes data by means of node and edge elements;*
- *provides built-in graph traversal functions;*
- *(hopefully) provides graph data integrity;*

which is very similar to the definition provided in "An Introduction to Graph Data Management" by Angles and Gutierrez (2018):

[48] Sɪɴɪᴄᴏ (2017)
Luca Sinico. *Graph databases and their application to the Italian Business Register for efficient search of relationships among companies.* Master's thesis. University of Padua, Apr. 2017. 208 pages. ᴜʀʟ: http://tesi.cab.unipd.it/54610/.

☰ **Definition 2.15 (of graph databases,** by Angles and Gutierrez**):** *A graph database is a database where the data structures for the schema and/or instances are modeled as a graph, where data manipulation is expressed by graph-oriented operations and appropriate integrity constraints can be defined over the graph structure.*[46]

Another definition focused on database requirements is the one that follows:

☰ **Definition 2.16 (of graph databases,** with requirements**):** *A graph database is a system with Create, Read, Update, and Delete (CRUD) methods that exposes a graph data model. Graph databases are generally built for use with transactional (OLTP) systems. Accordingly, they are normally optimized for transactional performance and engineered with transactional integrity and operational availability in mind.*[45]

One last definition shall be reported and that revolves around the idea a graph database has to provide index-free adjacency:

☰ **Definition 2.17 (of graph databases,** requiring index-free adjacency**):** *A graph database is a database that uses graph structures with vertices, edges and properties to represent and store data providing index-free adjacency.*[49,50]

On data retrieval, some new concepts and generally a new query language are required, if compared to the standard query language of relational databases. There is not, yet, a standard query language for this domain: each graph database product comes with its own query language and, up to now, none of these query languages has risen to prominence in the same fashion as SQL did for relational databases. Some standardization efforts are however taking place, leading to systems like *Gremlin* (which works with a variety of graph engines that use the property graph model) and *SPARQL* (which is used in RDF graphs).[51]

In a more general sense, a database model (or just data model) is a conceptual tool used to model representations of real world entities and the relationships among them. As is well known, a data model can be characterized by three basic components, namely:

1. data structures;
2. query and transformation language;
3. integrity constraints.

Following this definition:

☰ **Definition 2.15 (of graph databases,** by Angles and Gutierrez**, continued):** *A graph database model is a model where data structures for the schema and/or instances are modeled as graphs (or generalizations of them), where the data manipulation is expressed by graph-oriented operations (i.e. a graph query language) and appropriate integrity constraints can be defined over the graph structure.*[46]

Having given some definitions of what are commonly considered to be graph databases, in the next subsection graph data models shall be described.

## 2.2.2. Graph data models

When it comes to graph databases, there are a variety of features that may be considered. The data model that they use is one of them. Property graphs, RDF graphs, and hypergraphs are the three most used graph data models.[45]

[49] IBM System G (2018)

IBM System G. *Introduction of Graph Database*. Online. May 2018. URL: https://web.archive.org/web/20180521155557/http://systemg.research.ibm.com/database.html.

[50] Rodriguez (2010)

Marko Rodriguez. *Graph Databases: Trends in the Web of Data*. Online. Slides. Sept. 2010. URL: https://www.slideshare.net/slidarko/graph-databases-trends-in-the-web-of-data/.

[51] Wikimedia Foundation, Inc. (2021)

Wikimedia Foundation, Inc. *Wikipedia - Graph Database*. Online. Aug. 2021. URL: https://en.wikipedia.org/wiki/Graph_database.

The property graph is more of an industry standard than the RDF model, which is a W3C standard.[52]

In the following subsections, each of these models is presented in detail.

### 2.2.2.1. Property graph

▤ **Definition 2.18 (of property graphs):** *A property graph (see Figure 2.10[53]) is a one that has the following properties:*

- *It is made up of vertices, edges, attributes and labels - or some similar categorization form.*
- *Vertices can have an arbitrary amount of attributes, typically key-value pairs.*
- *Vertices are connected by edges which structure the graph. Each edge has a direction, a single name, a start vertex and an end vertex. There are no dangling edges.*
- *One or more categories can be applied to vertices and edges, by means of classes or labels. Vertices are grouped together by categories, which describe the functions they perform in the dataset. Edges can be categorized as well. Edge direction and type, category work together to provide semantic meaning to the network structure.*
- *Relationships too can have attributes. The ability to add attributes to them is especially useful for improving the graph's semantics and for limiting the results at query-time.[45]*

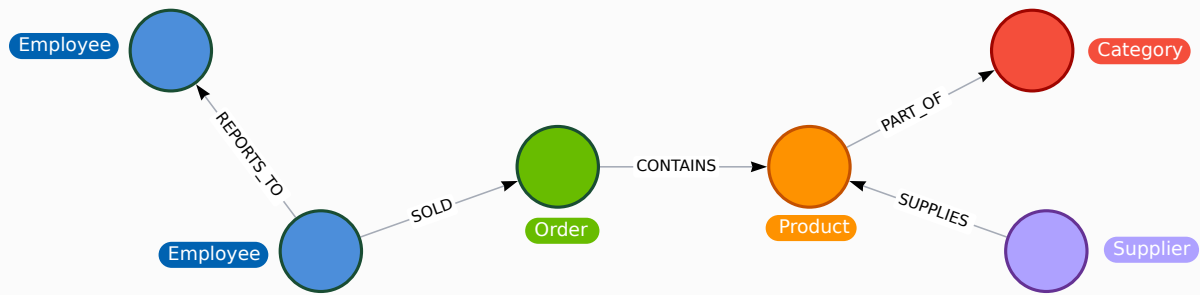

▤ Figure 2.10.: An example of a property graph[53]

This type of model has gained a lot of traction and it is used by the majority of the graph DBMSs today. In addition, the Apache TinkerPop project has sought to create the groundwork for standardization.[54,55]

### 2.2.2.2. RDF graphs, Ontologies, Inferences, SPARQL, Semantic Web and Linked Data

The formal definition of RDF, given by W3C, is:

▤ **Definition 2.19 (of RDF):** *Resource Description Framework (RDF) is a standard model for data interchange on the Web. RDF has features that make data merging easier even when the underlying schemas are different and it supports schema evolution over time without needing all data consumers to be updated.[56]*

To put it another way, RDF expands the Web's linking structure by allowing URIs to be used to name both relation-

[52] Seaborne (2015)

Andy Seaborne. *Two graph data models: RDF and Property Graphs*. Slides. Talk given at ApacheConEU Big Data 2015. 2015. url: https://www.slideshare.net/andyseaborne/two-graph-data-models-rdf-and-property-graphs.

[53] Neo4j, Inc. (2021)

Neo4j, Inc. *Neo4j Documentation*. Online. Documentation. Aug. 2021. url: https://neo4j.com/.

[54] Apache TinkerPop (2021)

Apache TinkerPop. *Apache TinkerPop Documentation*. Online. Documentation. Aug. 2021. url: https://tinkerpop.apache.org/docs/current/reference/.

[55] Apache TinkerPop (2021)

Apache TinkerPop. *Apache TinkerPop Website*. Online. Aug. 2021. url: https://tinkerpop.apache.org/.

[56] W3C (2014)

The World Wide Web Consortium W3C. *W3C Semantic Web - Resource Description Framework (RDF)*. Online. Feb. 2014. url: https://www.w3.org/RDF/.

ships between objects and the two ends of the link. This model is built on a set of triples, each of which consists of a subject, a predicate and an object. An RDF graph, also known as a triples store, is a collection of such triples.

In an RDF graph, there are three types of vertices:

- URI vertices,
- literal vertices
- and blank vertices.[57]

URI stands for Universal Resource Identifier and is a string of characters used to identify a resource. URL (Uniform Resource Locator) is the most popular type of URI, which is used to identify Web resources. Values like as strings, numbers and dates are stored in literal vertices. Instead, blank vertices indicate anonymous resources, or those without a URI or literal value. A blank vertex can only be used as the subject or object of an RDF triple. The object might be a URI or a literal vertiex, while the subject is usually specified by a URI. Predicates are generally identified by URIs and can be interpreted as a definition of an attribute value (object vertex) for some subject vertex or as a relationship between the two vertices.[57]

From a low-level perspective, the RDF graph architecture is made up of arcs that connect entities with both their characteristics and other entities. From a high-level perspective, The RDF graph produces a directed labeled graph made up of edges between entities. By employing this basic paradigm, RDF allows structured and semi-structured data to be mixed, exposed and shared across multiple different applications.[56]

**RDF storage techniques**

The implementation of the underlying storage model of RDF is not unique. RDF schemas (metadata) and instances can be read and manipulated quickly in main memory, they can be serialized to files for persistence, or for huge volumes of data, a database management system can be used. For this reason, several existing RDF stores employ relational and object-relational database management systems (RDBMS and ORDBMS). However, storing RDF data in a relational database requires an appropriate table design. There are two major approaches to this:

- generic schema (or schema-carefree), which are schemas that do not depend on the ontology
- and ontology specific schema (or schema-aware).[58]

The paper "A survey of RDF storage approaches", Faye, Curé and Blin (2012)[59] suggests a classification of the RDF storage techniques as shown in Figure 2.11 (b). In the figure is highlighted the difference between native and non-native storage techniques, which also divides them into several sub-types based on the underlying storage used.

[57] W3C (2014)

The World Wide Web Consortium W3C. *W3C Recommendation - RDF 1.1 Concepts and Abstract Syntax, Data Model*. Online. Feb. 2014. URL: https://www.w3.org/TR/2014/REC-rdf11-concepts-20140225/#data-model.

[58] HERTEL, BROEKSTRA and STUCKENSCHMIDT (2009)

Alice Hertel, Jeen Broekstra and Heiner Stuckenschmidt. *RDF Storage and Retrieval Systems*. In: (2009). https://www.researchgate.net/publication/227215511_RDF_Storage_and_Retrieval_Systems, pages 489–508. DOI: http://dx.doi.org/10.1007/978-3-540-92673-3_22. URL: http://publications.wim.uni-mannheim.de/informatik/lski/Hertel08RDFStorage.pdf.

[59] FAYE, CURÉ and BLIN (2012)

David C. Faye, Olivier Curé and Guillaume Blin. *A survey of RDF storage approaches*. In: Revue Africaine de la Recherche en Informatique et Mathématiques Appliquées 15 (2012). Pdf: https://hal.inria.fr/hal-01299496/file/Vol.15.pp.11-35.pdf, pages 11–35. URL: https://hal.inria.fr/hal-01299496.

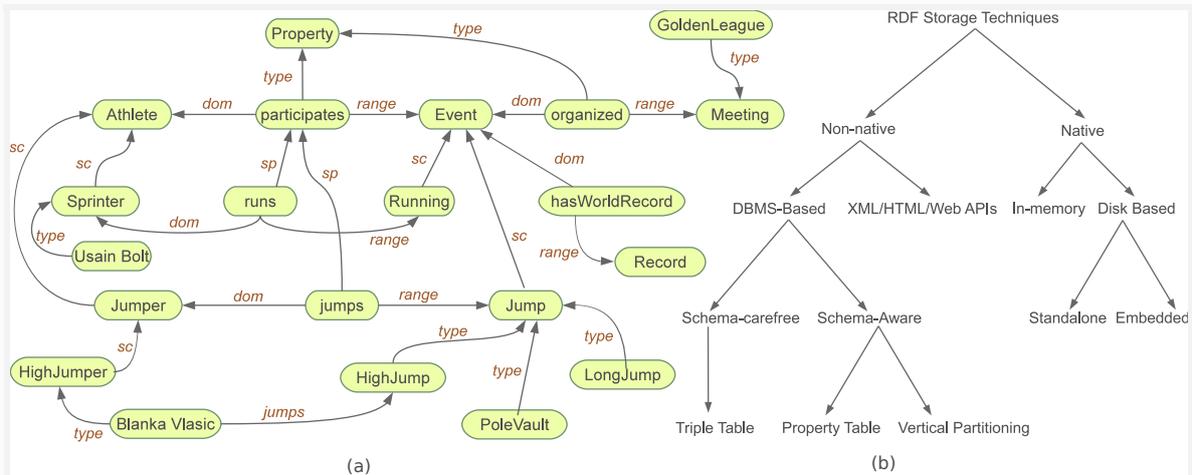

Figure 2.11.: RDF graphs. (a) An RDF graph describing athletics competitions. (b) A classification of RDF data storage approaches.[59]

RDF is supported by a variety of serialization formats, such as:

- RDF/XML,
- Turtle,
- JSON-LD,
- N-Triples,
- N3,
- N-Quads.

They are all human-readable and easy-to-parse. Some even allow the encoding of inference rules. In the example shown in Code listing 2.1 is brought how RDF triples may be encoded using the RDF/XML format. One of the existing ontologies (see the next paragraph for more on Ontologies) is used in the example, that is FOAF (Friend Of A Friend). FOAF was written to describe people, their activities, and their relationships to other people and objects.[59]

```
1 <rdf:RDF xmlns:rdf="http://www.w3.org/1999/02/22-rdf-syntax-ns#"
2          xmlns:foaf="http://xmlns.com/foaf/0.1/">
3    <foaf:Person rdf:nodeID="mark">
4        <foaf:name>Mark</foaf:name>
5        <foaf:mbox rdf:resource="mailto:mark@email.com"/>
6        <foaf:knows>
7            <foaf:Person>
8                <foaf:name>John</foaf:name>
9            </foaf:Person>
10       </foaf:knows>
11   </foaf:Person>
12
13   <foaf:Person rdf:nodeID="peter">
14       <foaf:name>Peter</foaf:name>
15           <rdfs:seeAlso rdf:resource="http://www.peter.com/peter.rdf"/>
16       <foaf:knows rdf:nodeID="mark"/>
17   </foaf:Person>
18 </rdf:RDF>
```

Code listing 2.1: RDF storage techniques example

The example presents a situation in which there are three vertices of type Person, one of which is named Mark, one is named John, and the third of them is named Peter. Mark has an email address, while Peter has an RDF description for itself. Furthermore, Mark knows John and Peter knows Mark.

Returning on RDF graphs known as triples stores, some triples stores have been developed as database engines from scratch. Others, on the other hand, have been developed on top of commercial relational database engines or NoSQL document-oriented database engines already in use. This method of using existing databases allows for the creation of database engines with minimal programming work. However, while triples may be stored in relational databases, implementing efficient querying mechanisms (e.g., mapping from SPARQL) into SQL queries for data domains made up of triples is difficult. As a result, a well-designed native triples store may provide better performance than non-native alternatives.[60]

**Ontologies**

☰ **Definition 2.20 (of ontologies, vocabularies):** *Because RDF uses URIs to encode information, concepts become more than just words on a page. They are linked to a unique definition that anybody can locate on the Web. Ontology or vocabulary is the term for this type of shared definition.*

> *"Imagine having access to a variety of databases with information about people, including their addresses. If there was an interest in finding the people living in a specific zip code, it would be required to know which fields in each database represent names and which represent zip codes. RDF can specify that '(field 5 in database A) (is a field of type) (zip code)', using URIs rather than phrases for each term. However, two databases may use different identifiers for what is in fact the same concept, such as zip code. A program that wants to compare or combine information across the two databases has to know that these two terms are being used to mean the same thing. Ideally, the program must have a way to discover such common meanings for whatever databases it encounters. A solution to this problem is provided by the third basic component of the Semantic Web, collections of information called ontologies."* - Berners-Lee, Hendler and Lassila in "The Semantic Web" (2001)[61]

These ontologies (or vocabularies) are thus one of RDF's most distinguishing features. Ontologies' primary role is to help data integration when there are ambiguities between the terminology used in different datasets, or when a little extra information can lead to the identification of new relationships. The other role of ontologies is to organize knowledge, which may come from multiple organizations' datasets, in order to disseminate standard and shared terminology.[62]

**Inferences**

The possibility to perform inferences (or reasoning) among data is another significant feature of RDF.

☰ **Definition 2.21 (of inferences):** *Inference is the ability to generate, using by inference engines (or "reasoners") automated procedures - new associations based on data and some additional information in the form of an ontology. The*

---

[60] Wikimedia Foundation, Inc. (2021)

Wikimedia Foundation, Inc. *Wikipedia - Triplestore*. Online. Aug. 2021. URL: https://en.wikipedia.org/wiki/Triplestore.

[61] Berners-Lee, Hendler and Lassila (2001)

Tim Berners-Lee, James Hendler and Ora Lassila. *The Semantic Web*. In: Scientific American 284.5 (May 2001). https://www.w3.org/2001/sw/, pages 34–43. URL: https://www.jstor.org/stable/26059207.

[62] W3C (2021)

The World Wide Web Consortium W3C. *W3C Semantic Web - Ontologies, Vocabolaries*. Online. Aug. 2021. URL: https://www.w3.org/standards/semanticweb/ontology.html.

*new resulting relationships can be returned at query time or explicitly added to the dataset.*[63] *This way, in addition to retrieving data, the database can also be used to deduce new information by analyzing facts in the data.*

Inferences, by automatically identifying new relationships while evaluating the data's content, can be used to enhance the quality of data integration on the Web or to manage knowledge on the Web in general. Inference-based approaches are also useful for detecting inconsistencies in the (integrated) data.[63]

Consider the following classical syllogism as an illustration of what inferences are: Imagine the information:

*"all men are mortal"*

and

*"Socrates is a man"*

is encoded in the RDF graph. The following conclusion may thus be drawn:

*"Socrates is mortal."*

This is important when working with entities that belong to classes and subclasses; in fact, inferences would make the fact that a record belonging to a subclass also belongs to its superclass relevant. This is useful when working with entities that belong to classes and subclasses. In fact, inferences would consider relevant the fact that a record belonging to a subclass also belongs to its superclass. The ability to infer is provided by rules specified on metadata, which can be stored with the data itself. They are usually expressed in OWL (Web Ontology Language), which provides the ability to express notions like:

- equality, e.g. *sameAs*;
- class relationships, e.g. *disjointWith*;
- richer properties, e.g. *symmetrical*;
- class property restrictions, e.g. *allValuesFrom*.[64]

It is therefore commonly used by applications that traverse the graph and must deal with issues that occur when using many different data sources.[65]

Considering another well-known example:

*"Epimenides says that all Cretans are liars"*

and

*"Epimenides is a Cretan"*

If the database is programmed to avoid infinite loops, it may point out that there is a contradiction in the data. Other graph databases usually do not handle such cases by default.[66] Nevertheless, some libraries and frameworks are moving in this direction - see *Apache TinkerPop Website*, Apache TinkerPop (2021)[55].

[63] W3C (2021)

The World Wide Web Consortium W3C. *W3C Semantic Web - Inference*. Online. Aug. 2021. URL: https://www.w3.org/standards/semanticweb/inference.

[64] LOVINGER (2009)

Rachel Lovinger. *RDF and OWL - A simple overview of the building blocks of the Semantic Web*. Slides. Presented at the Semantic Web Affinity Group at Razorfish. 2009. URL: https://www.slideshare.net/rlovinger/rdf-and-owl.

[65] W3C (2004)

The World Wide Web Consortium W3C. *W3C Recommendation - OWL Web Ontology Language*. Online. Feb. 2004. URL: https://www.w3.org/TR/owl-features/.

[66] BLOOR (2015)

Robin Bloor. *The Graph Database and the RDF Database*. Online. Jan. 2015. URL: https://web.archive.org/web/20150206002908/http://insideanalysis.com/2015/01/the-graph-database-and-the-rdf-database/.

Computers, in order for the semantic web to "work" must have access to structured collections of information and sets of inference rules that they may employ to conduct automated 'reasoning'. As a result, one of the Semantic Web's challenges has been to provide a language that conveys both data and reasoning rules, as well as the ability to export rules onto the Web from any currently existing knowledge representation system.[61]

**SPARQL**

**Definition 2.22 (of SPARQL):** *SPARQL, pronounced as "sparkle"*[67] *(in IPA phonetic transcription: /ˈspɑːkl/), is a recursive acronym for "SPARQL Protocol And RDF Query Language". It is an RDF query language, i.e. a semantic query language for databases able to retrieve and manipulate data stored in Resource Description Framework (RDF) format.*[68]

It is not the only RDF query language that exists, but it is the W3C Recommendation in that regard.

SPARQL can express queries on data stored in RDF format natively as well as data viewed in RDF format via middleware. It supports aggregation, subqueries, negation and limitation on the query by RDF source graph. SPARQL query results can provide result sets or RDF graphs.[69]

(Triple) patterns are the basis for SPARQL searches. The only difference between triple patterns and RDF triples is that one or more of the constituent resource references are variables. All triples that meet the patterns are returned by a SPARQL engine.

The following is an example of a query that asks for John's friends:

```
1 PREFIX sn: <http://www.socialnetwork.org/>
2 SELECT ?N
3 WHERE {
4     ?X sn:type sn:Person . ?X sn:firstName ?N .
5     ?X sn:knows ?Y . ?Y sn:firstName "John"
6 }
```

Code listing 2.2: SPARQL query asking for John's friends

The Semantic Web, an expanded version of the actual Web is possible by the combination of the RDF standard format for data and the SPARQL standard query language.[48]

**Semantic Web and Linked Data**

The key notion of Semantic Web can be summarized in these words:

"*A new form of Web content that is meaningful to computers*" - Tim Berners-Lee[61] referring to the Semantic Web

The Semantic Web is an effort to:
- give meaning to web page content;
- to enable guided exploration of the contents through the introduction of semantic information;
- to provide better processing support for other types of data, aside from web page documents, such as media and structured data;

[67] BECKETT (2011)
Dave Beckett. *What does SPARQL stand for?* Online. Semantic web mailing list. Oct. 2011. URL: https://lists.w3.org/Archives/Public/semantic-web/2011Oct/0041.html.
[68] WIKIMEDIA FOUNDATION, INC. (2021)
Wikimedia Foundation, Inc. *Wikipedia - SPARQL*. Online. Aug. 2021. URL: https://en.wikipedia.org/wiki/SPARQL.
[69] W3C (2008)
The World Wide Web Consortium W3C. *W3C Recommendation - SPARQL Query Language for RDF*. Online. Jan. 2008. URL: https://www.w3.org/TR/rdf-sparql-query/.

• and more broadly, to provide a means of integrating disparate sources of information.[69]

The term *Linked Data* refers to a set of guidelines for publishing and linking structured data on the Internet.[70] To make the Web of Data a reality, both nodes and relationships must be published on the Internet, allowing inbound and outbound connections and connecting the local graph to other graphs.

Semantic Web technologies can be used for a wide range of applications, such as:

• in resource discovery and classification, to improve domain-specific search engine capabilities;
• by intelligent software agents to facilitate knowledge sharing and exchange;
• in cataloging to describe the content and content relationships available at a Web site, page, or digital library;
• for describing collections of pages that represent a single logical "document";
• for content rating;
• for describing intellectual property rights of Web pages;
• in data integration, where data from multiple sources and formats can be combined into one;
• and in many others.[71]

DBPedia, which effectively makes Wikipedia available in RDF, is one example of a large Linked Dataset. The importance of DBPedia is not just because it contains Wikipedia data, but also because it provides linkages to other Web databases, such as Geonames. Applications may exploit the extra knowledge from other datasets by using those extra links (in terms of RDF triples).[72]

### 2.2.2.3. Hypergraph

A hypergraph is another way to describe graph data. When datasets contain a high number of many-to-many relationships, hypergraphs can be useful. However, with hyperedges might be not easy to define fine-grained edge characteristics for the relationships.[48]

Hypergraph models are more generic than property graphs due to the multi-dimensionality of their hyperedges. However, because the two are isomorphic, a hypergraph may always be represented as a property graph (albeit with more relationships and nodes). The opposite, on the other hand, is not so immediate and depends on the information stored.[73]

While property graphs are widely regarded as having the finest combination of pragmatism and modeling efficiency, hypergraphs stand out for their ability to capture meta-intent. Hypergraphs, for example, need fewer primitives than property graphs when qualifying one connection with another, like in[45]:

*"I like the fact that you loved that automobile"*

The hypergraph data model did not gain as much traction as the other two data models discussed in the previous subsubsections and only a few graph DBMSs use it to manage data. HypergraphDB is one example.[74]

---

[70] HEATH (2019)

Tom Heath. *Linked Data - Connect Distributed Data across the Web*. Online. Aug. 2019. URL: https://web.archive.org/web/20190802052922/http://linkeddata.org/.

[71] W3C and HERMAN (2021)

The World Wide Web Consortium W3C and Ivan Herman. *W3C Semantic Web Frequently Asked Questions*. Online. Aug. 2021. URL: https://www.w3.org/RDF/FAQ.

[72] W3C (2021)

The World Wide Web Consortium W3C. *W3C Semantic Web - Linked Data*. Online. Aug. 2021. URL: https://www.w3.org/standards/semanticweb/data.

[73] SASAKI and NEO4J, INC. (2018)

Bryce Merkl Sasaki and Neo4j, Inc. *Graph Databases for Beginners: Other Graph Technologies*. Online. Blog post. Dec. 2018. URL: https://neo4j.com/blog/other-graph-database-technologies/.

[74] KOBRIX SOFTWARE (2010)

Kobrix Software. *HypergraphDB*. Online. 2010. URL: http://www.hypergraphdb.org/.

## 2.2.3. Graph database characteristics

Graph databases can be considered from two points of view:

· the underlying storage mechanism according which they organize and store graph data,

· the processing engine, which itself can be divided:

  – with index-free adjacency,

  – without index-free adjacency.

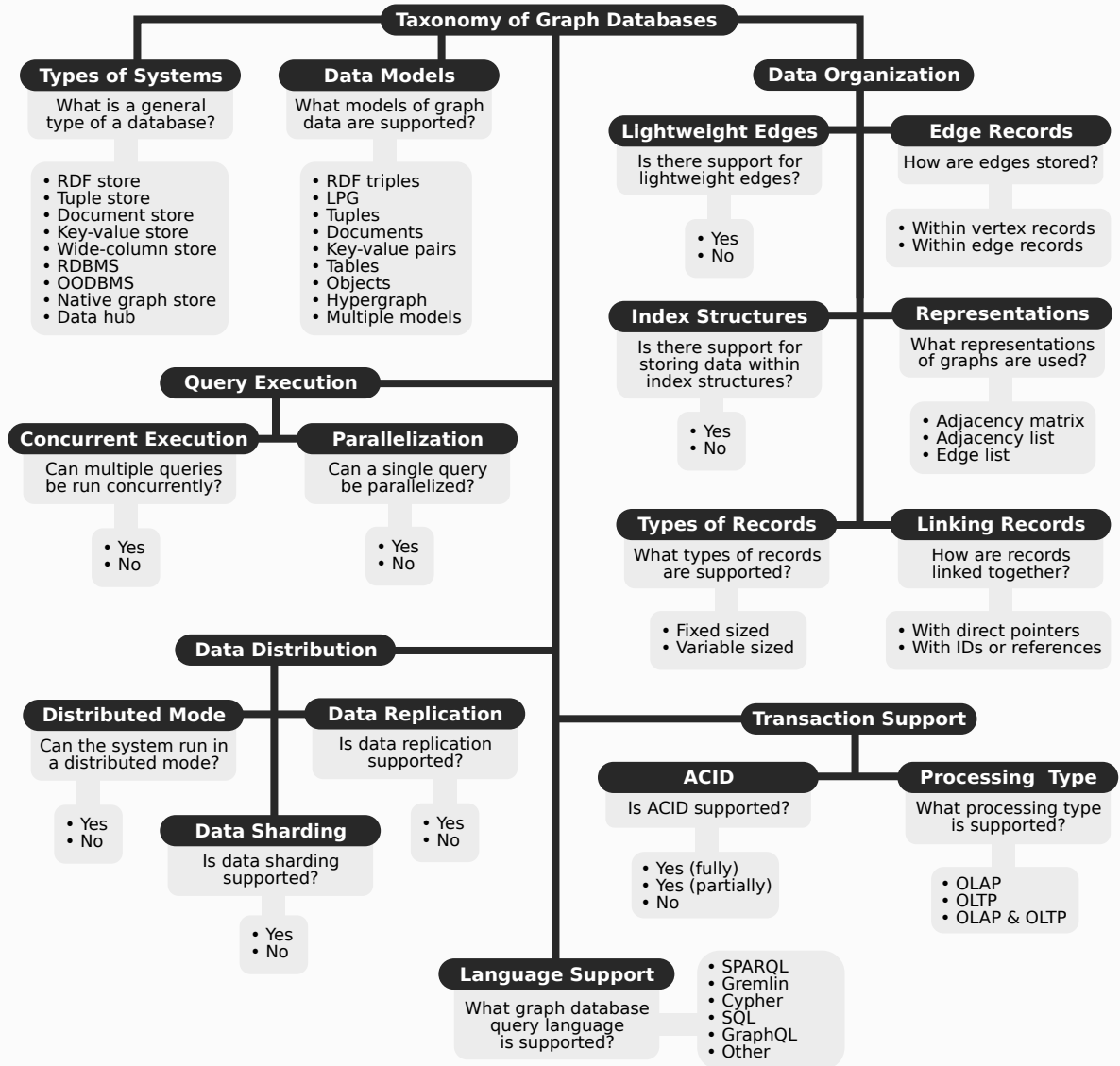

Figure 2.12.: Taxonomy of Graph Databases[14]

From the perspective of the underlying storage mechanism:

**Definition 2.23 (of native and non-native graph storages):** *A graph database is said to be a native graph storage if it organizes and stores graph data with an expressly built graph model architecture. It is said to be a non-native graph storage if it stores graph data with some already existing technology (e.g. relations for a relational database, files, documents, etc).*

Whereas from the perspective of the processing engine, whether they provide or not index-free adjacency:

≣ **Definition 2.24 (of native and non-native graph processing engine):** *A graph database is said to be a native graph processing engine if it provides index-free adjacency; otherwise it is said to be a non-native graph processing engine.*

In a database engine with index-free adjacency, each vertex stores references to its adjacent vertex or of the edges through which the adjacent vertices are reachable. As a result of this, each vertex serves as a "micro-index" of nearby vertices. Therefore this approach is in contrast with any system that relies on indexes to discover relations among data and is bound to the *locality principle* of graph queries. The locality principle says that, for non global graph analytics, during a standard querying only the part of the graph that is reachable by the specified node has to be examined.[48]

Wanting to evaluate the potential advantages of an index-free adjacency approach: Assuming a query starts with a specific node. The aim is to find the descendants of that node that have a specific attribute value. Assuming that the database that stores such data uses a traditional binary tree index. If the database does not support index-free adjacency, it will take $O\left(m \log n\right)$ time to traverse the graph, where $m$ is the number of hops and $n$ is the total number of nodes indexed in the graph. On the other hand, the traversal cost in an implementation with index-free adjacency, would only be $O\left(m\right)$.[45] This happens because the information about the edges is already stored in each node - and because the cost to find and traverse each connected edge is $O\left(1\right)$.

This is the common rationale given by the graph databases that support it within their engine. This motivation however, is as valid as the assumption that traditional databases with no index-free adjacency implement indexes with an algorithmic complexity of $O\left(\log n\right)$, such as the ones with a binary tree structure. If a Hash index were to be used, the lookup complexity would be $O\left(1\right)$. Should be said the one drawback of Hash indexes, is that while exhibiting good performance for direct text matching, they cannot be directly used for partial text matching or range searches.[48]

There is some disagreement among different graph database product providers on whether a database system must make use of index-free adjacency to be considered a graph database; For further details on this, see *Index Free Adjacency or Hybrid Indexes for Graph Databases* - ArangoDB and Weinberger (2016).[75]

### 2.2.3.1. Motivations for the adoption of graph databases

It should be clear by now that graph databases are a useful instrument when the real-world data to be modeled can be described with a graph-like structure and when the required properties are persistence and reliability.

> *"We live in a connected world. There are no isolated pieces of information, but rich, connected domains all around us. Only a database that embraces relationships as a core aspect of its data model is able to store, process and query connections efficiently."* - Neo4j's introduction line

The more connected the data domain is, the more real value it derives through the relationships. Facebook, was founded on the premise that while discrete information about individuals - their names, what they do, etc. - is valuable, the relationships between them are far more valuable. Similarly, Larry Page and Sergey Brin figured out not only how to store and analyze discrete online documents, but also how those web documents are linked: Google captured the web graph.[45]

Social graphs, business relationships, recommendation systems, geospatial applications (road maps and route planning for also rail or logistical networks), dependencies (network impact analysis), telecommunication or energy distribution networks, management of distributed data sources and access management are all common use cases for graph databases.[45,76]

---

[75] ARANGODB and WEINBERGER (2016)

ArangoDB and Claudius Weinberger. *Index Free Adjacency or Hybrid Indexes for Graph Databases*. Online. Blog post. Apr. 2016. URL: https://www.arangodb.com/2016/04/index-free-adjacency-hybrid-indexes-graph-databases/.

[76] AMAZON (2021)

Amazon. *What Is a Graph Database?* Online. Aug. 2021. URL: https://aws.amazon.com/nosql/graph/.

Fraud detection is another area where graph databases have had a lot of success. This is accomplished by identifying unusual or properly fraudulent relationship patterns on a graph. Patterns that deviate from the norm for a particular user can be identified on the graph and reported for further investigation.

In the following paragraphs will be presented a short explanation of the reasons why businesses consider the adoption of graph databases.

**Performance**

The hoped-for performance boost when working with related data is one reason why businesses look for and choose a graph database. This is due to the ability to retrieve the vertices that contain only certain types of incoming edges and not others; or the built-in features to work with weighted graphs and compute shortest or least-cost paths. The locality principle of graph queries, which was presented in § 2.2.3 - Graph database characteristics, contributes to better performance thus making graph database more appetible to businesses.[48]

**Scalability**

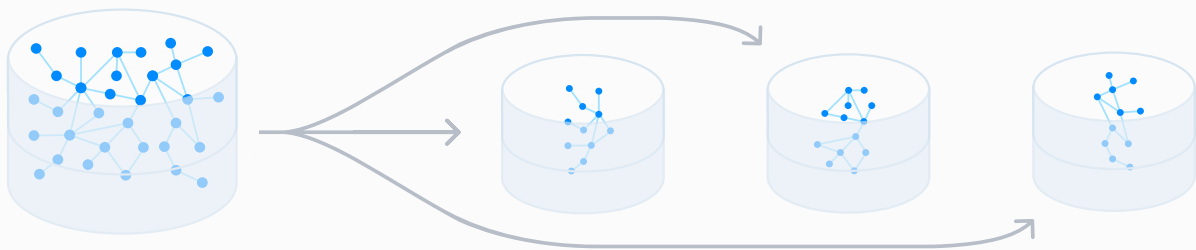

Some graph databases provide horizontal scalability at the data level, allowing large graphs to be saved as one across several devices. Horizontal scalability can help save on hardware, but in graphs comes with higher complexity of setup.

**Developer agility and schema flexibility**

New types of relationships, vertices, labels and, in general, subgraphs can be added to an existing structure without disrupting existing searches and application functionality. These factors may have beneficial effects on developer productivity and project maintenance. Because of the graph model's flexibility, it is not necessary to represent the domain in great detail ahead of time in order to limit future modifications, both in terms of structure (connection types, labels) and vertex and relationship attributes. Furthermore, the graph data model's schema-free nature eliminates the need for the relational world's schema-oriented data governance operations.[45]

However, this does not imply that schema flexibility is only beneficial; it may also bring problems later in the database life cycle.

**Intuitive data model**

Both users and developers are comfortable with and intuitive with graphs, and a well-designed query language or API saves time in both data model definition, development and maintenance.

### 2.2.3.2. Differences from graph computing engines

Graph databases and graph computing engines are the two major types of graph data systems.
- Graph databases, as described so far, are technologies that are largely used for transactional online graph persistence and are often accessed directly in real time from applications. They are the equivalent of the OLTP databases of the relational world.

- Graph computing engines (also known as graph analytics engines) are technologies that are largely used for offline graph analytics, which is often done in batches. These fall under the same category as other technologies for large-scale data analysis, such as data mining and online analytical processing (OLAP). Graph computing engines typically operate on a global graph level and make use of numerous processors to complete batch tasks.[45,46]

Graph computing engines are used to do tasks such as data clustering, like community detection or to answer questions like "how many relationships does anyone in a social network have on average?" While some graph analytics engines incorporate a graph storage layer, others (and likely the majority) focus solely on processing data from an external source and then deliver the results to store elsewhere.[45]

In some circumstances, graph databases are not the best tool for graph analytics tasks. The majority of graph processing use random data access. Random disk access becomes a performance bottleneck for huge graphs that cannot be stored in memory. Even for smaller graphs that might be entirely kept in memory, graph databases lack the computational capability of distributed and parallel systems since they are centralized systems that must ensure ACID features. When some computations must be performed on the complete graph in a single-thread technique, computation times quickly become prohibitive. The implementation of batch processes that conduct several queries with graph-local scope on the database and then do various computations on the intermediate results in order to obtain the graph global result might be one method for computing graph analytics using graph databases.[48]

The following are some products that perform distributed computation operations:

- **Apache Hadoop** is an open source distributed-processing framework for large data sets. It includes a `MapReduce` implementation. Commodity computer clusters can be programmed to execute large-scale data processing in a single pass using Hadoop and `MapReduce`. `MapReduce` is designed for analytics on big data volumes distributed over hundreds of machines.[77,78]

- **Pregel** is a scalable framework for implementing graph algorithms that was introduced by Google in 2010. Graphs, in particular, are intrinsically recursive data structures because vertices' properties are dependent on the qualities of their neighbors, who are dependent on the properties of their neighbors. As a result, many important Pregel graph algorithms recompute the properties of each vertex iteratively until a fixed-point condition is reached.[45,79]

- **GraphX** is an Apache Spark-based distributed graph processing framework. It provides APIs for implementing massively parallel algorithms, like PageRank. GraphX fully supports property graphs and is part of Apache Spark.[80]

### 2.2.3.3. Differences from relational databases

Relational databases were created with the intent of encoding tabular structures, and they excel at it. They may, however, struggle to simulate the web of relationships that emerge in various domains of the real world. One of the strongest arguments for graph databases is:

[77] Dean and Ghemawat (2008)

Jeffrey Dean and Sanjay Ghemawat. *MapReduce: Simplified Data Processing on Large Clusters*. In: Commun. ACM 51.1 (Jan. 2008). https://research.google.com/archive/mapreduce-osdi04.pdf, pages 107–113. ISSN: 0001-0782. DOI: 10.1145/1327452.1327492. URL: https://static.googleusercontent.com/media/research.google.com/en//archive/mapreduce-osdi04.pdf.

[78] Sakr (2013)

Sherif Sakr. *Processing large-scale graph data: A guide to current technology*. Online. June 2013. URL: https://developer.ibm.com/articles/os-giraph/.

[79] Apache Spark (2021)

Apache Spark. *Spark - GraphX Programming Guide, Pregel API*. Online. Documentation. Aug. 2021. URL: https://spark.apache.org/docs/latest/graphx-programming-guide.html#pregel-api.

[80] Wikimedia Foundation, Inc. (2021)

Wikimedia Foundation, Inc. *Wikipedia - Apache Spark*. Online. Aug. 2021. URL: https://en.wikipedia.org/wiki/Apache_Spark.

*"Relationships are first-class citizens of the graph data model"*

Graph databases do, in fact, value both vertices and relationships equally, storing them with dedicated constructs. Relationships are the fundamental descriptors for the web of connections among the stored elements and they are the primary information carriers for any programs that need to operate with graph data.d to work with graph data.[48] This is not the case in other database management systems, particularly relational ones: there are no dedicated constructs for treating relationship information differently from other forms of information and connections between entities must be inferred by performing join operations on values of fields with referential integrity constraints (like foreign keys); or by using out-of-band data processing such as `MapReduce`.[45]

The relational paradigm is "value-based" by definition: relationships between two separate relations are expressed by the values that exist within both relations' tuples. Graph databases, on the other hand, do not need to check and link entities by comparing the values of a given property in both of them; they already have the linking information stored and all of the operations that relational databases and document databases perform at query time are completely avoided. The database implementation determines how relationships are kept and linked to the vertices. When considering those that use index-free adjacency, the basic idea is that the node "record" stores a "pointer" (either logical or physical) to the linked nodes; or it stores a pointer to some relationship object that stores the information of the other vertex involved, as well as any other possible attribute. By directly storing relationships, the database relieves the system of those operations that can become burdensome such as `JOINS`, repeated index searches, temporary tables and so on, especially when dealing with huge amounts of data.[48]

Relationships are not just valued at the storage layer; they are also treated with full importance at the conceptual layer. This fact brings up another crucial feature of graph databases: the dedicated and simpler handling of relationships in any operations that the users or the programs want to execute on the underlying database. Making the relationship manageable as a standalone object, rather than being indirectly identified as a secondary object by the two vertexes connected by that edge - simplifies its manipulation in terms of:

- property updates,
- edge removal or construction
- and querying.[48]

Another compelling argument for graph databases is:

*"The true value of the graph approach becomes evident when one performs searches that are more than one level deep."*

Nested JOINs are the most common mean by which SQL performs hierarchical queries. When querying "two levels deep" in a recursive relationship, the query consists of a `JOIN` inside the `FROM` clause of an outer `SELECT` clause where another `JOIN` is executed. This means that for each desired depth level, would be needed a query consisting of many nested `SELECT` + `JOIN` statements as the required depth levels to be traversed. The more levels deep the querying into the network, the more complex and expensive computations become. Though the question "who are my friends-of-friends-of-friends?" can be answered in a reasonable amount of time, queries with four, five, or six degrees of friendship degrade dramatically due to the computational and space difficulty of recursively linking tables.[45]

Most relational databases provide a way to overcome situations of execution of nested `JOIN`s on sub-queries by using a special type of CTE (Common Table Expression) that is invoked by the WITH RECURSIVE clause.[48]

A last distinction between the relational and the graph model is that the separation between schema and data (instances) in graph databases is less pronounced than in relational databases. The translation between the "conceptual schema" and the "logical schema," which is part of the usual database development process, is typically imperceptible in graph databases. In contrast, the transformation of an ER schema to a relational "logical schema" is less straightforward. This is due to the normalization processes that must be applied to the ER schema in order to map it to a relational "logical schema". Furthermore, such normalization stages frequently tend to push the developer move away from the reality aimed to represent. Also, often this stage may cause a reduction of performance. This is also why denormalization steps are sometimes implemented afterwards.[46,48]

### 2.2.3.4. Differences from document stores

When dealing with graph data, relational databases are not the only ones to struggle. Most NoSQL databases indeed store sets of disconnected documents, values or columns. This makes it difficult to use them for connected data and graphs. In document stores, the simplest way for defining relationships between two document records is by embedding in the outer document, the document which is to be linked. Hierarchical situations can be represented using this approach; However it is immediate to see that if more documents are linked to the same document record, then such record would be replicated for both the outer documents; This would be a problem with many bad side effects.[48]

Another well-known method for adding relationships in such stores is to embed a record identifier in the other record rather than entirely embedding it. This way is resolved the issue of duplication of records and foreign keys may be applied. However, since with this second approach relationships are built from these flat, disconnected data structures, this approach presents the necessity to join records at the application level - which becomes quickly expensive. In addition, to ensure that the application updates or deletes these foreign record references in the same way that the rest of the data is updated or deleted, certain consistency methods must be provided. If this were not to happen, the store would accumulate dangling references which lower query performance and data quality.[45]

Another flaw in this method is that there are no reverse directed, backward pointing identifiers; the foreign "links" are not reflexive - therefore it is not possible to perform queries in the opposite direction.[48]

### 2.2.4. Graph Query Languages

When working with graph databases, the querying is though differently from the traditional relational database style queries. Instead of reasoning with tables, columns, joins and so on like in "SQL form", the main focus here are nodes and traversing the graph by walking the right edges.[45]

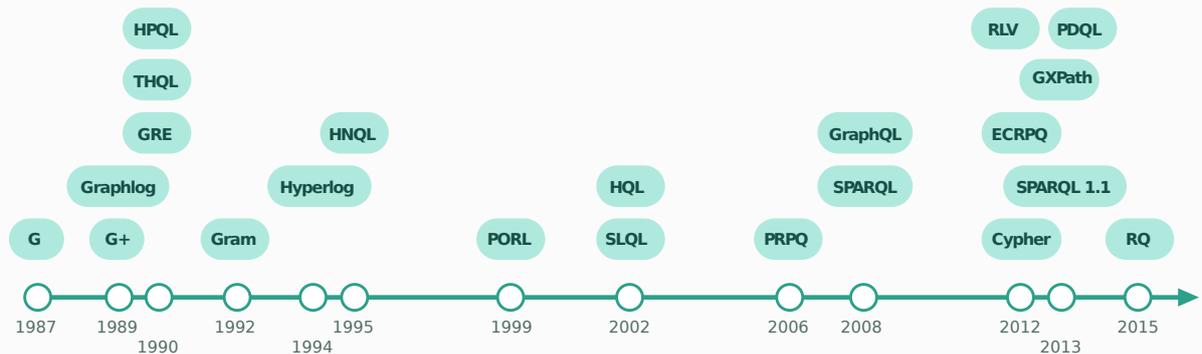

≡ Figure 2.14.: Evolution of graph query languages[46]

Graph queries usually are made of two parts:

1. the landing phase,
2. the exploration phase.

During the landing phase, generally is searched for the start vertex on the graph. This usually makes use of indices, labels or classes and properties for fast node look-up. The start node represents the starting point for the exploration phase that shall be performed right-after. There are situations in which during the landing phase more than one start node is searched. The same is valid for edges. On large graphs, the landing phase represents the most expensive operation. For queries with complex exploration phase, the landing phase could represent a fraction of the total query time. Following the landing phase is the exploration phase of the querying; During this phase in a certain sense happens the actual querying, the graph traversal - walking. The exploration phase makes use of a variety of graph theory navigation algorithms. The separation of landing and exploratory phase is noticeable also in the query formulation of many graph query languages. While syntactically this may not be the case in a few languages, their query execution nonetheless follows the same philosophy.[48]

In the following subsubsections a couple of Graph Query Languages are briefly described.

### 2.2.4.1. Cypher

Cypher[81] is a graph query language that is used to conduct graph traversals and Create Retrieve Update Delete (CRUD) operations on the Neo4j graph database. Cypher is declarative in nature and very expressive in terms of visual and logical representation of graph query patterns and relationship dependencies.[82]

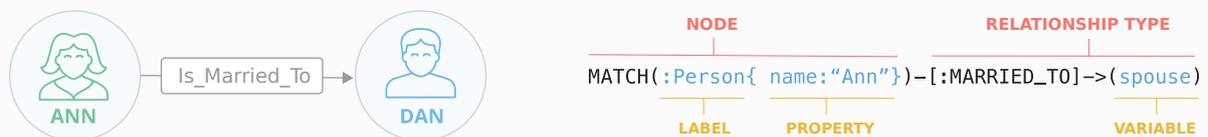

Figure 2.15.: Cypher syntax[81]

### 2.2.4.2. Gremlin

Gremlin[83,84] is a graph query language that is supported by Apache TinkerPop, a graph framework for different graph databases. Gremlin is based on Groovy, an OOP language for Java Platform. It supports both imperative and declarative queries and both Online Transactional Processing (OLTP) and Online Transactional Processing (OLTP).[82]

A Gremlin query example:

```
gremlin> g.V().has('Person', 'name', 'Ann')
gremlin> valueMap()with(WithOptions.tokens).unfold()
```

### 2.2.4.3. ArangoDB Query Language (AQL)

Arango Query Language (AQL) is used to carry out data modification operations in ArangoDB as well as graph traversals. It is a declarative language, written in Java and C. Because AQL is used to retrieve all sorts of datasets stored in ArangoDB, it is not limited to graph traversal queries. AQL supports the complex query patterns and the different data model (documents, graphs, and key-values). but does not support creating and dropping data entities. Furthermore, AQL also supports different traversal functionalities such as searching path using inbound and out-

[81] Neo4j, Inc. (2021)

Neo4j, Inc. *Neo4j Documentation - Cypher Query Language Manual*. Online. Documentation. Aug. 2021. URL: https://neo4j.com/docs/cypher-manual/current/.

[82] SmartM2M Technical Committee, Liquori and Guillemin (2020)

SmartM2M Technical Committee, Luigi Liquori and Patrick Guillemin. *SmartM2M (Smart Machine-to-Machine Communications) TR 103 715, Study for oneM2M, Discovery and Query solutions analysis & selection*. Technical report. V1.1.1. France: ETSI, Nov. 2020. 88 pages. DOI: 10.5445/IR/1000097104. URL: https://www.etsi.org/deliver/etsi_tr/103700_103799/103715/01.01.01_60/tr_103715v010101p.pdf.

[83] Apache TinkerPop (2021)

Apache TinkerPop. *Apache TinkerPop - The Gremlin Graph Traversal Machine and Language*. Online. Documentation. Aug. 2021. URL: https://tinkerpop.apache.org/gremlin.html.

[84] Chan and Neubauer (2013)

Harold Chan and Peter Neubauer. *Should I learn Cypher or Gremlin for operating a Neo4j database?* Online. May 2013. URL: https://www.quora.com/Should-I-learn-Cypher-or-Gremlin-for-operating-a-Neo4j-database.

bound relationships, result filtrations, shortest path traversals, community detection and many more Pregel custom algorithms.[82]

An AQL query example:

```
FOR p IN Person
    FILTER p.name == 'Ann'
    RETURN p
```

### 2.2.4.4. Query Language summary

In Table 2.3[82] is brought a summary of the features of the Graph Query Languages presented above.

| | Cypher | Gremlin | AQL |
|---|---|---|---|
| **Query Language Type** | Declarative | Declarative/Imperative | Declarative/data manipulation |
| **Effort (Query formation)** | Low | Medium | Low |
| **Readability** | Medium | Low | Medium |
| **Expressiveness** | High | Medium | High |
| **Access** | Embedded, WebSocket, REST | Embedded, Websockets, HTTP | REST & GraphQL |
| **OLTP** | Yes | Yes | Yes |
| **OLAP** | Yes | Yes | No |
| **Shortest Path Search** | Yes | Yes | Yes |

☰ Table 2.3.: Feature-wise comparison of selected Graph Query Languages[82]

## 2.2.5. GDBMSs comparison

It is possible to find in literature many benchmarking studies carried out in the last years, easily accessible online. However, it should be noted that the validity of the comparisons of Graph DBMSs is very short since it is a highly in-development field. A feature of a GDBMS that is not be available in a certain version of the software, might already be implemented in a successive version. Or viceversa, a feature available in a certain version of the software, might be deprecated in later versions. Since the changes are rather dynamic, the comparisons are to be taken with a grain of salt.

Furthermore, some of the benchmarking works are made from GDBMS product companies' insiders, or in certain ways related. Because of this there might also be some bais, cherrypicking in those works.

Below is reported a list of GDBMSs benchmarking works:

- "Demystifying Graph Databases: Analysis and Taxonomy of Data Organization, System Designs, and Graph Queries - Towards Understanding Modern Graph Processing, Storage, and Analytics" - Besta, Peter, Gerstenberger, Fischer, Podstawski, Barthels, Alonso and Hoefler (2019)[14]
- "XGDBench: A benchmarking platform for graph stores in exascale clouds" - Dayarathna and Suzumura (2012)[85]

[85] Dayarathna and Suzumura (2012)

Miyuru Dayarathna and Toyotaro Suzumura. *XGDBench: A benchmarking platform for graph stores in exascale clouds*. In: 4th IEEE International Conference on Cloud Computing Technology and Science Proceedings. Taipei, Taiwan, Dec. 2012, pages 363–370. DOI: 10.1109/CloudCom. 2012.6427516.

- "Benchmarking Traversal Operations over Graph Databases" - Ciglan, Averbuch and Hluchy (2012)[86]
- "Benchmarking Graph Databases on the Problem of Community Detection" - Beis, Papadopoulos and Kompatsiaris (2015)[34]
- "An Empirical Comparison of Graph Databases" - Jouili and Vansteenberghe (2013)[87]
- *Performance comparison between ArangoDB, MongoDB, Neo4j and OrientDB* - ArangoDB and Weinberger (2015)[88]
- *Rdf Store Benchmarking* - W3C (2018)[89]
- *Large Triple Stores* - W3C (2021)[90]
- Graph vs. Relational comparison: *Benchmark: PostgreSQL, MongoDB, Neo4j, OrientDB and ArangoDB* - ArangoDB and Weinberger (2015)[91]

In the next subsubsection, an in-detail view of major Graph DBMSs is given.

### 2.2.5.1. Major Graph DBMSs

Since by now the number of GDBMS products available in the market is of more than 30, doing a hand-by-hand manual analysis of each product continuously each month is not feasible. A list of the most known DBMSs is ranked monthly by popularity on the website `https://db-engines.com`[47]. The different products are categorized and there is a category for Graph Database Systems too.

At the beginning of § 2.2 was presented a chart (see Figure 2.9) from `https://db-engines.com`[47] showing the trends of popularity growth between the different categories of DBMSs. While the growth is staggering, Graph DBMSs have still a long way to do till they catch up the relational DBMSs. In Table 2.4 this is evident.

`https://db-engines.com`[47] calculates the popularity score of a system by standardizing and averaging different parameters. The score for each DBMS is obtained using the following parameters[48]:

- The number of times the system has been mentioned on the web, as measured by the number of times it has come up in search engine queries.
- Frequency of searches in Google Trends.
- Frequency of technical discussions about the system, measured as the number of questions on Stack Overflow and DBA Stack Exchange.
- Number of job offers, in which the system is mentioned, on the leading job search engines.
- Number of LinkedIn and Upwork profiles in which the system is mentioned.
- Number of Twitter tweets in which the system is mentioned.

[86] CIGLAN, AVERBUCH and HLUCHY (2012)
Marek Ciglan, Alex Averbuch and Ladialav Hluchy. *Benchmarking Traversal Operations over Graph Databases*. In: 2012 IEEE 28th International Conference on Data Engineering Workshops. `http://ups.savba.sk/~marek/papers/gdm12-ciglan.pdf`. 2012, pages 186–189. DOI: `10.1109/ICDEW.2012.47`. URL: `https://ieeexplore.ieee.org/document/6313678`.
[87] JOUILI and VANSTEENBERGHE (2013)
Salim Jouili and Valentin Vansteenberghe. *An Empirical Comparison of Graph Databases*. In: 2013 International Conference on Social Computing. Sept. 2013, pages 708–715. DOI: `10.1109/SocialCom.2013.106`. URL: `https://ieeexplore.ieee.org/document/6693403`.
[88] ARANGODB and WEINBERGER (2015)
ArangoDB and Claudius Weinberger. *Performance comparison between ArangoDB, MongoDB, Neo4j and OrientDB*. Online. Blog post. Discussion: `https://news.ycombinator.com/item?id=9699102`. June 2015. URL: `https://www.arangodb.com/2015/06/performance-comparison-between-arangodb-mongodb-neo4j-and-orientdb/`.
[89] W3C (2018)
The World Wide Web Consortium W3C. *Rdf Store Benchmarking*. Online. Oct. 2018. URL: `https://www.w3.org/wiki/RdfStoreBenchmarking`.
[90] W3C (2021)
The World Wide Web Consortium W3C. *Large Triple Stores*. Online. Apr. 2021. URL: `https://www.w3.org/wiki/LargeTripleStores`.
[91] ARANGODB and WEINBERGER (2015)
ArangoDB and Claudius Weinberger. *Benchmark: PostgreSQL, MongoDB, Neo4j, OrientDB and ArangoDB*. Online. Blog post. Oct. 2015. URL: `https://www.arangodb.com/2015/06/performance-comparison-between-arangodb-mongodb-neo4j-and-orientdb/`.

To be noted however, the disclaimer on `https://db-engines.com`:

*"Ranking does not measure the number of installations of the systems, or their use within IT systems. It can be expected, that an increase in the popularity of a system as measured by the* `https://db-engines.com` *Ranking (e.g. in discussions or job offers) precedes a corresponding broad use of the system by a certain time factor. Because of this, the* `https://db-engines.com` *Ranking can act as an early indicator"[47]*

| Rank Sep 2021 | DBMS | Database Model | Score Sep 2021 |
|---|---|---|---|
| 1. | Oracle | Relational, Multi-model | 1271.55 |
| 2. | MySQL | Relational, Multi-model | 1212.52 |
| 3. | Microsoft SQL Server | Relational, Multi-model | 970.85 |
| 4. | PostgreSQL | Relational, Multi-model | 577.50 |
| 5. | MongoDB | Document, Multi-model | 496.50 |
| 6. | Redis | Key-value, Multi-model | 171.94 |
| 7. | IBM Db2 | Relational, Multi-model | 166.56 |
| 8. | Elasticsearch | Search engine, Multi-model | 160.24 |
| 9. | SQLite | Relational | 128.65 |
| 10. | Cassandra | Wide column | 118.99 |
| | . . . | | |
| 17. | Teradata | Relational, Multi-model | 69.68 |
| 18. | Neo4j | Graph | 57.63 |
| 19. | SAP HANA | Relational, Multi-model | 56.24 |

| Rank Sep 2021 | DBMS | Database Model | Score Sep 2021 |
|---|---|---|---|
| 25. | Google BigQuery | Relational | 43.92 |
| 26. | Microsoft Azure Cosmos DB | Multi-model | 38.52 |
| 27. | PostGIS | Spatial DBMS, Multi-model | 30,71 |
| | . . . | | |
| 74. | Ignite | Multi-model | 4.85 |
| 75. | ArangoDB | Multi-model | 4.79 |
| 76. | MaxDB | Relational | 4.72 |
| 77. | Derby | Relational | 4.69 |
| 78. | Adabas | Multivalue | 4.63 |
| 79. | Virtuoso | Multi-model | 4.42 |
| 80. | Google Cloud Datastore | Document | 4.38 |
| 81. | UniData,UniVerse | Multivalue | 4.27 |
| 82. | OrientDB | Multi-model | 4.24 |
| 83. | Oracle NoSQL | Multi-model | 4.23 |

Table 2.4.: Rank by popularity of GDBMSs in top 100 DBMSs[47]

In Figure 2.16 is displayed a chart of the trends of selected Graph DBMSs only. Note the score axis in logarithmic scale. At the moment Neo4j is ranked as the most popular Graph Database Management System. It is followed by Microsoft Azure Cosmos Database and ArangoDB.

Whereas in Figure 2.17 are reported by Besta, Peter, Gerstenberger, Fischer, Podstawski, Barthels, Alonso and Hoefler (2019)[14] the various types of graph database systems with an indication of the data model, way of storing vertices and edges, a drawing of the data structures and examples of which GDBMS belongs to that category.

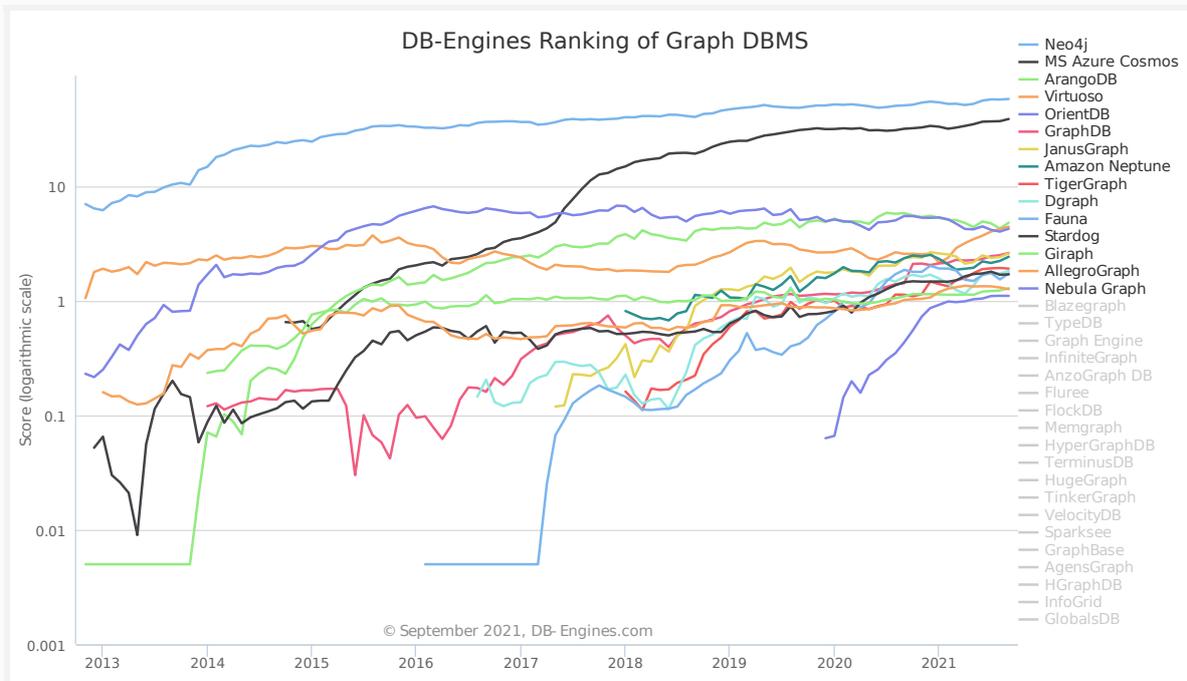

Figure 2.16.: Trend of Graph DBMS Popularity[47]

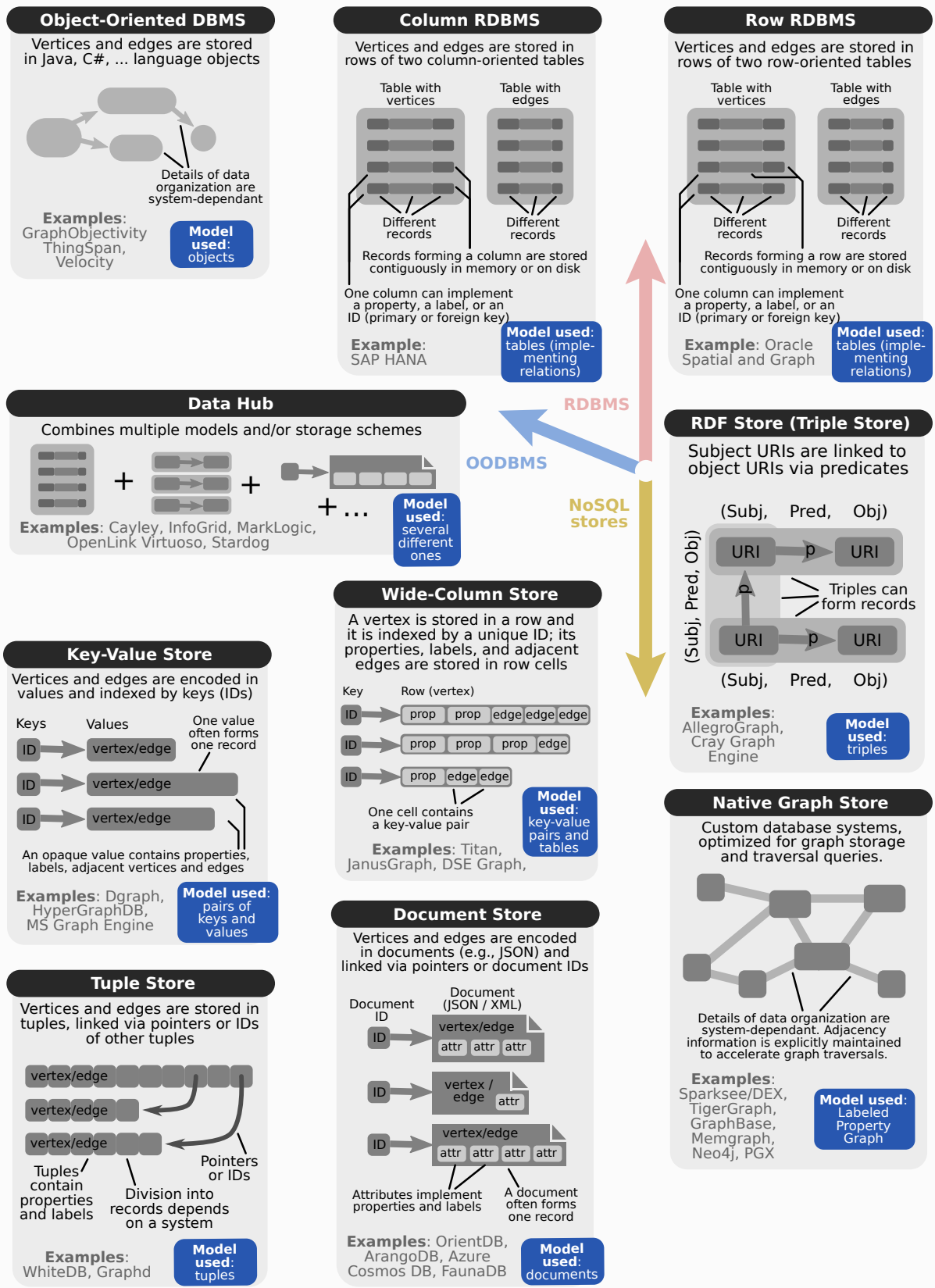

Figure 2.17.: Overview of different categories of graph database systems, with examples[14]

In the next subsubsection a comparison of the major GDBMSs summarized in two tables is presented.

### 2.2.5.2. Comparison tables

In Table 2.5 is brought an exhaustive comparison of all known graph database systems with indications on data models, storing mechanisms, operation mode, kind of transactions and other remarks. The various GDBMSs are divided in categories according to the data model used. To consult the original work, see "Demystifying Graph Databases: Analysis and Taxonomy of Data Organization, System Designs, and Graph Queries - Towards Understanding Modern Graph Processing, Storage, and Analytics" - Besta, Peter, Gerstenberger, Fischer, Podstawski, Barthels, Alonso and Hoefler (2019)[14].

| Graph Database System | Model | | | Records | | Storing Edges | | | | | Data Distribution & Query Execution | | | | | | | | Additional remarks |
|---|---|---|---|---|---|---|---|---|---|---|---|---|---|---|---|---|---|---|---|
| | MM | LPG | RDF | FS | VS | DP | AL | SE | SV | LW | MN | RP | SH | CE | PE | TR | OLTP | OLAP | |

**❶ RDF STORES (TRIPLE STORES)** — The main data model used: **RDF triples**

**BlazeGraph** — *BlazeGraph uses RDF*, an extension of RDF

**Cray Graph Engine** — *RDF triples are stored in hashtables.

AllegroGraph — *Triples are stored as integers (RDF strings are mapped to integers).

Amazon Neptune — *LPG is enabled via Gremlin

AnzoGraph — *Data model and schema support SPARQL.

Apache Marmotta — *The structure of data records is based on that of different RDBMS systems (H2, PostgreSQL, MySQL).

BrightstarDB — —

Ontotext GraphDB — —

Profium Sense — Profium Sense is "AI powered". *The format used is called JSON-LD: JSON for vertices and RDF for edge.

TripleBit — The data organization uses compression. *Strings are mapped to variable size integers. ‡Described as future work.

**❷ TUPLE STORES** — The main data model used: **tuples**

**WhiteDB** — *Implicit support for triples of integers. ‡Implementable by the user. †Transactions use a global shared/exclusive lock.

Graphd — Backend of Google Freebase. *Implicit support for triples. ‡Subset of ACID.

**❸ DOCUMENT STORES** — The main data model used: **documents**

**ArangoDB** — *Uses a hybrid index for retrieving edges.

**OrientDB** — *AL contains RIDs (i.e., physical locations) of edge and vertex records. ‡Sharding is defined by the user. OrientDB supports JSON and it offers certain object-oriented capabilities.

Azure Cosmos DB — —

Bitsy — The system is disk based and uses JSON files. The storage only allows for appending data.

FaunaDB — *Document, RDBMS, graph, and "time series". ‡Adjacency lists are separately precomputed.

**MM**: A system is **multi model**. **LPG**, **RDF**: A system supports, respectively, the **Labeled Property Graph** and **RDF** without prior data transformation. **FS**, **VS**: Data records are **fixed size** and **variable size**, respectively. **DP**: A system can use **direct pointers** to link records. This enables storing and traversing adjacency data without maintaining indices. **AL**: Edges are stored in the **adjacency list** format. **SE**: Edges can be **stored in a separate edge record**. **SV**: Edges can be **stored in a vertex record**. **LW**: Edges can be **lightweight** (containing just a vertex ID or a pointer, both stored in a vertex record). **MN**: A system can operate in a **Multi Server** (distributed) mode. **RP**: Given a distributed mode, a system enables **Replication** of datasets. **SH**: Given a distributed mode, a system enables **Sharding** of datasets. **CE**: Given a distributed mode, a system enables **Concurrent Execution** of multiple queries. **PE**: Given a distributed mode, a system enables **Parallel Execution** of single queries on multiple nodes/CPUs. **TR**: Support for **ACID Transactions**. **OLTP**: Support for **Online Transaction Processing**. **OLAP**: Support for **Online Analytical Processing**. 👍: A system offers a given feature. 👌: A system offers a given feature in a limited way. 👎: A system does not offer a given feature. ⓘ: Unknown.

☰ Table 2.5: Comparison of graph databases[14]. Continues on next page ...

... continued from previous page

| Graph Database System | Model | | | | Records | | | | | | Data Distribution & Query Execution | | | | | | | | Additional remarks |
|---|---|---|---|---|---|---|---|---|---|---|---|---|---|---|---|---|---|---|---|
| | MM | LPG | RDF | FS | VS | DP | AL | SE | SV | LW | MN | RP | SH | CE | PE | TR | OLTP | OLAP | |

**❶ KEY-VALUE STORES** — The main data model used: **key-value pairs**

| | | | | | | | | | | | | | | | | | | | |
|---|---|---|---|---|---|---|---|---|---|---|---|---|---|---|---|---|---|---|---|
| **HyperGraphDB** | 👎 | 👍* | 👎 | 👎 | 👍 | 👍 | 👍‡ | 👎 | 👎 | 👎 | 👍 | 👎 | 👎 | 👍 | 👍 | ✋† | 👍 | | *A Hypergraph model. ‡The system uses an incidence index to retrieve edges of a vertex. †Support for ACI only. |
| **MS Graph Engine** | 👍 | 👍 | 👎 | 👍 | 👍* | 👍 | 👍‡ | 👎 | 👍 | 👎 | 👍 | 👍 | 👍 | 👍 | 👍 | 👎 | 👍 | | *Schema is defined by Trinity Specification Language (TSL). ‡AL contains IDs of edges and/or vertices. |
| Dgraph | 👎 | 👍 | 👎 | 👎 | 👍 | 👎 | 👍 | 👎 | 👎 | 👎 | 👍 | 👍 | 👍 | 👍 | 👍 | 👍 | 👍 | | Dgraph is based on Badger. |
| RedisGraph | 👎 | 👍 | 👎 | 👎 | 👍 | 👎 | 👍* | 👎 | 👎 | 👍 | 👎 | 👍 | 👎 | 👎 | 👎 | 👎 | 👍‡ | | RedisGraph is based on Redis. ‡The OLAP part uses GraphBLAS. |

**❷ RELATIONAL DBMS (RDBMS)** — The main data model used: **tables (implementing relations)**.

| | | | | | | | | | | | | | | | | | | | |
|---|---|---|---|---|---|---|---|---|---|---|---|---|---|---|---|---|---|---|---|
| **Oracle Spatial & Graph** | 👍* | 👍 | 👍 | ○* | ○* | 👎 | 👍 | 👍* | 👎 | 👎 | 👍 | 👍 | 👍 | 👍 | 👍 | 👍 | 👍 | 👍 | *LPG and RDF use row-oriented storage. The system can also run on top of PGX (effectively as a **native graph database**). |
| AgensGraph | 👍 | 👍 | 👍 | ○ | ○ | 👎 | 👍 | ○ | 👎 | 👎 | 👍 | 👍 | 👍 | 👍 | 👍 | 👍 | 👍 | | AgensGraph is based on PostgreSQL. |
| FlockDB | 👎 | 👎 | 👎 | ○ | ○ | 👎 | 👎 | ○ | 👎 | 👎 | 👍 | 👍 | 👍 | 👍 | 👍 | 👎 | 👍 | 👎 | FlockDB is based on MySQL. The system focuses on "shallow" graph queries, such as finding mutual friends. |
| MS SQL Server 2017 | 👍 | 👍 | 👎 | ○ | ○ | 👎 | 👎 | ○ | 👎 | 👎 | 👍 | 👍 | 👍 | 👍 | 👍 | 👍 | 👍 | | The system uses an SQL graph extension. |
| OQGRAPH | 👎 | 👎 | 👎 | ○* | ○* | 👎 | ○ | 👍* | 👎 | 👎 | 👍 | 👍 | 👍 | 👍 | 👍 | 👍 | | ○ | OQGRAPH uses MariaDB. *OQGRAPH uses row-oriented storage. |
| SAP HANA | 👎 | 👍 | 👎 | 👎 | 👍* | 👍* | 👎 | 👎 | 👍* | 👎 | 👍 | 👍 | 👍 | 👍 | 👍 | 👍 | 👍 | | *SAP HANA is column-oriented, edges and vertices are stored in rows. SAP HANA can be used with a dedicated graph engine and it offers certain capabilities of a JSON document store |

**❸ OBJECT-ORIENTED DATABASES (OODBMS)** — The main data model used: **objects**.

| | | | | | | | | | | | | | | | | | | | |
|---|---|---|---|---|---|---|---|---|---|---|---|---|---|---|---|---|---|---|---|
| **VelocityGraph** | 👍 | 👍 | 👎 | 👎 | 👍 | 👍 | 👍 | 👍 | 👎 | 👎 | 👍 | 👍 | 👍 | 👍 | 👍 | 👍 | 👍 | | The system is based on VelocityDB. |
| Objectivity ThingSpan | 👎 | 👍 | 👎 | ○ | ○ | ○ | ○ | 👍 | ○ | 👎 | 👍 | 👍 | 👍 | 👍 | 👍 | 👍 | 👍 | | The system is based on ObjectivityDB. |

**❹ NATIVE GRAPH DATABASES** — The main data model used: **LPG**.

| | | | | | | | | | | | | | | | | | | | |
|---|---|---|---|---|---|---|---|---|---|---|---|---|---|---|---|---|---|---|---|
| **Neo4j** | 👎 | 👍 | 👎 | 👎 | 👍 | 👍 | 👎 | 👎 | 👎 | 👍 | 👍 | 👍 | 👎 | 👍 | 👍 | 👍 | 👍 | | Neo4j is provided as a cloud service by a system called Graph Story. |
| GBase | 👎 | 👍* | 👎 | 👎 | 👍 | 👎 | 👍 | 👍‡ | 👎 | 👎 | 👍 | ○ | ○ | ○ | ○ | ○ | 👎 | | *GBase supports simple graphs only. ‡GBase stores the AM sparsely. |
| **Sparksee/DEX** | 👎 | 👍 | 👎 | 👎 | 👍* | 👍* | 👍* | 👎 | 👎 | 👍‡ | 👍 | 👍 | 👎 | 👍 | 👎 | 👍 | 👍 | | *The system uses maps only. ‡Bitmaps are used for connectivity. |
| GraphBase | 👎 | 👍* | 👎 | 👎 | 👍 | 👍 | ○ | ○ | ○ | ○ | 👍 | 👍 | ○ | 👍 | ○ | ○ | 👍 | ○ | *No support for edge properties, only two types of edges available. |
| Memgraph | 👎 | 👍 | 👎 | 👎 | ○ | ○ | ○ | 👍 | ○ | 👍 | 👍 | 👍 | ✋* | ✋‡ | 👍 | 👍 | 👍 | | *This feature is under development. ‡Available only for some algorithms. |
| TigerGraph | 👎 | 👍 | 👎 | 👎 | ○ | ○ | ○ | ○ | ○ | 👍 | 👍 | 👍 | 👍 | 👍 | 👍 | 👍 | 👍 | | — |
| Weaver | 👎 | 👍 | 👎 | 👎 | ○ | ○ | ○ | ○ | ○ | 👍 | 👍 | 👍 | 👍 | 👍 | 👍 | 👍 | 👍 | | — |

<u>MM</u>: A system is **multi model**. <u>LPG</u>, <u>RDF</u>: A system supports, respectively, the **Labeled Property Graph** and **RDF** without prior data transformation. <u>FS</u>, <u>VS</u>: Data records are **fixed size** and **variable size**, respectively. <u>DP</u>: A system can use **direct pointers** to link records. This enables storing and traversing adjacency data without maintaining indices. <u>AL</u>: Edges are stored in the **adjacency list** format. <u>SE</u>: Edges can be **stored in a separate edge record**. <u>SV</u>: Edges can be **stored in a vertex record**. <u>LW</u>: Edges can be **lightweight** (containing just a vertex ID or a pointer, both stored in a vertex record). <u>MN</u>: A system can operate in a **Multi Server** (distributed) mode. <u>RP</u>: Given a distributed mode, a system enables **Replication** of datasets. <u>SH</u>: Given a distributed mode, a system enables **Sharding** of datasets. <u>CE</u>: Given a distributed mode, a system enables **Concurrent Execution** of multiple queries. <u>PE</u>: Given a distributed mode, a system enables **Parallel Execution** of single queries on multiple nodes/CPUs. <u>TR</u>: Support for **ACID Transactions**. <u>OLTP</u>: Support for **Online Transaction Processing**. <u>OLAP</u>: Support for **Online Analytical Processing**. 👍: A system offers a given feature. ✋: A system offers a given feature in a limited way. 👎: A system does not offer a given feature. ○: Unknown.

⊞ Table 2.5: Comparison of graph databases[14]. Continues on next page ...

... continued from previous page

| Graph Database System | Model | | | | Records | | | | | | Data Distribution & Query Execution | | | | | | | | Additional remarks |
|---|---|---|---|---|---|---|---|---|---|---|---|---|---|---|---|---|---|---|---|
| | MM | LPG | RDF | FS | VS | DP | AL | SE | SV | LW | MN | RP | SH | CE | PE | TR | OLTP | OLAP | |

| **❾ WIDE-COLUMN STORES** | | | | | | | | | | | | | | | | | | | The main data model used: **key-value pairs** and **tables**. |
| **JanusGraph** | 👎 | 👍 | 👎 | 👍 | 👍 | 👎 | 👍 | 👎 | 👍 | 👍 | 👍 | 👍 | 👍 | 👍 | 👍 | 👍 | 👍 | 👍 | JanusGraph is the continuation of Titan. |
| **Titan** | 👍 | 👍 | 👎 | 👍 | 👍 | 👎 | 👍 | 👎 | 👍 | 👍 | 👍 | 👍 | 👍 | 👍 | 👍 | 👍 | 👍 | 👍 | Titan enables various backends (e.g., Cassandra). |
| DSE Graph (DataStax) | 👎 | 👍* | 👎 | 👍 | 👍 | 👎 | 👍 | 👍 | 👎 | 👎 | 👍 | 👍 | 👍 | ? | ? | ◐‡ | 👍 | 👍 | DSE Graph is based on Cassandra. *Enabled by Gremlin query language. ‡Support for AID, Consistency is configurable. |
| HGraphDB | 👎 | 👍 | 👎 | 👍 | 👎 | 👎 | 👍 | 👎 | 👍 | 👍 | 👍 | ? | ? | ? | 👍* | 👍 | 👍 | | HGraphDB uses TinkerPop3 with HBase. *ACID is supported only within a row. |

| **❿ DATA HUBS** | | | | | | | | | | | | | | | | | | | The main data model used: **several different ones**. |
| **MarkLogic** | 👍 | 👎* | 👍 | ? | 👍 | 👍 | 👎 | 👍 | 👍* | 👎 | 👍 | 👍 | 👍 | 👍 | ? | 👍 | 👍 | 👍 | Supported storage/models: relational tables, RDF, various documents. *Vertices are stored as documents, edges are stored as RDF triples. |
| **OpenLink Virtuoso** | 👍 | 👎* | 👍 | ? | ? | 👎 | 👍 | 👎 | 👍 | 👎 | 👍 | 👍 | 👍 | ◐‡ | ? | 👍 | 👍 | 👍 | Supported storage/models: relational tables and RDF triples. ‡This feature can be used with relational data only. |
| Cayley | 👍 | 👍 | 👍 | ? | ? | ? | 👎 | ? | ? | ? | 👍 | 👍 | 👎 | 👍 | 👍* | ? | 👍 | ? | Supported storage/models: relational tables, RDF, document, key-value. *This feature depends on the backend. |
| InfoGrid | 👍 | 👍 | 👎 | ? | 👍 | ? | 👍 | 👎 | 👍 | 👍 | 👍 | ? | ? | ? | ? | ◐* | ? | ? | Supported storage/models: relational tables, Hadoop's filesystem, grid storage. *A weaker consistency model is used instead of ACID. |
| Stardog | 👍 | 👎* | ◐* | ? | 👎 | 👎 | 👍* | 👎 | 👍 | 👎 | 👍 | 👎 | ? | 👍 | 👍 | | 👍 | | Supported storage/models: relational tables, documents. *RDF is simulated on relational tables. |

__MM__: A system is **multi model**. __LPG__, **RDF**: A system supports, respectively, the **Labeled Property Graph** and **RDF** without prior data transformation. __FS__, __VS__: Data records are **fixed size** and **variable size**, respectively. __DP__: A system can use **direct pointers** to link records. This enables storing and traversing adjacency data without maintaining indices. __AL__: Edges are stored in the **adjacency list** format. __SE__: Edges can be **stored in a separate edge record**. __SV__: Edges can be **stored in a vertex record**. __LW__: Edges can be **lightweight** (containing just a vertex ID or a pointer, both stored in a vertex record). __MN__: A system can operate in a **Multi Server** (distributed) mode. __RP__: Given a distributed mode, a system enables **Replication** of datasets. __SH__: Given a distributed mode, a system enables **Sharding** of datasets. __CE__: Given a distributed mode, a system enables **Concurrent Execution** of multiple queries. __PE__: Given a distributed mode, a system enables **Parallel Execution** of single queries on multiple nodes/CPUs. __TR__: Support for **ACID Transactions**. __OLTP__: Support for **Online Transaction Processing**. __OLAP__: Support for **Online Analytical Processing**. 👍: A system offers a given feature. ◐: A system offers a given feature in a limited way. 👎: A system does not offer a given feature. ?: Unknown.

Table 2.5.: Comparison of graph databases[14]

In the next table is presented the current state of support for different graph QLs in the various GDBMSs.

| Graph Database System | Graph database query language | | | | | | Other languages and additional remarks |
|---|---|---|---|---|---|---|---|
| | SPARQL | Gremlin | Cypher | SQL | GraphQL | Progr. API | – |
| **⓫ TUPLE STORES** | | | | | | | |
| Graphd | – | – | – | – | – | – | Graphd uses MQL. |
| WhiteDB | – | – | – | – | – | 👍 (C, Python) | |

"**Progr. API**" determines whether a system supports formulating queries using some native prog. language such as C++. "👍": A system supports a given language. "◐": A system supports a given language in a limited way. "–": A system does not support a given language.

Table 2.6: Support for different graph query languages in different GDBMSs[14]. Continues on next page ...

... continued from previous page

| Graph Database System | Graph database query language | | | | | | Other languages and additional remarks |
|---|---|---|---|---|---|---|---|
| | SPARQL | Gremlin | Cypher | SQL | GraphQL | Progr. API | – |
| **❶ RDF STORES (TRIPLE STORES)** | | | | | | | |
| AllegroGraph | 👍 | – | – | – | – | – | – |
| Amazon Neptune | – | 👍 | – | – | – | – | – |
| AnzoGraph | – | – | 👍 | – | – | – | – |
| Apache Marmotta | 👍 | – | – | – | – | – | Apache Marmotta also supports its native LDP and LDPath languages. |
| BlazeGraph | 👍 | 👍 | – | – | – | – | – |
| BrightstarDB | 👍 | – | – | – | – | – | – |
| Cray Graph Engine | 👍 | – | – | – | – | – | – |
| Ontotext GraphDB | 👍 | – | – | – | – | – | – |
| Profium Sense | 👍 | – | – | – | – | – | – |
| TripleBit | 👍 | – | – | – | – | – | – |
| **❷ DOCUMENT STORES** | | | | | | | |
| OrientDB | 👍 | 👍 | 👍 | 👍* | – | 👍** | *An SQL extension for graph queries. **OrientDB offers bindings to Java, C, JS, PHP, .NET, Python, and others. |
| ArangoDB | – | 👍 | – | – | – | – | ArangoDB uses AQL (ArangoDB Query Language). |
| Azure Cosmos DB | – | 👍(limited) | – | 👍 | – | – | – |
| Bitsy | – | 👍 | – | – | – | – | Bitsy also supports other Tinkerpop-compatible languages such as SQL2Gremlin and Pixy. |
| FaunaDB | – | – | – | – | 👍 | – | – |
| **❸ KEY-VALUE STORES** | | | | | | | |
| MS Graph Engine | – | – | – | – | – | – | MS Graph Engine uses LINQ. |
| HyperGraphDB | – | – | – | – | – | 👍 (Java) | – |
| Dgraph | – | – | – | – | 👍* | – | *A variant of GraphQL. |
| RedisGraph | – | – | 👍 | – | – | – | – |
| **❹ WIDE-COLUMN STORES** | | | | | | | |
| Titan | – | 👍 | – | – | – | – | – |
| JanusGraph | – | 👍 | – | – | – | – | – |
| DSE Graph (DataStax) | – | 👍 | – | – | – | – | DSE Graph also supports CQL. |
| HGraphDB | – | 👍 | – | – | – | – | – |
| **❺ RELATIONAL DBMS (RDBMS)** | | | | | | | |
| Oracle Spatial and Graph | 👍 | – | – | 👍* | – | – | *PGQL [191], an SQL-like graph query language. |
| MS SQL Server 2017 | – | – | – | 👍* | – | – | *Transact-SQL. |
| SAP HANA | – | – | – | 👍* | – | 👍** | *SAP HANA offers bindings to Rust, ODBC, and others. **GraphScript, a domain-specific graph query language. |
| FlockDB | – | – | – | – | – | – | FlockDB uses the Gizzard framework and MySQL. |
| AgensGraph | – | – | 👍* | 👍** | – | – | *A variant called openCypher. **ANSI-SQL. |
| OQGRAPH | – | – | – | 👍 | – | – | – |
| **❻ OBJECT-ORIENTED DATABASES (OODBMS)** | | | | | | | |
| Objectivity ThingSpan | – | – | – | – | – | – | Objectivity ThingSpan uses a native DO query language. |
| Objectivity ThingSpan VelocityGraph | – | – | – | – | – | 👍 (.NET) | |
| **❼ NATIVE GRAPH DATABASES** | | | | | | | |
| Neo4j | – | 👍(limited)* | 👍 | – | 👍** | 👍(limited)*** | *Gremlin is supported as a part of TinkerPop integration. **GraphQL supported with the GRANDstack layer. ***Neo4j can be embedded in Java applications. |
| Gbase | – | – | – | – | 👍 | – | – |
| GraphBase | – | – | – | – | – | – | GraphBase uses its native query language. |
| Memgraph | – | – | 👍* | – | – | – | *openCypher. |
| Sparksee/DEX | – | 👍 | – | – | 👍 (.NET)* | – | *Sparksee/DEX also supports C++, Python, Objective-C, and Java APIs. |
| TigerGraph | – | – | – | – | – | – | TigerGraph uses GSQL. |
| Weaver | – | – | – | – | – | 👍 (Python) | |

"**Progr. API**" determines whether a system supports formulating queries using some native prog. language such as C++. "👍": A system supports a given language. "👍(limited)": A system supports a given language in a limited way. "–": A system does not support a given language.

▤ Table 2.6: Support for different graph query languages in different GDBMSs[14]. Continues on next page ...

... continued from previous page

| Graph Database System | Graph database query language | | | | | | Other languages and additional remarks |
| | SPARQL | Gremlin | Cypher | SQL | GraphQL | Progr. API | – |
| **⦾ DATA HUBS** | | | | | | | |
| Cayley | – | 👍* | – | – | 👍 | – | *Cayley supports Gizmo, a Gremlin dialect . Cayley also uses MQL . |
| MarkLogic | – | – | – | – | – | – | MarkLogic uses XQuery . |
| OpenLink Virtuoso | 👍 | – | – | 👍 | – | – | OpenLink Virtuoso also supports XQuery , XPath v1.0 [54], and XSLT v1.0. |
| Stardog | 👍* | 👍 | – | – | 👍 | – | *Stardog supports the Path Query extension. |

"**Progr. API**" determines whether a system supports formulating queries using some native prog. language such as C++. "👍": A system supports a given language. "👌": A system supports a given language in a limited way. "–": A system does not support a given language.

In the next subsection a GDBMS is taken into consideration, specifically ArangoDB - and its characteristics are presented in detail.

## 2.2.6. A GDBMS in detail: ArangoDB

ArangoDB is a NoSQL multi-model database management system supporting graphs, documents and key/value pairs. It was first released in 2012 and is developed by ArangoDB GmbH and triAGENS GmbH (Germany). In September 2021, according to `https://db-engines.com`[47], it ranks:

· 3rd in Graph DBMS category,
· 10th in Document Stores category,
· 11th in Key-Value Stores category,
· 75th overall the DBMS tracked by the site.

It is a schema-free database management system written in C/C++ and JavaScript that runs on a variety of operating systems (Linux, OS X, Windows, Raspbian and Solaris). ArangoDB is designed to serve documents to clients. These documents are sent in JSON format over HTTP. A REST API is available to interact with the database system. A web interface and an interactive shell are also available for accessing the database.[48]

It works with a variety of programming languages: C#, Clojure, Dart, Elixir, Erlang, Go, Java, JavaScript, PHP, Python, R, Ruby, Rust, Scala and so on - and allows the definition of stored procedures in JavaScript. Many community integrations and libraries exist for ArangoDB, worth mentioning: Camel, Ecto, Feathers, GORM, GraphQL, Gremlin, Hemera, Jakarta NoSQL, Kafka, Laravel, Mirconaut, Pydantic, Symfony2, Testcontainers and many others. It states to have a native multi-model approach, rather to, say, a graph database implemented as an abstraction layer on top of a document store. In this approach, when executing queries, it "does not switch" between the models behind the scenes.[92]

### 2.2.6.1. Logical data organization

ArangoDB organizes data into databases, collections and documents. Databases are sets of collections. Collections are sets of records, which are also known as documents. To have collection isolation, multiple databases can be defined. A special database called `_system` is always generated by default, it cannot be deleted and it is used as the administration database to execute tasks such as user and collection management.[48]

Documents can be thought of as rows in a table, while collections are the equivalent of tables in RDBMS. Because ArangoDB is schema-less, there is no need to define what attributes a collection - and consequently its documents

[92] ARANGODB (2021)
ArangoDB. *ArangoDB - Why use ArangoDB, Advantages of Native Multi-Model*. Online. Aug. 2021. URL: `https://www.arangodb.com/community-server/native-multi-model-database-advantages/`.

- can have before inserting data; Instead, each document can have a completely different structure while yet being stored with other documents in the same collection.[93,94]

There are two types of collections:

- document collections, also used as vertex collections in the context of graphs,
- and edge collections.

Edge collections also store documents, but they also have two special attributes, _from and _to, that are used to form relationships between vertex documents.

Documents follow the JSON standard. Internally they are stored in a binary format called `VelocyPack`. `VelocyPack` is described in § 2.2.6.2 - Physical data organization on page 55.

A document can have zero or more attributes, each with a value. A value can be:

- an atomic type, such as a number, string, boolean or null;
- or a compound type, such as an array;
- or an embedded document.

Each document has a unique primary key that allows it to be identified both within its collection and across all collections in the same database. The document handle is formed by the combination of the collection name and the document key.[93]

Three special attributes are part of every document:

- the document handle is saved as a string in `_id`,
- the primary key of the document is stored in `_key`,
- and the document revision is recorded in `_rev`.

When creating a document, the user can provide the value of the key attribute. However, after the document has been produced, the `_id` and `_key` values remain immutable, while the rev value is automatically maintained by ArangoDB.[93]

The direction of an edge depends on the fields `_from` and `_to`. It is possible to indicate within a query which direction the edge should be traversed. The possible directions are:

- `OUTBOUND`: `_from` → `_to`;
- `INBOUND`: `_from` ← `_to`;
- `ANY`: `_from` ↔ `_to`.[95]

In ArangoDB a named graph is used to define graphs. To define a graph, the edge collections to include are specified. The related vertex documents are automatically discovered by the `_from` `_to` attributes of each of the edge documents.[95]

## 2.2.6.2. Physical data organization

JSON is ArangoDB's 'default data format.[96] It can store a nested JSON object as a data entry inside a collection natively, So there is no need to unwrap the generated JSON objects for storage and the stored data simply inherits the

[93] ARANGODB (2021)

ArangoDB. *ArangoDB - Data Modeling and Operational Factors Documentation*. Online. Documentation. Aug. 2021. URL: https://www.arangodb.com/docs/stable/data-modeling-operational-factors.html.

[94] ARANGODB (2021)

ArangoDB. *ArangoDB - Getting Started Documentation*. Online. Documentation. Aug. 2021. URL: https://www.arangodb.com/docs/stable/getting-started.html.

[95] ARANGODB (2021)

ArangoDB. *ArangoDB - Graphs Documentation*. Online. Documentation. Aug. 2021. URL: https://www.arangodb.com/docs/stable/graphs.html.

[96] WIESE (2015)

Lena Wiese. *Advanced Data Management: For SQL, NoSQL, Cloud and Distributed Database*. Walter de Gruyter GmbH, Oct. 2015. 374 pages. ISBN: 978-3110441406. URL: https://www.oreilly.com/library/view/advanced-data-management/9783110433074/.

document's tree structure.[97]

ArangoDB, internally, uses VelocyPack. VelocyPack is a compact binary format for serialization and storage of documents, query results, and temporarily computed values.[98] VelocyPack is a (unsigned) byte-oriented serialization format whose values are platform-independent bytes sequences.[99] Its main purpose is to reduce the amount of storage space needed for "small" values such as booleans, integers, and short strings in order to speed up querying operations. VelocyPack document entries stored on disk are self-contained - each stored document contains all of the data type and attribute name definitions. While this may require a little more storage space, it eliminates the overhead of retrieving attribute names and document layout via shared structures. It also simplifies the code paths for document storage and reading.[100]

> "These days, JSON (JavaScript Object Notation) is used in many cases where data has to be exchanged. Lots of protocols between different services use it, databases store JSON (document stores naturally, but others increasingly as well). It is popular, because it is simple, human-readable and yet surprisingly versatile, despite its limitations. ArangoDB developed this - referring to VelocyPack - format because none of the several known JSON formats used by other applications (e.g. Universal Binary JSON, MongoDB's BSON, MessagePack, BJSON, Apache Thrift, Google's Protocol Buffers, etc.) manages to combine compactness, platform independence, fast access to subobjects and rapid conversion from and to JSON" - on the reason why ArangoDB developed VelocyPack[98]

ArangoDB stores graph models by using these particular forms of JSON documents, of the vertex or edge type, that are stored using the optimized VelocyPack format. Thus, there is no particular data structure that models a graph (as Neo4j has), but rather a specialized use of the JSON format. At server startup, indexes are built over the edge elements to rebuild the relationships between them and their referred node elements. These indexes, named Edge indexes, are a type of index that allows for fast document access based on their _from or _to attributes. The Edge Index internally is a hash index that stores the union of all _from and _to attributes. They are thus used every time an edge is walked during a traversal operation and they point to where the information related to the nodes is stored.[48]

ArangoDB does not appear to have index-free adjacency property. Getting from a vertex to an edge and viceversa requires an index lookup. However, ArangoDB uses a Hash index for these operations and the lookup complexity is $O(1)$. So, up to a small constant factor, the time it takes to travel from a vertex to the edge (and viceversa) is not all that different from the case when the edge's address were directly stored in the vertex itself, perhaps as a property and serialized to secondary memory.[101]

ArangoDB has also provided additional options for optimizing graph traversals, such as Vertex centric Indexes. Its main concept is to index a combination of node, direction of the associated edge and any other set of edge properties.

[97] Agoub, Kunde and Kada (2015)

Amgad Agoub, Felix Kunde and Martin Kada. *Potential of Graph Databases in Representing and Enriching Standardized Geodata*. In: Dreiländertagung der DGPF, der OVG und der SGPF. Bern, Switzerland, June 2015, pages 208–216. URL: https://www.dgpf.de/src/tagung/jt2016/proceedings/papers/20_DLT2016_Agoub_et_al.pdf.

[98] ArangoDB (2021)

ArangoDB. *ArangoDB - Velocity Pack (VPack) GitHub Repository*. Online. Aug. 2021. URL: https://github.com/arangodb/velocypack.

[99] ArangoDB (2021)

ArangoDB. *ArangoDB Velocity Pack (VPack) GitHub Repository VelocyPack.md*. Online. Aug. 2021. URL: https://github.com/arangodb/velocypack/blob/main/VelocyPack.md.

[100] ArangoDB (2021)

ArangoDB. *ArangoDB Release Notes Documentation - What's New, Changelogs*. Online. Aug. 2021. URL: https://www.arangodb.com/docs/stable/release-notes.html.

[101] ArangoDB Google Groups (2014)

ArangoDB Google Groups. *Discussion on ArangoDB and index-free adjacency*. Online. Mar. 2014. URL: https://groups.google.com/forum/#!topic/arangodb/xO0qIcZ6h60.

Considering a social networking situation in which users have different types of relationships, such as `friend_of` or follows, likes etc. - on the edges, there would be a Type attribute. ArangoDB can quickly find the list of all edges attached to the node by using the built-in Edge Index, but it must still walk through the result list and check if all of them contain the attribute Type == "`friend_of`". ArangoDB, using a Vertex centric Index, locates all edges for a vertex having the attribute Type == "`friend_of`" at the same time, without any need to check the condition on all the result edges.[102]

Memory-mapped files are used to store documents to disk. These memory-mapped files by default are synchronized to disk on a regular basis, but not instantly. This is a trade-off between data durability and storage performance. If this level of durability is deemed too low for an application, the server can instantly sync all modifications to disk; this would provide full durability but at the cost of performance because each data alteration would initiate a sync I/O operation.[103]

Lastly, ArangoDB creates a new version for each updated document (MVCC - Multi-Version Concurrency Control) and this is also the case when a document is deleted. This way objects can be kept in main memory coherently and compactly, and that isolated writing and reading transactions allow for parallel access to these objects.[48]

### 2.2.6.3. Data Integrity

Data integrity is unquestionably one of the most crucial features of a database; it is much more critical in scenarios involving complicated data models. In this subsubsection is presented a brief presentation of data and graph integrity, which refers to the consistency of the state of the edges of a graph.

Because ArangoDB is a NoSQL schema-less DBMS, it does not expect a schema to be defined before data is inserted, hence data constraints are usually not imposed. Within the same edge collection, there may be documents with different properties, attributes, or fields, as well as edges. Automatic indexes on system attributes, such as `_key`, `_from` and `_to`, ensure unique constraints. Users or applications can also impose indexes on additional fields.[48]

*Insert*, *update* and *delete* are the operations that pose the greatest danger to graphs and in general data integrity. Because the collections of a graph can still be modified using traditional INSERT, UPDATE, DELETE operations ways even after it has been created, graph inconsistency can occur. In fact, deleting a node should not be taken lightly because it may result in dangling edges. If the collections are accessible by graph module functions, however, the following assurances are guaranteed:

- All changes are carried out transactionally.
- If a node is deleted, all connecting edges are removed as well avoiding loose/dangling ends.
- When an edge is inserted, it is checked to see if it matches the edge definitions, ensuring that edge collections contain only valid edges.[95]

Because ArangoDB's named graphs are graph definitions done on collections that can exist independently of the graph, the same node collection could be used by two different named graphs at the same time. However, the graph module handles this situation as well avoiding dangling edges in this case too. This necessitates extra database operations, which do not obviously come not free.

When it comes to data consistency and concurrent activities, transactions are used to preserve data integrity. ArangoDB transactions are not the same as SQL transactions. A SQL transaction, in particular, begins with an explicit command, such as BEGIN or START TRANSACTION, then proceeds via a series of data retrieval or modification activi-

---

[102] ArangoDB (2021)

ArangoDB. *ArangoDB - Indexing Documentation - Vertex Centric Indexes*. Online. Documentation. Aug. 2021. URL: https://www.arangodb.com/docs/stable/indexing-vertex-centric.html.

[103] ArangoDB (2021)

ArangoDB. *ArangoDB - Architecture Documentation*. Online. Documentation. Aug. 2021. URL: https://www.arangodb.com/docs/stable/architecture.html.

ties and concludes with a COMMIT command, or a rollback - a ROLLBACK command.[95] There are no specific BEGIN, COMMIT, or ROLLBACK transaction commands in ArangoDB. Instead, a transaction is launched by passing a transaction description to the JavaScript method db._executeTransaction(description). This function will then initiate a transaction, perform all necessary data retrieval and/or modification actions and finally commit the transaction. If an error occurs during transaction execution, the transaction will be aborted and all modifications will be rolled back.[104]

Furthermore, a transaction is always a server-side operation that is executed in one go on the server, with no client interaction. However, ACID properties will be provided throughout transaction execution by exploiting different techniques such as:

- document revision,
- collections locking for those involved in the transaction,
- transactions interruptions deactivation
- and others.[104]

WAL (Write-Ahead Log) files are used by transactions, which are files where all alterations are appended before they are applied and persisted to disk. After a failure or server crash, this strategy allows you to simply search for a file's valid start-section. Instead of this, the DBMS would overwrite existing data, would have to validate each block before reenabling access to the database, as other databases generally do. This approach is used to run data recovery after a crash and it may also be utilized in a replication configuration where slaves need to replay the same sequence of actions as the master.[103] However, this technique is based on the assumption that the server is subject to fewer crashes.[105]

The fact that transaction operation information (record pointers, revision numbers, and rollback information) must fit into main memory is a limitation of ArangoDB. Furthermore, transactions should be kept as small as possible to ensure the Write Ahead Log trash collection progresses and large transactions are broken into numerous smaller transactions. In ArangoDB transactions cannot be nested. Attempting to call a transaction from within an ongoing transaction will result in an error.[48]

Depending on how the database is structured, transactions have different guarantees. Multi-document and multi-collection queries are guaranteed to be fully ACID while using a single database instance and single-document actions are also fully ACID in cluster mode. Multidocument and multi-collection queries in a cluster, on the other hand, may not be ACID.[104]

### 2.2.6.4. Query language and graph functions

SQL has made history in the databases' world. It was created to address the needs of relational models. However, if used to query graph structures, SQL does not excel - it was not designed for that. It is possible with SQL to query or walk graph data mapped to a relational database and get desired data; but, the complexity of these queries increases really fast.

As a result, most graph DBMSs have created their own graph-oriented query language. The query language used by ArangoDB is known as AQL, which stands for ArangoDB Query Language. AQL is a declarative language for managing all three data models supported by its database management system. It supports reading and modifying collection data, but it does not support creating and dropping databases, collections and indexes. As a result, it is a pure data

[104] ARANGODB (2021)

ArangoDB. *ArangoDB - Transactions Documentation*. Online. Documentation. Aug. 2021. URL: https://www.arangodb.com/docs/stable/transactions.html.

[105] EVERETT and MALYSHEV (2015)

Nik Everett and Stas Malyshev. *Investigate ArangoDB for Wikidata Query*. Online. Feb. 2015. URL: https://phabricator.wikimedia.org/T88549.

manipulation language (DML) and not a data definition (DDL) or data control language (DCL).[106]

Even though certain terms are identical, AQL queries have a different syntax from SQL queries. It provides various aggregating, ordering, filtering and sub-querying capabilities, as well as an EXPLAIN clause for getting insights about query the execution plan. It offers some support for JOIN operations between documents, despite the fact that, due to its schema-less nature, a null value could be delivered if one of the involved documents has a missing attribute.[107] It does not have a SELECT clause like SQL for selecting the data to be returned; instead, it uses the keywords FOR and RETURN to select what to return as a result.

In Code listing 2.3 is shown the query syntax for graph traversals.[108]

```
1  [WITH vertexCollection1[, vertexCollection2[, ...vertexCollectionN]]]
2  FOR vertex[, edge[, path]]
3     IN [min[..max]]
4     OUTBOUND | INBOUND | ANY startVertex
5     GRAPH graphName
6     [PRUNE pruneCondition]
7     [OPTIONS options]
8     // RETURN [...]
9     // UPDATE vertex WITH {...} IN vertexCollectionN
10    // REMOVE vertex IN vertexCollectionN
11    // INSERT newVertex INTO vertexCollectionN
```

Code listing 2.3: ArangoDB Traversal Query

The start vertex and edge directions are specified for a named graph. Inputting min and max depth parameters gives the possibility to explore surrounding vertices. The FOR line can be used to give names to graph elements, which can then be used in the RETURN clause. Before invoking RETURN, the OPTIONS clause allows the specification of policies for graph traversal. Additionally, other clauses like FILTER or LIMIT can be specified before invoking RETURN. AQL can be invoked via the Aardvark web interface, the arangosh shell, via HTTP API/REST/GraphQL or Foxx Services.[92,106]

Apart from simple document retrieval and graph traversals, ArangoDB also supports AQL requests for shortest path and geospatial functions.[106]

A number of JavaScript functions are also readily available, for example:[109]

· common descendants of two vertices,

· nodes betweenness,

· distance between two vertices,

· nodes with the same common properties,

· nodes eccentricity,

· graph radius and diameter,

· all paths between two vertices,

· nodes closeness,

[106] ARANGODB (2021)

ArangoDB. *ArangoDB Query Language (AQL) Documentation*. Online. Documentation. Aug. 2021. URL: https://www.arangodb.com/docs/stable/aql/.

[107] ARANGODB (2021)

ArangoDB. *ArangoDB - Joins Documentation*. Online. Documentation. Aug. 2021. URL: https://www.arangodb.com/docs/stable/aql/examples-join.html.

[108] ARANGODB (2021)

ArangoDB. *ArangoDB - Traversals Documentation*. Online. Documentation. Aug. 2021. URL: https://www.arangodb.com/docs/stable/graphs-traversals.html.

[109] ARANGODB (2021)

ArangoDB. *ArangoDB - Graph Functions Documentation*. Online. Documentation. Aug. 2021. URL: https://www.arangodb.com/docs/stable/graphs-general-graphs-functions.html.

- minimum weight path between two vertices,
- and many more.

It is also possible to specify additional functions and create new visitor or expander methods using JavaScript.

ArangoDB does (did) not natively support Gremlin for query specification[110] - but there are some projects making it possible.[111,112]

### 2.2.6.5. Caching

For database management systems caching is a fundamental feature. In general, caching means that important or useful data is kept in main memory and may be accessed more quickly. As a result, a great percentage of queries run faster because the graph data needed to answer them are already in memory. Caching generally has three key goals:

1. reduction of disk access,
2. reduction of CPU utilization,
3. and faster response times for the user[48]

Databases usually implement caching in the following three forms: Three types of caching are generically possible for databases:

1. **caching of query results**: the exact output of a read-only query is saved in main memory for the next time the same query is run. It reduces the number of accesses to the disk, uses less CPU time and returns the result quicker. It is especially useful when there are a lot of data-reading queries but fewer write queries.
2. **caching of query plans**: it stores the optimizer's results, which are responsible for determining how the database will obtain the requested data. With the exception of one or more "placeholders", this type of caching usually requires a pre-prepared query. The optimizer does not need to be called because the plan is already known, which saves CPU and time.
3. **caching of data itself**: it stores graph data structures, tables or indexes in memory for quicker readings. This reduces the number of disk accesses, which saves time.[113]

Each of the three types of caching might be used in conjunction with the others - a query may be able to use one, two, or all three types of caches.

ArangoDB uses the filesystem page cache. ArangoDB has a query caching system in place. Query cache can be requested/disabled on demand by AQL queries, or it can be enabled and disabled for the entire server service. Hash table is used to implement the query cache.[48]

If two queries have exactly the same query string, including whitespaces, the query cache mechanism considers them identical. The query string is hashed and used as the cache lookup key. Bind parameters are hashed and used as the cache lookup key if a query uses them, this way there will not be a hit for the same query with new parameters.

[110] ARANGODB (2013)

ArangoDB. *ArangoDB - Gremlin graph queries for REST*. Online. GitHub Repository Issue #392. Feb. 2013. URL: https://github.com/arangodb/arangodb/issues/392.

[111] ARANGODB, BRANDT, HOYOS and MCCOY (2020)

ArangoDB, Achim Brandt, Horacio Hoyos and Nathan McCoy. *Using ArangoDB with Gremlin*. Online. GitHub Repository. https://github.com/ArangoDB-Community/arangodb-tinkerpop-provider/wiki/Server. May 2020. URL: https://github.com/ArangoDB-Community/arangodb-tinkerpop-provider.

[112] ARANGODB (2021)

ArangoDB. *ArangoDB - TinkerPop Provider for ArangoDB, GitHub Repository*. Online. An implementation of the Apache TinkerPop OLTP Provider API for ArangoDB.. Aug. 2021. URL: https://github.com/ArangoDB-Community/arangodb-tinkerpop-provider.

[113] MULLANE (2002)

Greg Sabino Mullane. *PostgreSQL - Database Caching*. Online. PostgreSQL mailing list (pgsql-hackers). Feb. 2002. URL: https://www.postgresql.org/message-id/E16gYpD-0007KY-00@mclean.mail.mindspring.net.

To be eligible for caching, a query needs to meet the following criteria:[114]

- is a read-only query;
- no warnings were generated;
- only uses deterministic functions.

If queries modify the data of collections that were used during the computation of the cached query results, the query cache results are entirely or partially invalidated automatically; So there are extra cache invalidation checks for each data-modification operation.

### 2.2.6.6.  Scalability approach and data partitioning

Graph data domains can be quite large. Think of social networks, energy distribution networks or railway/transportation networks. Scalability issues arise. While scaling vertically could be a way to temporarily postpone the problem, it usually comes with some drawbacks. It will not be discussed here. NoSQL databases brought the ability to grow horizontally and did it in a cost-effective manner. Is horizontal scaling applicable in graph databases too?[48]

The capacity to disperse workload over several clusters/servers/locations is known as horizontal scalability. Rather than putting all of the load on one server, it is possible to delegate at least some of the work to other servers to speed up things even when there are a lot of requests. It also improves resilience through replication and fail-over management. Sharding is a method of achieving horizontal scalability by dividing the data storage into multiple partitions, each of which is stored on a different machine. Clusters of machines are set up so that all instances appear as a single database. Being able to manage a database that is split and distributed across multiple machines provides a significant opportunity of horizontal scalability, allowing vast amounts of data to be handled. This feature is one of the strongest points of NoSQL DBMSs; When it comes to graph databases, however, the situation is more complicated. The challenge derives from the fact that, according to graph theory, partitioning a graph into distinct partitions is an NP-complete problem[74,115,116] and thus impractical. A naïve solution to the problem can result in unpredictable query times because of graph traversals jumping between machines over the (slow) network.[48]

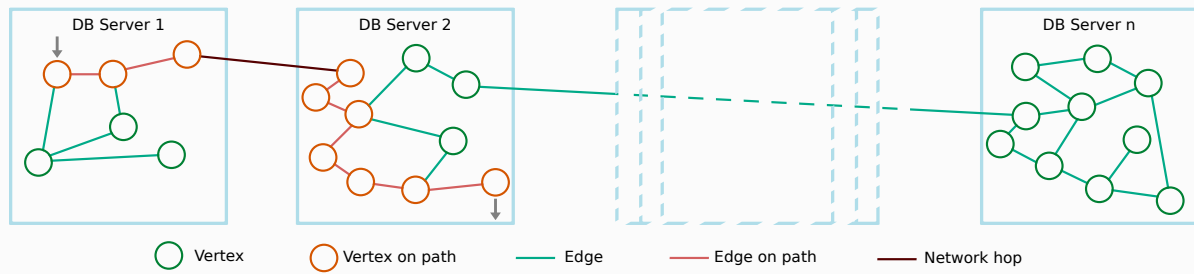

Figure 2.18.: ArangoDB's SmartGraphs for sharding

ArangoDB may be horizontally scaled by using many servers. Because of their nature, the data models created by ArangoDB offer different opportunities for scalability. Particularly, the ability to scale decreases from key/value over documents (documents with joins) to graphs. The key/value store data model is the most scalable since a document collection always has a primary key (_key attribute) and document collections behave like simple key/value stores in

[114] ArangoDB (2021)

ArangoDB. *ArangoDB - The AQL query results cache Documentation*. Online. Documentation. Aug. 2021. URL: https://www.arangodb.com/docs/stable/aql/execution-and-performance-query-cache.html.

[115] Savage and Wloka (1991)

John E. Savage and Markus G. Wloka. *Heuristics for Parallel Graph-Partitioning*. Technical report. Technical Report No. CS-89-41, Revised Version. Providence, Rhode Island 02912, USA, Jan. 1991.

[116] Wikimedia Foundation, Inc. (2021)

Wikimedia Foundation, Inc. *Wikipedia - Graph partition*. Online. Aug. 2021. URL: https://en.wikipedia.org/wiki/Graph_partition.

absence of additional secondary indexes. Single key lookups and key/value pair insertions and updates are the only logical in this situation. If the key attribute is the only sharding attribute, sharding is done with respect to the primary key and all operations scale linearly.[48] Even in presence of secondary indexes, the same is valid for the document store, because an index for a sharded collection is simply the same as a local index for each shard. Each shard only stores the part of an index that it needs. As a result, single document operations continue to scale linearly with cluster size. However, because the AQL query language permits queries to employ multiple collections, secondary indexes and joins, scalability might be difficult if the data to be combined is spread over numerous machines, as a lot of communication is required. When working with graph data, the same thing happens.[48]

It is therefore, crucial to configure the graph data distribution across the shards in a well-studied manner in order to provide good performance at scale. ArangoDB asks users to specify which attributes to utilize for the graph data to be sharded. The suggestion is to make sure that edges originating at a node are located on the same cluster machine as the node itself. ArangoDB Enterprise Edition, includes the SmartGraph feature, which implements graph partitioning based on community detection in order to reduce network communication. The distributed architecture is managed by a multi-master model and a number of ArangoDB instances communicate with one another via the network and play different roles (Agents, Coordinators, Primary and Secondary DBservers).[92]

#### 2.2.6.7. Graph visualization

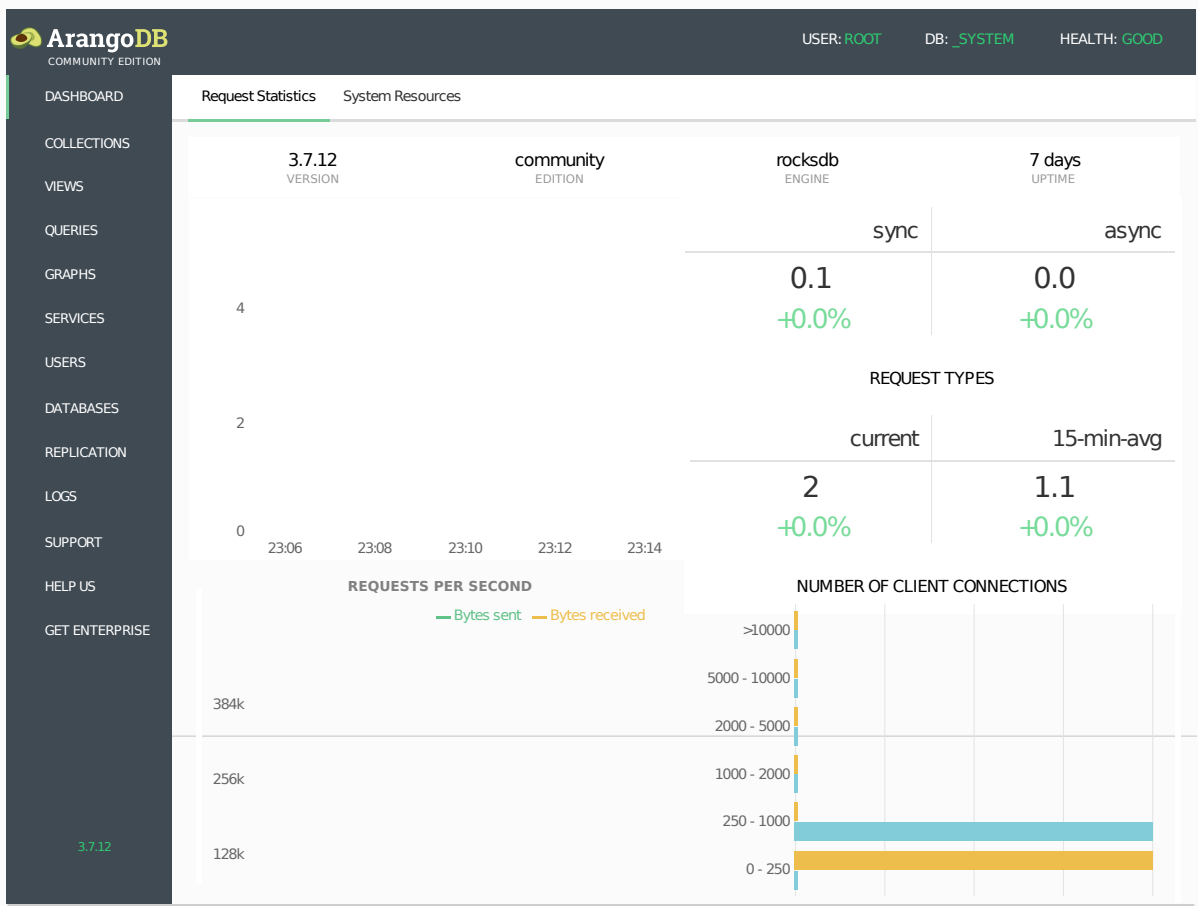

≡ Figure 2.19.: ArangoDB's Aardvark web interface

ArangoDB comes with a useful web interface called Aardvark. With this interface the database administrator can view some live statistics on the dashboard and perform operations like:

- manage databases, collections and documents;
- execute and save AQL queries;
- manage the database schema, impose constraints and create indexes;
- store and launch services/procedures;
- get a graph-representation of the data contained in the form of a graph model, as shown in ;
- view the database logs.
- create, manage and explore graphs click-by-click;
- create new vertices and edges in an interactive way directly in the graph visualizer;

### 2.2.6.8. Licensing

ArangoDB offers three different levels of subscription[92,117]:

- The Community Edition is an open and free license without direct support and without custom advanced features. It is released under the Apache v2 license. Of all the major GDBMSs Community Editions, ArangoDB's CE is the one that provides the most features of its upgraded Enterprise Edition.
- The Enterprise Edition is a commercial license with dedicated support and contact to experts of the core team. It includes all features of the Community version plus SmartJoins, SmartGraphs, Encryption, Auditing and training.
- The Oasis Edition offers fully managed service with physical or cloud machines.

With this ends the in-detail description of ArangoDB and this chapter altogether. In the next one shall be presented the Label Propagation Community Detection algorithm and the usage of ArangoDB for the its application to cluster academic collaboration communities.

[117] ArangoDB (2021)
ArangoDB. *ArangoDB Commercial Subscriptions*. Online. Aug. 2021. URL: https://www.arangodb.com/subscriptions/.

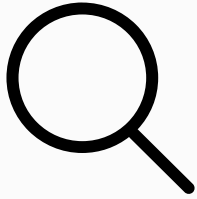

# 3. Community Detection

**This chapter's contents:**

In the previous chapter the literature on graph theory concepts, problems, algorithms and on graph database systems was reviewed. With these concepts fresh in mind it is now possible to go more in depth with graph clustering, community detection in graphs using graph databases. The chosen GDBMS for storing the data is ArangoDB since it comes with ready to execute algorithms on graphs, has very few feature limitations on its Community Edition license and is generally well documented. The graph dataset that is going to be used contains information of computer science academic publications obtained from dblp.org[1].

The goal of the community detection applied to the dblp.org[1] dataset is to find groups of scientific collaboration between researchers that are more densely connected between them than others. The dataset contains typical citation data, like authors, title of publications, date, publishers, editors, affiliations etc. In this chapter, how community detection is carried out - is investigated in the literature. Then, having seen what are the options on the table, will be chosen an algorithm and described the practical part of actually executing it on the graph data.

## 3.1. Literature review of clustering methodologies and algorithms

In this section different graph clustering algorithms are presented.

Some common community detection algorithms are:

- **Triangle Count** and **Clustering Coefficient** for overall relationship density
- **Strongly Connected Components (SCC)** and **Weakly Connected Components (WCC)** for finding connected clusters
- **Label Propagation**[118] for quickly inferring groups based on node labels

[118] Raghavan, Albert and Kumara (2007)
Usha Nandini Raghavan, Réka Albert and Soundar Kumara. *Near linear time algorithm to detect community structures in large-scale networks*. In: Physical Review E 76.3 (Sept. 2007). ISSN: 1550-2376. DOI: 10.1103/physreve.76.036106. URL: https://arxiv.org/abs/0709.2938.

- **Louvain Modularity** for looking at grouping quality and hierarchies
- **InfoMap** discovers communities by applying the Random Walk technique
- **Leading Eigenvector** separates the vertices into communities by using the eigenvector of the modularity matrix of the graph
- **Walktrap** computes the community structure of a network based on a similarity metric among vertices
- other more specific algorithms[119] such as:
  - **Hierarchical Clustering**
  - **Girvan-Newman Algorithm (GNA)**
  - **Algorithm of Duch and Arenas**
  - **Algorithm of Clauset et. al.**
  - **Newman's Spectral Algorithm**
  - **Genetic Algorithms** like:
    * **Traditional GA**
    * **Falkanuer's Grouping Genetic Algorithm**
    * **Grouping Genetic Algorithm** of Tasgin et. al.

In Table 3.1 are brought descriptions and usecases for some of the algorithms listed above - whereas in Figure 3.2 is brought a graphical visualization of the distributions of the size of the communities detected with various clustering algorithms.[16] After that, the complexity of some of the community detection algorithms is brought in Code listing 3.1. algorithms.[120,121]

| Algorithm type | What it does | Usecases |
|---|---|---|
| **Triangle Count and Clustering Coefficient** | Measures how many nodes form triangles and the degree to which vertices tend to cluster together | Estimating group stability and whether the network might exhibit "small-world" behaviors seen in graphs with tightly knit clusters |
| **Strongly Connected Components** | Finds groups where each vertex is reachable from every other vertex in that same group following the direction of relationships | Making product recommendations based on group affiliation or similar items |
| **Weakly Connected Components** | Finds groups where each node is reachable from every other node in that same group, regardless of the direction of relationships | Performing fast grouping for other algorithms and identify islands |

Table 3.1: Overview of community detection algorithms[41]. Continues on next page …

[119] Öztürk (2014)
Koray Öztürk. *Community detection in social networks*. Master's thesis. Middle East Technical University, Dec. 2014. 105 pages. URL: https://open.metu.edu.tr/bitstream/handle/11511/24245/index.pdf.
[120] Leão, Brandão, Vaz de Melo and Laender (2018)
Jeancarlo C. Leão, Michele A. Brandão, Pedro O. S. Vaz de Melo and Alberto H. F. Laender. *Who is really in my social circle? Mining social relationships to improve detection of real communities*. In: Journal of Internet Services and Applications 9.1 (Oct. 2018). ISSN: 1869-0238. DOI: 10.1186/s13174-018-0091-6. URL: https://www.researchgate.net/publication/327824385.
[121] Wagenseller III and Wang (2017)
Paul Wagenseller III and Feng Wang. *Size Matters: A Comparative Analysis of Community Detection Algorithms*. 2017. eprint: 1712.01690. URL: https://arxiv.org/abs/1712.01690.

... continued from previous page

| Algorithm type | What it does | Usecases |
|---|---|---|
| **Label Propagation** | Infers clusters by spreading labels based on neighborhood majorities | Understanding consensus in social communities or finding dangerous combinations of possible co-prescribed drugs |
| **Louvain Modularity** | Maximizes the presumed accuracy of groupings by comparing relationship weights and densities to a defined estimate or average | In fraud analysis, evaluating whether a group has just a few discrete bad behaviors or is acting as a fraud ring |

Table 3.1.: Overview of community detection algorithms[41]

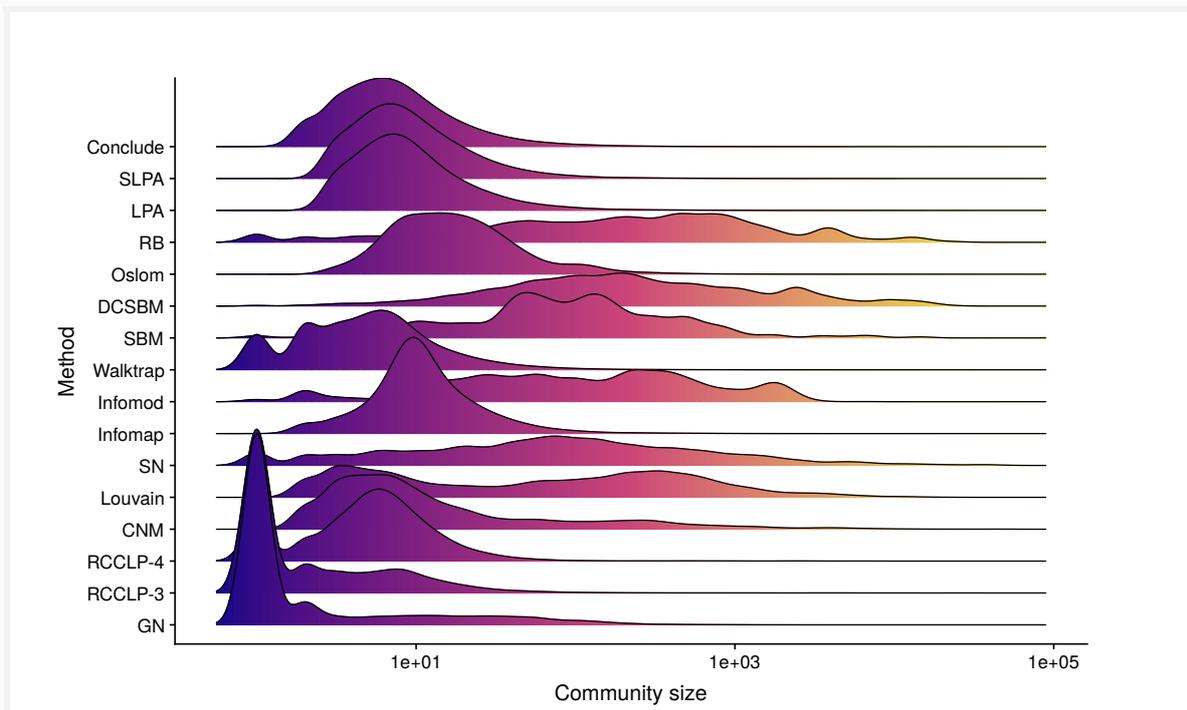

Figure 3.2.: Detected community size distributions using different methods, alogrithms[16]

| Abbr. | Algorithm name | Deterministic or non | Complexity |
|---|---|---|---|
| LPA | Label Propagation Algorithm | Non-deterministic | $O(V)$; $O(E)$[121] |
| LMA | Louvain Modularity Algorithm | Deterministic | $O(V \log V)$ |
| IMA | InfoMap Algorithm | Non-deterministic | $O(V \log V)$; $O(E)$[121] |
| GOMA | Greedy Optimization Of Modularity Algorithm | Deterministic | $O\left(V \log^2 (V)\right)$ |
| LEA | Leading Eigenvector Algorithm | Deterministic | $O\left(V^2 \log V\right)$; $O\left(V(V+E)\right)$[121] |
| WTA | WalkTrap Algorithm | Non-deterministic | $O\left(V^2 \log V\right)$ |
| EBA | Edge Betweenness Algorithm | Deterministic | $O\left(V^3\right)$ |

Table 3.2.: Complexity of some of the community detection algorithms[120] ($V$, $E$: number of vertices, edges)

In the following subsections is brought a description of the algorithms and some usage examples.

### 3.1.1. Triangle Count and Clustering Coefficient

Triangle Count and Clustering Coefficient measures how many nodes form triangles and the degree of community formation between the nodes.[41]

▤ **Formula 3.1 (for the calculation of the clustering coefficient for a node):** [41]

$$CC\left(u\right) = \frac{2R_u}{k_u\left(k_u - 1\right)}$$

Expression 3.1

**Usecases**

It is principally used to evaluate the stability of groups and if networks show some form of "small-world" behavior. Some concrete examples are for classifying spam content, community structure of online social networks, detection of webpages related to common topics.[41]

where:

- $u$ is a node.
- $R\left(u\right)$ is the number of edges through the neighbors of $u$.
- $k\left(u\right)$ is the degree of $u$.

### 3.1.2. Strongly Connected Components (SCC) and Weakly Connected Components (WCC)

Strongly Connected Components finds groups of vertices where each vertex is reachable from all other other vertices in that same group traversing the edges in their direction. In the case of Weakly Connected Components edge direction is irrelevant, so the traversal is done regardless of the direction of the relationships.[41]

**Usecases**

Strongly Connected Components and Weakly Connected Components are generally used for generation of recommendations, identification of islands or for fast grouping in early steps of graph analysis, since they scale nicely. Concrete usecases are the detection of groups of companies whose stakeholders have shares on the other companies of the same group or the analysis of citation networks.[41]

### 3.1.3. Label Propagation

Label Propagation[118] tries to infer communities by propagating labels on neighbor vertices and detect neighborhood majorities. It is a fast algorithm.[41]

The idea behind the algorithm is that in a densely connected community of vertices, a label quickly becomes dominant whereas it is more difficult to do so in a sparsely connected region of the graph.[41]

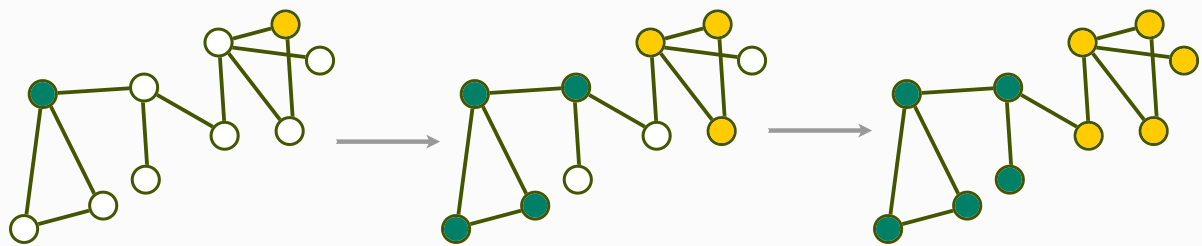

▤ Figure 3.3.: Label propagation algorithm's spread of label to or from neighbors

There are two ways of propagating the labels, one is the push method and the other is the pull. The pull method is a better choice when parallelization is involved.

**The steps of the algorithms**

The pull method of the LPA works as follows:

1. Every vertex is labeled with a unique label.
2. The labels are propagated from vertex to vertex throughout the graph.
3. At the end of each propagation iteration, each vertex updates its label to match the one with the maximum weight, which is calculated based on the weights of neighbor vertices and their relationships. Ties are broken uniformly and randomly.
4. Convergence is reached when each vertex is labeled as the majority of its neighbors. It may happen that convergence on a single solution is not reached in a reasonable time. In such cases, a trade-off between accuracy and execution time would be setting a maximum number of iterations to avoid maybe never-ending execution. This way consensus is reached in groups of vertices.

**Note**

Should be noted that the algorithm may produce for the same graph different community structures each time it is run. This happens because the order in which LPA evaluates the vertices, influences the communities it detects. A way to reduce the range of possible solutions is by setting preliminary label named "seed labels" to some vertices while others are not labeled. This way of initialization of the LPA can be considered a semi-supervised learning to clustering graphs.

**Usecases**

LPA is mainly used to get an understanding of community consensus in social groups, or for semantic analysis of tweets to classify their polarity as positive or negative. It finds use also in pharmaceutical fields when dealing with chemical components and their combinations.[41]

## 3.1.4. Louvain Modularity

Louvain Modularity makes use of relationship weights and densities comparisons to maximize the accuracy of groupings. It is one of the fastest modularity-based algorithms for clustering. Apart from community detection, it reveals also a hierarchy of communities at different scales.[41]

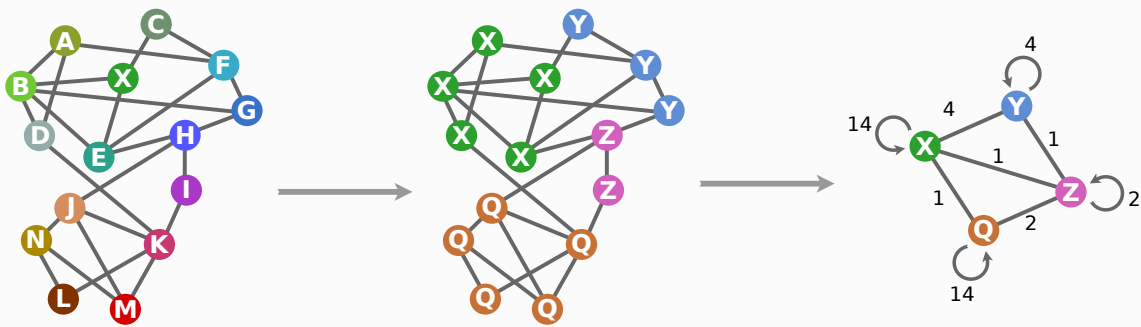

Figure 3.4.: Louvain Modularity algorithm moving of nodes into higher relationship density groups and aggregating

To quantify how well a vertex is assigned to a cluster, LMA looks at the density of connections within a community and compares it to an average or random sample. This measure is called *modularity*. Modularity algorithms optimize detected clusters locally and globally.[41]

**The steps of the algorithms**

1. "Greedy" assignment of nodes to communities, with local optimizations of modularity.
2. Definition of a more fine-grained network structure based on the clusters found in the first step.

3. Repeat iteratively the first two steps until no reassignments of communities increase the modularity.

**☰ Formula 3.2 (for the modularity of a group in Louvain Modularity Algorithm):** [41]

$$M = \Sigma_{c=1}^{n_c} \left[ \frac{L_c}{L} - \left( \frac{k_c}{2L} \right)^2 \right]$$

Expression 3.2

where:

- $L$ is the number of relationships in the entire group.
- $L_c$ is the number of relationships in a partition.
- $k_c$ is the total degree of nodes in a partition.

**☰ Formula 3.3 (for the first optimization step of Louvain Modularity Algorithm):** [41]

$$Q = \frac{1}{2m} \Sigma_{u,v} \left[ A_{uv} - \frac{k_u k_v}{2m} \right] \delta \left( c_u, c_v \right)$$

Expression 3.3

where:

- $u$ and $v$ are vertices.
- $m$ is the total relationship weight across the entire graph
- $2m$ is a common normalization value in modularity formulas.
- $A_{uv} - \frac{k_u k_v}{2m}$ is the strength of the relationship between $u$ and $v$ compared to what it is expected with a random assignment of those vertices in the network.
  - $A_{uv}$ is the weight of the relationship between $u$ and $v$.
  - $k_u$ is the sum of relationship weights for $u$.
  - $k_v$ is the sum of relationship weights for $v$.
- $\delta \left( c_u, c_v \right)$ is the Kronecker delta function which is equal to 1 if $u$ and $v$ are assigned to the same community and 0 otherwise.

**Usecases**

It is used in fraud detection or evaluation of unusual behaviors. Similar usecase is the detection of cyberattacks. Another area of application is the detection of hierarchical community structures within the brain's functional network.[41]

## 3.2. Chosen algorithm

The algorithm chosen for clustering is the Label Propagation Community Detection Algorithm. ArangoDB provides a custom implementation of it, so it is directly applicable on the graph data of the database.

In general, ArangoDB offers a variety of graph processing algorithms, implemented with a custom Pregel. Pregel is a Distributed Iterative Graph Processing algorithm developed at Google by Malewicz, Austern, Bik, Dehnert, Horn, Leiser and Czajkowski in "Pregel: A System for Large-Scale Graph Processing" (2009).[122] It is generally used for graph computations that involve local data, with sparse connections between vertices and whose data is not able to fit all in one computing node. Some example problems include:

- **Transportation Routes**: Shortest Path algorithms
- **Web**: PageRank
- **Social Networks**: Clustering Techniques

[122] MALEWICZ, AUSTERN, BIK, DEHNERT, LEISER and CZAJKOWSKI (2009)

Grzegorz Malewicz, Matthew H. Austern, Aart J. C. Bik, James C. Dehnert, Ilan Horn, Naty Leiser and Grzegorz Czajkowski. *Pregel: A System for Large-Scale Graph Processing.* In: SPAA 2009: Proceedings of the 21st Annual ACM Symposium on Parallelism in Algorithms and Architectures, Calgary, Alberta, Canada, August 11-13, 2009. Jan. 2009, pages 135–145. DOI: 10.1145/1582716.1582723. URL: https://www.researchgate.net/publication/221257383_Pregel_A_system_for_large-scale_graph_processing.

• **Citations Relationships**: Connected Components

A common Pregel algorithm's very high level pseudocode is reported in Code listing 3.1.

```
while not converged do
    /* Superstep */
    communicate:
        send(msg: own_value, to: neighbors)
    compute:
        own_value ← max(value_from_all_messages ∪ own_value)
end while
```

Code listing 3.1: A generic Pregel algorithm's pseudocode[123]

Basically it involves two supersteps:

• **Diffusion**: information is propagated from vertex to neighbors

• **Fusion**: information is aggregated from neighbors to a set of entities

Pregel is inspired by the Bulk Synchronous Parallel model[124], is vertex centric, computes a sequence of iterations (supersteps) in a master-worker model that involves communication between nodes. The communication is organized via the C++ API with message passing, no guaranteed order of delivery but knowing that messages are delivered exactly once. Pregels synchronization machanism ensures the absense of deadlocks and data races[125].

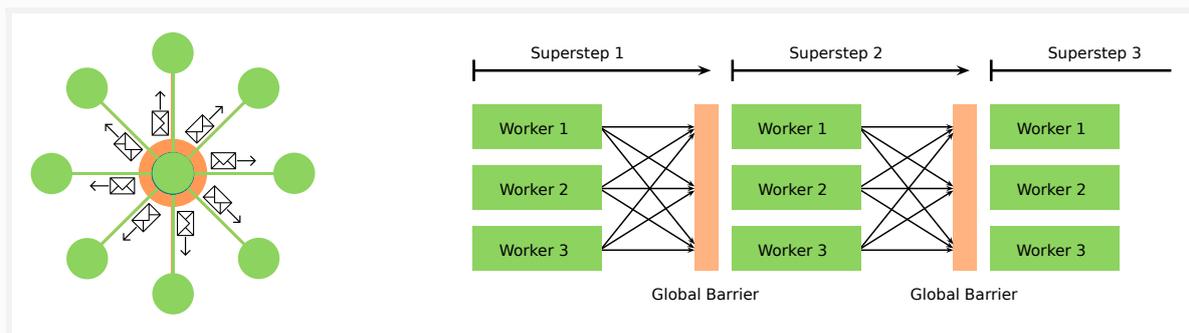

Figure 3.5.: Pregel's diffusion and fusion computation supersteps[123]

ArangoDB offers the following custom Pregel ready-to-run algorithms:[26,123]

• PageRank

• Seeded PageRank

• Single-Source Shortest Path

• Connected Components

    – Weakly Connected Components (WCC)

    – Strongly Connected Components (SCC)

• Hyperlink-Induced Topic Search (HITS)

[123] ARANGODB, SCHAD and KERNBACH (2020)
ArangoDB, Jörg Schad and Heiko Kernbach. *Custom Pregel Algorithms in ArangoDB*. Online. Slides. Docs: https://www.arangodb.com/docs/stable/graphs-pregel.html. Nov. 2020. URL: https://www.slideshare.net/arangodb/custom-pregel-algorithms-in-arangodb.
[124] VALIANT (1990)
Leslie G. Valiant. *A bridging model for parallel computation*. In: Commun. ACM 33 (Aug. 1990). https://dl.acm.org/doi/10.1145/79173.79181, pages 103–111. DOI: 10.1145/79173.79181. URL: https://www.semanticscholar.org/paper/A-bridging-model-for-parallel-computation-Valiant/8665c9b459e4161825baf1f25b5141f41a5085ff.
[125] AGUILAR GONZALES (2018)
Ezequiel Aguilar Gonzales. *Pregel: A System for Large-Scale Graph Processing Presentation*. Online. Slides. 2018. URL: https://ranger.uta.edu/~sjiang/CSE6350-spring-18/12-pregel-slides.pdf.

- Vertex CentralityPermalink
- Effective Closeness
- LineRank
- **Community Detection**
  - **Label Propagation**
  - Speaker-Listener Label Propagation

As mentioned above, of the two available algorithms for community detection, Label Propagation is chosen to be used.

At initialization, random Community IDs are assigned to the vertices. Then during each iteration, a vertex sends its current community ID to all its neighbor vertices. Each vertex adopts the community ID it received most frequently during the iteration. The algorithm ideally runs until it converges, but this is very unlikely to happen for large graphs. There is therefore, a need to specify a maximum iteration bound.[26]

In the next section are shown all the details of the setup and parameters for the execution of the algorithm.

## 3.3.  Clustering collaboration communities - Algorithm execution

In this section is presented how the clustering is done, with a hands-on approach showing the setup and launch of the algorithm execution and the results afterwards.

### 3.3.1.  Execution setup and parameters

Before launching the algorithm execution, it is necessary a first step of data preparation in collections of vertices, edges and a graph made out of these. A brief look at the parameters of the algorithm is given too.

#### 3.3.1.1.  Environment setup

The setup phase is conducted and described in § 4.2 - The data, during which the data is prepared, imported, distributed in vertices and edges are defined between them. In the end, a graph is defined using the vertex and edges collections.

Once ready, `arangosh` is launched from terminal.  It is now time to insert the commands with the parameters indicated in the following subsection.

#### 3.3.1.2.  Parameters used

Regarding the parameters used in the commands, their values follow:

The parameter regarding the name of the algorithm to use is

```
"labelpropagation"
```

The name of the graph on which to execute the algorithm is:

```
"author_publisher_editor_journal_publication_series_affiliation_school_cited_crossreffed"
```

The graph contains vertices and edges on authors, publications, affiliation institutions and schools, publishers, editors, journals, cited publications and crossref-fed ones. It also contains data on the year of publication and publication type for the vertices but edges connecting these data are not included in the graph.

The maximum iteration bound chosen is:

```
maxGSS: 100
```

The attribute field of each vertex where to place the result of the computation is:

```
resultField:  "community"
```

### 3.3.2. Algorithm launch

After having inserted the parameters from the above section, the command is launched.

The execution is fairly straightforward, with just a few lines of code:

First of it is necessary to have shell access to the machine(s) where ArangoDB is installed. Start by launching arangosh with the shown parameter.

```
1 // launch arangosh in terminal with
2 /usr/bin/arangosh --javascript.execute
```
≣ Code sublisting 3.2 (a): Launch arangosh

Then proceed to setting which database is going to be used. It should have all the vertex and edge collections as well as the graph on which the community detection algorithm is going to be run.

```
3 // choose database
4 db._useDatabase('academic_db')
```
≣ Code sublisting 3.2 (b): Select the database to use

After that it is time to import Pregel defined algorithms from @arangodb

```
5 // import pregel
6 const pregel = require("@arangodb/pregel");
```
≣ Code sublisting 3.2 (c): Import Pregel

and go on with the instruction that was discussed in § 3.3.1.2 - Parameters used. That will start the execution of the algorithm.

```
7 // start execution
8 const handle = pregel.start(
9     "labelpropagation", // type of algorithm
10    "nameOfTheGraph", // name of the graph on which to execute the algorithm
11    {
12        maxGSS: 100, // maximum iteration bound, default is 500
13        resultField: "community" // attribute field where to place the result for each vertex's
     detected community
14    }
15 );
```
≣ Code sublisting 3.2 (d): Start the execution of the algorithm

In order to check the progress of the computation, a status check can be repeatedly obtained with the command in line 17 of Code sublisting 3.2 (e).

```
16 // check the status periodically for completion
17 pregel.status(handle);
```
≣ Code sublisting 3.2 (e): Periodically check for the completion progress

≣ Code listing 3.2: Commands to run to execute the community detection algorithm

With the machines involved in this project (see § 4.1.1 - The machines on page 79) and the chosen number of iterations, it takes about two hours to compute the clusters of collaboration communities.

### 3.3.3. Results

In the end the result of the detected community of a specific vertex is as shown in Figure 3.6, by means of a new attribute field of named community.

```
_id:    author/40474848
_rev:   _cszBrjq--C
_key:   40474848
```

```
1 ▾ {
2      "name": "Marcel Thaens",
3      "orcid": "",
4      "bibtex": "",
5      "aux": "",
6      "other_names": [],
7 ▾    "affiliation": [
8 ▾      {
9          "value": "Department of Public Administration, Erasmus University Rotterdam, The Netherlands",
10         "label": "",
11         "type": "affiliation"
12       }
13     ],
14     "note": [],
15 ▾   "url": [
16 ▾     {
17         "address": "homepages/57/1144",
18         "label": "Home Page"
19       }
20     ],
21     "graph_name": "Marcel Thaens",
22     "community": 2886711
23 }
```

Figure 3.6.: Collection document representing a vertex of the graph whose membership to a community has been identified

Each detected community has an identifier, a distinct number. To all the vertices found as members of a community is added the `community` property and its value is set to the identifier of the belonging community.

In the table below are reported some numbers on the results:

| Vertex type | Number of vertices | Number of detected communities |
|---|---|---|
| author | 2786113 | 177592 |
| editor | 43644 | 9837 |
| institution | 56918 | 25415 |
| journal | 1905 | 1896 |
| publication | 5662747 | 141939 |
| publisher | 2292 | 1437 |
| school | 2098 | 1677 |
| series | 1742 | 934 |
| all types | 8557459 | 187451 |

Table 3.3.: Summary of the community detection results

In Figure 3.7 (the query) and Figure 3.8 (the result) are shown in a graph vertices of hop distance 1 to 4 from the start vertex displayed in Figure 3.6.

```
1   FOR vertex, edge, path IN 1..4 ANY "author/40474848"
2   GRAPH author_publisher_editor_journal_publication_series_affiliation_school_cited_crossreffed
3     FILTER vertex.community == 2886711
4     RETURN { vertices: path.vertices[*], edges: path.edges[*] }
```

Figure 3.7.: Querying about the neighborhood of a vertex after community detection

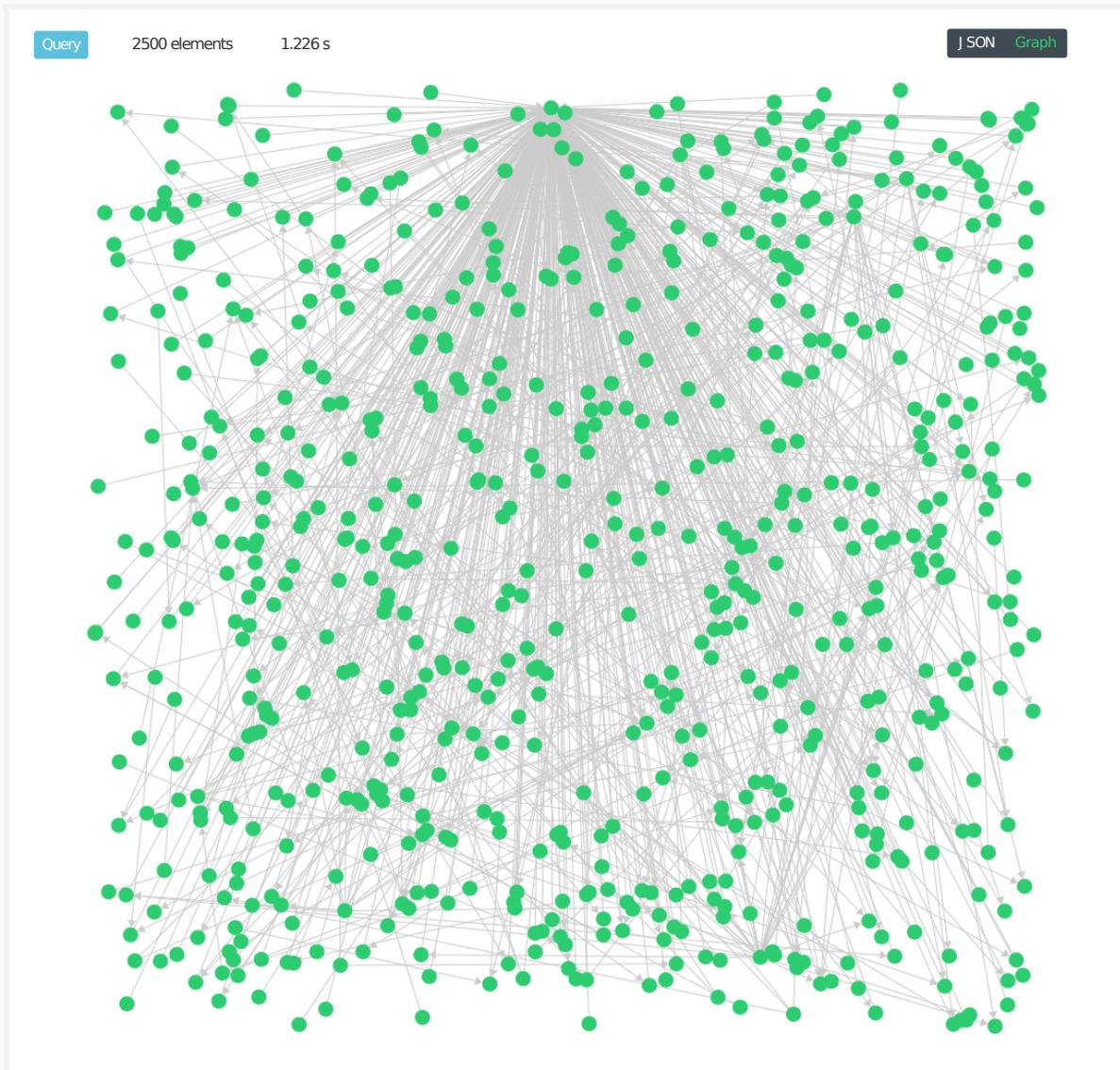

Figure 3.8.: Result of the query (in Figure 3.7) on the neighborhood of a vertex after community detection

As can be seen from the graph it is not clear which vertices belong to which communities, even though information on the detected communities is present in the database. Because of these limitations on graphical coloring or grouping the nodes on the ArangoDB web interface, in the next chapter is going to be developed an ad-hoc Web Application. The goal is to display in a graphically understandable manner, groups of nodes and to which community they belong to.

# Implementation of a WebApp with Graph Databases, GraphQL API and graph rendering libraries in frontend

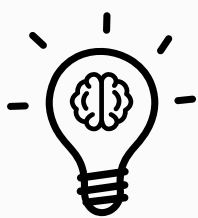

# 4. Implementing the WebApp

**This chapter's contents:**

## 4.1.  The hardware and the OS

In order to host the application with all its components, different machines were bought. These include 5 general purpose used computers, 1 programmable new router, a new network switch (after the first one stopped working during the development phase), 2 used monitors and lots of cables.

### 4.1.1.  The machines

The bought computers were 2 Lenovo ThinkCentre M73 Tiny Form Factor - each with 8 GB of RAM, CPU Intel Core i3-4130 4 cores, 3.4GHz and an SSD harddrive of 250 GB. Apart from the Lenovo-s, 2 HP Compaq 8200 Elite Small

Farm Factor were bought - each with 8 GB of RAM, CPU Intel Core i5-2400, 4 cores, 3.1 GHz and an SSD of 120 GB. Another last machine with similar specifications was bought but was never used because it was decided to put the database in a single machine, thus not creating a cluster. A photo of the machines is diplayed in Figure 4.2.

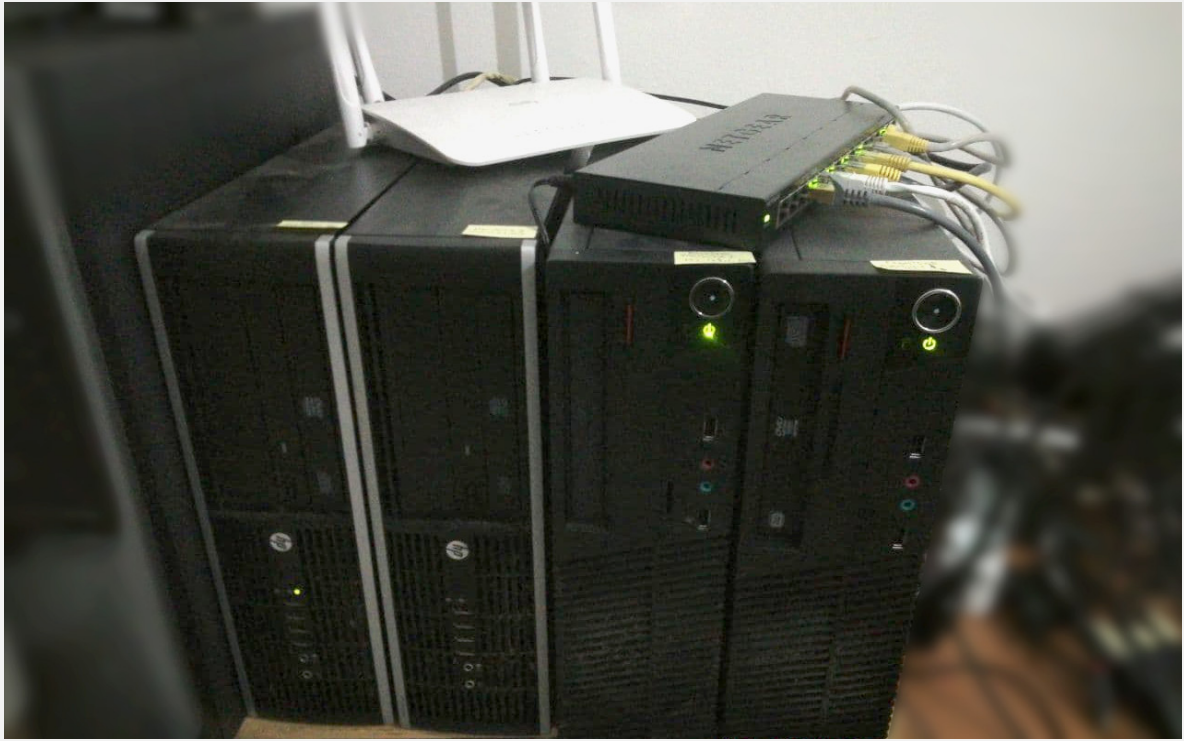

≡ Figure 4.2.: Photo of the computers, router and switch used for hosting the Web App API, frontend UI and the database

One machine is being used as a server for the frontend. A second machine is being used as a server of the API, the backend. A third machine hosts the database, an ArangoDB on single node. The fourth machine is used for development, deployment and remote access purposes.

### 4.1.2. The network

The network is setup in a home environment. The CUDY WR1300 router of MediaTek MT7621 architecture routes the connection from the main (and distant) home router to the network switch. The CUDY router is connected in client (bridge) mode to the main home router. Before setting up the network, CUDY's firmware was changed from the stock one to a new version and than to OpenWRT 19.07. This was made to have full control of the configurations of the router.

The switch, a NETGEAR Switch Ethernet Gigabit GS316, switches the connection from CUDY to all the machines.

### 4.1.3. Setup

In the home router some configurations are made to expose CUDY directly to the Internet. In CUDY static IP addresses and symbolic hostnames are assigned to DHCP clients (the computers). In the routers firewall port forwarding is setup to allow access to the server of the frontend of the website. Also a Dynamic DNS service is setup to help with the dynamic IP changes.

On the computers CentOS Linux 7 (Core) is installed, kernel version: `Linux 3.10.0-1160.el7.x86_64`. On the frontend server the usual configurations for serving a website are made. On the backend server Nodejs and `pm2` is installed. On the DB host ArangoDB is installed.

ArangoDB GDBMS is installed on one of the machines.

A dummy application is deployed on the machines - some example data is imported on an ArangoDB collection, a ready-made Rest API is setup on the backend server and a frontend interface is placed in the frontend server. A simple test visit from local network and from external network (using mobile data connection tethering) is done to the website's dyndns provided domain address. Then, on the dummy application's frontend some DB communication via API requiring actions are undertaken to see if everything is working as should between the machines, the API and the DB. Having everything setup, working and briefly tested, the environment is ready to host and serve.

## 4.2. The data

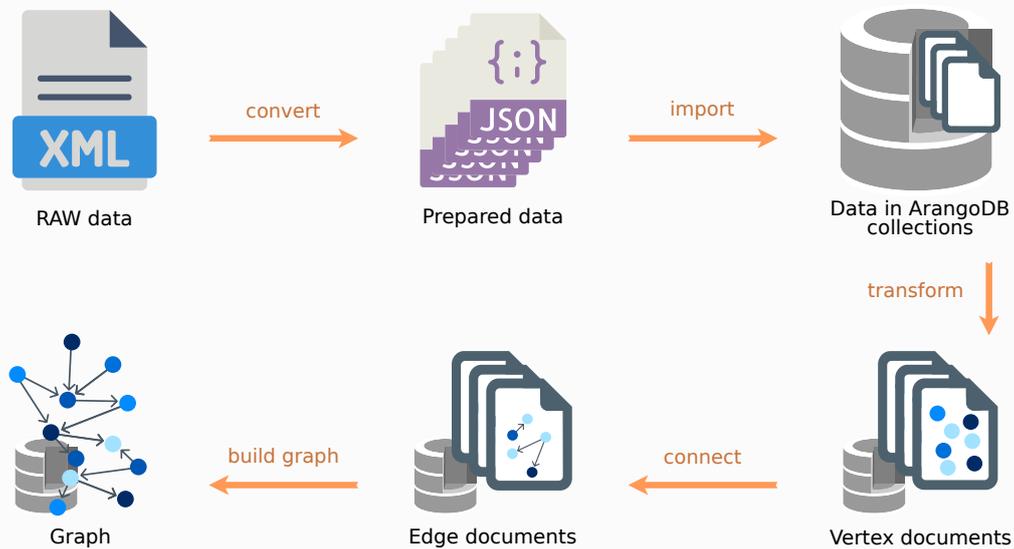

Figure 4.3.: The process of transformations of the data from dblp.org[1] downloading to graph building in ArangoDB DBMS

In this section are described all the steps carried from download of the data, their conversion, import in the database and reorganizing into vertices and edges afterwards. In the end the graph is built. See Figure 4.3 for a graphical representation of the process.

In the following subsection dblp.org[1] dataset download and data conversion is presented.

### 4.2.1. dblp.org[1] dataset download and data conversion

As mentioned a few times now, the dataset is downloaded from dblp.org[1] or more precisely from dblp.uni-trier.de[1]

The data, version of 2021-07-01 is a single .gz compressed archive of size 611MB. Once extracted, the output is a single XML file *dblp.xml* of 3.2GB in size. From the website is possible to download also some text files on changes, README snd documentation.

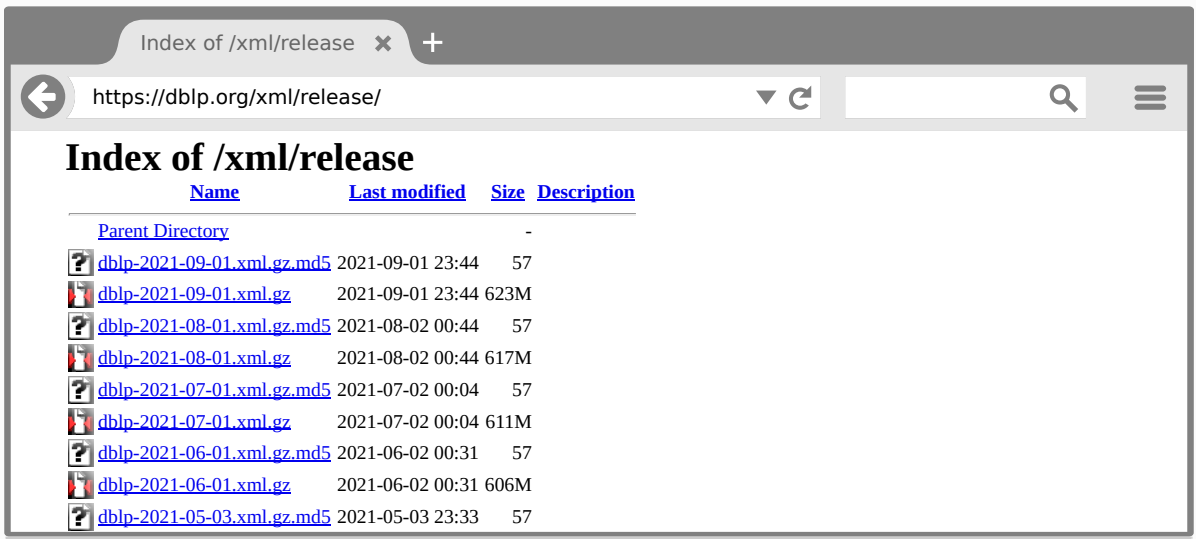

☰ Figure 4.4.: dblp.org[1] dataset download

As ArangoDB requires the data to be imported as JSON, more specifically single line JSON for a single document in a collection, the data has to be converted. In the following paragraph data conversion is explained.

**Data conversion**

Because the dataset format does not correspond to the one accepted by ArangoDB for data import, the data needs to be converted from XML to JSON line format.

```
1  <?xml version="1.0" encoding="ISO-8859-1"?>
2  <!DOCTYPE dblp SYSTEM "dblp-2019-11-22.dtd">
3  <dblp>
4  <article mdate="2017-06-08" key="dblpnote/error" publtype="informal">
5  <title>(error)</title>
6  </article><article mdate="2017-06-08" key="dblpnote/ellipsis" publtype="informal">
7  <title>…</title>
8  </article><article mdate="2017-06-08" key="dblpnote/neverpublished" publtype="informal">
9  <title>(was never published)</title>
10 </article><phdthesis mdate="2002-01-03" key="phd/Turpin92">
11 <author>Russell Turpin</author>
12 <title>Programming Data Structures in Logic.</title>
13 <year>1992</year>
14 <school>University of Texas, Austin</school>
15 </phdthesis>
```

☰ Code listing 4.1: An extract of dblp.org[1] dataset, top 15 lines

From a first analysis of the contents of the extracted XML is possible to notice that the file in many (significantly) is not well formatted. Many XML entries start right at the end of the previous ones without any empty line, or even without spaces - which makes it a hassle for parsers to directly parse the XML entries.

Another issue is that the extracted XML file is a little too large for a one go conversion. A splitting into smaller chunks is required too.

To solve these little problems, is proceeded in the following manner:

1. A Python script is written to clearly separate each XML entry from the previous and the next one with empty newlines.

2. Another script is written to split the big XML file into smaller chunks of valid (not cut in half) XML entries. The script splits to a new file every 1000000 lines (arbitrarily chosen). If the millionth line cuts an XML entry in half, it will include the rest of the lines needed to complete that last XML entry for that chunk.

3. A last script is written to convert each small chunk from XML to JSON. `xmltodict` library is used to do the parsing and dumping of the converted JSON string.

In Table 4.1 are shown the various scripts written to perform the needed data transformations. For more, in the appendices see § A.1 and § A.2.1 on page 117.

| Activity | Script's filename |
|---|---|
| XML format correction | `1_xml_to_xml_newlines.py`[a] |
| XML splitting into smaller chunks | `2_split_xml_to_many_smaller_xml_files.py`[b] |
| XML chunks conversion to line JSON entry chunks | `3_convert_xml_blocks_to_json_blocks.py`[c] |

Table 4.1.: Scripts written for format correction, data splitting and conversion from XML to line JSON

After having done the intermediate transformations and completed the conversion, is proceeded to the next phase: the importation of the line JSON data in an ArangoDB database' collection.

## 4.2.2. Importing the data

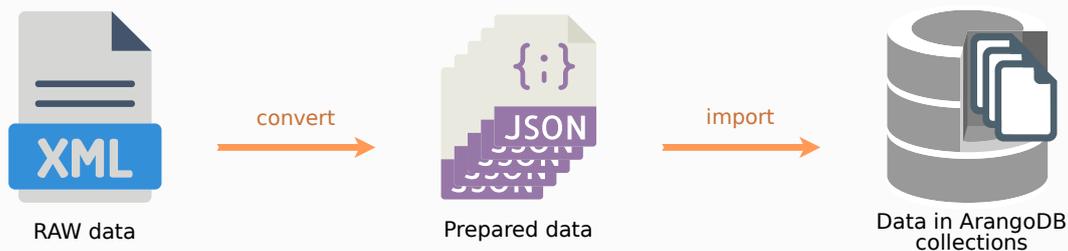

RAW data → convert → Prepared data → import → Data in ArangoDB collections

Figure 4.5.: Import of data in the database after conversion

In order to import the data in ArangoDB, firstly all the files containing JSON entries are copied to the DB host machine.

In Table 4.2 are shown the various scripts written to perform the movement of data files and then the import. For more, in the appendices see § A.1 on page 117 and § A.2.2 on page 118.

| Activity | Script's filename |
|---|---|
| Copy JSON files to database host machine | `1_copy_json_to_DB_host_machine.sh`[d] |
| Import in ArangoDB | `2_import_json_DB_to_arangodb.sh`[e] |

Table 4.2.: Scripts written for line JSON data import in ArangoDB

Firstly a new database is created in ArangoDB - if the data will not be imported in `_system`, the default autocreated DB during ArangoDB installation. The ArangoDB user declared in the command for the importing of the data (see Code

[a] github.com/A-Domain-that-Rocks/convert_large_xml_to_json/blob/main/1_xml_to_xml_newlines.py
[b] github.com/A-Domain-that-Rocks/convert_large_xml_to_json/blob/main/2_split_xml_to_many_smaller_xml_files
[c] github.com/A-Domain-that-Rocks/convert_large_xml_to_json/blob/main/3_convert_xml_blocks_to_json_blocks.py
[d] github.com/A-Domain-that-Rocks/arangodb_import_json_data/blob/main/1_copy_json_to_DB_host_machine.sh
[e] github.com/A-Domain-that-Rocks/arangodb_import_json_data/blob/main/2_import_json_DB_to_arangodb.sh

listing 4.2) must have permissions setup correctly to do the said operations. After that a collection is also created. All this is information is given as arguments to the `arangoimport` command in the terminal.

```bash
1 #!/bin/bash
2 # SSH to DB host machine and run:
3 for file in /remote/path/in/DB/host/where/to/store/json/data/*; do arangoimport --server.endpoint tcp
      ://127.0.0.1:8529 --server.username DB_USERNAME --server.database DB_NAME --file "$file" --type
      jsonl --collection COLLECTION_NAME --create-collection false --progress true --threads 4; done
```

Code listing 4.2: Bash command to import the JSON data ArangoDB using arangosh[126]

The command is executed and once finished importing, the data in the collection is in the state shown in Code listing 4.2. About 8.45 million documents are imported in total.

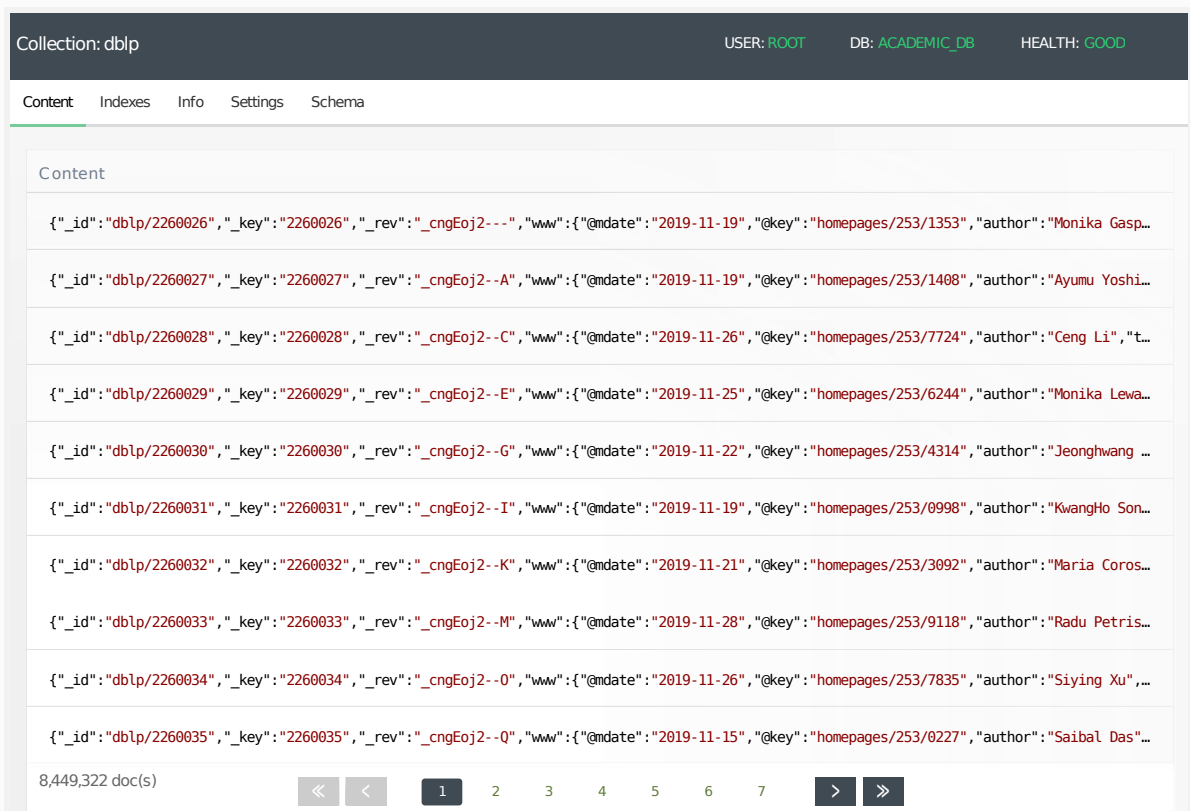

Figure 4.6.: Data in the dblp.org[1] collection after the importation has finished

For a graph to be built with the data at hand, vertices and edges have to be defined. In the next subsection are presented the data manipulations performed to distribute the documents in collection into separate collections of vertices and edges.

[126] ArangoDB (2021)
ArangoDB. *ArangoDB - arangoimport Documentation*. Online. Documentation. Aug. 2021. URL: https://www.arangodb.com/docs/stable/programs-arangoimport-details.html.

### 4.2.3. Transformations for vertex definitions

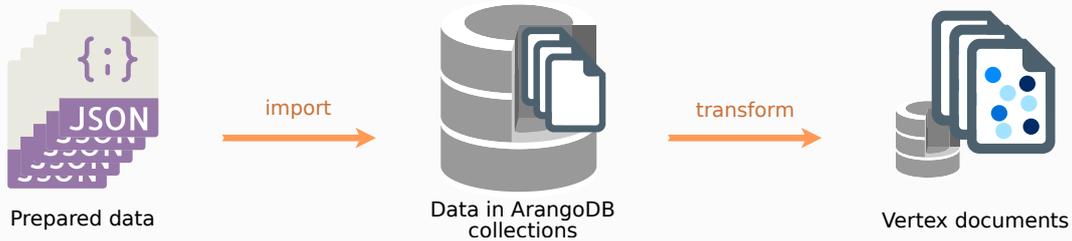

Figure 4.7.: Data transformations after import in order to obtain vertices

Documents in the dblp.org[1] collection have to be separated and distributed in different collections, each representing the kind of vertex they semantically are. For example, publication documents will be collected in a collection called Publication, being vertices of that type. Authors shall be collected in a collection called Author, affiliation institutions in Institution and so on.

Most of the vertices aren't directly extractable from the data in the current form (see Figure 4.8). Some preprocessing has to be done in order to get separate collections for each kind of vertex, without duplicates and with the correct attributes in a well structured JSON document format.

```
_id:   dblp/10709800
_rev:  _cngHde0--o
_key:  10709800
```

```
1  {
2    "article": {
3      "@mdate": "2020-02-21",
4      "@key": "tr/ibm/RJ1200",
5      "@publtype": "informal",
6      "author": [
7        "Dines Bjørner",
8        "E. F. Codd",
9        "Kenneth L. Deckert",
10       "Irving L. Traiger"
11     ],
12     "title": "The Gamma-0 n-ary Relational Data Base Interface Specifications of Objects and Operations.",
13     "journal": "Research Report / RJ / IBM / San Jose, California",
14     "volume": "RJ1200",
15     "month": "April",
16     "year": "1973",
17     "note": "republished on \"ACM SIGMOD Anthology\""
18   }
19 }
```

Figure 4.8.: Structure of documents from the imported data

The list of the files with the AQL queries for the transformations and vertex document creations are reported in Table 4.3). For more, in the appendices see § A.1 on page 117 and § A.2.3 on page 118.

During this process, in separated collections are created vertices of types: author, publication, affiliated_institution, journal, school, editor and publisher. Information on location was maintained as attributes of vertices that have it already. It was not transformed into a separate type of vertex since very rarely present and even when there was information on location, it was in unstructured format. Even in the case of affiliated institutions and schools, the case was somehow similar - even though less degenerated than locations with addresses and building numbers or only town names without any indication of countries...

| Vertex type | AQL query filename |
|---|---|
| Author | 02_create_author_nodes.aql[(f)] |

Table 4.3: AQL queries for the creation of vertex documents from the imported data. Continues on next page ...

(f) github.com/A-Domain-that-Rocks/distribute_data_in_arangodb/blob/main/02_create_author_nodes.aql

... continued from previous page

| Vertex type | AQL query filename |
|---|---|
| Publication | `03_create_publication_nodes.aql`[(g)] |
| Editor | `08_create_editor_nodes.aql`[(h)] |
| Publisher | `10_create_publisher_nodes.aql`[(i)] |
| Series | `12_create_series_nodes.aql`[(j)] |
| School | `14_create_school_nodes.aql`[(k)] |
| Journal | `24_create_journal_nodes.aql`[(l)] |

⊞ Table 4.3.: AQL queries for the creation of vertex documents from the imported data

⊞ **Remark 4.1 (on affiliated institutions and schools):** *Should be noted that affiliated institutions and schools are created as different vertex types. This is done on purpose because in many publications with numerous authors, some authors have attributes of affiliations different from the institution where the main author teaches. Other reason is for example when a conference is organized in an institution and its proceedings affiliated to that institutions - but all the authors participating that conference are affiliated to other institutions different from the conference organizer's one. Because of this, these two vertices were maintained separate even though semantically school and institution are interchangeable.*

Once all the transformations have been performed, special ArangoDB edge collections will be used to store the edge documents linking millions of vertices. This is the argument of the next subsection.

### 4.2.4. Connecting the vertices with edges

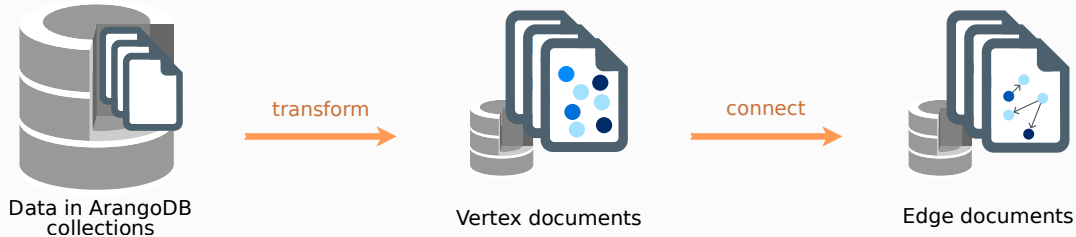

Data in ArangoDB        transform        Vertex documents        connect        Edge documents
collections

⊞ Figure 4.9.: Linking vertices with edges

In order to produce a graph, sets of vertices and edges linking them are needed. Till now, with the various data transformations, only sets (of different types) of vertices have been produced. To create connections between them, like "Author `HAS_PUBLISHED` Publication" or "Publication `WAS_CITED_IN` Publication" and similar - special ArangoDB edge collection are used. These collection force the use of documents with attributes specific to edges. An edge document part of an edge collection has `_to` and `_from` attributes, used for creating a connection between the `_id` of a vertex with one of another vertex.

The edge documents linking vertices are easily produced using AQL queries. The data involved are both the new vertex collections's produced in the previous subsection and the dblp.org[1] collection documents's.

The list of the files with the queries for the edge creations are reported in Table 4.4).

---
(g) github.com/A-Domain-that-Rocks/distribute_data_in_arangodb/blob/main/03_create_publication_nodes.aql
(h) github.com/A-Domain-that-Rocks/distribute_data_in_arangodb/blob/main/08_create_editor_nodes.aql
(i) github.com/A-Domain-that-Rocks/distribute_data_in_arangodb/blob/main/10_create_publisher_nodes.aql
(j) github.com/A-Domain-that-Rocks/distribute_data_in_arangodb/blob/main/12_create_series_nodes.aql
(k) github.com/A-Domain-that-Rocks/distribute_data_in_arangodb/blob/main/14_create_school_nodes.aql
(l) github.com/A-Domain-that-Rocks/distribute_data_in_arangodb/blob/main/24_create_journal_nodes.aql

| Vertex linked | to other vertex | AQL query filename |
| --- | --- | --- |
| Author | Publication | `16_create_edges_between_author_and` `_publication_nodes.aql`[m] |
| Editor | Publication | `17_create_edges_between_editor_and` `_publication_nodes.aql`[n] |
| Publisher | Publication | `20_create_edges_between_publisher_and` `_publication_nodes.aql`[o] |
| School | Publication | `21_create_edges_between_school_and` `_publication_nodes.aql`[p] |
| Journal | Publication | `36_create_edges_between_journal_and` `_publication_nodes.aql`[q] |
| Series | Publication | `37_create_edges_between_series_and` `_publication_nodes.aql`[r] |
| Affiliation institution | Author | `38_create_edges_between_affiliation_institution_and` `_author_nodes.aql`[s] |
| Publication | Cited publication | `39_create_edges_between_publication_and` `_cited_publication_nodes.aql`[t] |
| Publication | Crossreffed publication | `40_create_edges_between_publication_and` `_crossref-fed_publication_nodes.aql`[u] |

≡ Table 4.4.: AQL queries for the creation of edge documents linking pairs of vertex types

After having produced all the edges linking the vertices, the view of the vertex and edge collections in the ArangoDB's web interface is as shown in Figure 4.10. On the upper part of the figure are listed the vertex collections, with the exception of dblp.org[1], `pub_type` and `year` that are not used as vertices. Whereas on the lower part of the figure are listed the edge collections. The collections `author-author_isnot`, `pub_type-publication`, `year-publication` and `year-year` are not used. These edge collections were produced but their value for the detection of communities was deemed neglectable. For example, `author-author_isnot` links an author to another author whose names are similar. In the dblp.org[1] dataset this information is provided with an attribute of `isnot` taking as value the name of another author.

In the next subsection, making use of the vertices and edges produced till now, the graph construction is presented.

[m] github.com/A-Domain-that-Rocks/distribute_data_in_arangodb/blob/main/
16_create_edges_between_author_and_publication_nodes.aql
[n] github.com/A-Domain-that-Rocks/distribute_data_in_arangodb/blob/main/
17_create_edges_between_editor_and_publication_nodes.aql
[o] github.com/A-Domain-that-Rocks/distribute_data_in_arangodb/blob/main/
20_create_edges_between_publisher_and_publication_nodes.aql
[p] github.com/A-Domain-that-Rocks/distribute_data_in_arangodb/blob/main/
21_create_edges_between_school_and_publication_nodes.aql
[q] github.com/A-Domain-that-Rocks/distribute_data_in_arangodb/blob/main/
36_create_edges_between_journal_and_publication_nodes.aql
[r] github.com/A-Domain-that-Rocks/distribute_data_in_arangodb/blob/main/
37_create_edges_between_series_and_publication_nodes.aql
[s] github.com/A-Domain-that-Rocks/distribute_data_in_arangodb/blob/main/
38_create_edges_between_affiliation_institution_and_author_nodes.aql
[t] github.com/A-Domain-that-Rocks/distribute_data_in_arangodb/blob/main/
39_create_edges_between_publication_and_cited_publication_nodes.aql
[u] github.com/A-Domain-that-Rocks/distribute_data_in_arangodb/blob/main/
40_create_edges_between_publication_and_crossref-fed_publication_nodes.aql

Add Collection

affiliation_institution `loaded`

all_nodes `loaded`

author `loaded`

dblp `loaded`

editor `loaded`

journal `loaded`

pub_type `loaded`

publication `loaded`

publisher `loaded`

school `loaded`

series `loaded`

year `loaded`

affiliation_institution-author `loaded`

author-author_isnot `loaded`

author-publication `loaded`

editor-publication `loaded`

journal-publication `loaded`

pub_type-publication `loaded`

publication-publication_cited `loaded`

publication-publication_crossref `loaded`

publisher-publication `loaded`

school-publication `loaded`

series-publication `loaded`

year-publication `loaded`

year-year `loaded`

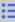 Figure 4.10.: Collections of vertices and edges in the ArangoDB web interface

### 4.2.5. Building the scientific publications' graph

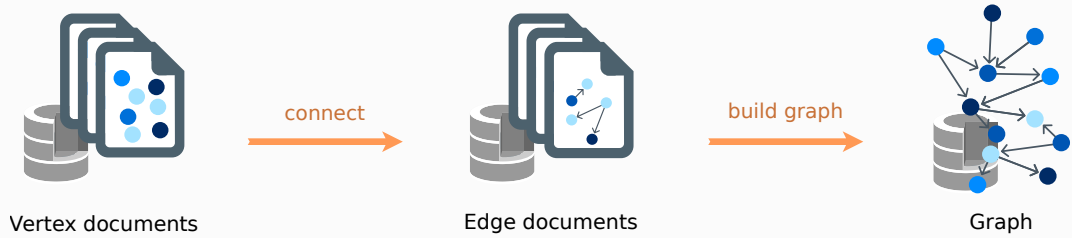

Using the edge collections produced in the previous subsection, it is possible to build the graph of the scientific publications provided by dblp.org[1]. ArangoDB obtains the set of vertices of the graph from the _to and _from fields of each edge document.

The final graph is shown in § 4.12. Arrived at this point, it is possible to detect the communities as described in § 3.3 on page 72 or go on with the development of the Web Application's backend API (see § 4.3 on page 90) and frontend UI (see § 4.4 on page 95).

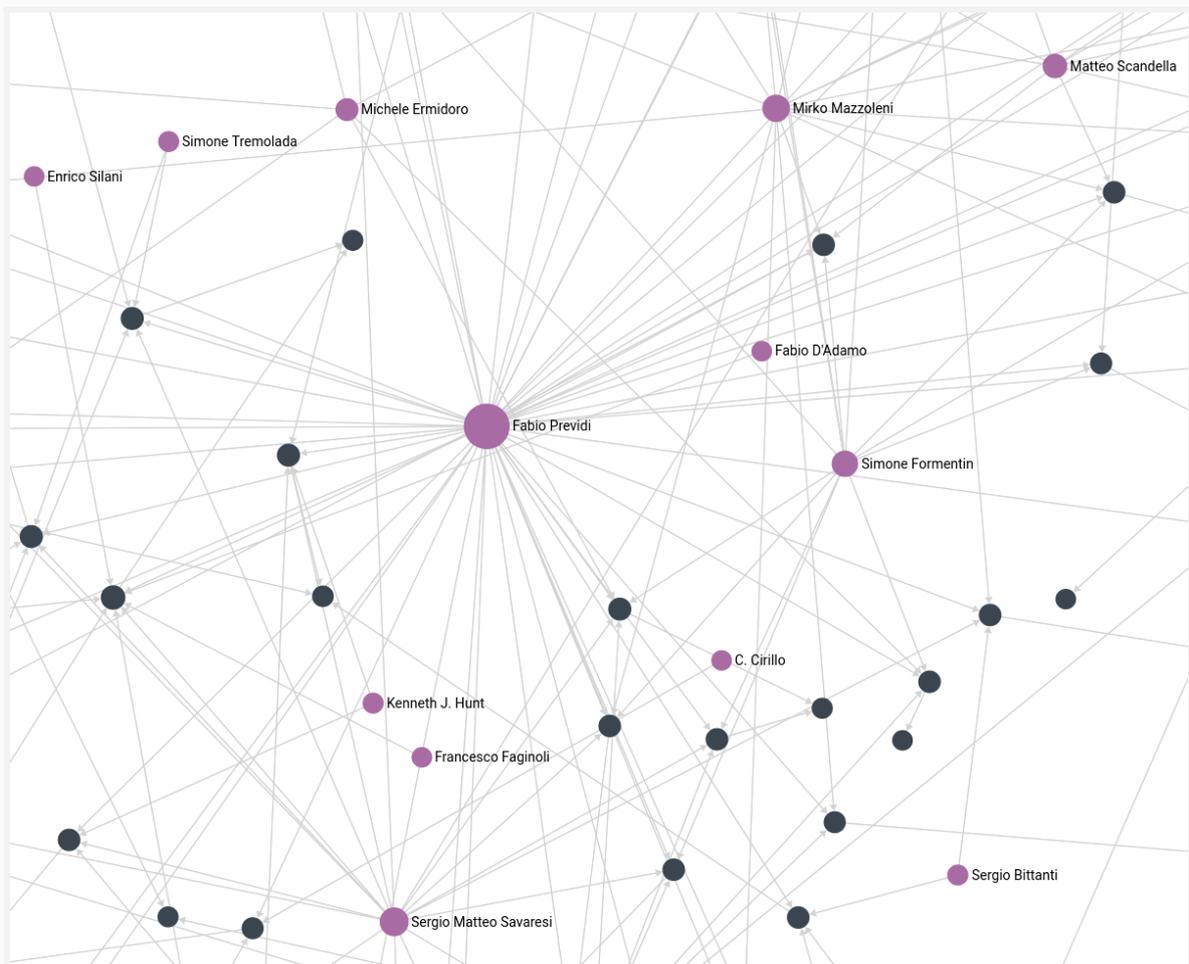

Figure 4.12.: The graph built after all the data manipulations - Vertices shown: Prof. Previdi, Mazzoleni, Ermidoro, Scandella and Prof. Bittanti

## 4.3. The API

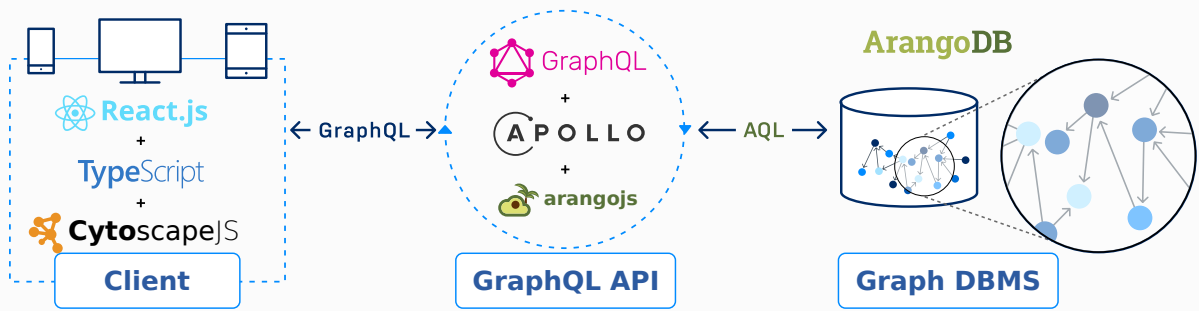

Figure 4.13.: Architecture of the Web Application

In this section is described the implementation of an initially Rest and then GraphQL API for querying the ArangoDB Graph Database from the website's frontend, or even other clients.

At its final state, the implemented server uses NodeJS, Express, GraphQL and Apollo Server to serve its functionalities.

### 4.3.1. Implementing a NodeJS & Express simple Rest API server

Initially, the backend of the Web Application is developed as a traditional Rest API. It is done this way because of the greater familiarity with this type of API and for getting then started straightahead with the development of the frontend interface.

NodeJS and Express are used. See § A.1 for source code and § A.2.4 for instructions on how to get the source code, run and deploy it.

The built API, at user request from the frontend or from any API querying client, responds with vertex IDs in the case of the search form with autocomplete suggestions - and with vertex and edge data when a specific graph is requested.

```
1 import { Database, aql } from "arangojs";
2 import dbConfig from '../config/db.js';
3
4 const db = new Database({
5     url: dbConfig.url,
6     databaseName: dbConfig.database,
7     auth: {username: dbConfig.username, password: dbConfig.password}
8 });
9
10 try { /* query db */ } catch (e) { /* handle exception */ };
11
12 export default service;
```
Code listing 4.3: Rest API's service

At this stage, no information on detected communities is sent in responses. Once the frontend shall be at a development stage to display compound nodes, detected communities' data will be included in the responses.

Once developed a minimal Web App consisting of DB + backend with Rest API + basic frontend, the API is upgraded to a GraphQL + Apollo Server API. This is presented in the next subsection.

## 4.3.2. Moving towards a GraphQL API

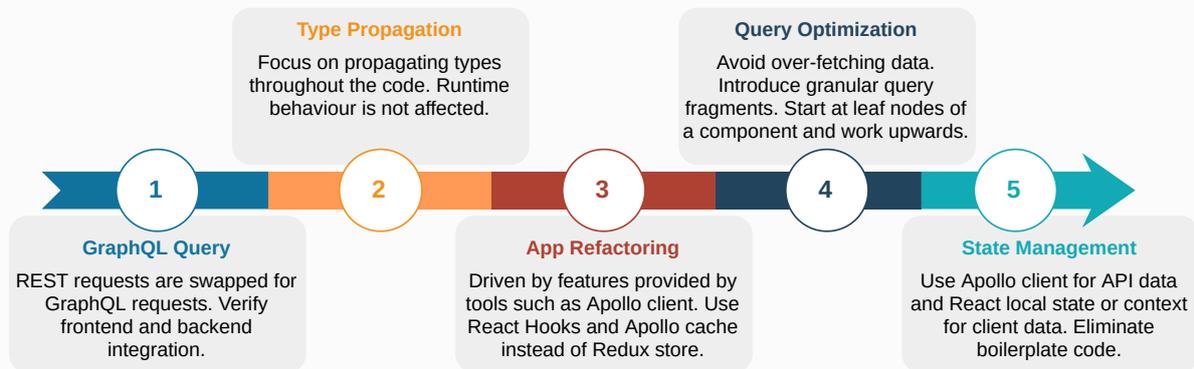

**Type Propagation**
Focus on propagating types throughout the code. Runtime behaviour is not affected.

**Query Optimization**
Avoid over-fetching data. Introduce granular query fragments. Start at leaf nodes of a component and work upwards.

**GraphQL Query**
REST requests are swapped for GraphQL requests. Verify frontend and backend integration.

**App Refactoring**
Driven by features provided by tools such as Apollo client. Use React Hooks and Apollo cache instead of Redux store.

**State Management**
Use Apollo client for API data and React local state or context for client data. Eliminate boilerplate code.

Figure 4.14.: A typical transition from Rest to an API with GraphQL and Apollo

GraphQL is an open-source data query and manipulation language for APIs. It also serves as a runtime for fulfilling queries. GraphQL was developed at Facebook and was publicly released in 2015. It is chosen to develop the API with it because of a variety of (good) reasons - having the word "graph" in its name is not one of them.

*"Ask for what you need, get exactly that"* - GraphQL

While Rest APIs require querying multiple URLs/routes for different data, GraphQL APIs get all the data in a single request. Differently from the many endpoints of a typical Rest API, GraphQL APIs make use of a type system consisting of types and fields to ensure an app only asks for what's offered and provide clear and helpful errors. Also using types avoids the need to write manual parsing functionalities.

Along with GraphQL, Apollo is used. Apollo Server is an open-source, spec-compliant GraphQL server that is compatible with any GraphQL client. It's the best way to build a production-ready, self-documenting GraphQL API that can use data from any source.

```
1  import express from "express";
2  import dotenv  from "dotenv";
3  import cors from 'cors';
4  import { resolvers } from './graphql/resolvers.js';
5  import { ApolloServer } from 'apollo-server-express';
6  import { db } from './config/db.js';
7  import expressPlaygroundMiddleware from 'graphql-playground-middleware-express';
8  import { typeDefs } from './graphql/schema.js';
9
10 dotenv.config();
11 const app = express();
12 app.use(cors());
13
14 const server = new ApolloServer({ introspection: false, playground: true, typeDefs, resolvers, db });
15 await server.start();
16 server.applyMiddleware({ app, path: '/graphql' });
17 app.get('/', (req, res) => res.end('Welcome to the API'));
18 app.get('/playground', expressPlaygroundMiddleware.default({ endpoint: '/graphql' }));
19 app.listen(process.env.PORT, () =>
20     console.log('API up and running, listening on port ${process.env.PORT}!'));
```

Code listing 4.4: Part of the code of the API server with GraphQL and Apollo

A GraphQL and Apollo API was implemented (with some basic resolver functions) and their integration with ArangoDB, a graph DBMS (at least for this thesis purposes) - was tested. Everything communicates flawlessly and the responses are as intended.

At migration concluded from the Rest API to the GraphQL API with Apollo, the code is as shown in Code listing 4.4. For the GraphQL schema with the type (and their fields) definitions, see Code listing B.1 in appendix § B on page 125. For the resolvers, implemented functions' source code and how to get them running, see appendix § A on page 118.

### 4.3.3. Resolvers

The API needs to handle two types of requests:

1. request for the ids (and names) of nodes named similarly to a given string;
2. graph data request (nodes, edges communities) given a vertex id and min, max number of hops.

Follows why and how these requests are handled.

```
1  import { getNodesIDByName, getNodeGraph } from './resolverFunctions.js';
2
3  const resolvers = {
4      Query: {
5          nodesID: async (root, args) => await getNodesIDByName(args.name),
6          nodeGraph: async (root, args) => await getNodeGraph(args.node_id, args.minDepth, args.maxDepth)
7      }
8  };
9
10 export { resolvers }
```

Code listing 4.5: GraphQL API's resolvers changed version after frontend implementation

#### 4.3.3.1. Autocomplete suggestion query for string to ID translation

In short, the user from the frontend inserts three inputs that indicate which entity's collaboration graph to display and how far (how many hops) from that entity to look for other entities (min and max depth). In terms of information to include in the query, this translates to a string representing the name or title of the entity/vertex (author, publication, institution etc.), a number for the minimum depth and another number for the maximum depth.

Since naming is a difficult problem, the string representing the name or title of the start vertex has to refer to exactly one vertex of the graph - otherwise the graph traversal cannot be performed since the startnode's identifying string would not be deterministic.

To solve this, it is needed to make a translation from the user's input string - that is a name, be it an author's name or a publication title or some institutions name - into an id corresponding exactly to one document in the database collections, the startNode vertex document. Making use of autocomplete suggestions in the search form, a user's input in the form field queries the API for vertices named in a similar way to that string. The API then queries the database sending a specific AQL query to obtain possible vertices that match the user's input. In the response to the frontend, the API sends the full names/titles of the vertices similar to the string the user wrote - and these vertices IDS. The search form then displays these elements to the user as a list of suggestions (selectable vertices) so he/she can select one. The selected vertex's ID is used, once the user has inserted the min and max depth too, to form an exact (deterministic) query that the API sends to the DB. That query is non-ambiguous since IDs are unique. It would not have been like that if names/title were used to refer to vertices.

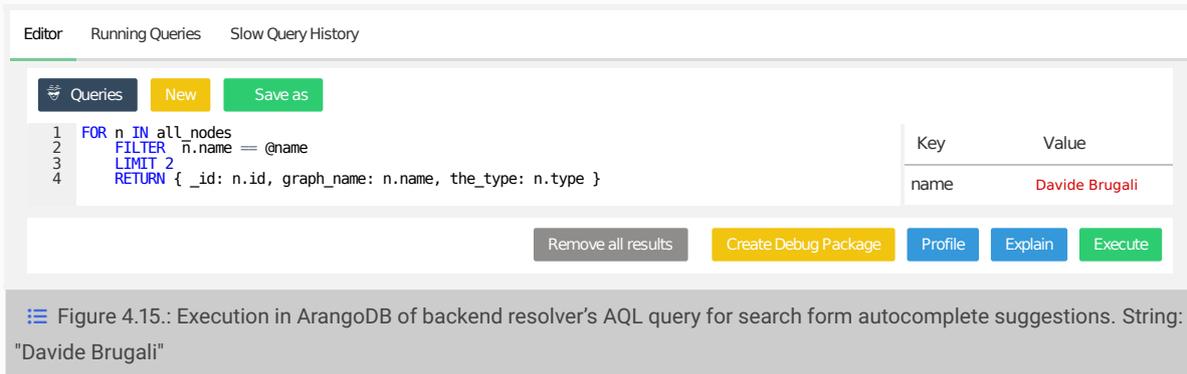

Figure 4.15.: Execution in ArangoDB of backend resolver's AQL query for search form autocomplete suggestions. String: "Davide Brugali"

In Figure 4.15 is shown the API's AQL query for autocomplete suggestions (translations from string name/title to unique ID of a vertex of the graph) executed directly in the ArangoDB database's web interface. The found results shown in Figure 4.16, in the same fashion are sent to the API - which then forwards them to the frontend interface.

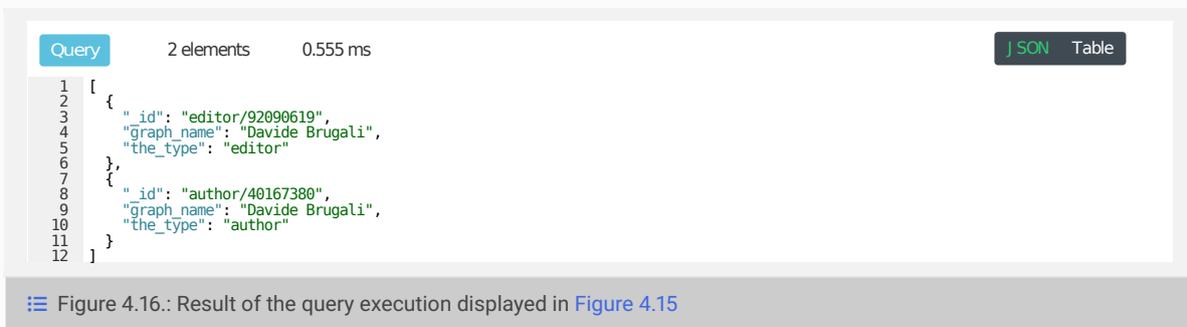

Figure 4.16.: Result of the query execution displayed in Figure 4.15

It is now clear why the needed queries (thus resolver functions) are two: one for name to ID translation just presented and the second one for requesting collaboration graph nodes, vertices and communities data. The second resolver is reported below.

### 4.3.3.2. Querying for graph data

With two resolvers and many types (see Code listing B.1 in appendix § B on page 125) defined as the different vertices (`author`, `publication`, `affiliation_institution`, `publisher`, `editor`, ...) - the API manages to get queries from the frontend, create AQL queries to send to the DB and forward the response data back to the frontend.

The graph data requiring query is handled by a resolver function that builds an AQL query using the translated ID and the depth inputs of the user and sends it to the database in order to perform a graph traversal.

In Figure 4.17, the ID of the author found in the previous query is used as input in this new query along with the min and max depth. The query returns the graph data consisting of the nodes, edges and communities (compound nodes) the derived number of hops distant from the start node. In the ArangoDB's web interface, when a query's result is made of nodes and edges, that result's data is sufficient for ArangoDB to automatically render a graph (see Figure 4.18). The response returned is of course in JSON format consisting of:

- a startNode vertex, corresponding to the vertex id given in input;
- a list of distinct vertices composing the graph;
- a list of the edges connecting the vertices of the graph;
- a list of the communities the nodes of the graph belong to.

These resolver functions are sufficient to provide the data needed to build graphs in the frontend.

Editor      Running Queries      Slow Query History

```
1   LET graph_data = (
2       FOR vertex, edge, path IN TO_NUMBER(@minD)..TO_NUMBER(@maxD)
3           ANY @aID GRAPH
              author_publisher_editor_journal_publication_series_affilia
              n_school_cited_crossreffed
4           RETURN {vertices: path.vertices[*], edges: path.edges[*]}
5   )
6   LET allVertices = UNIQUE(FLATTEN(
7       FOR el IN graph_data
8           RETURN el.vertices
9   ))
10  LET startN = (
11      FOR el IN allVertices
12          FILTER el._id == @aID
13          LIMIT 1
14          RETURN el
15  )
16  LET allEdges = UNIQUE(FLATTEN(
17      FOR el IN graph_data
18          RETURN el.edges
19  ))
20  LET allCommunities = UNIQUE(FLATTEN(
21      FOR el IN allVertices
22          RETURN { number: el.community }
23  ))
24  RETURN { startNode: startN[0],
25          vertices: allVertices,
26          edges: allEdges,
27          communities: allCommunities }
```

| Key   | Value           |
|-------|-----------------|
| minD  | 1               |
| maxD  | 2               |
| aID   | author/40167380 |

Remove all results      Create Debug Package      Profile      Explain      Execute

Figure 4.17.: Execution in ArangoDB of backend resolver's AQL query for graph data retrieval

Query      1 elements      41.316 ms                    JSON  Graph

Figure 4.18.: Result of the query execution displayed in Figure 4.17

In the next section is presented the development of the frontend interface and how the data from the API are used.

## 4.4.  The frontend

In this section are presented the implementation details of the frontend of the WebApp from its creation from a Typescript template to rendering graphs and communities within it.

### 4.4.1.  Creating a simple React & TypeScript project

As shown in , the frontend is made using React, TypeScript and Apollo Client - even though TypeScript is used not the way it is intended to be used. `Cytoscape.JS` library is used for graph rendering and `Bootstrap` for styling. See  for source code and  for instructions on how to get the source code, run, build and deploy it.

The features/functionalities/components the frontend interface needs are two:

- a search form component to give the user the possibility to search for some entities collaboration network,
- and a graph rendering component in order to display the graph that was searched for.

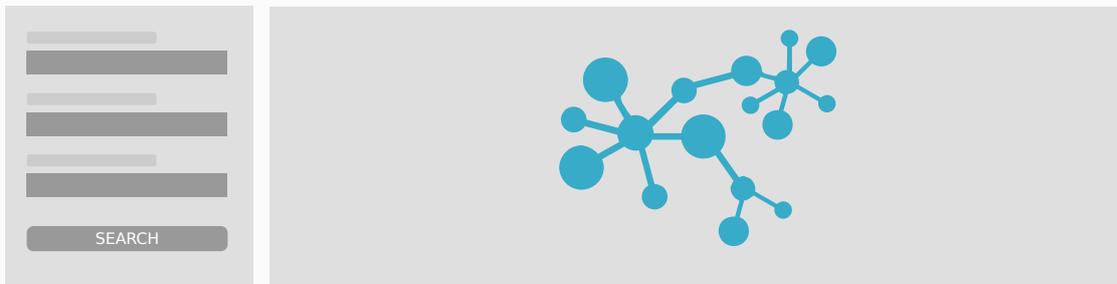

**Academic Graph Connections**

SEARCH

Made with ❤

Figure 4.19.: Layout of the frontend interface with Bootstrap: Header row, Content row made of Search Form column component and the Graph column component, Footer row

By making use of `Bootstrap` for React, `Header`, `Content`, `Footer` are defined and the `Content` is divided into two columns:

- one that will host the search form component,
- and the other - the area where the graph is going to be displayed;

as shown in . The ratios of the columns are about 1/4 for the search form and 3/4 for the graph area.

Having created the basis of the website with the placeholders for the components, it is time to present the development of the Search Form component.

### 4.4.2.  Search Form component

In explaining the implementation of the GraphQL API in  is briefly mentioned that the user will have to insert three values in order to perform a search and thus request the collaboration graph of an entity. These values are:

- a string text referred to the name or the title of the start node, the focal entity of the graph around which the collaboration community is built.
- the number of minimum depth/hops, that is the number of edge traversals from the start node and further - distant nodes to include in the graph.

- the number of maximum depth/hops, that is the number of edge traversals from the start node and nearer - distant nodes to include in the graph. The intersection of min and max depth makes it possible to include in the graph only the nodes belonging to that intersection.

For each of these three values there is a distinct form field to be filled. In the following paragraphs each of these fields is explained in detail.

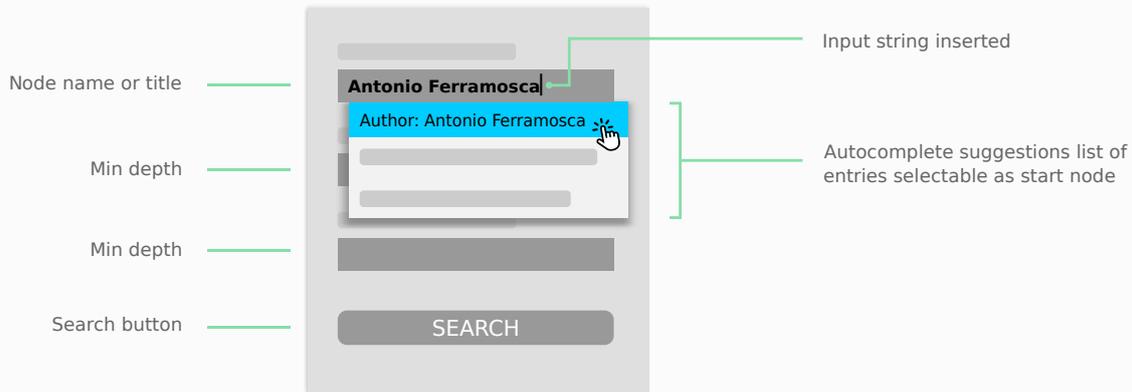

Figure 4.20.: Search Form with autocomplete suggestions: search string "Antonio Ferramosca"

### 4.4.2.1. Start vertex

Previously, presenting the implementation of the API, were mentioned the reasons why there is a need to translate the input string to an id. See § 4.3.3.1 on page 92.

The start vertex can be an author e.g.: "Antonio Ferramosca" (as shown in Figure 4.20) or an institution e.g.: "University of Bergamo" or whatever vertex type like title of a publication, publisher, journal, editor, series, institution etc. . To obtain the mapping of the input string from a humanly meaningful string to an id, an autocomplete suggestion feature is implemented. The user inputs the intended string representing the name or title of the vertex. That string is sent to the backend API, which performs a request to the database with a filter of all the nodes named that way. In Code listing 4.6 is shown the GraphQL query the frontend's Search Form Component builds and sends to the backend API.

```
103        return gql`
104            query {
105                nodesID${num}(name: "${searchV}") {
106                    ... on SuggestedNode {
107                        _id
108                        graph_name
109                        the_type
110                    }
111                }
112            }
113        `;
```

Code listing 4.6: GraphQL query in Search Form component for autocomplete suggestions

Once done, a list of possible vertices is brought back to the frontend and shown as a suggestion list. The list is a dictionary of the name of the vertex and its id. For the visual result, see Figure 5.2 in § 5.1 - Frontend search form, page 101. The user selects an item from the list and that gets saved as an id the state of the search form.

#### 4.4.2.2. Hops, minimum and maximum depth

The hops range is inserted as minimum and maximum depth of traversal. The default minimum depth is set to 1 and maximum depth is set to 2. A minimum and maximum depth of both selected as 1 requests all vertices directly connected to the start vertex. Whereas a minimum and maximum depth of both selected as 2 requests all vertices that are exactly at 2 hops distance from the start vertex. In the results of this search would not be present the vertices directly connected to the start vertex. If the minimum depth is set to 1 and the maximum to 2, then both the 1 hop and 2 hops distant vertices would be included in the results. The same logic follows for higher minimum and maximum depth values.

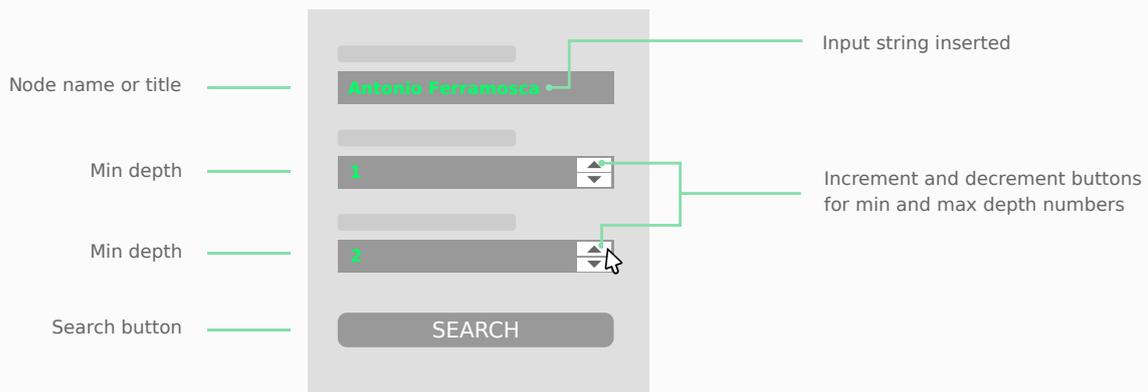

Figure 4.21.: Search Form's min and max depth values

**Remark 4.2 (on big depth values):** *Should be noted that a maximum depth of 3 produces a graph of about 250 thousand vertices. Users client machine might not be able to render or fit a graph with that many nodes. A maximum depth of 4 produces a graph with about 1 million nodes. It takes about 30 seconds for ArangoDB to perform that traversal. A maximum depth of 5 produces a graph with many million nodes. For ArangoDB it takes from 15 minutes to an hour to perform that traversal. It produces a response a few hundreds of MB total size, if not some GB. The WebApp is not designed to handle these cases. This situation is indicated as in need of a possible improvement in* § 6 - Conclusions ▸ § 6.2 - Possible improvements, *specifically in* § 6.2.1 - Limit maximum number of nodes in the frontend graph, *on page 114. Imposing a limit on the maximum number of nodes to obtain, some kind of an upper bound, would be a trade-off between completeness and usability but might still be a reasonable improvement in order to cover also the display of graphs with high values of depths.*

#### 4.4.2.3. Search inputs validation

The inputs are validated client side with some basic rules. If an item from the suggestion list is not selected, thus no id is indicated, the form is considered invalid - warnings in red are shown to the user.

If the minimum depth is not 1 or greater, the form is invalid. If the maximum depth is not greater or equal to the minimum depth, the form is invalid. In all other cases the search request for the graph gets sent to the backend, provided that an element is already selected from the autocomplete of the name/title. If the form is valid borders of the fields are colored in green.

**Remark 4.3 (on server side validation):** *Should be noted that unvalidated queries from the frontend can be sent to the API through other clients and no check is performed in the backend. For example in the case of very big values of depths, the database received the query and tries to do its best to compute a result and return it. But sometimes, especially when the query computation overloads the DB host machine, this leads to an automatic interruption of the* *connection between the API and the ArangoDB DBMS and the cancellation of all ongoing query computations.*

*It is outside the goals of this work to implement security features, thus server side validation was neglected.*

Having described the features of the search form, it is time to present the graph rendering component that would display the collaboration networks. Before that, how the collaboration graph data get requested is presented.

### 4.4.3. Requesting collaboration graph data

In a similar fashion to how the GraphQL API was queried for vertex suggestions, is queries also for the collaboration graph data. The only change is that now the response data shall contain types of elements that are not anymore simple or primitive.

When querying about suggestions, the response was an array of strings. While for graph data the response shall be composed of arrays of objects having diffrent attributes. To this end, GraphQL comes to help with its type system. In Code listing 4.7, lines 22-24 inline fragments (spread operator) are (is) used to access data on the underlying concrete type.

```
16        const getGraphDataQuery = gql'
17            query {
18                nodeGraph(node_id: "${nodeId}",
19                          minDepth: "${String(minDepth)}",
20                          maxDepth: "${String(maxDepth)}") {
21                    startNode   { _id graph_name }
22                    vertices    { ... on SlimNode  { _id graph_name community }}
23                    edges       { ... on SlimEdge  { _from _to label}}
24                    communities { ... on Community { number }}
25                }
26            }
27        ';
28        setIsLoadingGraph(true);
29        const result = await myApolloClient.query({ query: getGraphDataQuery });
```

≡ Code listing 4.7: GraphQL query to request graph data from the API

With the graph data obtained from the backend, it is now possible to start drawing graphs in the dedicated area. In the following subsection the graph rendering component's features are brought.

### 4.4.4. Graph rendering component

There exist many different JavaScript graph visualization libraries with different features. To show a network graph with simple and compound nodes (used for communities), Cytoscape.JS `Cytoscape` was deemed appropriate.

Cytoscape.JS is an open-source graph theory library written in JavaScript. It makes developer's life easier for displaying and manipulating rich, interactive graphs.

Cytoscape is domain independent but usually when used by social scientists, it is employed to visualize and analyze large social networks of interpersonal relationships. It has many extensions and is able to perform analytical computations. It even provides different clustering algorithms, such as the ones shown in Table 4.5

| Attribute Cluster Algorithms | Network Cluster Algorithms |
| --- | --- |
| | – Affinity Propagation |
| – AutoSOME Clustering | – Connected Components |
| – Creating Correlation Networks | – Community Clustering (GLay) |
| – Hierarchical Clustering | – MCODE |
| – K-Means Clustering | – MCL |
| – K-Medoid Clustering | – SCPS (Spectral Clustering of Protein Sequences) |
| | – Transitivity Clustering |

Table 4.5.: Clustering algorithms provided by Cytoscape

Cytoscape.JS includes all the gestures expected out-of-the-box, such as pinch-to-zoom, box selection, panning, etc.[27,127]

In order to render a graph, it needs just - as one would expect - a set of vertices and a set of edges. More details follow in the next subsubsections.

## 4.4.5.  Sets of vertices and edges

```
107    let constructed_node = {
108        data: {
109            id: n._id,
110            label: the_name,
111            parent: the_community
112        },
```

Code sublisting 4.8 (a): Vertices with community stated as parent node

With Cytoscape, a vertex in the set of vertices has to have an `id` property and a `label`. The `id` must be distinct. If it belongs to or is inside a compound node as shown in Code sublisting 4.8 (a) it also has a parent attribute. For each node, styling of the shape, size and color or text format - is applicable. On vertex click, vertex information is shown in a pop-up.

An edge on the other hand has a `source`, a `target` and a `label` attribute as shown in Code sublisting 4.8 (b). The `source` and `target` values have to be `ids` present in the set of vertices. Edges too can be styled with properties like color, thickness etc.

```
137    let constructed_edge = {
138        data: {
139            source: e._from,
140            target: e._to,
141            label: 'edge'
142        },
```

Code sublisting 4.8 (b): Edges with source and target vertices

## 4.4.6.  Sets of communities and compound nodes

Using the `CoseBilken` layout of Cytoscape.JS, it is possible to properly fit and render compound nodes, that is the community nodes acting as parents for the vertices that belong to them. Compound nodes have as their properties an `id` and a `label` as shown in Code sublisting 4.8 (c). The `id` must be distinct between all nodes. Since the same community generally appears many times (as parent of different vertices) in a collaboration graph, it may happen that the communities the API provides are included multiple times in the response, as many as the number of vertices belonging to that community. This generates errors with Cytoscape.JS (`ids` must be unique) and is avoided by changing the APIs AQL query to the database in order to request a list of distinct communities.

[127] PLOTLY (2021)
Plotly. *React Cytoscape react-cytoscapejs*. Online. Documentation. Aug. 2021. URL: https://github.com/plotly/react-cytoscapejs.

```
149          let constructed_community = {
150              data: {
151                  id: c.number,
152                  label: 'Community ' + c.number
153              },
```

Code sublisting 4.8 (c): Communities, the compound nodes

In the end, all elements of the collaboration graph are feeded together to Cytoscape like in Code sublisting 4.8 (d) and the library handles the positioning of the elements in the drawing area. Initially set to random positions, using gravity and repelling forces between the elements, it is possible to fit them all in a nice graph.

```
166
167          const graphElements = [...mappedNodes, ...mappedEdges, ...mappedCommunities];
168
```

Code sublisting 4.8 (d): All graph elements joined using spread operator expansion

Code listing 4.8: Cytoscape Graph Component source code snippets for the definition of the constituting elements of the collaboration graph

Having seen the community detection algorithm execution and now the implementation of the Web App, in the next chapter are going to be presented some concrete views of different collaboration graphs visualized in the Web App's frontend UI.

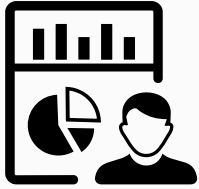

# 5. Display of the results

**This chapter's contents:**

In this chapter are presented the results post implementation of the WebApp.

## 5.1. Frontend search form

This first section is about the searching of a start vertex from the user inputting node names or titles and the number of hops from that vertex to all the other vertices that shall be shown in the graph.

In Figure 5.2 is displayed the search process of a node, in the specific case of the name of professor Gargantini, University of Bergamo.

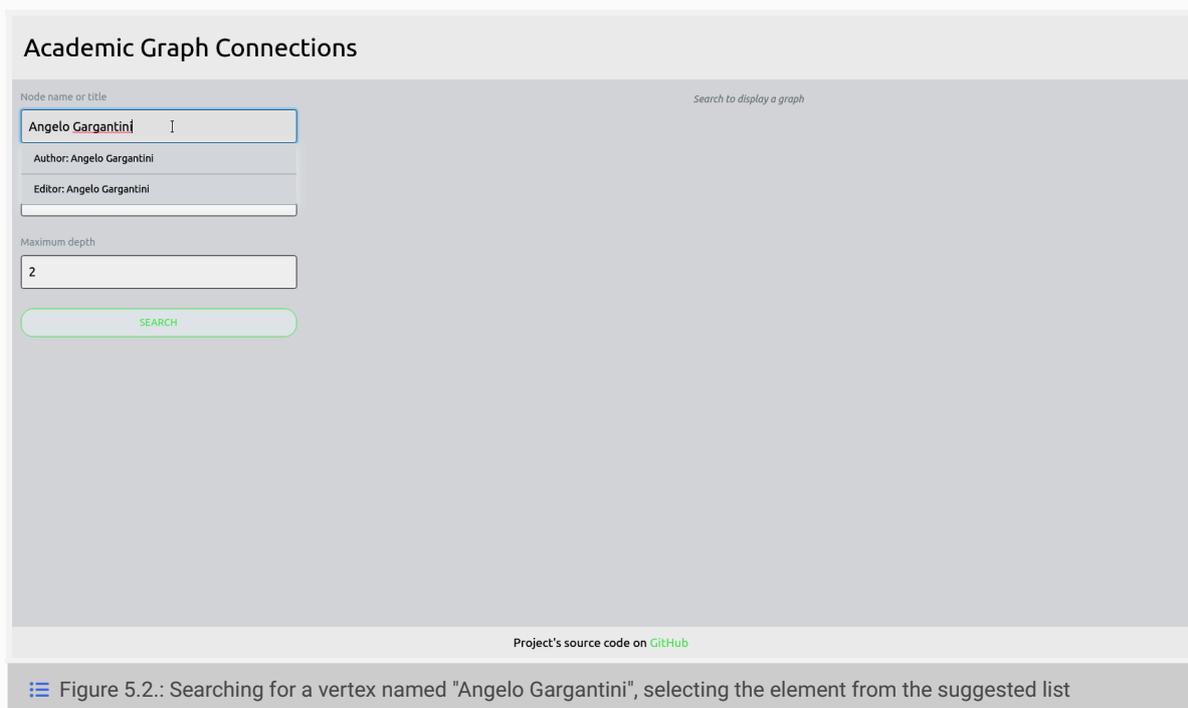

Figure 5.2.: Searching for a vertex named "Angelo Gargantini", selecting the element from the suggested list

The user inputs a string and in a moment a suggestion list of nodes with that name or title pops up. Supposing to have selected the author element of the pop up list, it is possible to input the values for the other two fields of the search form.

By default are set to 1 and 2 the number of hops, that is the minimum depth and maximum depth. These values indicate the distance from the start node (chosen in the first field of the search form) of the vertices to be included in the graph.

In Figure 5.3 is visualized the webpage just after having clicked on the search button. At this precise moment the GraphQL API is queried and the database is performing the needed calculations.

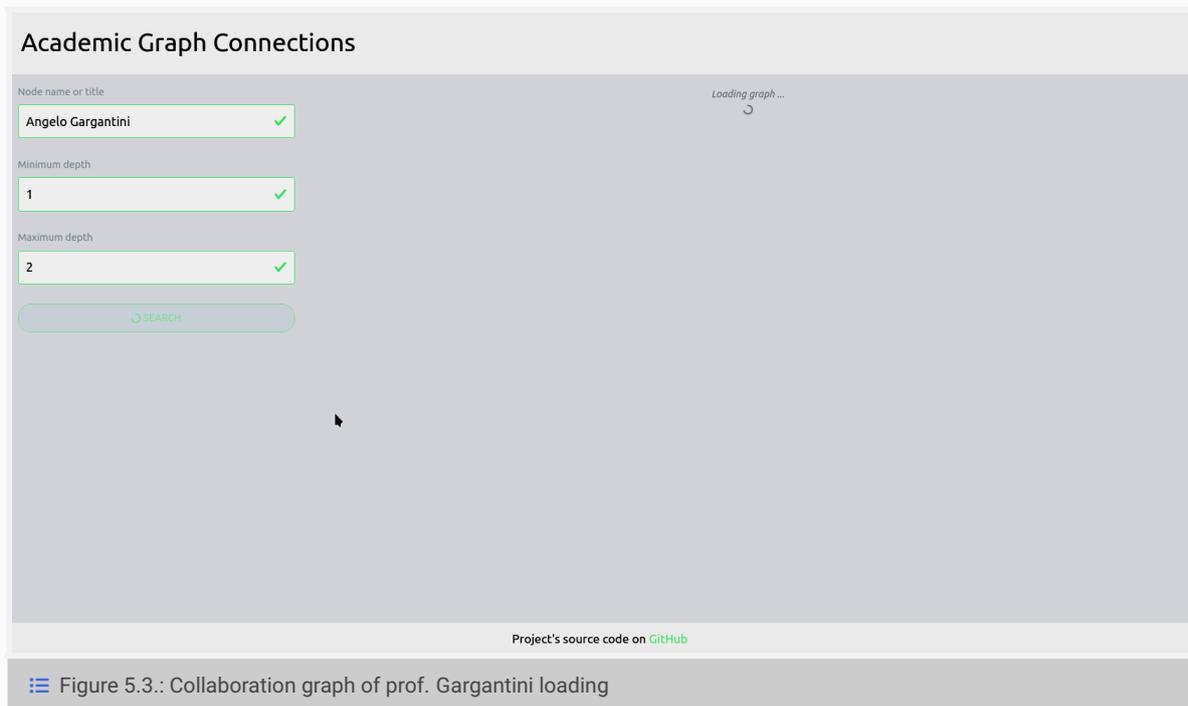

Figure 5.3.: Collaboration graph of prof. Gargantini loading

In the next section the loading graph's results are visualized. The results are limited to just a frontend, client interface point of view since that is the main usecase of the developed WebApp. What happens behind the scenes with the querying and the traversal computation to show the results, is not treated here. For a more detailed explanation of how everything is integrated together see § 3 - Community Detection on page 65 and § 4 - Implementing the WebApp on page 79.

## 5.2. Frontend graph visualization

Once the database has traversed the graph from the start vertex to all the other nodes at distance 1 to 2, the list of the visited nodes, of the edges connecting them and of their communities gets sent back from the API to the frontend. The graph rendering libraries use this data to fit all the nodes in a canvas for the user to easily navigate on.

In Figure 5.4 (and in Figure 5.5 focused on the canvas) is rendered the collaboration graph searched for, with the detected collaboration communities highlighted in rectangular compound nodes.

There are many communities shown because some of the researchers with whom prof. Gargantini has collaborated are more tightly connected to other groups of researchers, which means they belong to some other collaboration community.

Some communities are displayed having only one node. This happens because many of the other nodes of those communities are not in the range 1 to 2 - minimum and maximum depth, distance from the start node that was searched at the beginning in the shown example.  While it makes sense from the theoretical point of view to have 1-member communities, this is not what is happening in this case.

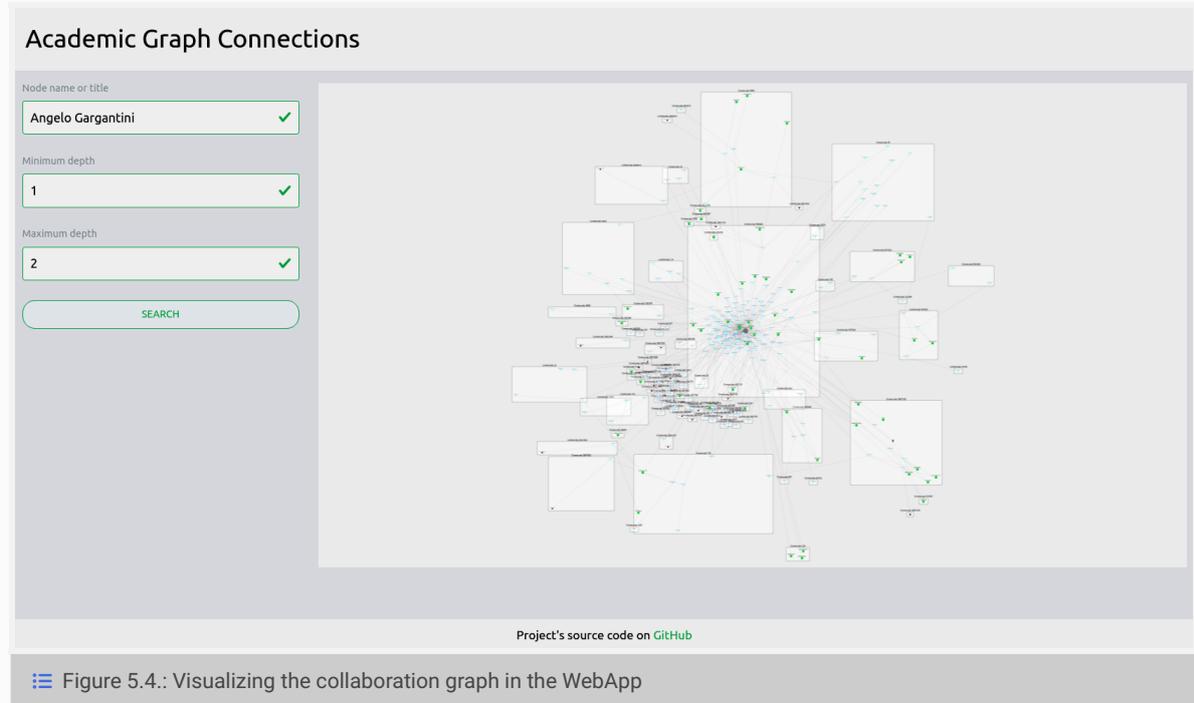

Figure 5.4.: Visualizing the collaboration graph in the WebApp

A consideration worthy to be made, in view of possible improvements treated in § 6.2.2 on page 114 - is the possibility of having the compound nodes (community nodes), collapsible and expandable.

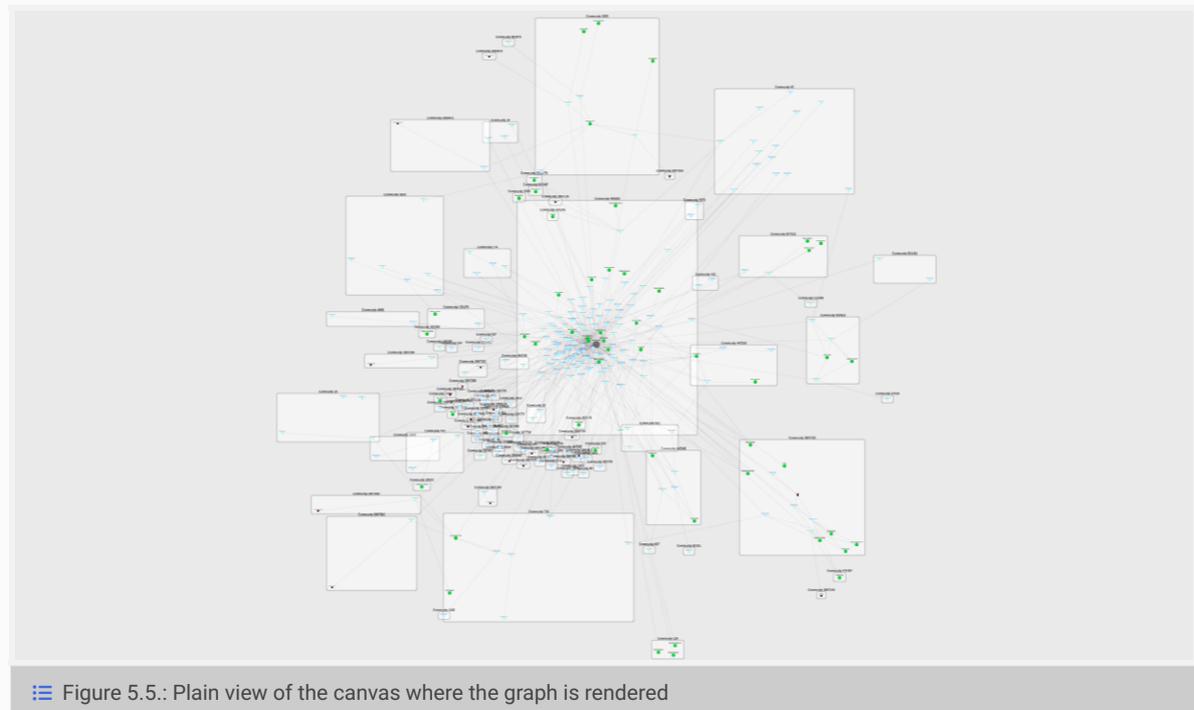

Figure 5.5.: Plain view of the canvas where the graph is rendered

In the next section detail level zoomed graphs are shown and their features are interpreted.

## 5.3. Detailed view of some examples of graphs of collaboration communities

### 5.3.1. Prof. Angelo Gargantini's cluster of research collaboration

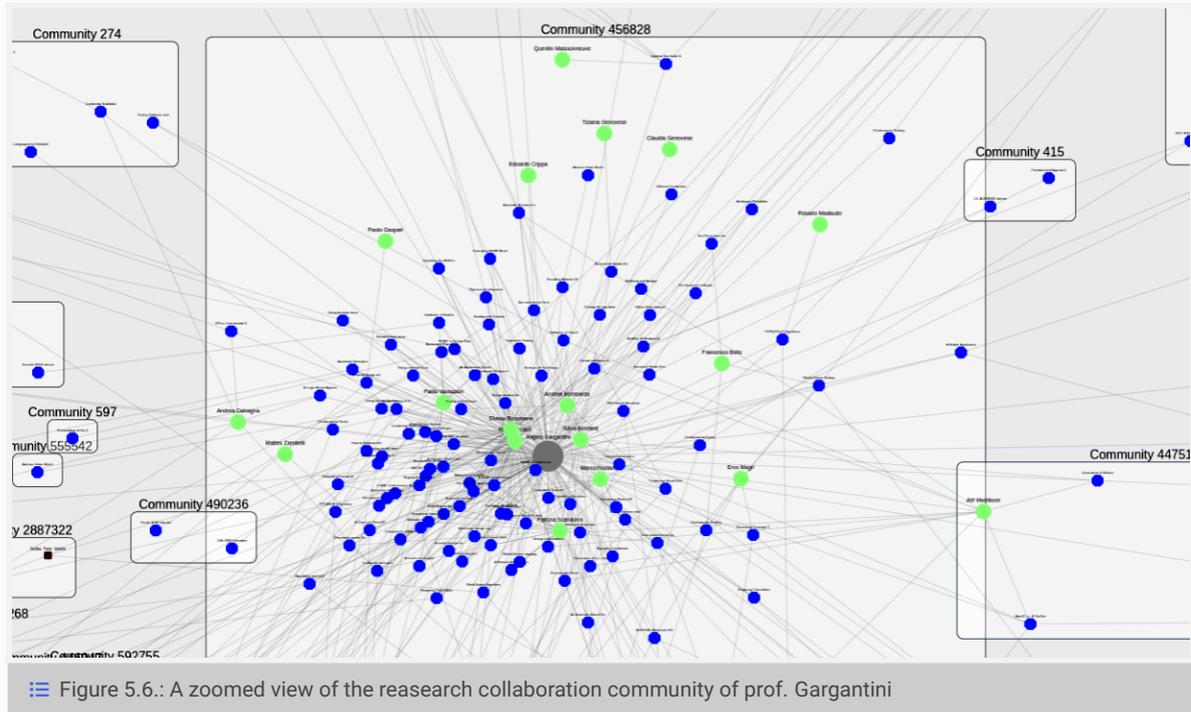

Figure 5.6.: A zoomed view of the reasearch collaboration community of prof. Gargantini

In the last searched graph, zooming in again in the middle where the start node is placed (author: Angelo Gargantini) - like in Figure 5.6, it is possible to see the dense relations (edges) between the vertices. The dark blue nodes are publication works, while the light green ones are other peer researchers.

This panoramic is interesting because it makes it possible to grasp a sense of boundary of the community even without the rectangle highlighting it. The transition from densely to sparsely linked is noticeable around the main group of nodes. The further is gone from its gravity center, the fewer the number of connections (edges) between this group's nodes and the others around it.

Zooming in again, this time to a level of detail able to read the names of the nodes like in Figure 5.7, some other considerations can be made. It is possible to see that some of the most tightly linked nodes of prof. Gargantini are indeed his real life collaborators, like his PhD students or his university colleagues, professors.

In the next subsection is considered one of the nodes that appeared in prof. Gargantini's collaboration graph, his colleague prof. Patrizia Scandurra.

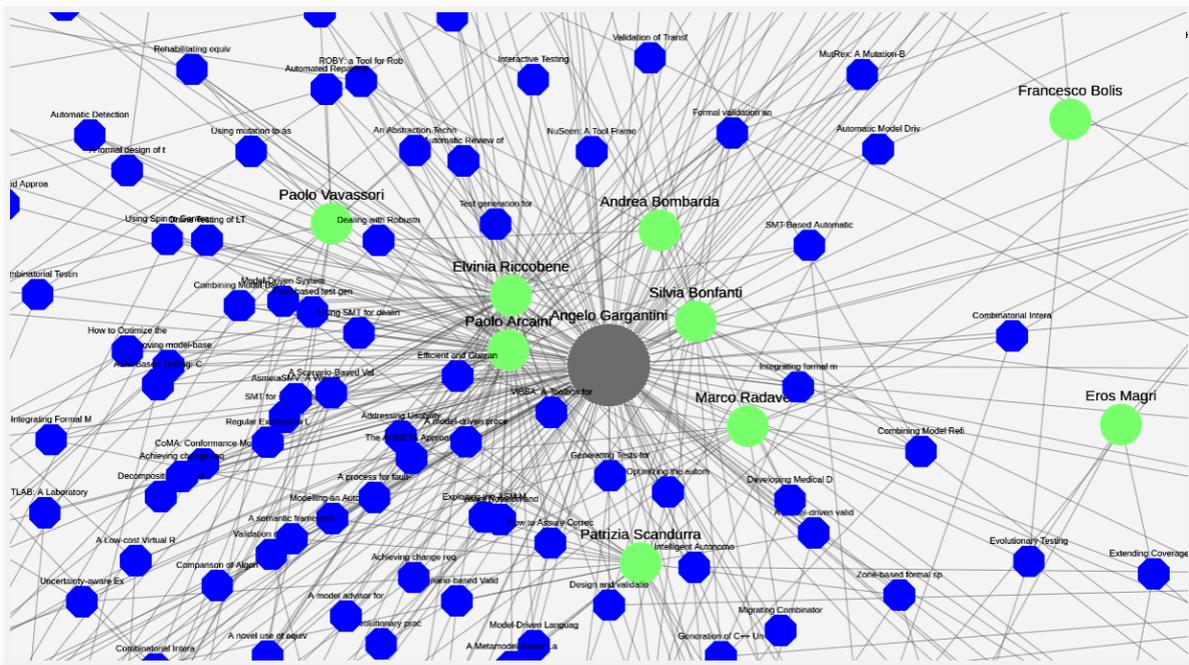

Figure 5.7.: Detail level zoom of the collaboration graph of prof. Gargantini and its peers, related publications

## 5.3.2. Prof. Patrizia Scandurra's research collaboration cluster

After having searched for the graph of prof. Scandurra, the zoomed view is the one shown in Figure 5.8. As expected, since prof. Scandurra appeared in prof. Gargantini's collaboration community graph, also the viceverca is true.

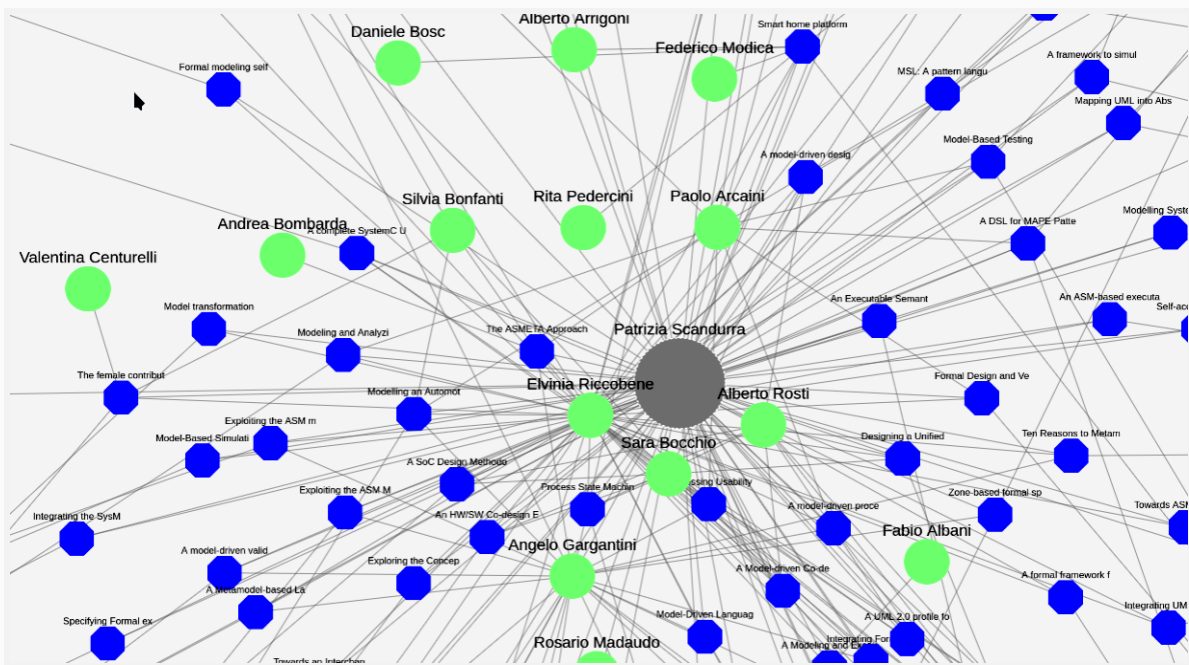

Figure 5.8.: Detailed view of the collaboration graph of prof. Scandurra and of its peers, publications

Furthermore, both of them have many researchers in common in their research collaboration graphs. In a sense,

collaborating together and having many connections in common leads to being detected by the algorithm as part of
the same community.

Worth noting here is the fact that in the graph of prof. Gargantini the dominating color was the dark blue, which
means a lot of publications (dark blue vertices) and fewer collaborations with other authors (light green vertices) -
while in the graph of prof. Scandurra the situation is more balanced. Qualitatively it might be possible to say that the
number of prof. Scandurra's connections to other authors over her total number of connections is greater than that
of prof. Gargantini's. While the opposite can be said about the number of connections to publication works.

In the next subsection is considered the case of prof. Psaila, who even though is part of the same educational
community (University of Bergamo) as prof. Gargantini and prof. Scandurra, is not detected as being part of the same
community of academic collaborations of the two.

### 5.3.3. Prof. Giuseppe Psaila's graph of collaboration community

In this section is presented the collaboration graph of prof. Giuseppe Psaila. The graph of the community he
belongs to is shown in Figure 5.9

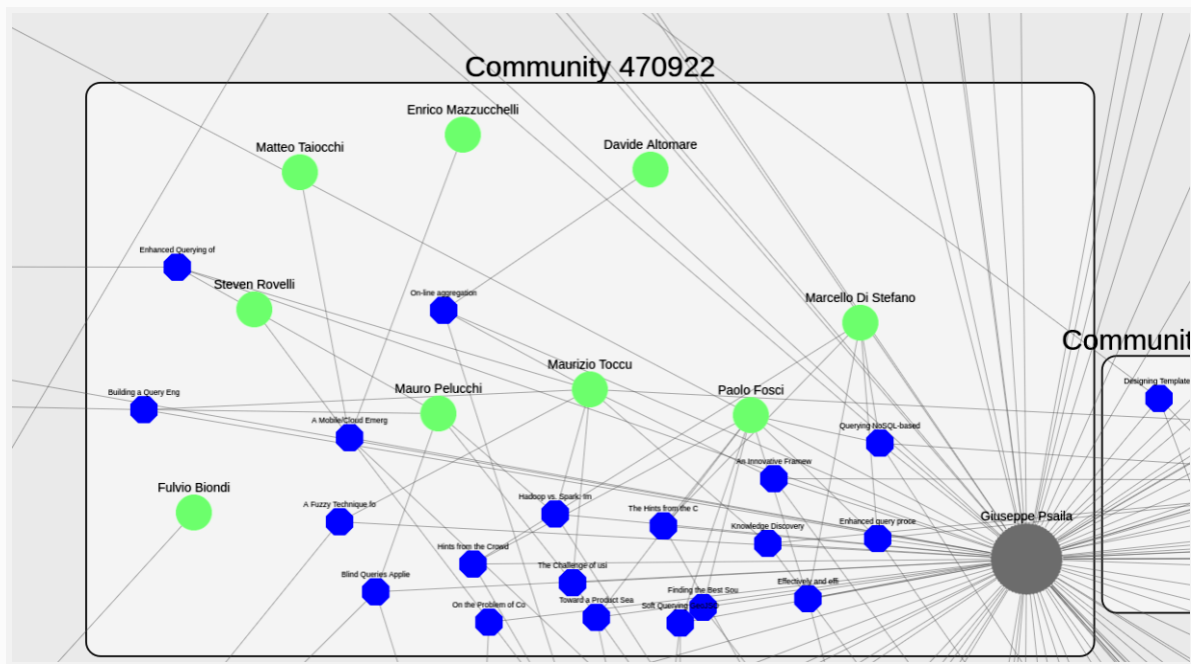

Figure 5.9.: Detailed view of the collaboration graph of prof. Psaila and of its peers, publications

Prof. Psaila is detected to be part of an academic collaboration community totally different from those of pro-
fessors Gargantini and Scandurra. Even though all three are part of the same institutional environment, that is the
University of Bergamo and also part of the same faculty, departments, prof. Psaila's community is a separate one.
This has to do with fewer connections, in other words collaborations between him and the other two professor's com-
munity vertices. Prof. Psaila collaborates with a totally different group of vertices, thus it makes sense from the point
of view of the interconnections density between different groups, to be detected as part of a distinct cluster.

Something interesting to mention here is the concept of overlapping communities that will be considered in § 6
- Conclusions ▸ § 6.3 - Further avenues of exploration, specifically in § 6.3.3 - Exploring overlapping communities
detection on page 115. Even though prof. Psaila's community is a different one from the other two professors's, it
might be interesting to consider the possibility of having overlapping parts of the different detected communities.

In the next subsection is shown an example of a vast graph of collaboration, that of prof. Paraboschi's.

### 5.3.4. Prof. Stefano Paraboschi's community of research collaboration

In this subsection is considered the situation of a pretty vast graph of academic collaborations that are detected as fragmented in many smaller communities. That is the case of the graph of prof. Paraboschi, shown in Figure 5.10.

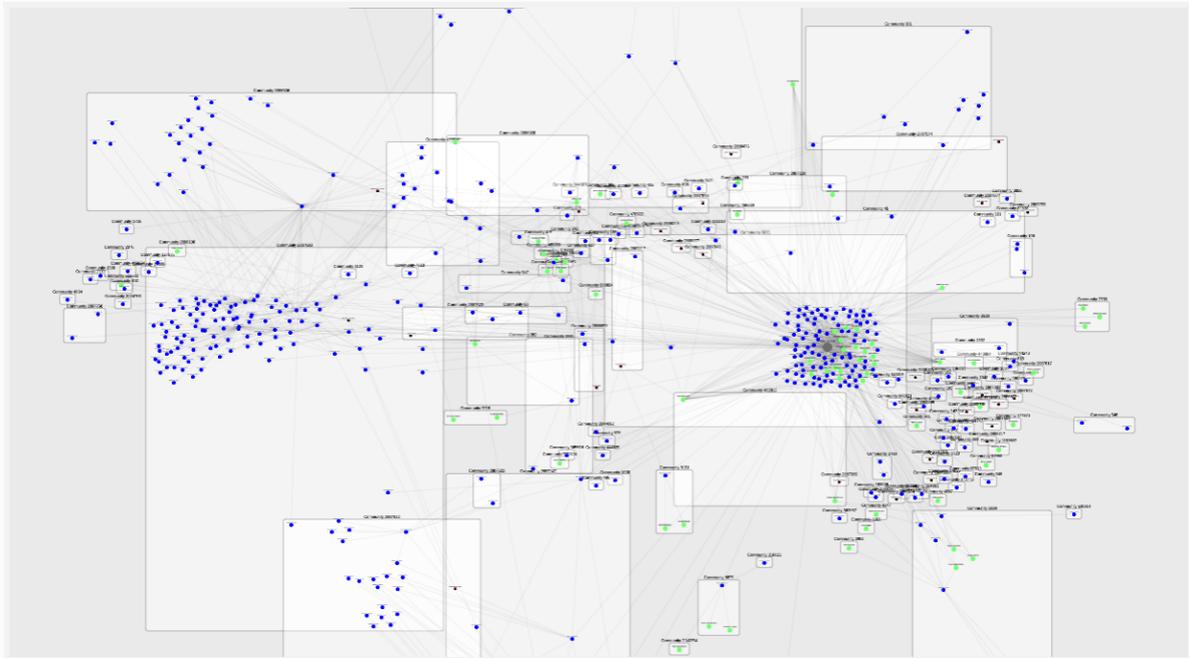

Figure 5.10.: Panoramic view of the collaboration graph of prof. Paraboschi

It is visually possible to see how wide the graph is, with lots of locally sparsed groups of subcommunities.

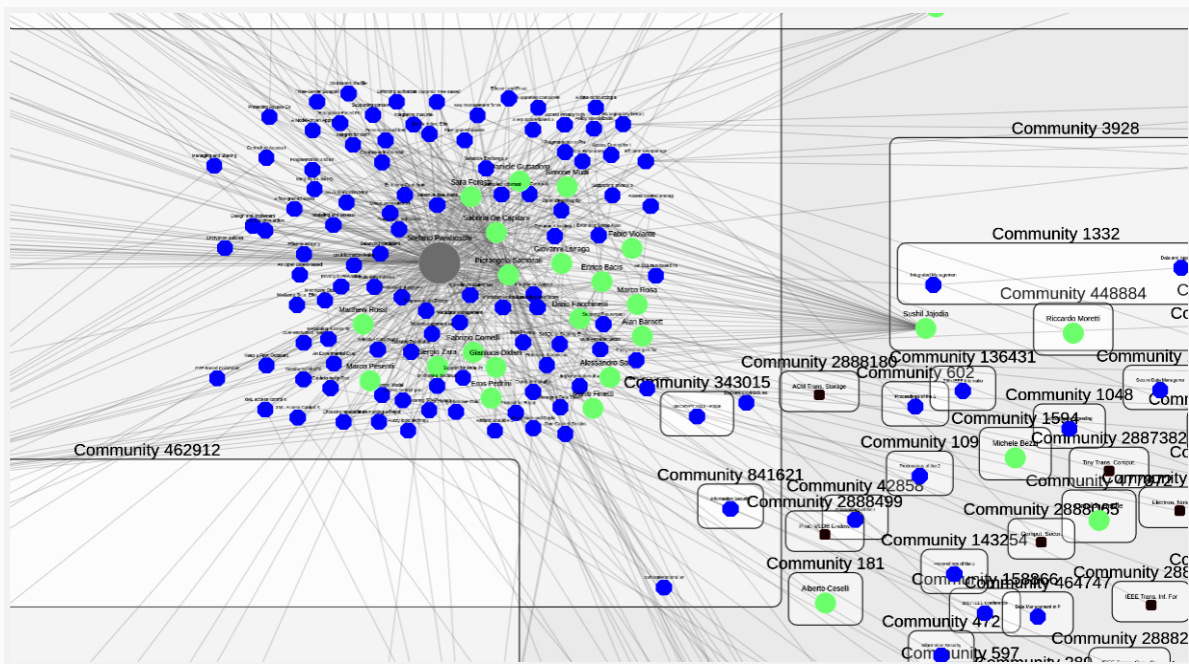

Figure 5.11.: A zoom in of the collaboration graph of prof. Paraboschi

Zooming in the graph, other features are noticeable - like for example the concentrated high density of subcommunities. The difference in interconnection densities between groups of vertices, or between a group and some sparsely linked vertices leads to this fragmentation of detected communities. The relative difference of densities of links between different communities is an important parameter to consider in order to have a detection of clusters as much of a reflection of reality as possible. This is considered in § 6 - Conclusions ▸ § 6.3 - Further avenues of exploration, specifically in § 6.3.1 - Detection of communities and their hierarchy at different scales on page 114.

In Figure 5.12 is possible to see the highly connected vertices to the point where the area between nodes gets colored by the dense edges.

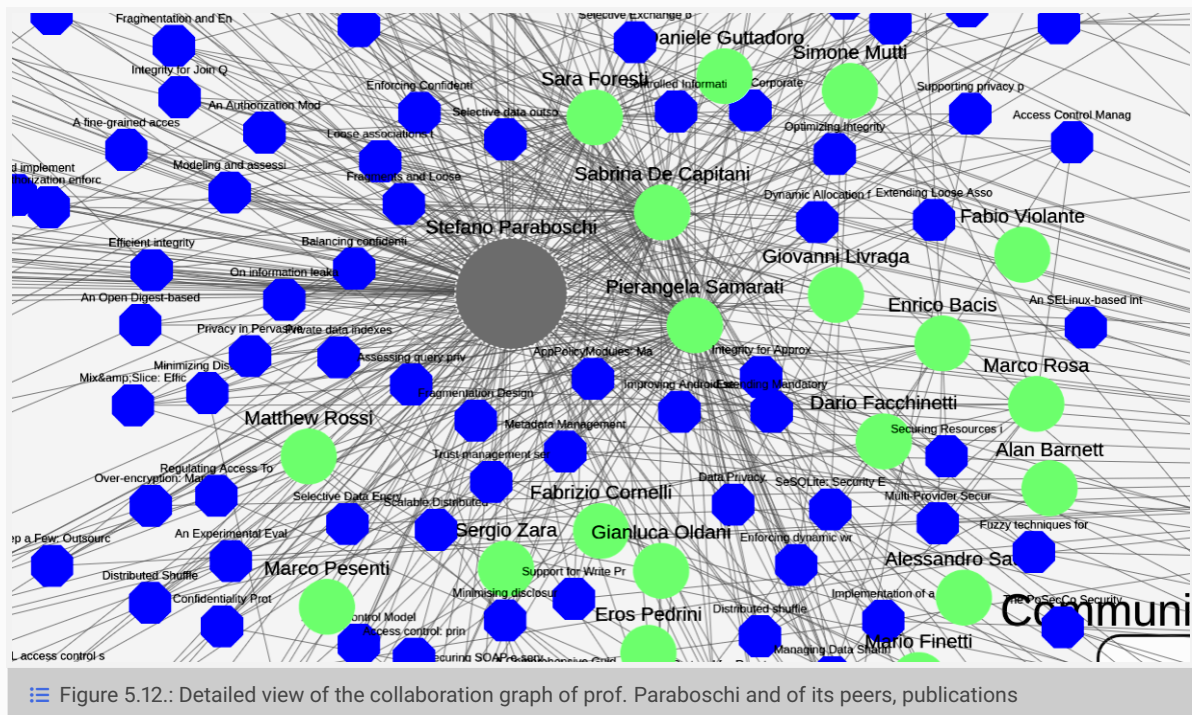

⋮≡ Figure 5.12.: Detailed view of the collaboration graph of prof. Paraboschi and of its peers, publications

In the next two subsections are shown graphs of academic collaborations build around a vertex that is not an author, such as for example having a start node set as a journal or as an institution.

### 5.3.5. Graph of the community of the *Statistical Methods & Applications* journal and prof. Alessandro Fassò node

In this subsection is shown a graph of academic collaborations formed around a journal, *Statistical Methods & Applications* - that is the journal of the *Italian Statistical Society*.[128]

Figure 5.13 and Figure 5.14 offer a lot of graphically accessible information for an interested user. The graph is associated to only one focal input, that is the name of the journal. The same can be done with a series[a] for example.

---

(a) https://en.wikipedia.org/wiki/Serial_(publishing)

[128] ITALIAN STATISTICAL SOCIETY (2021)
Italian Statistical Society. *Statistical Methods & Applications*. Edited by Carla Rampichini, Alessio Farcomeni and Stefano Campostrini. Journal of the Italian Statistical Society. Officially cited as: Stat. Methods Appl. Aug. 2021. URL: https://www.springer.com/journal/10260.

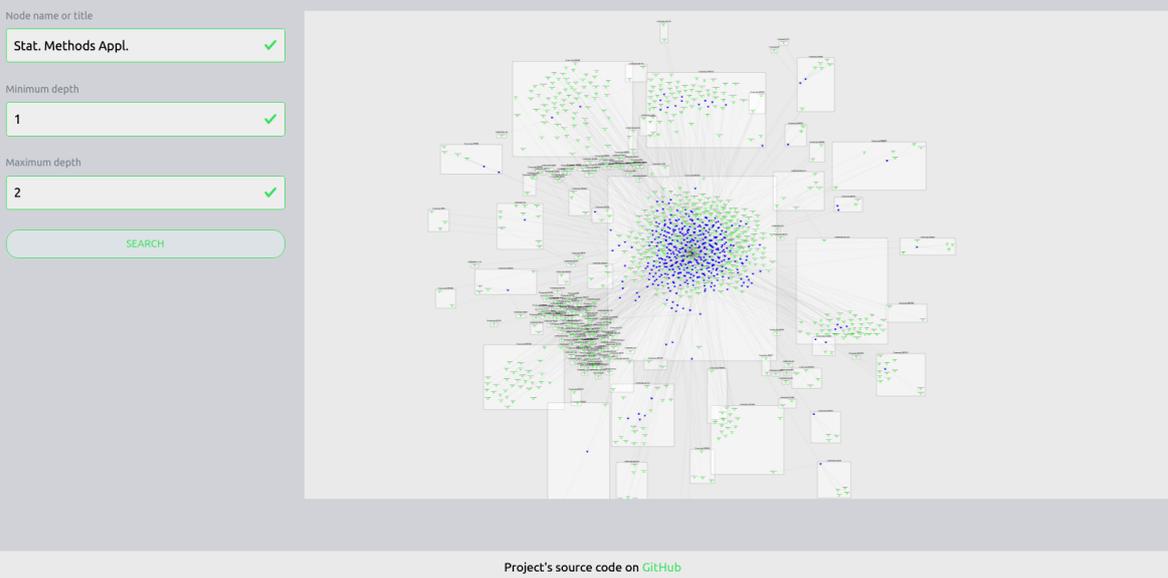

☰ Figure 5.13.: Searching for the collaboration communities built around *Statistical Methods & Applications journal*

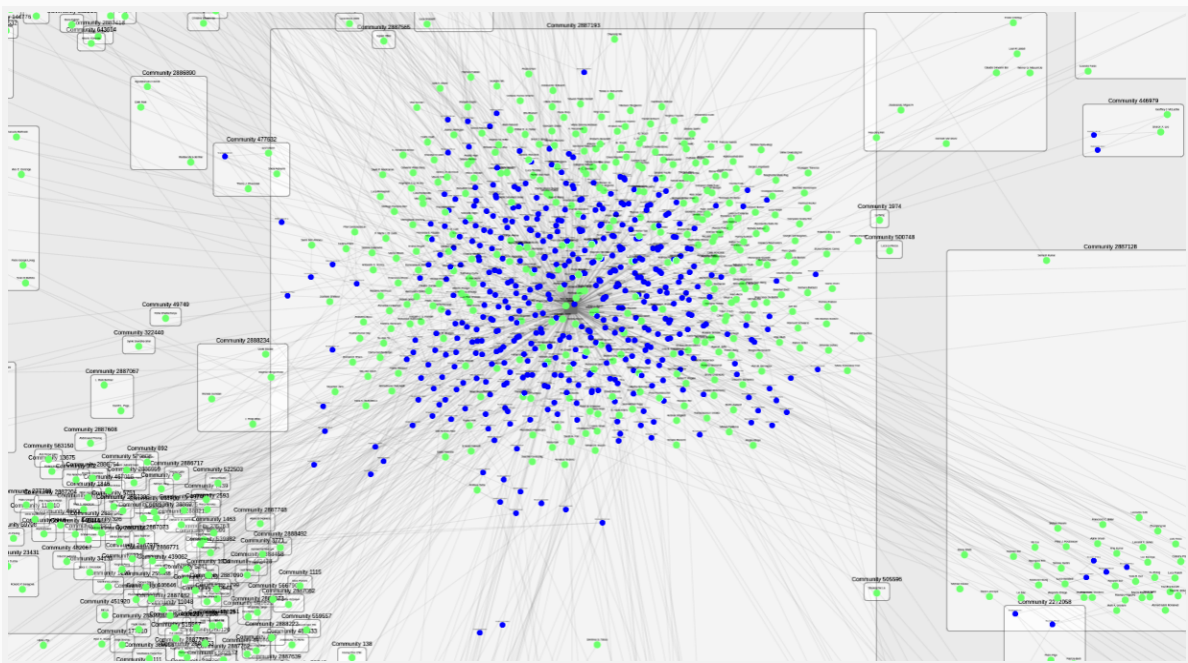

☰ Figure 5.14.: Zoomed in view of the graph of collaboration communities built around Statistical Methods & Applications journal

At the bottom part of Figure 5.15, in the middle is found the professor of statistics at the University of Bergamo, Alessandro Fassò - part of the enormous academic collaboration community of researchers who publish in *Statistical Methods & Applications journal*.

In the next subsection is presented the graph of the communities affiliated to the most prestigious computer science university in Europe, ETH Zurich, Switzerland.

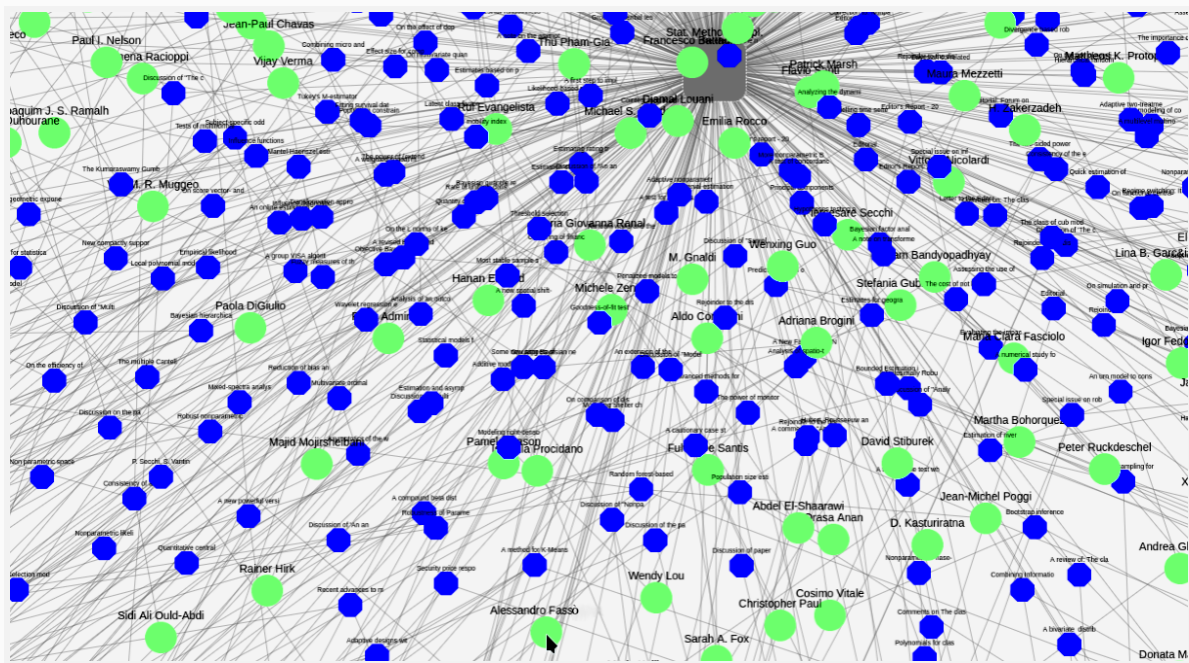

☰ Figure 5.15.: Detailed view of the graph of communties of Statistical Methods & Applications journal and prof. Alessandro Fassò node

### 5.3.6. Graph of researchers affiliated to ETH Zurich

A special case of graphs regarding academic collaboration communities are the ones built around universities, schools and research institutions in general. In this last subsection are shown two graphs. In Figure 5.16 is presented a maximum 1 hop graph of the communities in ETH Zurich, Switzerland. It shows only the researchers directly affiliated to that institution.

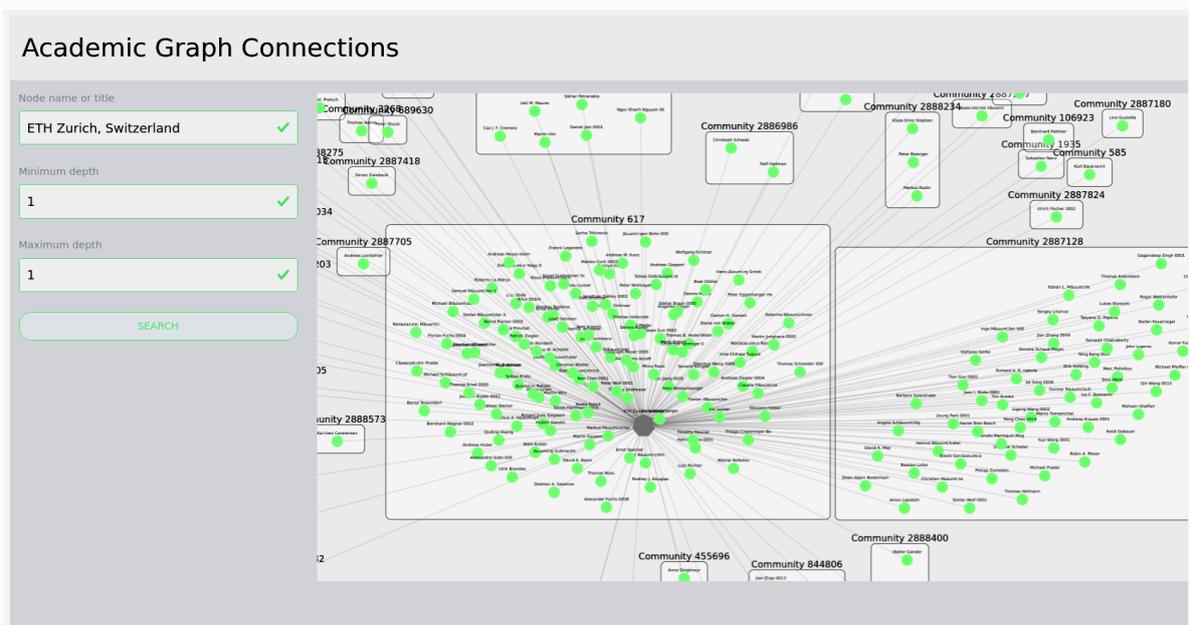

☰ Figure 5.16.: Detailed view of the graph of the authors of the academic communities at ETH Zurich, Switzerland

Whereas in Figure 5.16 is presented a maximum 2 hops graph of the colalboration communities in ETH Zurich. Since the maximum depth is set to 2, the publications of the researchers (the dark blue nodes) are also shown as part of the communities.

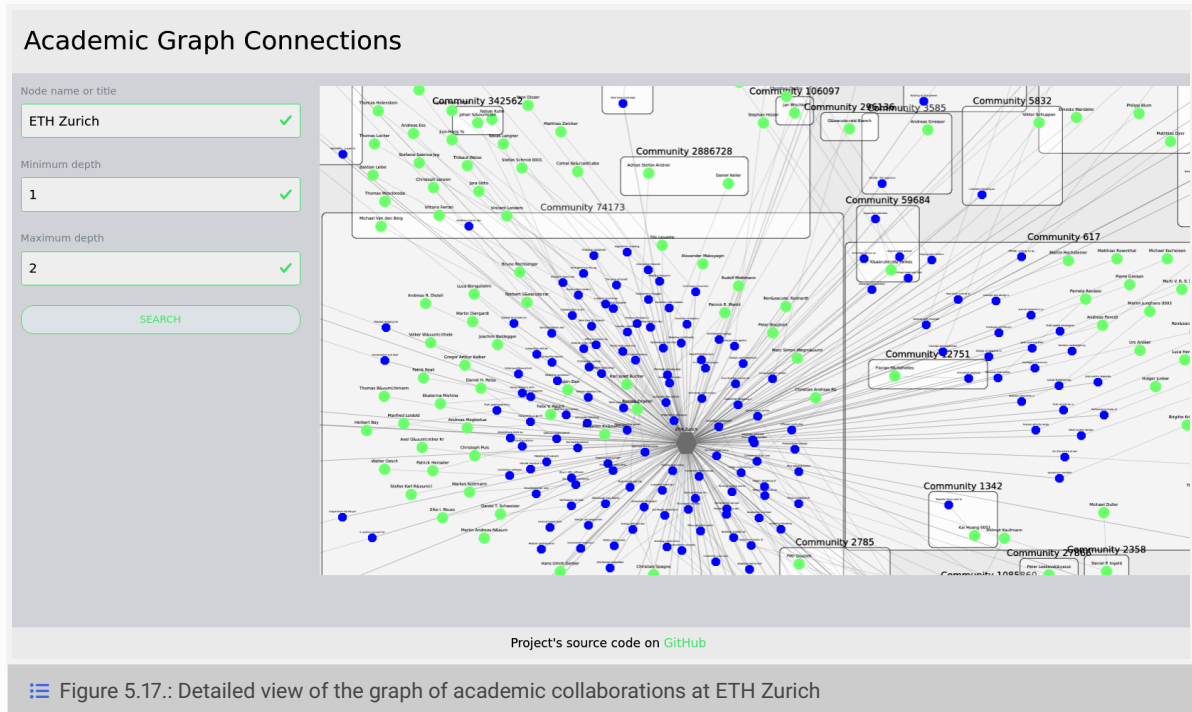

Figure 5.17.: Detailed view of the graph of academic collaborations at ETH Zurich

For more exploration, visit http://adomainthat.rocks/.[129]

[129] FERHATI (2021)

Andi Ferhati. *Clustering Graphs - Applying a Label Propagation Algorithm to Detect Communities in Graph Databases*. Master's Degree in Computer Science & Engineering. GitHub repo: https://github.com/A-Domain-that-Rocks/. Master's thesis. Viale G. Marconi, 5, 24044 Dalmine, BG, Italy: University of Bergamo, Sept. 2021. 182. URL: http://adomainthat.rocks/.

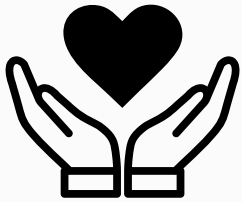

## 6. Conclusions

"Jumping to conclusions, may be an indication of Borderline Personality Disorder... I thought you should know."

*— Emma Paul*

**This chapter's contents:**

## 6.1. Summary

In this thesis was presented how new technologies like graph databases applying graph theory concepts, algorithms and novel data models can be used to infer information from highly interconnected data such as academic networks of (computer science) scientific publications to the end of detecting collaboration communities between researchers. In addition, it was shown how these results can be queried by and delivered to the user in a graphically beautiful fashion making good use of graph rendering libraries like Cytoscape.JS, of state-of-the-art frontend web development technologies as React and TypeScript - and of modern API development paradigms such as GraphQL and Apollo.

Specifically:

- Academic publications data from a dblp.org[1] dataset were used;
- The data was stored in a graph database management system like ArangoDB, was distributed into documents representing graph vertices and edges of relationships were drawn between them;
- A graph of about 8.5 million vertices and around 24 million edges was built;
- A Pregel Label Propagation Community Detection was applied to the graph and 187k clusters were discovered;
- Using the developed Web Application, results of the clustering were displayed graphically for a number of University of Bergamo's professors, the Statistical Methods & Applications journal and affiliated researchers of an ETH Zurich.

While this thesis may have accomplished the goals stated at the beginning of the work, there is always space to improve what is done and/or find new and better paths in the exploration of the graph of this topic. To this end, in the next section are shown some improvements that can be made to the developed WebApp, while in § 6.3 - Further avenues of exploration are presented some possible directions for improving the quality of the detected clusters.

## 6.2.  Possible improvements

### 6.2.1.  Limit maximum number of nodes in the frontend graph

One of the improvements that can be made to the Web App is to limit the maximum number of nodes to display. While it may be better to draw each and every vertex returned by the API in response to user queries, sometimes that is simply not possible with current commodity client computers. For example, when depth is 3 hops from the start vertex, the client has to render 50k to 250k vertices on average. When the depth becomes 4, the number of vertices becomes something like 800k (and millions of edges) which with current browsers and advancements of client computers is not possible to efficiently render.

Therefore, a way to render graphs with large number of hops from startnode would be to limit (maybe dynamically, or in percentage of the limit) the number of vertices displayable from the 3$^{rd}$ hop and deeper.

### 6.2.2.  Collapsible and expandable community compound nodes

As briefly mentioned in § 5.2 on page 103 another possible improvement is the possibility of having the compound nodes (community nodes) be collapsible and expandable.

A collapsed community node (say community 1) would have edges:

- to all other linked collapsed communities (say community 2) - this would happen if at least one vertex member of community 1 is linked to at least one vertex member of community 2.
- to all vertices of other expanded (non collapsed) community nodes (say community 3) - this would happen for each member of community 3 that is linked to at least one vertex member of community 1.

A side effect of this feature would also be the possibility to view a larger graph of many more collaboration communities, for example in a regional or national level. Having all community compound nodes collapsed might give valuable information on macrocollaborations of scientific communities.

## 6.3.  Further avenues of exploration

### 6.3.1.  Detection of communities and their hierarchy at different scales

As briefly mentioned in § 5.3.4 on page 108 a possible topic for further exploration might be the analysis of how the detected communities vary at different scales or how they are structured hierarchically. One algorithm that may be employed is the Louvain Modularity presented in § 3.1.4 on page 69.

### 6.3.2.  Performance analysis of the Pregel Community Detection algorithm run on a graph in clustered, sharded database

Another direction of exploration might be the analysis of the performance difference in terms of time and memory between the execution of the Pregel Label Propagation Community Detection algorithm on a graph in a database hosted on a single machine vs. the execution of the algorithm on a graph in a database hosted on a cluster of nodes, a sharded database.

### 6.3.3. Exploring overlapping communities detection

As briefly mentioned in § 5.3.3 on , would be interesting to perform cluster detection with overlapping communities.[10,130–139] If the same GDBMS was to be used again, possible algorithms might the Speaker-Listener Label Propagation (SLPA) or the Disassortative Degree Mixing and Information Diffusion (DMID) - both already present in ArangoDB.

[130] AMELIO and PIZZUTI (2014)

Alessia Amelio and Clara Pizzuti. *Overlapping Community Discovery Methods: A Survey*. In: CoRR abs/1411.3935 (2014). DOI: 10.1007/978-3-7091-1797-2. eprint: 1411.3935. URL: http://arxiv.org/abs/1411.3935.

[131] FONSECA VIEIRA, XAVIER and EVSUKOFF (2020)

Vinícius da Fonseca Vieira, Carolina Ribeiro Xavier and Alexandre Gonçalves Evsukoff. *A comparative study of overlapping community detection methods from the perspective of the structural properties*. In: Applied Network Science 5.1 (2020), pages 1–42. DOI: 10.1007/s41109-020-00289-9; URL: https://appliednetsci.springeropen.com/articles/10.1007/s41109-020-00289-9.

[132] GREGORY (2010)

Steve Gregory. *Finding overlapping communities in networks by label propagation*. In: New Journal of Physics 12.10 (Feb. 2010). ISSN: 1367-2630. DOI: 10.1088/1367-2630/12/10/103018. URL: https://arxiv.org/abs/0910.5516.

[133] LANCICHINETTI, FORTUNATO and KERTÉSZ (2009)

Andrea Lancichinetti, Santo Fortunato and János Kertész. *Detecting the overlapping and hierarchical community structure in complex networks*. In: New Journal of Physics 11.3 (Mar. 2009). ISSN: 1367-2630. DOI: 10.1088/1367-2630/11/3/033015. URL: https://arxiv.org/abs/0802.1218.

[134] LI, HE, BINDEL and HOPCROFT (2015)

Yixuan Li, Kun He, David Bindel and John Hopcroft. *Overlapping Community Detection via Local Spectral Clustering*. 2015. eprint: 1509.07996. URL: https://arxiv.org/abs/1509.07996.

[135] SHAHRIARI (2018)

Mohsen Shahriari. *Detection and Analysis of Overlapping Community Structures for Modelling and Prediction in Complex Networks*. PhD thesis. Aachen, Germany: RWTH Aachen University, July 2018. 229 pages. DOI: 10.18154/RWTH-2018-226325. URL: https://www.researchgate.net/publication/330601257_Detection_and_Analysis_of_Overlapping_Community_Structures_for_Modelling_and_Prediction_in_Complex_Networks.

[136] SHEN, CHENG, CAI and HU (2009)

Huawei Shen, Xueqi Cheng, Kai Cai and Mao-Bin Hu. *Detecting overlapping and hierarchical community structure in networks*. In: Physica A: Statistical Mechanics and its Applications 388.8 (Apr. 2009), pages 1706–1712. ISSN: 0378-4371. DOI: 10.1016/j.physa.2008.12.021. URL: https://arxiv.org/abs/0810.3093.

[137] WANG, TANG, GAO and LIU (2010)

Xufei Wang, Lei Tang, Huiji Gao and Huan Liu. *Discovering Overlapping Groups in Social Media*. In: 2010 IEEE International Conference on Data Mining. Dec. 2010, pages 569–578. DOI: 10.1109/ICDM.2010.48.

[138] XIE, KELLEY and SZYMANSKI (2013)

Jierui Xie, Stephen Kelley and Boleslaw K. Szymanski. *Overlapping community detection in networks*. In: ACM Computing Surveys 45.4 (Aug. 2013), pages 1–35. ISSN: 1557-7341. DOI: 10.1145/2501654.2501657. URL: https://arxiv.org/abs/1110.5813.

[139] ZHU, ZHOU, JIA, LIU and CAO (2020)

Ziqing Zhu, Tao Zhou, Chenghao Jia, Weijia Liu, Bo Liu and Jiuxin Cao. *Community detection across multiple social networks based on overlapping users*. In: Transactions on Emerging Telecommunications Technologies (Mar. 2020). ISSN: 2161-3915. DOI: 10.1002/ett.3928. URL: https://arxiv.org/abs/1909.09007.

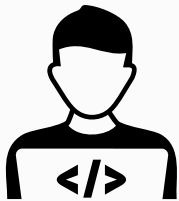

# A. Source Code

**This appendix's contents:**

## A.1. Project repositories

The work for this thesis is carried in different steps. Each phase and conceptually different workspace's source code is put in its own repository. All the repositories of the project are shown in Table A.1.

| Description of the repo | URL to the repo |
|---|---|
| XML to JSON conversion scripts | `github.com/A-Domain-that-Rocks/convert_large_xml_to_json` |
| Commands to import the JSON data in ArangoDB | `github.com/A-Domain-that-Rocks/arangodb_import_json_data` |
| AQL Queries to edit collections, create nodes, edges and graphs from the imported data | `github.com/A-Domain-that-Rocks/distribute_data_in_arangodb` |
| Backend's source code | `github.com/A-Domain-that-Rocks/adomainthat-rocks_backend` |
| Frontend's source code | `github.com/A-Domain-that-Rocks/adomainthat-rocks_frontend` |
| Thesis's LaTeX source code and PDF | `github.com/A-Domain-that-Rocks/masters_thesis` |

Table A.1.: All the repositories of the project.

All the code is public, open and freely downloadable, or clonable/pullable with git.

## A.2. Instructions on how to run, build and deploy

### A.2.1. XML to JSON conversion scripts

Open a terminal, change directory to where to store repo's source code and clone it:

```
$ git clone https://github.com/A-Domain-that-Rocks/convert_large_xml_to_json.git
```

Run the scripts with Python or open and run them from an IDE.

### A.2.2. Commands to import the JSON data in ArangoDB

Open a terminal, change directory to where to store repo's source code and clone it:

```
$ git clone https://github.com/A-Domain-that-Rocks/arangodb_import_json_data.git
```

Run in bash the commands included in the files.

### A.2.3. AQL Queries to edit collections, create nodes, edges and graphs from the imported data

Open a terminal, change directory to where to store repo's source code and clone it:

```
$ git clone https://github.com/A-Domain-that-Rocks/distribute_data_in_arangodb.git
```

Copy commands in the ArangoDB web interface query editor and execute them.

For files `44_detecting_communities.sh` and `45_community_detection_with_pregel.js`, execute those instructions in a terminal.

### A.2.4. Backend's source code

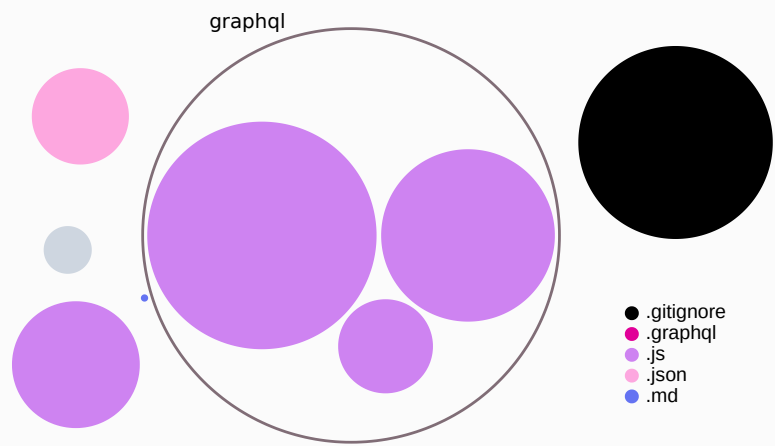

graphql

- ● .gitignore
- ● .graphql
- ● .js
- ● .json
- ● .md

Figure A.2.: Backend's repo files

Open a terminal, change directory to where to store repo's source code and clone it:

```
$ git clone https://github.com/A-Domain-that-Rocks/adomainthat-rocks_backend.git
```

**Run the app's backend/API in localhost**:

Install NodeJS, instructions: https://nodejs.org/en/download/. NodeJS verion used 14.17.1 .

Change directory to inside the cloned repo:

```
$ cd adomainthat-rocks_backend
```

Install all the project libraries:

```
$ npm install
```

Create a `.env` environment file to communicate with the database:

```
$ echo -e "HOST=localhost \nPORT=5000 \n\nDB_HOST=localhost \nDB_NAME=db_name \nDB_USER=root
\nDB_PASS=pass \n\nAPI_KEY=1234567890" >> .env
```

Launch the app:

```
$ npm start
```

The API endpoint is `http://localhost:5000/graphql`.

**Deploy the backend/API on a web server.**

Copy the files to remote web server:

```
$ ssh root@mydomain.com -p port "mkdir -p /root/api/" && scp -P port -r
/path/to/my/local/repositories/adomainthat-rocks_backend/* root@mydomain.com:/root/api
```

Install NodeJS on remote API server machine, instructions: `https://nodejs.org/en/download/`. NodeJS version used: 14.17.1 .

Change directory to inside the cloned repo (run command on remote machine):

```
# cd /root/api/
```

Install all the project libraries (run command on remote machine):

```
# npm install
```

Create a `.env` environment file on remote API server machine to communicate with the database host (run command on remote machine):

```
# echo -e "HOST=databasehost.com \nPORT=5000 \n\nDB_HOST=databasehost.com \nDB_NAME=db_name
\nDB_USER=root \nDB_PASS=pass \n\nAPI_KEY=1234567890" >> .env
```

Install `PM2` Process Management daemon (run command on remote machine)

```
# npm install -g pm2
```

Start the API server (run command on remote machine)

```
# pm2 start /root/api/server.js
```

```
# pm2 save
```

The API endpoint is `http://mydomain.com:5000/graphql`.

## A.2.5.  Frontend's source code

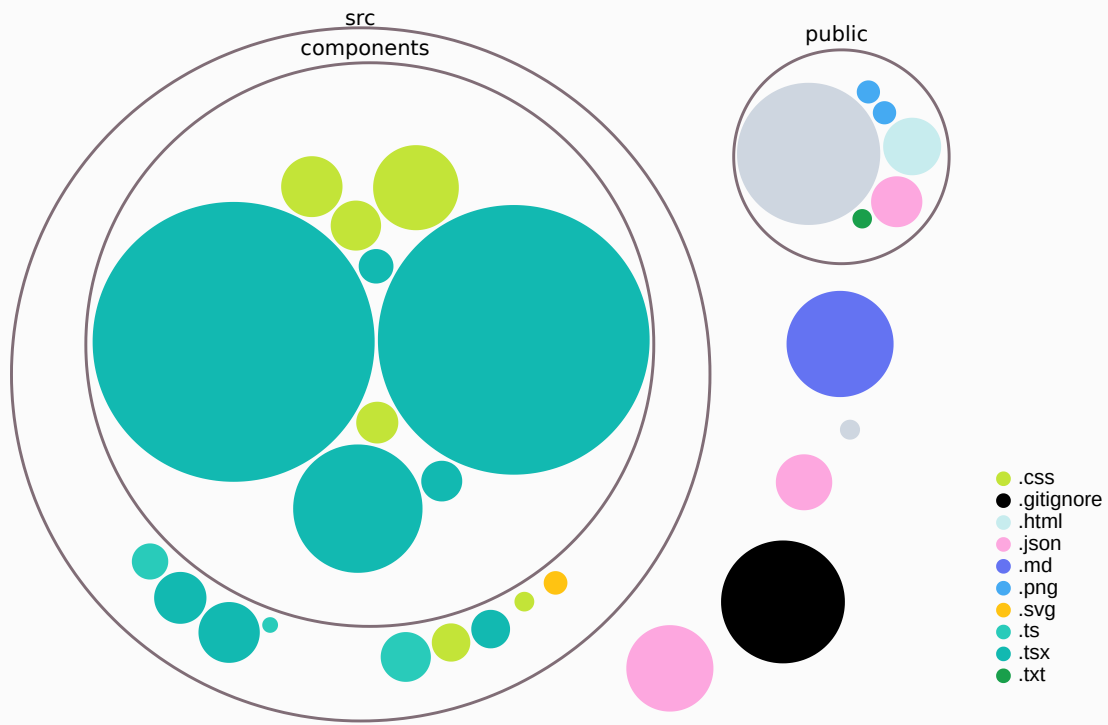

**Figure A.3.:** Frontend's repo files

Open a terminal, change directory to where to store repo's source code and clone it:

```
$ git clone https://github.com/A-Domain-that-Rocks/adomainthat-rocks_frontend.git
```

Install NodeJS, instructions: https://nodejs.org/en/download/. NodeJS verion used 14.17.1 .
Change directory to inside the cloned repo:

```
$ cd adomainthat-rocks_frontend
```

Install all the project libraries:

```
$ npm install
```

Create a `.env` environment file to communicate with the API.
If the API is in a remote machine:

```
$ echo 'REACT_APP_API_URL=http://mydomain.com:5000/graphql' >> .env
```

If the API is in localhost:

```
$ echo 'REACT_APP_API_URL=http://localhost:5000/graphql' >> .env
```

**Run the app's client interface in localhost**:
Launch the app:

```
$ npm start
```

If the app did not automatically open in the browser, visit http://localhost:3000/.

**Build the Web App interface and deploy on a web server.**
Build the app:

```
$ npm run build
```

Copy the built app to remote web server:

```
$ scp -P port -r /path/to/my/local/repositories/adomainthat-rocks_frontend/build/*
root@mydomain.com:/var/www/html/
```

Open firewall ports (run commands on remote machine):

```
# firewall-cmd --permanent --zone=public --add-service=http
```

```
# firewall-cmd --permanent --zone=public --add-service=https
```

```
# firewall-cmd --reload
```

Enable Apache to start automatically on boot (run commands on remote machine):

```
# systemctl enable httpd
```

```
# systemctl restart httpd.service
```

```
# apachectl restart
```

Visit http://mydomain.com.

## A.2.6. Thesis's LaTeX source code

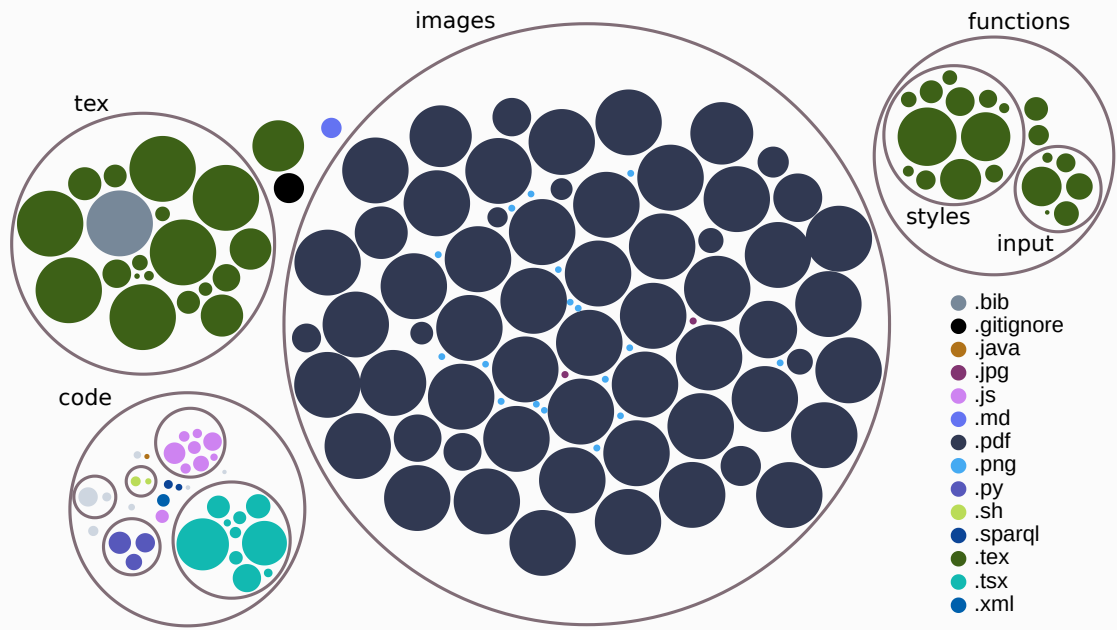

Figure A.4.: Thesis's repo files

**Compile with Overleaf**

Open a terminal, change directory to where to store repo's ZIP archive and download it:

```
$ curl -L https://github.com/A-Domain-that-Rocks/masters_thesis/archive/refs/heads/main.zip
--output masters_thesis_Andi_Ferhati.zip
```

Open https://www.overleaf.com/project ▸ **New Project** ▸ Upload Project ▸ *Upload the previously downloaded* masters_thesis_Andi_Ferhati.zip *file* ▸ **⟳ Recompile**

**Compile locally**

Install a TeX or LaTeX compilers. Clone the repo:

```
$ git clone https://github.com/A-Domain-that-Rocks/masters_thesis.git
```

Install the packages used in the project. Compile the project locally.

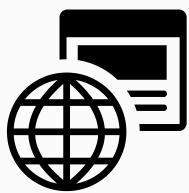

# B. API Docs

**This appendix's contents:**

## B.1. GraphQL API Schema

The GraphQL schema is composed of the type definitions shown in Code listing B.1.

Specifically types are defined for `Author`: Code sublisting B.1 (a), `Publication`: Code sublisting B.1 (b), `Institution`: Code sublisting B.1 (c), `Community`: Code sublisting B.1 (d) and others that are not presented here. For a complete list of type definitions of the GraphQL schema, see § A.2.4 on page A.2.4.

```
12    type Author {
13        _id: String!
14        name: String!
15        orcid: String
16        bibtex: String
17        aux: String
18        graph_name: String!
19        kind: String
20        other_orcid: String
21        isnot: String
22        url: [Url]
23        other_names: [String]
24        affiliation: [Affiliation]
25        note: [Note]
26    }
```
Code sublisting B.1 (a): The Author type

Even though not shown here for brevity, `Affiliation` and `Note` types are defined in previous lines of code.

```
27    type Publication {
28        _id: String!
29        title: String
30        author: [String]
31        year: String
32        graph_name: String!
33    }
```
Code sublisting B.1 (b): The Publication type

The `Publication`'s author field is of type `[String]`. This is so because, not all author's attributes are included when asking for a publication, just the names in the form of an array of strings.

```
34    type Institution {
35        _id: String!
36        name: String!
37        graph_name: String!
38    }
```

≣ Code sublisting B.1 (c): The Institution type

All attributes having as value types, types that are followed by an exclamation mark (!), have to be included in the requested fields of a GraphQL query. The other attributes, whose value types are not followed by an exclamation mark, are optional. It is within the developer's discretion to include them or not.

```
39    type Community {
40        number: String!
41    }
```

≣ Code sublisting B.1 (d): The Community type

The `Community` type has an attribute `number` of type `String` even though it effectively is a number. This is done for query building purposes.

```
50    type SuggestedNode {
51        _id: String!
52        graph_name: String
53        the_type: String
54        appearances: Int
55    }
```

≣ Code sublisting B.1 (e): The SuggestedNode type, a Vertex/Node with fewer attributes used for autocomplete suggestions

`SuggestedNode` displayed in Code sublisting B.1 (e) represents a type of Vertex/Node that is sent in response when querying the API for autocomplete suggestions (see § 4.3.3.1).

Whereas the `SlimNode`, `SlimEdge` and `SlimGraph` displayed respectively in Code sublisting B.1 (f), Code sublisting B.1 (g) and Code sublisting B.1 (h) - are Vertex/Node, Edge and Graph types with reduced number of attributes.

```
56    type SlimNode {
57        _id: String!
58        graph_name: String!
59        community: String
60    }
```

≣ Code sublisting B.1 (f): The SlimNode type, a Vertex/Node with fewer attributes

When querying for `SlimEdge`, is required the request of `_to` and `from` attributes while is optional the `label` of an edge.

```
61    type SlimEdge {
62        _from: String!
63        _to: String!
64        label: String
65    }
```

≣ Code sublisting B.1 (g): The SlimEdge type, an Edge with fewer attributes

When querying for a collaboration graph, the `SlimGraph` type is used. It has an attribute on the start node, of type `SlimNode` from where to start the traversal. `SlimGraph` also has an attribute for vertices (and edges). It takes as value an array of `SlimNode` (`SlimEdge`). An attribute on the request of communities is also part of the attributes of SlimGraph. It takes as value an array of elements of type `Community`.

```
66    type SlimGraph {
67        startNode: SlimNode!
68        vertices: [SlimNode!]
69        edges: [SlimEdge]
70        communities: [Community]
71    }
```

≣ Code sublisting B.1 (h): The SlimGraph type, a Graph with fewer attributes

When building a GraphQL query, the API can be queried for `nodesID` which translates a name/title to an `_id` - and `nodeGraph`, which returns a collaboration graph with the communities detected in it. The `nodesID` query requires an argument of type `String`, that would be the name/title of the node whose `_id` is being requested. It returns a `SuggestedNode`.

The `nodeGraph` query requires an argument of type `node_id`, that would be the `_id` obtained above - and `minDepth`, `maxDepth` parameters which indicate the range of hop bounds for the the collaboration graph. `minDepth` and `maxDepth` are optional parameters - their default values are 1 and 2 respectively. It returns a `SlimGraph`.

```
72    type Query {
73        nodesID(name: String!): [SuggestedNode]
74        nodeGraph(node_id: String!, minDepth: String = "1", maxDepth: String = "2"): SlimGraph!
75    }
```

≣ Code sublisting B.1 (i): The Query type

≣ Code listing B.1: GraphQL API's schema, type definitions

## B.2. A GraphQL API query example

For the example shown below, GraphQL Playground is used. `Postman` of whatever other favorite software for API querying works just fine for sending GraphQL requests to the API.

GraphQL Playground is a library that gives the developer the possibility to test and run GraphQL queries from a GUI interface like the browser or the desktop app, enabling better (local) development workflows. It supports GraphQL Subscriptions, configuration of HTTP headers, interactive, multi-column schema documentation and many other features. More on https://github.com/graphql/graphql-playground [140].

In Figure B.2 is shown the querying of the API using a GraphQL query requesting all the attributes needed for a collaboration graph to be built.

The `node_id` value given in input to the `nodeGraph` query type, is the `_id` in ArangoDB of the vertex of Prof. Davide Brugali in the `author` collection. That means that his collaboration graph is being requested. In order to obtain a specific node's `_id`, see § 4.3.3.1 on page 92 and specifically Figure 4.15 on page 93.

Apart from the `node_id` value, `minDepth` and `maxDepth` values are also given in input, even though are same as the default ones.

The start node is requested, along with the lists of vertices, edges and communities composing the graph. The GraphQL syntax for their request requires the use of inline fragments, spread operators. This is done this way in order to access data on the underlying concrete type, since the requested types are not made of only primitive attribute types.

[140] Suchanek, Huvar, Schickling and Schulte (2021)
Tim Suchanek, Lukáš Huvar, Johannes Schickling and Rikki Schulte. *GraphQL Playground IDE*. Aug. 2021. URL: https://github.com/graphql/graphql-playground.

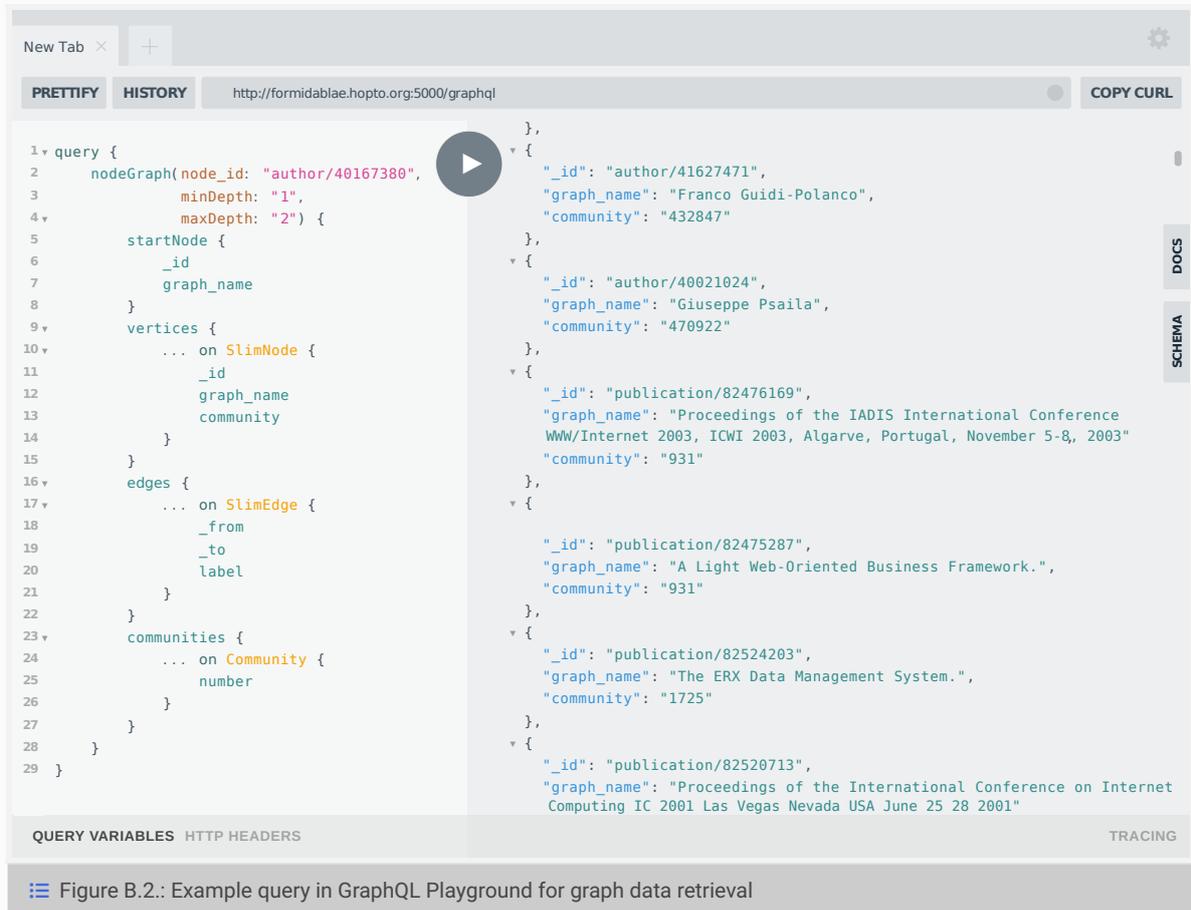

Figure B.2.: Example query in GraphQL Playground for graph data retrieval

Once the request is sent, the response shown on the right of Figure B.2 is obtained.

# Glossary

Index of all terms and their pages of appearance in the thesis.

# Bibliography

## 1. Graph theory and algorithms related references

[13] BOLLOBÁS (1998)
Béla Bollobás. *Modern Graph Theory*. 1st edition. Graduate Texts in Mathematics 184. Springer-Verlag New York, 1998. ISBN: 978-0-387-98488-9. URL: https://www.springer.com/gp/book/9780387984889
Cited on pages 12, 20, 22.

[35] DIESTEL (2000)
Reinhard Diestel. *Graph Theory*. Volume 173. Graduate Texts in Mathematics. Heidelberg, Germany: Springer-Verlag, 2000. 447 pages. ISBN: 0-387-98976-5. URL: https://diestel-graph-theory.com/index.html
Cited on pages 20, 21, 22.

[36] GROSS, YELLEN and ZHANG (2014)
Jonathan L. Gross, Jay Yellen and Ping Zhang, editors. *Handbook of Graph Theory*. Discrete Mathematics and it's Applications. 6000 Broken Sound Parkway NW, Suite 300: CRC Press, 2014. 1633 pages. ISBN: 9781439880180. DOI: https://doi.org/10.1201/b16132. URL: https://www.taylorfrancis.com/books/mono/10.1201/b16132/handbook-graph-theory-jonathan-gross-jay-yellen-ping-zhang
Cited on page 21.

[37] CORMEN, LEISERSON, RIVEST and STEIN (2009)
Thomas H. Cormen, Charles E. Leiserson, Ronald L. Rivest and Clifford Stein. *Introduction to Algorithms*. 3rd Edition. MIT Press, 2009. 1313 pages. ISBN: 978-0-262-03384-8. URL: http://mitpress.mit.edu/books/introduction-algorithms
Cited on pages 21, 23.

[38] ALDOUS and WILSON J. (2004)
Joan M. Aldous and Robin Wilson J. *Graph and Applications*. Walton Hall, Milton Keynes MK7 6AA, UK: Springer-Verlag, 2004. 458 pages. ISBN: 1-85233-259-X. URL: https://www.springer.com/gp/book/9781852332594
Cited on pages 21, 22.

[43] WIKIMEDIA FOUNDATION, INC. (2021)
Wikimedia Foundation, Inc. *Wikipedia - Breadth-first search*. Online. Aug. 2021. URL: https://en.wikipedia.org/wiki/Breadth-first_search
Cited on page 28.

[44] WIKIMEDIA FOUNDATION, INC. (2021)
Wikimedia Foundation, Inc. *Wikipedia - Depth-first search*. Online. Aug. 2021. URL: https://en.wikipedia.org/wiki/

## 2. Graph databases related references

[12] BECHBERGER and PERRYMAN (2020)
Dave Bechberger and Josh Perryman. *Graph Databases in Action*. In Action. Manning Publications, 2020. 338 pages.
ISBN: 9781617296376. URL: https://books.google.it/books?id=kWIFEAAAQBAJ
Cited on pages 12, 13.

[14] BESTA, PETER, GERSTENBERGER, FISCHER, PODSTAWSKI, BARTHELS, ALONSO and HOEFLER (2019)
Maciej Besta, Emanuel Peter, Robert Gerstenberger, Marc Fischer, Michał Podstawski, Claude Barthels, Gustavo Alonso
and Torsten Hoefler. *Demystifying Graph Databases: Analysis and Taxonomy of Data Organization, System Designs, and
Graph Queries - Towards Understanding Modern Graph Processing, Storage, and Analytics*. In: (Oct. 2019). URL: https:
//arxiv.org/abs/1910.09017
Cited on pages 13, 23, 39, 46, 48, 49, 50, 51, 52, 53, 54.

[34] BEIS, PAPADOPOULOS and KOMPATSIARIS (2015)
Sotirios Beis, Symeon Papadopoulos and Yiannis Kompatsiaris. *Benchmarking Graph Databases on the Problem of Com-
munity Detection*. In: New Trends in Database and Information Systems II. Edited by Nick Bassiliades, Mirjana Ivanovic,
Margita Kon-Popovska, Yannis Manolopoulos, Themis Palpanas, Goce Trajcevski and Athena Vakali. https://github.
com/socialsensor/graphdb-benchmarks. Springer International Publishing, 2015, pages 3–14. ISBN: 978-3-319-10518-
5. DOI: 10.1007/978-3-319-10518-5_1. URL: https://link.springer.com/chapter/10.1007/978-3-319-10518-5_1
Cited on pages 15, 47.

[40] SCIFO (2020)
Estelle Scifo. *Hands-On Graph Analytics with Neo4j - Perform graph processing and visualization techniques using con-
nected data across you entreprise*. Livery Place, 35 Livery Street, Birmingham, B3 2PB, UK: Packt, Aug. 2020. 510 pages.
ISBN: 978-1-83921-261-1. URL: https://www.packtpub.com/product/hands-on-graph-analytics-with-neo4j/
9781839212611
Cited on pages 25, 26, 27, 28.

[41] NEEDHAM and HODLER (2021)
Mark Needham and Amy E. Hodler. *Graph Algorithms - Practical Examples in Apache Spark & Neo4j*. 1005 Gravenstein
Highway North, Sebastopol, CA 95472, USA: O'Reilly, 2021. 300 pages. ISBN: 978-1-492-05781-9. URL: https://neo4j.
com/graph-algorithms-book/
Cited on pages 26, 27, 28, 66, 67, 68, 69, 70.

[45] ROBINSON, WEBBER and EIFREM (2015)
Ian Robinson, Jim Webber and Emil Eifrem. *Graph Databases*. 2nd Edition. O'Reilly Media, Inc., June 2015. 224 pages.
ISBN: 9781491930892. URL: https://www.oreilly.com/library/view/graph-databases-2nd/9781491930885/
Cited on pages 29, 31, 32, 38, 40, 41, 42, 43, 44.

[46] ANGLES and GUTIERREZ (2018)
Renzo Angles and Claudio Gutierrez. *An Introduction to Graph Data Management*. In: Graph Data Management (2018),
pages 1–32. ISSN: 2197-974X. DOI: 10.1007/978-3-319-96193-4_1. URL: https://arxiv.org/abs/1801.00036
Cited on pages 29, 30, 31, 42, 44.

[47] SOLID IT GMBH (2021)
solid IT GmbH. *db-engines.com website*. Online. Created and maintained by solid IT GmbH, an Austrian IT consulting
company with a special focus on software development, consulting and training for database-centric applications. Aug.
2021. URL: https://db-engines.com
Cited on pages 29, 30, 47, 48, 54.

[49] IBM SYSTEM G (2018)
IBM System G. *Introduction of Graph Database*. Online. May 2018. URL: https://web.archive.org/web/
20180521155557/http://systemg.research.ibm.com/database.html
Cited on page 31.

[50] RODRIGUEZ (2010)
Marko Rodriguez. *Graph Databases: Trends in the Web of Data*. Online. Slides. Sept. 2010. URL: https://www.

slideshare.net/slidarko/graph-databases-trends-in-the-web-of-data/
Cited on page 31.

[51]  WIKIMEDIA FOUNDATION, INC. (2021)
Wikimedia Foundation, Inc. *Wikipedia - Graph Database*. Online. Aug. 2021. URL: https://en.wikipedia.org/wiki/
Graph_database
Cited on page 31.

[58]  HERTEL, BROEKSTRA and STUCKENSCHMIDT (2009)
Alice Hertel, Jeen Broekstra and Heiner Stuckenschmidt. *RDF Storage and Retrieval Systems*. In: (2009). https://
www.researchgate.net/publication/227215511_RDF_Storage_and_Retrieval_Systems, pages 489–508. DOI:
http://dx.doi.org/10.1007/978-3-540-92673-3_22. URL: http://publications.wim.uni-mannheim.de/
informatik/lski/Hertel08RDFStorage.pdf
Cited on page 33.

[59]  FAYE, CURÉ and BLIN (2012)
David C. Faye, Olivier Curé and Guillaume Blin. *A survey of RDF storage approaches*. In: Revue Africaine de la Recherche
en Informatique et Mathématiques Appliquées 15 (2012). Pdf: https://hal.inria.fr/hal-01299496/file/Vol.15.
pp.11-35.pdf, pages 11–35. URL: https://hal.inria.fr/hal-01299496
Cited on pages 33, 34.

[60]  WIKIMEDIA FOUNDATION, INC. (2021)
Wikimedia Foundation, Inc. *Wikipedia - Triplestore*. Online. Aug. 2021. URL: https://en.wikipedia.org/wiki/
Triplestore
Cited on page 35.

[66]  BLOOR (2015)
Robin Bloor. *The Graph Database and the RDF Database*. Online. Jan. 2015. URL: https://web.archive.org/web/
20150206002908/http://insideanalysis.com/2015/01/the-graph-database-and-the-rdf-database/
Cited on page 36.

[67]  BECKETT (2011)
Dave Beckett. *What does SPARQL stand for?* Online. Semantic web mailing list. Oct. 2011. URL: https://lists.w3.org/
Archives/Public/semantic-web/2011Oct/0041.html
Cited on page 37.

[68]  WIKIMEDIA FOUNDATION, INC. (2021)
Wikimedia Foundation, Inc. *Wikipedia - SPARQL*. Online. Aug. 2021. URL: https://en.wikipedia.org/wiki/SPARQL
Cited on page 37.

[73]  SASAKI and NEO4J, INC. (2018)
Bryce Merkl Sasaki and Neo4j, Inc. *Graph Databases for Beginners: Other Graph Technologies*. Online. Blog post. Dec.
2018. URL: https://neo4j.com/blog/other-graph-database-technologies/
Cited on page 38.

[74]  KOBRIX SOFTWARE (2010)
Kobrix Software. *HypergraphDB*. Online. 2010. URL: http://www.hypergraphdb.org/
Cited on pages 38, 61.

[76]  AMAZON (2021)
Amazon. *What Is a Graph Database?* Online. Aug. 2021. URL: https://aws.amazon.com/nosql/graph/
Cited on page 40.

[77]  DEAN and GHEMAWAT (2008)
Jeffrey Dean and Sanjay Ghemawat. *MapReduce: Simplified Data Processing on Large Clusters*. In: Commun. ACM 51.1
(Jan. 2008). https://research.google.com/archive/mapreduce-osdi04.pdf, pages 107–113. ISSN: 0001-0782.
DOI: 10.1145/1327452.1327492. URL: https://static.googleusercontent.com/media/research.google.com/en/
/archive/mapreduce-osdi04.pdf
Cited on page 42.

[78]  SAKR (2013)
Sherif Sakr. *Processing large-scale graph data: A guide to current technology*. Online. June 2013. URL: https://
developer.ibm.com/articles/os-giraph/
Cited on page 42.

[80] WIKIMEDIA FOUNDATION, INC. (2021)
Wikimedia Foundation, Inc. *Wikipedia - Apache Spark*. Online. Aug. 2021. URL: https://en.wikipedia.org/wiki/Apache_Spark
Cited on page 42.

[82] SMARTM2M TECHNICAL COMMITTEE, LIQUORI and GUILLEMIN (2020)
SmartM2M Technical Committee, Luigi Liquori and Patrick Guillemin. *SmartM2M (Smart Machine-to-Machine Communications) TR 103 715, Study for oneM2M, Discovery and Query solutions analysis & selection*. Technical report. V1.1.1. France: ETSI, Nov. 2020. 88 pages. DOI: 10.5445/IR/1000007104. URL: https://www.etsi.org/deliver/etsi_tr/103700_103799/103715/01.01.01_60/tr_103715v010101p.pdf
Cited on pages 45, 46.

[84] CHAN and NEUBAUER (2013)
Harold Chan and Peter Neubauer. *Should I learn Cypher or Gremlin for operating a Neo4j database?* Online. May 2013. URL: https://www.quora.com/Should-I-learn-Cypher-or-Gremlin-for-operating-a-Neo4j-database
Cited on page 45.

[85] DAYARATHNA and SUZUMURA (2012)
Miyuru Dayarathna and Toyotaro Suzumura. *XGDBench: A benchmarking platform for graph stores in exascale clouds*. In: 4th IEEE International Conference on Cloud Computing Technology and Science Proceedings. Taipei, Taiwan, Dec. 2012, pages 363–370. DOI: 10.1109/CloudCom.2012.6427516
Cited on page 46.

[86] CIGLAN, AVERBUCH and HLUCHY (2012)
Marek Ciglan, Alex Averbuch and Ladialav Hluchy. *Benchmarking Traversal Operations over Graph Databases*. In: 2012 IEEE 28th International Conference on Data Engineering Workshops. http://ups.savba.sk/~marek/papers/gdm12-ciglan.pdf. 2012, pages 186–189. DOI: 10.1109/ICDEW.2012.47. URL: https://ieeexplore.ieee.org/document/6313678
Cited on page 47.

[87] JOUILI and VANSTEENBERGHE (2013)
Salim Jouili and Valentin Vansteenberghe. *An Empirical Comparison of Graph Databases*. In: 2013 International Conference on Social Computing. Sept. 2013, pages 708–715. DOI: 10.1109/SocialCom.2013.106. URL: https://ieeexplore.ieee.org/document/6693403
Cited on page 47.

[96] WIESE (2015)
Lena Wiese. *Advanced Data Management: For SQL, NoSQL, Cloud and Distributed Database*. Walter de Gruyter GmbH, Oct. 2015. 374 pages. ISBN: 978-3110441406. URL: https://www.oreilly.com/library/view/advanced-data-management/9783110433074/
Cited on page 55.

[97] AGOUB, KUNDE and KADA (2015)
Amgad Agoub, Felix Kunde and Martin Kada. *Potential of Graph Databases in Representing and Enriching Standardized Geodata*. In: Dreiländertagung der DGPF, der OVG und der SGPF. Bern, Switzerland, June 2015, pages 208–216. URL: https://www.dgpf.de/src/tagung/jt2016/proceedings/papers/20_DLT2016_Agoub_et_al.pdf
Cited on page 56.

[113] MULLANE (2002)
Greg Sabino Mullane. *PostgreSQL - Database Caching*. Online. PostgreSQL mailing list (pgsql-hackers). Feb. 2002. URL: https://www.postgresql.org/message-id/E16gYpD-0007KY-00@mclean.mail.mindspring.net
Cited on page 60.

[159] DATASTAX (2015)
DataStax. *DataStax Acquires Aurelius, The Experts Behind TitanDB*. Online. Feb. 2015. URL: https://web.archive.org/web/20150207074233/http://www.datastax.com/2015/02/datastax-acquires-aurelius-the-experts-behind-titandb
Not cited.

[165] YUHANNA, EVELSON, HOPKINS and JEDINAK (2014)
Noel Yuhanna, Boris Evelson, Brian Hopkins and Emily Jedinak. *IT Market Clock for Database Management Systems, 2014*. Technical report. In-Memory, NoSQL, And DBaaS Forge Ahead; Relational And Data Warehouses Focus On Innovation; And Market Starts Consolidating. Feb. 2014. URL: https://www.forrester.com/report/TechRadar+Enterprise+

DBMS+Q1+2014/RES106801
   Not cited.

[167]  HEUDECKER and BEYER (2015)
   Nick Heudecker and Mark Beyer. *Making Big Data Normal With Graph Analysis for the Masses*. ID: G00278415. July 2015.
   URL: https://www.gartner.com/en/documents/3100219/making-big-data-normal-with-graph-analysis-for-the-masse
   Not cited.

[168]  HEUDECKER and FEINBERG (2014)
   Nick Heudecker and Donald Feinberg. *IT Market Clock for Database Management Systems, 2014*. ID: G00261661. Sept.
   2014. URL: https://www.gartner.com/en/documents/2852717/it-market-clock-for-database-management-systems-2014
   Not cited.

[182]  MOMJIAN (2001)
   Bruce Momjian. *PostgreSQL - Introduction and Concepts*. Addison-Wesley, 2001. 490 pages. ISBN: 0-201-70331-9. URL:
   https://lab.demog.berkeley.edu/Docs/Refs/aw_pgsql_book.pdf
   Not cited.

[197]  NEO4J, INC., RATHLE and RANGANATHAN (2016)
   Neo4j, Inc., Philip Rathle and Keshav Ranganathan. *Large Scale Graph Processing with IBM Power Systems & Neo4j*.
   Online. Webinar. Oct. 2016. URL: https://www.slideshare.net/neo4j/webinar-large-scale-graph-processing-with-ibm-power-systems-neo4j
   Not cited.

[229]  RAMAMONJISON (2015)
   Rindra Ramamonjison. *Apache Spark Graph Processing*. Packt Publishing, 2015. 148 pages. ISBN: 978-1-78439-180-5.
   DOI: 10.5555/2886274
   Not cited.

[240]  TESORIERO (2013)
   Claudio Tesoriero. *Getting Started with OrientDB*. Packt Publishing, Aug. 2013. ISBN: 978-1782169956. URL: https://www.packtpub.com/product/getting-started-with-orientdb/9781782169956
   Not cited.

# 3.  Community Detection related references

[7]  NEWMAN and GIRVAN (2004)
   M. E. J. Newman and M. Girvan. *Finding and evaluating community structure in networks*. In: Physical Review E 69.2
   (Feb. 2004). ISSN: 1550-2376. DOI: 10.1103/physreve.69.026113. URL: https://arxiv.org/abs/cond-mat/0308217
   Cited on pages 10, 15.

[8]  GIRVAN and NEWMAN (2002)
   M. Girvan and M. E. J. Newman. *Community structure in social and biological networks*. In: Proceedings of the National
   Academy of Sciences 99.12 (June 2002). https://arxiv.org/abs/cond-mat/0112110, pages 7821–7826. ISSN: 1091-
   6490. DOI: 10.1073/pnas.122653799. URL: https://www.pnas.org/content/99/12/7821
   Cited on page 10.

[9]  PARTHASARATHY, SHAH and ZAMAN (2019)
   Dhruv Parthasarathy, Devavrat Shah and Tauhid Zaman. *Leaders, Followers, and Community Detection*. 2019. eprint:
   1011.0774. URL: https://arxiv.org/abs/1011.0774
   Cited on page 10.

[10]  PALLA, DERÉNYI, FARKAS and VICSEK (2005)
   Gergely Palla, Imre Derényi, Illés Farkas and Tamás Vicsek. *Uncovering the overlapping community structure of complex
   networks in nature and society*. In: Nature 435.7043 (June 2005), pages 814–818. ISSN: 1476-4687. DOI: 10.1038/nature03607. URL: https://www.nature.com/articles/nature03607
   Cited on pages 11, 115.

[11]  FORTUNATO (2010)
   Santo Fortunato. *Community detection in graphs*. In: Physics Reports 486.3-5 (Feb. 2010), pages 75–174. ISSN: 0370-

1573. DOI: 10.1016/j.physrep.2009.11.002. URL: https://arxiv.org/abs/0906.0612.
Cited on page 11.

[15] BRANDES, GAERTLER and WAGNER (2003)
Ulrik Brandes, Marco Gaertler and Dorothea Wagner. *Experiments on Graph Clustering Algorithms*. In: volume 2832.
Nov. 2003. ISBN: 978-3-540-20064-2. DOI: 10.1007/978-3-540-39658-1_52. URL: https://www.researchgate.net/
publication/2939172.
Cited on page 13.

[16] DAO, BOTHOREL and LENCA (2020)
Vinh Loc Dao, Cécile Bothorel and Philippe Lenca. *Community structure: A comparative evaluation of community detection methods*. In: Network Science 8.1 (Jan. 2020), pages 1–41. ISSN: 2050-1250. DOI: 10.1017/nws.2019.59. URL: https:
//arxiv.org/abs/1812.06598.
Cited on pages 13, 66, 67.

[17] FORTUNATO and CASTELLANO (2007)
Santo Fortunato and Claudio Castellano. *Community Structure in Graphs*. 2007. eprint: 0712.2716. URL: https://arxiv.
org/abs/0712.2716.
Cited on page 13.

[18] LANCICHINETTI and FORTUNATO (2009)
Andrea Lancichinetti and Santo Fortunato. *Community detection algorithms: A comparative analysis*. In: Physical Review
E 80.5 (Nov. 2009). ISSN: 1550-2376. DOI: 10.1103/physreve.80.056117. URL: https://arxiv.org/abs/0908.1062.
Cited on page 13.

[19] LIU, CHENG and ZHANG (2019)
Xin Liu, Hui-Min Cheng and Zhong-Yuan Zhang. *Evaluation of Community Detection Methods*. 2019. eprint: 1807.01130.
URL: https://arxiv.org/abs/1807.01130.
Cited on page 13.

[20] NEWMAN (2004)
M. E. J. Newman. *Detecting community structure in networks*. In: The European Physical Journal B 38.2 (Mar. 2004),
pages 321–330. ISSN: 1434-6036. DOI: 10.1140/epjb/e2004-00124-y. URL: http://www-personal.umich.edu/~mejn/
papers/epjb.pdf.
Cited on page 13.

[21] ORMAN and LABATUT (2009)
Günce Orman and Vincent Labatut. *A Comparison of Community Detection Algorithms on Artificial Networks*. In: volume 5808. Nov. 2009, pages 242–256. DOI: 10.1007/978-3-642-04747-3_20. URL: https://www.researchgate.net/
publication/224921426.
Cited on page 13.

[22] ROSVALL, DELVENNE, SCHAUB and LAMBIOTTE (2019)
Martin Rosvall, Jean-Charles Delvenne, Michael T. Schaub and Renaud Lambiotte. *Different Approaches to Community Detection*. In: Advances in Network Clustering and Blockmodeling (Nov. 2019), pages 105–119. DOI: 10.1002/
9781119483298.ch4. URL: https://arxiv.org/abs/1712.06468.
Cited on page 13.

[23] SHAI, STANLEY, GRANELL, TAYLOR and MUCHA (2017)
Saray Shai, Natalie Stanley, Clara Granell, Dane Taylor and Peter J. Mucha. *Case studies in network community detection*.
2017. eprint: 1705.02305. URL: https://arxiv.org/abs/1705.02305.
Cited on page 13.

[24] WU, WU, CHEN, LI and ZHANG (2021)
Sissi Xiaoxiao Wu, Zixian Wu, Shihui Chen, Gangqiang Li and Shengli Zhang. *Community Detection in Blockchain Social Networks*. 2021. eprint: 2101.06406. URL: https://arxiv.org/abs/2101.06406.
Cited on page 13.

[25] YAN, CHENG, XING, LU, NG and BU (2014)
Da Yan, James Cheng, Kai Xing, Yi Lu, Wee Keong Ng and Yingyi Bu. *Pregel Algorithms for Graph Connectivity Problems with Performance Guarantees*. In: Proceedings of the VLDB Endowment 7(14). Volume 7. Oct. 2014, pages 1821–
1832. DOI: 10.14778/2733085.2733089. URL: https://www.researchgate.net/publication/271020290_Pregel_

`Algorithms_for_Graph_Connectivity_Problems_with_Performance_Guarantees`
Cited on page 13.

[29]  CHEN and REDNER (2010)
P. Chen and S. Redner. *Community structure of the physical review citation network*. In: Journal of Informetrics 4.3 (July 2010), pages 278–290. ISSN: 1751-1577. DOI: `10.1016/j.joi.2010.01.001`. URL: `https://arxiv.org/abs/0911.0694`
Cited on page 15.

[30]  NEWMAN (2001)
M. E. J. Newman. *The structure of scientific collaboration networks*. In: Proceedings of the National Academy of Sciences 98.2 (Jan. 2001). `https://arxiv.org/abs/cond-mat/0007214`, pages 404–409. ISSN: 1091-6490. DOI: `10.1073/pnas.98.2.404`. URL: `https://www.pnas.org/content/98/2/404`
Cited on page 15.

[31]  NEWMAN (2001)
M. E. J. Newman. *Scientific collaboration networks. I. Network construction and fundamental results*. In: Physical review. E, Statistical, nonlinear, and soft matter physics 64 (Aug. 2001). DOI: `10.1103/PhysRevE.64.016131`
Cited on page 15.

[32]  NEWMAN (2001)
M. E. J. Newman. *Scientific collaboration networks. II. Shortest paths, weighted networks, and centrality*. In: Phys. Rev. E 64 (1 June 2001). DOI: `10.1103/PhysRevE.64.016132`. URL: `https://link.aps.org/doi/10.1103/PhysRevE.64.016132`
Cited on page 15.

[33]  RADICCHI, FORTUNATO and VESPIGNANI (2012)
Filippo Radicchi, Santo Fortunato and Alessandro Vespignani. *Citation Networks*. In: Models of Science Dynamics: Encounters Between Complexity Theory and Information Sciences. Edited by Andrea Scharnhorst, Katy Börner and Peter van den Besselaar. Berlin, Heidelberg - Germany: Springer Berlin Heidelberg, 2012, pages 233–257. ISBN: 978-3-642-23068-4. DOI: `10.1007/978-3-642-23068-4_7`. URL: `https://link.springer.com/chapter/10.1007/978-3-642-23068-4_7`
Cited on page 15.

[48]  SINICO (2017)
Luca Sinico. *Graph databases and their application to the Italian Business Register for efficient search of relationships among companies*. Master's thesis. University of Padua, Apr. 2017. 208 pages. URL: `http://tesi.cab.unipd.it/54610/`
Cited on pages 30, 37, 38, 40, 41, 42, 43, 44, 45, 47, 54, 56, 57, 58, 60, 61, 62.

[115]  SAVAGE and WLOKA (1991)
John E. Savage and Markus G. Wloka. *Heuristics for Parallel Graph-Partitioning*. Technical report. Technical Report No. CS-89-41, Revised Version. Providence, Rhode Island 02912, USA, Jan. 1991
Cited on page 61.

[116]  WIKIMEDIA FOUNDATION, INC. (2021)
Wikimedia Foundation, Inc. *Wikipedia - Graph partition*. Online. Aug. 2021. URL: `https://en.wikipedia.org/wiki/Graph_partition`
Cited on page 61.

[118]  RAGHAVAN, ALBERT and KUMARA (2007)
Usha Nandini Raghavan, Réka Albert and Soundar Kumara. *Near linear time algorithm to detect community structures in large-scale networks*. In: Physical Review E 76.3 (Sept. 2007). ISSN: 1550-2376. DOI: `10.1103/physreve.76.036106`. URL: `https://arxiv.org/abs/0709.2938`
Cited on pages 65, 68.

[119]  ÖZTÜRK (2014)
Koray Öztürk. *Community detection in social networks*. Master's thesis. Middle East Technical University, Dec. 2014. 105 pages. URL: `https://open.metu.edu.tr/bitstream/handle/11511/24245/index.pdf`
Cited on page 66.

[120]  LEÃO, BRANDÃO, VAZ DE MELO and LAENDER (2018)
Jeancarlo C. Leão, Michele A. Brandão, Pedro O. S. Vaz de Melo and Alberto H. F. Laender. *Who is really in my social circle? Mining social relationships to improve detection of real communities*. In: Journal of Internet Services and Applications 9.1 (Oct. 2018). ISSN: 1869-0238. DOI: `10.1186/s13174-018-0091-6`. URL: `https://www.researchgate.net/publication/`

327824385
  Cited on pages 66, 67.

[121] WAGENSELLER III and WANG (2017)
Paul Wagenseller III and Feng Wang. *Size Matters: A Comparative Analysis of Community Detection Algorithms*. 2017. eprint: 1712.01690. URL: https://arxiv.org/abs/1712.01690
  Cited on pages 66, 67.

[129] FERHATI (2021)
Andi Ferhati. *Clustering Graphs - Applying a Label Propagation Algorithm to Detect Communities in Graph Databases*. Master's Degree in Computer Science & Engineering. GitHub repo: https://github.com/A-Domain-that-Rocks/. Master's thesis. Viale G. Marconi, 5, 24044 Dalmine, BG, Italy: University of Bergamo, Sept. 2021. 182. URL: http://adomainthat.rocks/
  Cited on page 111.

[130] AMELIO and PIZZUTI (2014)
Alessia Amelio and Clara Pizzuti. *Overlapping Community Discovery Methods: A Survey*. In: CoRR abs/1411.3935 (2014). DOI: 10.1007/978-3-7091-1797-2. eprint: 1411.3935. URL: http://arxiv.org/abs/1411.3935
  Cited on page 115.

[131] FONSECA VIEIRA, XAVIER and EVSUKOFF (2020)
Vinícius da Fonseca Vieira, Carolina Ribeiro Xavier and Alexandre Gonçalves Evsukoff. *A comparative study of overlapping community detection methods from the perspective of the structural properties*. In: Applied Network Science 5.1 (2020), pages 1–42. DOI: 10.1007/s41109-020-00289-9;. URL: https://appliednetsci.springeropen.com/articles/10.1007/s41109-020-00289-9
  Cited on page 115.

[132] GREGORY (2010)
Steve Gregory. *Finding overlapping communities in networks by label propagation*. In: New Journal of Physics 12.10 (Feb. 2010). ISSN: 1367-2630. DOI: 10.1088/1367-2630/12/10/103018. URL: https://arxiv.org/abs/0910.5516
  Cited on page 115.

[133] LANCICHINETTI, FORTUNATO and KERTÉSZ (2009)
Andrea Lancichinetti, Santo Fortunato and János Kertész. *Detecting the overlapping and hierarchical community structure in complex networks*. In: New Journal of Physics 11.3 (Mar. 2009). ISSN: 1367-2630. DOI: 10.1088/1367-2630/11/3/033015. URL: https://arxiv.org/abs/0802.1218
  Cited on page 115.

[134] LI, HE, BINDEL and HOPCROFT (2015)
Yixuan Li, Kun He, David Bindel and John Hopcroft. *Overlapping Community Detection via Local Spectral Clustering*. 2015. eprint: 1509.07996. URL: https://arxiv.org/abs/1509.07996
  Cited on page 115.

[135] SHAHRIARI (2018)
Mohsen Shahriari. *Detection and Analysis of Overlapping Community Structures for Modelling and Prediction in Complex Networks*. PhD thesis. Aachen, Germany: RWTH Aachen University, July 2018. 229 pages. DOI: 10.18154/RWTH-2018-226325. URL: https://www.researchgate.net/publication/330601257_Detection_and_Analysis_of_Overlapping_Community_Structures_for_Modelling_and_Prediction_in_Complex_Networks
  Cited on page 115.

[136] SHEN, CHENG, CAI and HU (2009)
Huawei Shen, Xueqi Cheng, Kai Cai and Mao-Bin Hu. *Detecting overlapping and hierarchical community structure in networks*. In: Physica A: Statistical Mechanics and its Applications 388.8 (Apr. 2009), pages 1706–1712. ISSN: 0378-4371. DOI: 10.1016/j.physa.2008.12.021. URL: https://arxiv.org/abs/0810.3093
  Cited on page 115.

[137] WANG, TANG, GAO and LIU (2010)
Xufei Wang, Lei Tang, Huiji Gao and Huan Liu. *Discovering Overlapping Groups in Social Media*. In: 2010 IEEE International Conference on Data Mining. Dec. 2010, pages 569–578. DOI: 10.1109/ICDM.2010.48
  Cited on page 115.

[138] XIE, KELLEY and SZYMANSKI (2013)
Jierui Xie, Stephen Kelley and Boleslaw K. Szymanski. *Overlapping community detection in networks*. In: ACM Computing

Surveys 45.4 (Aug. 2013), pages 1–35. ISSN: 1557-7341. DOI: 10.1145/2501654.2501657. URL: https://arxiv.org/abs/1110.5813
    Cited on page 115.

[139]  ZHU, ZHOU, JIA, LIU, LIU and CAO (2020)
       Ziqing Zhu, Tao Zhou, Chenghao Jia, Weijia Liu, Bo Liu and Jiuxin Cao. *Community detection across multiple social networks based on overlapping users*. In: Transactions on Emerging Telecommunications Technologies (Mar. 2020). ISSN: 2161-3915. DOI: 10.1002/ett.3928. URL: https://arxiv.org/abs/1909.09007
       Cited on page 115.

[146]  BALL, KARRER and NEWMAN (2011)
       Brian Ball, Brian Karrer and M. E. J. Newman. *Efficient and principled method for detecting communities in networks*. In: Physical Review E 84.3 (Sept. 2011). ISSN: 1550-2376. DOI: 10.1103/physreve.84.036103. URL: https://arxiv.org/abs/1104.3590
       Not cited.

[148]  BRODER, KUMAR, MAGHOUL, RAGHAVAN, RAJAGOPALAN, STATA, TOMKINS and WIENER (2000)
       Andrei Broder, Ravi Kumar, Farzin Maghoul, Prabhakar Raghavan, Sridhar Rajagopalan, Raymie Stata, Andrew Tomkins and Janet Wiener. *Graph structure in the Web*. https://www.cis.upenn.edu/~mkearns/teaching/NetworkedLife/broder.pdf. 2000. URL: http://citeseerx.ist.psu.edu/viewdoc/summary?doi=10.1.1.517.3557
       Not cited.

[149]  CAPOCCI, SERVEDIO, CALDARELLI and COLAIORI (2005)
       A. Capocci, V.D.P. Servedio, G. Caldarelli and F. Colaiori. *Detecting communities in large networks*. In: Physica A: Statistical Mechanics and its Applications 352.2-4 (July 2005), pages 669–676. ISSN: 0378-4371. DOI: 10.1016/j.physa.2004.12.050. URL: https://arxiv.org/abs/cond-mat/0402499
       Not cited.

[150]  CHERIFI (2018)
       Hocine Cherifi. *Non-overlapping community detection*. 2018. eprint: 1805.11584. URL: https://arxiv.org/abs/1805.11584
       Not cited.

[151]  CLAUSET, NEWMAN and MOORE (2004)
       Aaron Clauset, M. E. J. Newman and Cristopher Moore. *Finding community structure in very large networks*. In: Phys. Rev. E 70 (6 Dec. 2004). ISSN: 1550-2376. DOI: 10.1103/PhysRevE.70.066111. URL: https://arxiv.org/abs/cond-mat/0408187
       Not cited.

[153]  CONTISCIANI (2019)
       Martina Contisciani. *A new approach for community detection in multilayer networks*. Master's thesis. University of Padua, Sept. 2019. 61 pages. URL: http://tesi.cab.unipd.it/63015/
       Not cited.

[154]  COPPENS, DE VENTER, MITROVI and DE WEERDT (2019)
       Lauranne Coppens, Jonathan De Venter, Sandra Mitrovi and Jochen De Weerdt. *A comparative study of community detection techniques for large evolving graphs*. In: LEG@ ECML: The third International Workshop on Advances in Managing and Mining Large Evolving Graphs collocated with ECML-PKDD. Springer. 2019, pages 1157–1165. DOI: 10.1016/j.procs.2020.04.124
       Not cited.

[155]  CRAMPES and PLANTIÉ (2014)
       Michel Crampes and Michel Plantié. *A Unified Community Detection, Visualization and Analysis method*. Feb. 2014. eprint: 1301.7006. URL: https://arxiv.org/abs/1301.7006
       Not cited.

[156]  CRANE and DEMPSEY (2015)
       Harry Crane and Walter Dempsey. *Community detection for interaction networks*. 2015. eprint: 1509.09254. URL: https://arxiv.org/abs/1509.09254
       Not cited.

[160]  DELLING, GAERTLER, GÖRKE, NIKOLOSKI and WAGNER (2007)
       Daniel Delling, Marco Gaertler, Robert Görke, Zoran Nikoloski and Dorothea Wagner. *How to Evaluate Clustering Tech-

*niques*. Technical report 24. Universität Karlsruhe (TH), 2007. 12 pages. DOI: 10.5445/IR/1000007104. URL: https://publikationen.bibliothek.kit.edu/1000007104

    Not cited.

[161] DEL PIA, KHAJAVIRAD and KUNISKY (2020)

Alberto Del Pia, Aida Khajavirad and Dmitriy Kunisky. *Linear Programming and Community Detection*. 2020. eprint: 2006.03213. URL: https://arxiv.org/abs/2006.03213

    Not cited.

[162] DE MEO, FERRARA, FIUMARA and PROVETTI (2011)

Pasquale De Meo, Emilio Ferrara, Giacomo Fiumara and Alessandro Provetti. *Generalized Louvain method for community detection in large networks*. In: 2011 11th International Conference on Intelligent Systems Design and Applications (Nov. 2011). DOI: 10.1109/isda.2011.6121636. URL: https://arxiv.org/abs/1108.1502

    Not cited.

[163] DU, WU, WANG and WANG (2008)

Nan Du, Bin Wu, Bai Wang and Yi Wang. *Overlapping Community Detection in Bipartite Networks*. 2008. eprint: 0804.3636. URL: https://arxiv.org/abs/0804.3636?context=physics.soc-ph

    Not cited.

[164] FANG, WANG, LIU, WU, TANG and ZHENG (2020)

Wenyi Fang, Xin Wang, Longzhao Liu, Zhaole Wu, Shaoting Tang and Zhiming Zheng. *Community Detection through Vector-label Propagation Algorithms*. 2020. eprint: 2011.08342. URL: https://arxiv.org/abs/2011.08342

    Not cited.

[166] FORTUNATO and HRIC (2016)

Santo Fortunato and Darko Hric. *Community detection in networks: A user guide*. In: Physics Reports 659 (Nov. 2016), pages 1–44. ISSN: 0370-1573. DOI: 10.1016/j.physrep.2016.09.002. URL: https://arxiv.org/abs/1608.00163

    Not cited.

[169] GULBAHCE and LEHMANN (2008)

Natali Gulbahce and Sune Lehmann. *The art of community detection*. 2008. eprint: 0807.1833. URL: https://arxiv.org/abs/0807.1833

    Not cited.

[171] HOLLOCOU, BONALD and LELARGE (2016)

Alexandre Hollocou, Thomas Bonald and Marc Lelarge. *Improving PageRank for Local Community Detection*. 2016. eprint: 1610.08722. URL: https://arxiv.org/abs/1610.08722

    Not cited.

[172] HE, ZHA, DING and SIMON (2001)

Xiaofeng He, Hongyuan Zha, Chris H. Q. Ding and Horst D. Simon. *Web document clustering using hyperlink structures*. Technical report. https://citeseerx.ist.psu.edu/viewdoc/download?doi=10.1.1.331.3652&rep=rep1&type=pdf, https://citeseerx.ist.psu.edu/viewdoc/download?doi=10.1.1.331.3652&rep=rep1&type=pdf. USA, 2001. 22 pages. DOI: 10.2172/815474. URL: https://www.osti.gov/biblio/815474

    Not cited.

[173] LABATUT (2015)

Vincent Labatut. *Generalised measures for the evaluation of community detection methods*. In: International Journal of Social Network Mining 2.1 (2015). ISSN: 1757-8493. DOI: 10.1504/ijsnm.2015.069776. URL: https://arxiv.org/abs/1303.5441

    Not cited.

[174] LAI and MCKENZIE (2018)

Ming-Jun Lai and Daniel Mckenzie. *A Compressive Sensing Approach to Community Detection with Applications*. 2018. eprint: 1708.09477. URL: https://arxiv.org/abs/1708.09477

    Not cited.

[175] LANCICHINETTI and FORTUNATO (2009)

Andrea Lancichinetti and Santo Fortunato. *Benchmarks for testing community detection algorithms on directed and weighted graphs with overlapping communities*. In: Physical Review E 80.1 (July 2009). ISSN: 1550-2376. DOI: 10.1103/physreve.80.016118. URL: https://arxiv.org/abs/0904.3940

    Not cited.

[176] LANCICHINETTI, FORTUNATO and RADICCHI (2008)
Andrea Lancichinetti, Santo Fortunato and Filippo Radicchi. *Benchmark graphs for testing community detection algorithms*. In: Physical Review E 78.4 (Oct. 2008). ISSN: 1550-2376. DOI: 10.1103/physreve.78.046110. URL: https://arxiv.org/abs/0805.4770
Not cited.

[177] LESKOVEC, LANG and MAHONEY (2010)
Jure Leskovec, Kevin J. Lang and Michael W. Mahoney. *Empirical Comparison of Algorithms for Network Community Detection*. 2010. eprint: 1004.3539. URL: https://arxiv.org/abs/1004.3539
Not cited.

[178] LI, CHIEN and MILENKOVIC (2019)
Pan Li, Eli Chien and Olgica Milenkovic. *Optimizing Generalized PageRank Methods for Seed-Expansion Community Detection*. 2019. eprint: 1905.10881. URL: https://arxiv.org/abs/1905.10881
Not cited.

[180] MALLIAROS and VAZIRGIANNIS (2013)
Fragkiskos D. Malliaros and Michalis Vazirgiannis. *Clustering and community detection in directed networks: A survey*. In: Physics Reports 533.4 (Dec. 2013), pages 95–142. ISSN: 0370-1573. DOI: 10.1016/j.physrep.2013.08.002. URL: https://arxiv.org/abs/1308.0971
Not cited.

[181] MA and NANDY (2021)
Zongming Ma and Sagnik Nandy. *Community Detection with Contextual Multilayer Networks*. 2021. eprint: 2104.02960. URL: https://arxiv.org/abs/2104.02960
Not cited.

[183] MOORE (2017)
Cristopher Moore. *The Computer Science and Physics of Community Detection: Landscapes, Phase Transitions, and Hardness*. 2017. eprint: 1702.00467. URL: https://arxiv.org/abs/1702.00467
Not cited.

[199] NEWMAN (2013)
M. E. J. Newman. *Community detection and graph partitioning*. In: EPL (Europhysics Letters) 103.2 (July 2013). ISSN: 1286-4854. DOI: 10.1209/0295-5075/103/28003. URL: https://arxiv.org/abs/1305.4974
Not cited.

[212] PANG, SHAO, SUN and LI (2009)
Chuanjun Pang, Fengjing Shao, Rencheng Sun and Shujing Li. *Detecting Community Structure in Networks by Propagating Labels of Nodes*. In: Proceedings of the 6th International Symposium on Neural Networks: Advances in Neural Networks - Part III. ISNN 2009. Wuhan, China: Springer-Verlag, 2009, pages 839–846. ISBN: 9783642015120. DOI: 10.1007/978-3-642-01513-7_91. URL: https://dl.acm.org/doi/10.1007/978-3-642-01513-7_91
Not cited.

[213] PEEL (2010)
Leto Peel. *Estimating Network Parameters for Selecting Community Detection Algorithms*. 2010. eprint: 1010.5377. URL: https://arxiv.org/abs/1010.5377
Not cited.

[214] PORTER, ONNELA and MUCHA (2009)
Mason A. Porter, Jukka-Pekka Onnela and Peter J. Mucha. *Communities in Networks*. In: Notices of the American Mathematical Society 56.9 (Feb. 2009). eprint: 0902.3788. URL: https://arxiv.org/abs/0902.3788
Not cited.

[226] POTHEN (1997)
Alex Pothen. *Graph Partitioning Algorithms with Applications to Scientific Computing*. Technical report. Norfolk, VA, USA, 1997. 43 pages. DOI: 10.5555/890609. URL: https://citeseerx.ist.psu.eu/viewdoc/download?doi=10.1.1.34.3131&rep=rep1&type=pdf
Not cited.

[227] RADICCHI (2014)
Filippo Radicchi. *A paradox in community detection*. In: EPL (Europhysics Letters) 106.3 (May 2014). ISSN: 1286-4854.

DOI: `10.1209/0295-5075/106/38001`. URL: https://arxiv.org/abs/1312.4224
  Not cited.

[231]  SCHAEFFER (2007)
Satu Elisa Schaeffer. *Graph clustering*. In: Computer Science Review 1.1 (2007), pages 27–64. ISSN: 1574-0137. DOI: `10.1016/j.cosrev.2007.05.001`. URL: http://www.sciencedirect.com/science/article/pii/S1574013707000020
  Not cited.

[232]  SCHAUB, DELVENNE, ROSVALL and LAMBIOTTE (2017)
Michael T. Schaub, Jean-Charles Delvenne, Martin Rosvall and Renaud Lambiotte. *The many facets of community detection in complex networks*. In: Applied Network Science 2.1 (Feb. 2017). ISSN: 2364-8228. DOI: `10.1007/s41109-017-0023-6`. URL: https://arxiv.org/abs/1611.07769
  Not cited.

[234]  SHI, LIU and ZHANG (2018)
Cheng Shi, Yanchen Liu and Pan Zhang. *Weighted community detection and data clustering using message passing*. In: Journal of Statistical Mechanics: Theory and Experiment 2018.3 (Mar. 2018). ISSN: 1742-5468. DOI: `10.1088/1742-5468/aaa8f5`. URL: https://arxiv.org/abs/1801.09829
  Not cited.

[236]  STEENSTRUP (2001)
Martha Steenstrup. *Cluster-Based Networks*. In: Ad Hoc Networking. USA: Addison-Wesley Longman Publishing Co., Inc., 2001, pages 75–138. ISBN: 0201309769. DOI: `10.5555/374547.374551`
  Not cited.

[237]  STEINHAEUSER and CHAWLA (2008)
Karsten Steinhaeuser and Nitesh V. Chawla. *Community Detection in a Large Real-World Social Network*. In: Social Computing, Behavioral Modeling, and Prediction. Edited by Huan Liu, John J. Salerno and Michael J. Young. Boston, MA: Springer US, Jan. 2008, pages 168–175. ISBN: 978-0-387-77672-9. DOI: `10.1007/978-0-387-77672-9_19`
  Not cited.

[238]  ŠUBELJ and BAJEC (2011)
L. Šubelj and M. Bajec. *Robust network community detection using balanced propagation*. In: The European Physical Journal B 81.3 (May 2011), pages 353–362. ISSN: 1434-6036. DOI: `10.1140/epjb/e2011-10979-2`. URL: https://arxiv.org/abs/1106.5524
  Not cited.

[243]  TU, ZENG, WANG, ZHANG, LIU, SUN, ZHANG and LIN (2018)
Cunchao Tu, Xiangkai Zeng, Hao Wang, Zhengyan Zhang, Zhiyuan Liu, Maosong Sun, Bo Zhang and Leyu Lin. *A Unified Framework for Community Detection and Network Representation Learning*. 2018. eprint: `1611.06645`. URL: https://arxiv.org/abs/1611.06645
  Not cited.

[244]  VENTURINI (2020)
Sara Venturini. *Methods for community detection in multi-layer networks*. Master's thesis. University of Padua, July 2020. 143 pages. URL: http://tesi.cab.unipd.it/64409/
  Not cited.

[247]  XIE, CHEN and SZYMANSKI (2013)
Jierui Xie, Mingming Chen and Boleslaw K. Szymanski. *LabelRankT: Incremental Community Detection in Dynamic Networks via Label Propagation*. 2013. eprint: `1305.2006`. URL: https://arxiv.org/abs/1305.2006
  Not cited.

[248]  XIE and SZYMANSKI (2011)
Jierui Xie and Boleslaw K. Szymanski. *Community detection using a neighborhood strength driven Label Propagation Algorithm*. In: 2011 IEEE Network Science Workshop (June 2011). DOI: `10.1109/nsw.2011.6004645`. URL: https://arxiv.org/abs/1105.3264
  Not cited.

[249]  YANG, ALGESHEIMER and TESSONE (2016)
Zhao Yang, René Algesheimer and Claudio J. Tessone. *A Comparative Analysis of Community Detection Algorithms on Artificial Networks*. In: Scientific Reports 6.1 (Aug. 2016). ISSN: 2045-2322. DOI: `10.1038/srep30750`. URL: https:

//arxiv.org/abs/1608.00763
 Not cited.

# 4. Technical documentation references

## 4.1. ArangoDB docs references

[26] ArangoDB (2021)
ArangoDB. *ArangoDB - Distributed Iterative Graph Processing (Pregel), Community Detection*. Online. Documentation. Aug. 2021. URL: https://www.arangodb.com/docs/devel/graphs-pregel.html#community-detection
 Cited on pages 13, 71, 72.

[92] ArangoDB (2021)
ArangoDB. *ArangoDB - Why use ArangoDB, Advantages of Native Multi-Model*. Online. Aug. 2021. URL: https://www.arangodb.com/community-server/native-multi-model-database-advantages/
 Cited on pages 54, 59, 62, 63.

[93] ArangoDB (2021)
ArangoDB. *ArangoDB - Data Modeling and Operational Factors Documentation*. Online. Documentation. Aug. 2021. URL: https://www.arangodb.com/docs/stable/data-modeling-operational-factors.html
 Cited on page 55.

[94] ArangoDB (2021)
ArangoDB. *ArangoDB - Getting Started Documentation*. Online. Documentation. Aug. 2021. URL: https://www.arangodb.com/docs/stable/getting-started.html
 Cited on page 55.

[95] ArangoDB (2021)
ArangoDB. *ArangoDB - Graphs Documentation*. Online. Documentation. Aug. 2021. URL: https://www.arangodb.com/docs/stable/graphs.html
 Cited on pages 55, 57, 58.

[98] ArangoDB (2021)
ArangoDB. *ArangoDB - Velocity Pack (VPack) GitHub Repository*. Online. Aug. 2021. URL: https://github.com/arangodb/velocypack
 Cited on page 56.

[99] ArangoDB (2021)
ArangoDB. *ArangoDB Velocity Pack (VPack) GitHub Repository VelocyPack.md*. Online. Aug. 2021. URL: https://github.com/arangodb/velocypack/blob/main/VelocyPack.md
 Cited on page 56.

[100] ArangoDB (2021)
ArangoDB. *ArangoDB Release Notes Documentation - What's New, Changelogs*. Online. Aug. 2021. URL: https://www.arangodb.com/docs/stable/release-notes.html
 Cited on page 56.

[102] ArangoDB (2021)
ArangoDB. *ArangoDB - Indexing Documentation - Vertex Centric Indexes*. Online. Documentation. Aug. 2021. URL: https://www.arangodb.com/docs/stable/indexing-vertex-centric.html
 Cited on page 57.

[103] ArangoDB (2021)
ArangoDB. *ArangoDB - Architecture Documentation*. Online. Documentation. Aug. 2021. URL: https://www.arangodb.com/docs/stable/architecture.html
 Cited on pages 57, 58.

[104] ArangoDB (2021)
ArangoDB. *ArangoDB - Transactions Documentation*. Online. Documentation. Aug. 2021. URL: https://www.arangodb.com/docs/stable/transactions.html
 Cited on page 58.

[106]  ARANGODB (2021)
ArangoDB. *ArangoDB Query Language (AQL) Documentation*. Online. Documentation. Aug. 2021. URL: `https://www.arangodb.com/docs/stable/aql/`
Cited on page 59.

[107]  ARANGODB (2021)
ArangoDB. *ArangoDB - Joins Documentation*. Online. Documentation. Aug. 2021. URL: `https://www.arangodb.com/docs/stable/aql/examples-join.html`
Cited on page 59.

[108]  ARANGODB (2021)
ArangoDB. *ArangoDB - Traversals Documentation*. Online. Documentation. Aug. 2021. URL: `https://www.arangodb.com/docs/stable/graphs-traversals.html`
Cited on page 59.

[109]  ARANGODB (2021)
ArangoDB. *ArangoDB - Graph Functions Documentation*. Online. Documentation. Aug. 2021. URL: `https://www.arangodb.com/docs/stable/graphs-general-graphs-functions.html`
Cited on page 59.

[111]  ARANGODB, BRANDT, HOYOS and MCCOY (2020)
ArangoDB, Achim Brandt, Horacio Hoyos and Nathan McCoy. *Using ArangoDB with Gremlin*. Online. GitHub Repository. `https://github.com/ArangoDB-Community/arangodb-tinkerpop-provider/wiki/Server`. May 2020. URL: `https://github.com/ArangoDB-Community/arangodb-tinkerpop-provider`
Cited on page 60.

[112]  ARANGODB (2021)
ArangoDB. *ArangoDB - TinkerPop Provider for ArangoDB, GitHub Repository*. Online. An implementation of the Apache TinkerPop OLTP Provider API for ArangoDB. Aug. 2021. URL: `https://github.com/ArangoDB-Community/arangodb-tinkerpop-provider`
Cited on page 60.

[114]  ARANGODB (2021)
ArangoDB. *ArangoDB - The AQL query results cache Documentation*. Online. Documentation. Aug. 2021. URL: `https://www.arangodb.com/docs/stable/aql/execution-and-performance-query-cache.html`
Cited on page 61.

[126]  ARANGODB (2021)
ArangoDB. *ArangoDB - arangoimport Documentation*. Online. Documentation. Aug. 2021. URL: `https://www.arangodb.com/docs/stable/programs-arangoimport-details.html`
Cited on page 84.

[143]  ARANGODB (2017)
ArangoDB. *ArangoDB Cookbook*. Recipes for ArangoDB. Dec. 2017. 147 pages. URL: `https://download.arangodb.com/arangodb33/doc/ArangoDB_Cookbook_3.3.0.pdf`
Not cited.

[144]  ARANGODB (2021)
ArangoDB. *ArangoDB - Naming Conventions Documentation*. Online. Documentation. Aug. 2021. URL: `https://www.arangodb.com/docs/stable/data-modeling-naming-conventions.html`
Not cited.

[145]  ARANGODB and WEINBERGER (2015)
ArangoDB and Claudius Weinberger. *Native multi-model can compete with pure document and graph databases*. Online. Blog post. June 2015. URL: `https://www.arangodb.com/2015/06/multi-model-benchmark/`
Not cited.

## 4.2. Other GDBMSs docs references

[53]  NEO4J, INC. (2021)
Neo4j, Inc. *Neo4j Documentation*. Online. Documentation. Aug. 2021. URL: `https://neo4j.com/`
Cited on pages 32, 41.

[81]   Neo4j, Inc. (2021)
       Neo4j, Inc. *Neo4j Documentation - Cypher Query Language Manual*. Online. Documentation. Aug. 2021. URL: `https://neo4j.com/docs/cypher-manual/current/`
       Cited on page 45.

[184]  Neo4j, Inc. (2021)
       Neo4j, Inc. *Neo4j APOC User Guide 4.1 (Awesome Procedures On Cypher)*. Online. Documentation. Aug. 2021. URL: `https://neo4j.com/labs/apoc/4.1/`
       Not cited.

[185]  Neo4j, Inc. (2021)
       Neo4j, Inc. *Neo4j Documentation - Clustering*. Online. Documentation. Aug. 2021. URL: `https://neo4j.com/docs/operations-manual/current/clustering/`
       Not cited.

[187]  Neo4j, Inc. (2021)
       Neo4j, Inc. *Neo4j Transactional HTTP API Endpoint*. Online. Documentation. Aug. 2021. URL: `https://neo4j.com/docs/http-api/current/introduction/#http-api-transactional`
       Not cited.

[188]  Neo4j, Inc. (2021)
       Neo4j, Inc. *Neo4j Documentation - Cypher Administration, Indexes, Security*. Online. Documentation. Aug. 2021. URL: `https://neo4j.com/docs/cypher-manual/current/`
       Not cited.

[189]  Neo4j, Inc. (2021)
       Neo4j, Inc. *Neo4j Java API documentation - org.neo4j.graphdb.index*. Online. Documentation. Aug. 2021. URL: `https://neo4j.com/docs/java-reference/current/javadocs/org/neo4j/graphdb/package-summary.html`
       Not cited.

[190]  Neo4j, Inc. (2021)
       Neo4j, Inc. *Neo4j Java API documentation - Relationship*. Online. Documentation. Aug. 2021. URL: `https://neo4j.com/docs/java-reference/current/javadocs/org/neo4j/graphdb/Relationship.html`
       Not cited.

[191]  Neo4j, Inc. (2021)
       Neo4j, Inc. *Neo4j Java API documentation - Transaction*. Online. Documentation. Aug. 2021. URL: `https://neo4j.com/docs/java-reference/current/javadocs/org/neo4j/graphdb/Transaction.html`
       Not cited.

[192]  Neo4j, Inc. (2021)
       Neo4j, Inc. *Neo4j Licensing*. Online. Aug. 2021. URL: `https://neo4j.com/licensing/`
       Not cited.

[193]  Neo4j, Inc. and Gordon (2021)
       Neo4j, Inc. and Dave Gordon. *Warm the cache to improve performance from cold start*. Online. Aug. 2021. URL: `https://neo4j.com/developer/kb/warm-the-cache-to-improve-performance-from-cold-start/`
       Not cited.

[194]  Neo4j, Inc. (2021)
       Neo4j, Inc. *Neo4j Documentation - Operational Performance*. Online. Documentation. Aug. 2021. URL: `https://neo4j.com/docs/operations-manual/current/performance/`
       Not cited.

[196]  Neo4j, Inc. (2021)
       Neo4j, Inc. *Neo4j Documentation - Query Tuning*. Online. Documentation. Aug. 2021. URL: `https://neo4j.com/docs/cypher-manual/current/query-tuning/`
       Not cited.

[198]  Neo4j, Inc. (2021)
       Neo4j, Inc. *Neo4j Documentation - Transactionsg*. Online. Documentation. Aug. 2021. URL: `https://neo4j.com/docs/cypher-manual/current/introduction/transactions/`
       Not cited.

[200]  ORIENTDB LTD. (2021)
OrientDB Ltd. *OrientDB - Caching*. Online. Documentation. Aug. 2021. URL: https://orientdb.com/docs/last/internals/Caching.html
    Not cited.

[201]  ORIENTDB LTD. (2020)
OrientDB Ltd. *OrientDB - Clusters tutorial*. Online. Documentation. Mar. 2020. URL: https://web.archive.org/web/20200301072455/http://orientdb.com/docs/last/Tutorial-Clusters.html
    Not cited.

[202]  ORIENTDB LTD. (2021)
OrientDB Ltd. *OrientDB - Command Cache*. Online. Documentation. Aug. 2021. URL: https://orientdb.com/docs/last/sql/Command-Cache.html
    Not cited.

[203]  ORIENTDB LTD. (2021)
OrientDB Ltd. *OrientDB - Data Modeling*. Online. Documentation. Aug. 2021. URL: http://orientdb.com/docs/last/datamodeling/Tutorial-Document-and-graph-model.html
    Not cited.

[205]  ORIENTDB LTD. (2020)
OrientDB Ltd. *OrientDB - Graph API*. Online. Documentation. June 2020. URL: https://web.archive.org/web/20200629171431/http://www.orientdb.com/docs/last/Graph-Database-Tinkerpop.html
    Not cited.

[206]  ORIENTDB LTD. (2020)
OrientDB Ltd. *OrientDB - Graph API, Graph Batch Insert*. Online. Documentation. June 2020. URL: https://web.archive.org/web/20200629171416/http://www.orientdb.com/docs/last/Graph-Batch-Insert.html
    Not cited.

[207]  ORIENTDB LTD. (2020)
OrientDB Ltd. *OrientDB - Gremlin API*. Online. Documentation. June 2020. URL: https://web.archive.org/web/20200629171506/http://www.orientdb.com/docs/last/Gremlin.html
    Not cited.

[208]  ORIENTDB LTD. (2021)
OrientDB Ltd. *OrientDB - PLocal Storage*. Online. Documentation. Aug. 2021. URL: http://orientdb.com/docs/3.1.x/internals/Paginated-Local-Storage.html
    Not cited.

[209]  ORIENTDB LTD. (2020)
OrientDB Ltd. *OrientDB - Support and Subscriptions*. Online. Documentation. June 2020. URL: https://web.archive.org/web/20200629040612/http://orientdb.com/support/
    Not cited.

[210]  ORIENTDB LTD. (2020)
OrientDB Ltd. *OrientDB - SQL*. Online. Documentation. June 2020. URL: https://web.archive.org/web/20200629174343/http://orientdb.com/docs/last/SQL.html
    Not cited.

[211]  ORIENTDB LTD. (2020)
OrientDB Ltd. *OrientDB - Transactions*. Online. Documentation. June 2020. URL: https://web.archive.org/web/20200629174742/http://www.orientdb.com/docs/last/Transactions.html
    Not cited.

## 4.3. Non GDBMSs docs references

[215]  THE POSTGRESQL GLOBAL DEVELOPMENT GROUP (2021)
The PostgreSQL Global Development Group. *PostgreSQL - SQL Conformance*. Online. Documentation. Aug. 2021. URL: https://www.postgresql.org/docs/current/features.html
    Not cited.

[216]  THE POSTGRESQL GLOBAL DEVELOPMENT GROUP (2014)
       The PostgreSQL Global Development Group. *PostgreSQL - Monitoring query plan cache*. Online. PostgreSQL discussion
       thread. Dec. 2014. URL: https://www.postgresql.org/message-id/549564E4.4060800%40aule.net
          Not cited.

[217]  THE POSTGRESQL GLOBAL DEVELOPMENT GROUP (2021)
       The PostgreSQL Global Development Group. *Overview of PostgreSQL Internals, Planner/Optimizer*. Online. Documenta-
       tion. Aug. 2021. URL: https://www.postgresql.org/docs/current/planner-optimizer.html
          Not cited.

[218]  THE POSTGRESQL GLOBAL DEVELOPMENT GROUP (2021)
       The PostgreSQL Global Development Group. *PostgreSQL - PL/pgSQL Under the Hood*. Online. Documentation. Aug.
       2021. URL: https://www.postgresql.org/docs/current/plpgsql-implementation.html
          Not cited.

[219]  THE POSTGRESQL GLOBAL DEVELOPMENT GROUP (2021)
       The PostgreSQL Global Development Group. *PostgreSQL - PREPARE*. Online. Documentation. Aug. 2021. URL: https:
       //www.postgresql.org/docs/current/sql-prepare.html
          Not cited.

[220]  THE POSTGRESQL GLOBAL DEVELOPMENT GROUP (2011)
       The PostgreSQL Global Development Group. *PostgreSQL Query Cache released*. Online. Mar. 2011. URL: https://www.
       postgresql.org/about/news/postgresql-query-cache-released-1296/
          Not cited.

[221]  THE POSTGRESQL GLOBAL DEVELOPMENT GROUP (2021)
       The PostgreSQL Global Development Group. *PostgreSQL - Resource Consumption*. Online. Documentation. Aug. 2021.
       URL: https://www.postgresql.org/docs/current/runtime-config-resource.html
          Not cited.

[222]  THE POSTGRESQL GLOBAL DEVELOPMENT GROUP (2021)
       The PostgreSQL Global Development Group. *Tuning Your PostgreSQL Server*. Online. Mar. 2021. URL: https://wiki.
       postgresql.org/wiki/Tuning_Your_PostgreSQL_Server
          Not cited.

[223]  THE POSTGRESQL GLOBAL DEVELOPMENT GROUP (2021)
       The PostgreSQL Global Development Group. *What is PostgreSQL?* Online. Documentation. Aug. 2021. URL: https:
       //www.postgresql.org/docs/current/intro-whatis.html
          Not cited.

[224]  THE POSTGRESQL GLOBAL DEVELOPMENT GROUP (2021)
       The PostgreSQL Global Development Group. *PostgreSQL - WITH Queries (Common Table Expressions)*. Online. Docu-
       mentation. Aug. 2021. URL: https://www.postgresql.org/docs/current/queries-with.html
          Not cited.

## 4.4. Web development docs references

[27]   DONNELLY CENTRE - UNIVERSITY OF TORONTO (2021)
       Donnelly Centre - University of Toronto. *Cytoscape cytoscape*. Online. Documentation. Aug. 2021. URL: https://js.
       cytoscape.org/
          Cited on pages 14, 99.

[127]  PLOTLY (2021)
       Plotly. *React Cytoscape react-cytoscapejs*. Online. Documentation. Aug. 2021. URL: https://github.com/plotly/
       react-cytoscapejs
          Cited on page 99.

[140]  SUCHANEK, HUVAR, SCHICKLING and SCHULTE (2021)
       Tim Suchanek, Lukáš Huvar, Johannes Schickling and Rikki Schulte. *GraphQL Playground IDE*. Aug. 2021. URL: https:
       //github.com/graphql/graphql-playground
          Cited on page 125.

[157] FACEBOOK INC. (2021)
Facebook Inc. *Create a New React App create-react-app*. Online. Documentation. Aug. 2021. URL: https://reactjs.org/docs/create-a-new-react-app.html
  Not cited.

[158] FACEBOOK INC. (2021)
Facebook Inc. *Create a New React App create-react-app - Adding TypeScript*. Online. Documentation. Aug. 2021. URL: https://create-react-app.dev/docs/adding-typescript/
  Not cited.

[225] SUCHANEK, HUVAR, SCHICKLING and SCHULTE (2021)
Tim Suchanek, Lukáš Huvar, Johannes Schickling and Rikki Schulte. *GraphQL Playground IDE*. Aug. 2021. URL: https://www.postman.com/
  Not cited.

[230] DOZOIS, SHEMETOVSKIY, JIA and SMITH (2021)
Hugo Dozois, Alexander Shemetovskiy, Jimmy Jia and Matt Smith. *React Bootstrap react-bootstrap*. Online. Documentation. Aug. 2021. URL: https://react-bootstrap.github.io/
  Not cited.

## 4.5. Other documentation references

[54] APACHE TINKERPOP (2021)
Apache TinkerPop. *Apache TinkerPop Documentation*. Online. Documentation. Aug. 2021. URL: https://tinkerpop.apache.org/docs/current/reference/
  Cited on page 32.

[79] APACHE SPARK (2021)
Apache Spark. *Spark - GraphX Programming Guide, Pregel API*. Online. Documentation. Aug. 2021. URL: https://spark.apache.org/docs/latest/graphx-programming-guide.html#pregel-api
  Cited on page 42.

[83] APACHE TINKERPOP (2021)
Apache TinkerPop. *Apache TinkerPop - The Gremlin Graph Traversal Machine and Language*. Online. Documentation. Aug. 2021. URL: https://tinkerpop.apache.org/gremlin.html
  Cited on page 45.

[141] APACHE TINKERPOP (2021)
Apache TinkerPop. *Apache TinkerPop Documentation - Neo4j Gremlin*. Online. Documentation. Aug. 2021. URL: https://tinkerpop.apache.org/docs/current/reference/#neo4j-gremlin
  Not cited.

[142] APACHE TINKERPOP (2021)
Apache TinkerPop. *Apache TinkerPop Documentation - Recipes*. Online. Documentation. Aug. 2021. URL: https://tinkerpop.apache.org/gremlin.html
  Not cited.

[147] BELAID (2015)
Rachid Belaid. *Introduction to PostgreSQL physical storage*. Online. Blog post. Nov. 2015. URL: http://rachbelaid.com/introduction-to-postgres-physical-storage/
  Not cited.

## 5. Other references

[1] SCHLOSS DAGSTUHL - LEIBNIZ CENTER FOR INFORMATICS (2021)
Schloss Dagstuhl - Leibniz Center for Informatics. *The dblp computer science bibliography*. Online. dblp.org. July 2021. URL: http://dblp.uni-trier.de/xml/
  Cited on pages 3, 11, 14, 15, 65, 81, 82, 84, 85, 86, 87, 89, 113.

[2]  UNIVERSITÀ DEGLI STUDI DI BERGAMO (2021)
     Università degli Studi di Bergamo. *Logo UniBG*. Online. July 2021. URL: https://unibg.it/
     Cited on page 4.

[3]  CLODE (2018)
     David Clode. *A couple of interesting frayed ropes, subtly lit in the shadow between two fishing boats at the Cairns prawn trawler base*. Online. Photographer's profile: https://unsplash.com/@davidclode. Jan. 2018. URL: https://unsplash.com/photos/hpQAUR9jkaM
     Cited on page 4.

[4]  UXWING (2020)
     UXWing. *Icons*. Online. May 2020. URL: https://uxwing.com/
     Cited on page 5.

[5]  KELLY (2005)
     Kevin Kelly. *We Are the Web*. Wired article. Jan. 2005. URL: https://www.wired.com/2005/08/tech/
     Cited on page 9.

[6]  MARX (2013)
     Vivien Marx. *The big challenges of big data*. In: Nature 498 (7453 June 2013), pages 255–260. DOI: 10.1038/498255a. URL: https://www.nature.com/articles/498255a
     Cited on page 9.

[28] PRICE (1965)
     Derek J. de Solla Price. *Networks of scientific papers*. In: Science (New York, N.Y.) 149.3683 (July 1965), pages 510–515. ISSN: 0036-8075. DOI: 10.1126/science.149.3683.510. URL: https://doi.org/10.1126/science.149.3683.510
     Cited on page 14.

[39] WIKIMEDIA FOUNDATION, INC. (2021)
     Wikimedia Foundation, Inc. *Wikipedia - List of Milan Metro stations*. Online. Sept. 2021. URL: https://en.wikipedia.org/wiki/List_of_Milan_Metro_stations
     Cited on pages 24, 25.

[42] GROWING WITH THE WEB and IMMS (2021)
     Growing with the Web and Daniel Imms. *growingwiththeweb.com website*. Online. Sept. 2021. URL: https://www.growingwiththeweb.com/
     Cited on page 27.

[52] SEABORNE (2015)
     Andy Seaborne. *Two graph data models: RDF and Property Graphs*. Slides. Talk given at ApacheConEU Big Data 2015. 2015. URL: https://www.slideshare.net/andyseaborne/two-graph-data-models-rdf-and-property-graphs
     Cited on page 32.

[55] APACHE TINKERPOP (2021)
     Apache TinkerPop. *Apache TinkerPop Website*. Online. Aug. 2021. URL: https://tinkerpop.apache.org/
     Cited on pages 32, 36.

[56] W3C (2014)
     The World Wide Web Consortium W3C. *W3C Semantic Web - Resource Description Framework (RDF)*. Online. Feb. 2014. URL: https://www.w3.org/RDF/
     Cited on pages 32, 33.

[57] W3C (2014)
     The World Wide Web Consortium W3C. *W3C Recommendation - RDF 1.1 Concepts and Abstract Syntax, Data Model*. Online. Feb. 2014. URL: https://www.w3.org/TR/2014/REC-rdf11-concepts-20140225/#data-model
     Cited on page 33.

[61] BERNERS-LEE, HENDLER and LASSILA (2001)
     Tim Berners-Lee, James Hendler and Ora Lassila. *The Semantic Web*. In: Scientific American 284.5 (May 2001). https://www.w3.org/2001/sw/, pages 34–43. URL: https://www.jstor.org/stable/26059207
     Cited on pages 35, 37.

[62] W3C (2021)
     The World Wide Web Consortium W3C. *W3C Semantic Web - Ontologies, Vocabolaries*. Online. Aug. 2021. URL: https:

//www.w3.org/standards/semanticweb/ontology.html
  Cited on page 35.

[63]  W3C (2021)
      The World Wide Web Consortium W3C. *W3C Semantic Web - Inference*. Online. Aug. 2021. URL: https://www.w3.org/standards/semanticweb/inference
      Cited on page 36.

[64]  LOVINGER (2009)
      Rachel Lovinger. *RDF and OWL - A simple overview of the building blocks of the Semantic Web*. Slides. Presented at the Semantic Web Affinity Group at Razorfish. 2009. URL: https://www.slideshare.net/rlovinger/rdf-and-owl
      Cited on page 36.

[65]  W3C (2004)
      The World Wide Web Consortium W3C. *W3C Recommendation - OWL Web Ontology Language*. Online. Feb. 2004. URL: https://www.w3.org/TR/owl-features/
      Cited on page 36.

[69]  W3C (2008)
      The World Wide Web Consortium W3C. *W3C Recommendation - SPARQL Query Language for RDF*. Online. Jan. 2008. URL: https://www.w3.org/TR/rdf-sparql-query/
      Cited on pages 37, 38.

[70]  HEATH (2019)
      Tom Heath. *Linked Data - Connect Distributed Data across the Web*. Online. Aug. 2019. URL: https://web.archive.org/web/20190802052922/http://linkeddata.org/
      Cited on page 38.

[71]  W3C and HERMAN (2021)
      The World Wide Web Consortium W3C and Ivan Herman. *W3C Semantic Web Frequently Asked Questions*. Online. Aug. 2021. URL: https://www.w3.org/RDF/FAQ
      Cited on page 38.

[72]  W3C (2021)
      The World Wide Web Consortium W3C. *W3C Semantic Web - Linked Data*. Online. Aug. 2021. URL: https://www.w3.org/standards/semanticweb/data
      Cited on page 38.

[75]  ARANGODB and WEINBERGER (2016)
      ArangoDB and Claudius Weinberger. *Index Free Adjacency or Hybrid Indexes for Graph Databases*. Online. Blog post. Apr. 2016. URL: https://www.arangodb.com/2016/04/index-free-adjacency-hybrid-indexes-graph-databases/
      Cited on page 40.

[88]  ARANGODB and WEINBERGER (2015)
      ArangoDB and Claudius Weinberger. *Performance comparison between ArangoDB, MongoDB, Neo4j and OrientDB*. Online. Blog post. Discussion: https://news.ycombinator.com/item?id=9699102. June 2015. URL: https://www.arangodb.com/2015/06/performance-comparison-between-arangodb-mongodb-neo4j-and-orientdb/
      Cited on page 47.

[89]  W3C (2018)
      The World Wide Web Consortium W3C. *Rdf Store Benchmarking*. Online. Oct. 2018. URL: https://www.w3.org/wiki/RdfStoreBenchmarking
      Cited on page 47.

[90]  W3C (2021)
      The World Wide Web Consortium W3C. *Large Triple Stores*. Online. Apr. 2021. URL: https://www.w3.org/wiki/LargeTripleStores
      Cited on page 47.

[91]  ARANGODB and WEINBERGER (2015)
      ArangoDB and Claudius Weinberger. *Benchmark: PostgreSQL, MongoDB, Neo4j, OrientDB and ArangoDB*. Online. Blog post. Oct. 2015. URL: https://www.arangodb.com/2015/06/performance-comparison-between-arangodb-mongodb-neo4j-and-orientdb/
      Cited on page 47.

[101]  ARANGODB GOOGLE GROUPS (2014)
       ArangoDB Google Groups. *Discussion on ArangoDB and index-free adjacency*. Online. Mar. 2014. URL: https://groups.
       google.com/forum/#!topic/arangodb/x0QqIcZ6h60
         Cited on page 56.

[105]  EVERETT and MALYSHEV (2015)
       Nik Everett and Stas Malyshev. *Investigate ArangoDB for Wikidata Query*. Online. Feb. 2015. URL: https://phabricator.
       wikimedia.org/T88549
         Cited on page 58.

[110]  ARANGODB (2013)
       ArangoDB. *ArangoDB - Gremlin graph queries for REST*. Online. GitHub Repository Issue #392. Feb. 2013. URL: https:
       //github.com/arangodb/arangodb/issues/392
         Cited on page 60.

[117]  ARANGODB (2021)
       ArangoDB. *ArangoDB Commercial Subscriptions*. Online. Aug. 2021. URL: https://www.arangodb.com/subscriptions/
         Cited on page 63.

[122]  MALEWICZ, AUSTERN, BIK, DEHNERT, HORN, LEISER and CZAJKOWSKI (2009)
       Grzegorz Malewicz, Matthew H. Austern, Aart J. C. Bik, James C. Dehnert, Ilan Horn, Naty Leiser and Grzegorz Czajkowski.
       *Pregel: A System for Large-Scale Graph Processing*. In: SPAA 2009: Proceedings of the 21st Annual ACM Symposium on
       Parallelism in Algorithms and Architectures, Calgary, Alberta, Canada, August 11-13, 2009. Jan. 2009, pages 135–145.
       DOI: 10.1145/1582716.1582723. URL: https://www.researchgate.net/publication/221257383_Pregel_A_system_
       for_large-scale_graph_processing
         Cited on page 70.

[123]  ARANGODB, SCHAD and KERNBACH (2020)
       ArangoDB, Jörg Schad and Heiko Kernbach. *Custom Pregel Algorithms in ArangoDB*. Online. Slides. Docs: https:
       //www.arangodb.com/docs/stable/graphs-pregel.html. Nov. 2020. URL: https://www.slideshare.net/arangodb/
       custom-pregel-algorithms-in-arangodb
         Cited on page 71.

[124]  VALIANT (1990)
       Leslie G. Valiant. *A bridging model for parallel computation*. In: Commun. ACM 33 (Aug. 1990). https://dl.acm.org/
       doi/10.1145/79173.79181, pages 103–111. DOI: 10.1145/79173.79181. URL: https://www.semanticscholar.org/
       paper/A-bridging-model-for-parallel-computation-Valiant/8665c9b459e4161825baf1f25b5141f41a5085ff
         Cited on page 71.

[125]  AGUILAR GONZALES (2018)
       Ezequiel Aguilar Gonzales. *Pregel: A System for Large-Scale Graph Processing Presentation*. Online. Slides. 2018. URL:
       https://ranger.uta.edu/~sjiang/CSE6350-spring-18/12-pregel-slides.pdf
         Cited on page 71.

[128]  ITALIAN STATISTICAL SOCIETY (2021)
       Italian Statistical Society. *Statistical Methods & Applications*. Edited by Carla Rampichini, Alessio Farcomeni and Stefano
       Campostrini. Journal of the Italian Statistical Society. Officially cited as: Stat. Methods Appl. Aug. 2021. URL: https:
       //www.springer.com/journal/10260
         Cited on page 108.

[152]  COBDEN, BLACK, GIBBINS, CARR and SHADBOLT (2011)
       Marcus Cobden, Jennifer Black, Nicholas Gibbins, Les Carr and Nigel Shadbolt. *A Research Agenda for Linked Closed
       Data*. In: Second International Workshop on Consuming Linked Data. Germany, Sept. 2011. URL: https://eprints.
       soton.ac.uk/272711/
         Not cited.

[170]  HEATH and BIZER (2011)
       Tom Heath and Christian Bizer. *Linked Data: Evolving the Web into a Global Data Space*. Synthesis Lectures on the
       Semantic Web: Theory and Technology. Morgan & Claypool Publishers, Feb. 2011. 136 pages. ISBN: 978-1608454303.
       DOI: https://doi.org/10.2200/S00334ED1V01Y201102WBE001. URL: https://www.morganclaypool.com/doi/abs/10.
       2200/S00334ED1V01Y201102WBE001
         Not cited.

[179] Madusudanan (2016)
B. N. Madusudanan. *Understanding caching in Postgres - An in-depth guide*. Online. Blog post. May 2016. URL: https://madusudanan.com/blog/understanding-postgres-caching-in-depth/
Not cited.

[186] Neo4j, Inc. (2021)
Neo4j, Inc. *Neo4j download page*. Online. Aug. 2021. URL: https://neo4j.com/download/
Not cited.

[195] Neo4j, Inc. (2021)
Neo4j, Inc. *Neo4j Pricing, Compare editions*. Online. Aug. 2021. URL: https://neo4j.com/pricing/
Not cited.

[204] OrientDB Ltd. (2020)
OrientDB Ltd. *OrientDB - Enterprise*. Online. May 2020. URL: https://web.archive.org/web/20200529225838/http://orientdb.com/orientdb-enterprise/
Not cited.

[228] Raghavendra (2012)
Rao Raghavendra. *Caching in PostgreSQL*. Online. Blog post. Apr. 2012. URL: https://raghavt.blogspot.com/2012/04/caching-in-postgresql.html
Not cited.

[233] Schmachtenberg, Bizer, Jentzsch and Cyganiak (2014)
Max Schmachtenberg, Christian Bizer, Anja Jentzsch and Richard Cyganiak. *Linking Open Data cloud diagram 2014*. Online. 2014. URL: https://lod-cloud.net/
Not cited.

[235] Smith (2008)
Greg Smith. *Inside the PostgreSQL Shared Buffer Cache*. Online. Slides. July 2008. URL: https://www.2ndquadrant.com/wp-content/uploads/2019/05/Inside-the-PostgreSQL-Shared-Buffer-Cache.pdf
Not cited.

[239] Suzuki (2021)
Hironobu Suzuki. *The Internals of PostgreSQL, for database administrators and system developers*. Online. Aug. 2021. URL: http://www.interdb.jp/pg/pgsql01.html
Not cited.

[241] Thomson Reuters, Baker, Hapuarachchi and Bailey (2016)
Thomson Reuters, Tim Baker, Tharindi Hapuarachchi and Bob Bailey. *The future is graph shaped*. Online. Oct. 2016. URL: https://web.archive.org/web/20201127053648/https://blogs.thomsonreuters.com/answerson/future-graph-shaped/
Not cited.

[242] Trylks and Travers (2013)
Trylks and Chris Travers. *PostgreSQL temporary table cache in memory?* Online. Stack Overflow question. Jan. 2013. URL: https://stackoverflow.com/questions/14162917/postgresql-temporary-table-cache-in-memory
Not cited.

[245] W3C (2014)
The World Wide Web Consortium W3C. *W3C Working Group Note - RDF 1.1 Primer*. Online. June 2014. URL: https://www.w3.org/TR/rdf11-primer/
Not cited.

[246] W3C (2014)
The World Wide Web Consortium W3C. *W3C Recommendation - RDF 1.1 XML Syntax*. Online. Feb. 2014. URL: https://www.w3.org/TR/rdf-syntax-grammar/
Not cited.

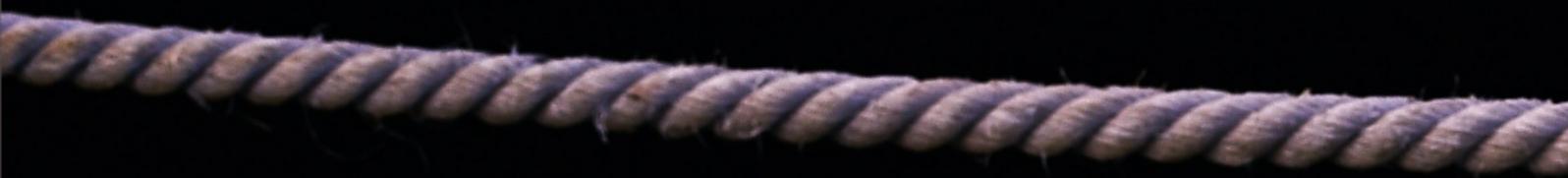
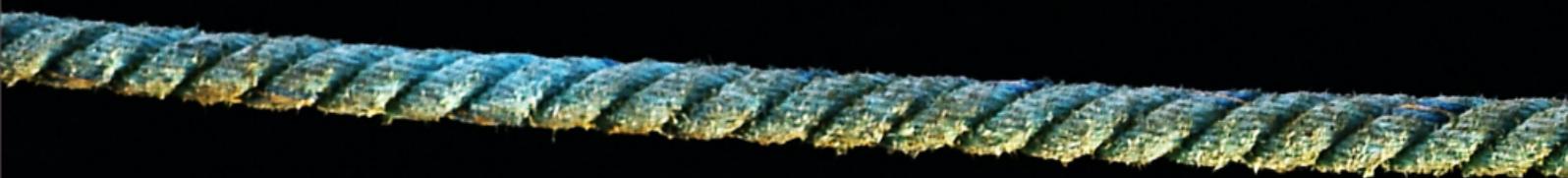